\documentclass[a4paper]{article}

\usepackage{icrc2013}
\usepackage[english]{babel}
\usepackage{amsmath}
\usepackage{amssymb}

\addto\captionsenglish{
  \renewcommand{\contentsname}%
    {\centerline{{\LARGE List of Contributions}}%
    \vskip 0.25cm%
    \hrule%
    }%
}

\newcommand{\etal}{\MakeLowercase{\textit{et al. }}} 

\newcommand{\comment}[1]{}

\newcommand{\ant}{\textsc{Antares}}

\newcounter{IdContrib}
\newcommand{\id}[1]{\refstepcounter{IdContrib}\label{#1}}

\begin{document}

\shorttitle{}

\begin{onecolumn}

\begin{center}
%

{
{\Huge \bf%

The {\sc Antares} Collaboration } \\
 \vskip 0.25cm
\hrule
\vskip 0.5cm
 {\Large Contributions to the\\
 33$^{st}$  International Cosmic Ray Conference (ICRC 2013) \\
 \vskip 0.25cm
 Rio de Janeiro, Brazil\\
 July 2013}
  \vskip 0.25cm
\hrule
 }
 \end {center}
  \vspace{2cm}
\centerline{\includegraphics[width=0.15\textwidth]{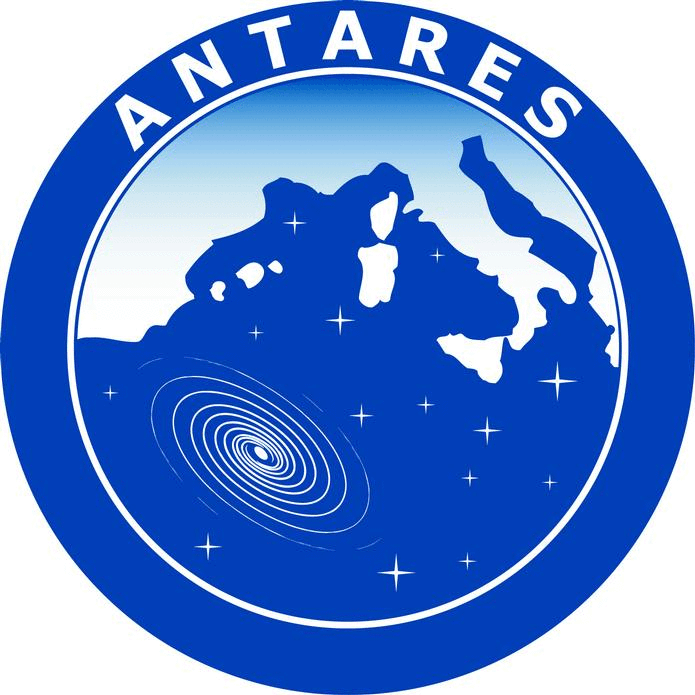}}
 \vspace{2cm}
 \begin{center}
 {\LARGE \bf Abstract}
 
  \end{center} 
{\large
The ANTARES detector, completed in 2008, is the largest neutrino telescope in the Northern hemisphere. 
Located at a depth of 2.5 km in the Mediterranean Sea, 40 km off the Toulon shore, its main goal is the 
search for astrophysical high energy neutrinos. 
In this paper we collect the 14 contributions of the ANTARES collaboration to the 33rd International Cosmic Ray Conference (ICRC 2013). 
The scientific output is very rich and the contributions included in these proceedings cover the main physics results,
ranging from steady point sources to exotic physics and multi-messenger analyses.

 }
 \vskip 0.125cm
\hrule
\end{onecolumn} 

\newpage
\null
\newpage
\shorttitle{{\sc Antares} Contributions to ICRC2013}
\begin{center}
{\large \bf The {\sc Antares} Collaboration
\phantomsection
 \addcontentsline{toc}{part}{The {\sc Antares} Collaboration, list of authors}}
\end{center}


\authors[{
\noindent
S.~Adri\'an-Mart\'inez$^a$,
M.~Ageron$^g$,
A.~Albert$^b$, M.~Andr\'e$^c$, M.~Anghinolfi$^d$,
I. Al Samarai$^g$,
G.~Anton$^e$, M.~Ardid$^a$, T.~Astraatmadja$^f$,
J.-J.~Aubert$^g$,
B.~Baret$^h$,
J.~Barrios-Mart\'{\i}$^i$,
S.~Basa$^j$,
V.~Bertin$^g$,
S.~Biagi$^{k,l}$,
C.~Bigongiari$^i$,
C.~Bogazzi$^f$,
B.~Bouhou$^h$,
M.C.~Bouwhuis$^f$,
J.~Brunner$^g$,
J.~Busto$^g$,
A.~Capone$^{m,n}$,
L.~Caramete$^o$,
C.~C$\mathrm{\hat{a}}$rloganu$^p$,
J.~Carr$^g$,
Ph.~Charvis$^q$,
T.~Chiarusi$^k$,
M.~Circella$^r$,
F.~Classen$^e$,
L.~Core$^g$,
H.~Costantini$^g$,
P.~Coyle$^g$,
A.~Creusot$^h$,
C.~Curtil$^g$,
I.~Dekeyser$^s$,
G.~Derosa$^{t,u}$,
A.~Deschamps$^q$,
G.~De~Bonis$^{m,n}$,
{C.~Distefano$^v$,
C.~Donzaud$^{h,w}$,
D.~Dornic$^g$,
Q.~Dorosti$^x$,
D.~Drouhin$^b$,
A.~Dumas$^p$,
T.~Eberl$^e$,
D.~Els\"asser$^y$,
U.~Emanuele$^i$,
A.~Enzenh\"ofer$^e$,
J.-P.~Ernenwein$^g$,
S.~Escoffier$^g$,
K.~Fehn$^e$,
I.~Felis$^a$,
P.~Fermani$^{m,n}$,
F.~Folger$^e$,
L.A.~Fusco$^{k,l}$,
S.~Galat\`a$^h$,
P.~Gay$^p$,
S.~Gei{\ss}els\"oder$^e$,
K.~Geyer$^e$,
V.~Giordano$^z$,
A.~Gleixner$^e$,
J.P.~ G\'omez-Gonz\'alez$^i$,
K.~Graf$^e$,
G.~Guillard$^p$,
H.~van~Haren$^{aa}$,
A.J.~Heijboer$^f$,
Y.~Hello$^q$,
J.J.~Hern\'andez-Rey$^i$,
B.~Herold$^e$,
J.~H\"o{\ss}l$^e$,
J.~Hofest\"adt$^e$,
C.~Hugon$^d$,
C.W~James$^e$,
M.~de~Jong$^f$,
M.~Kadler$^y$,
O.~Kalekin$^e$,
A.~Kappes$^e$,
U.~Katz$^e$,
D.~Kie{\ss}ling$^e$,
P.~Kooijman$^{f,ab,ac}$,
A.~Kouchner$^h$,
I.~Kreykenbohm$^{ad}$,
V.~Kulikovskiy$^{d,ae}$,
R.~Lahmann$^e$,
E.~Lambard$^g$,
G.~Lambard$^i$,
G.~Larosa$^a$,
D.~Lattuada$^v$,
D. ~Lef\`evre$^s$,
E.~Leonora$^{z,af}$,
H.~Loehner$^x$,
S.~Loucatos$^{ah}$,
S.~Mangano$^i$,
M.~Marcelin$^j$,
A.~Margiotta$^{k,l}$,
J.A.~Mart\'inez-Mora$^a$,
S.~Martini$^s$,
A.~Mathieu$^g$,
T.~Michael$^f$,
P.~Migliozzi$^t$,
L.~Moscoso$^{h,\dagger}$,
H.~Motz$^e$,
C.~Mueller$^{ad,y}$,
M.~Neff$^e$,
E.~Nezri$^j$,
D.~Palioselitis$^f$,
G.E.~P\u{a}v\u{a}la\c{s}$^o$,
P.~Payre$^{g,\dagger}$,
C.~Perrina$^{m,n}$,
P.~Piattelli$^v$,
V.~Popa$^o$,
T.~Pradier$^{ai}$,
C.~Racca$^{b}$,
G.~Riccobene$^v$,
R.~Richter$^e$,
C.~Rivi\`ere$^g$,
K.~Roensch$^e$,
A.~Rostovtsev$^{aj}$,
M.~Salda\~{n}a$^a$,
D.F.E.~Samtleben$^{f,ak}$,
M.~Sanguineti$^{d,al}$,
P.~Sapienza$^v$,
J.~Schmid$^e$,
J.~Schnabel$^e$,
S.~Schulte$^f$,
F.~Sch\"ussler$^{ah}$,
T.~Seitz$^e$,
R.~Shanidze$^e$,
C.~Sieger$^e$,
A.~Spies$^e$,
M.~Spurio$^{k,l}$,
J.J.M.~Steijger$^f$,
Th.~Stolarczyk$^{ah}$,
D.~Stransky$^e$,
A.~S{\'a}nchez-Losa$^i$,
M.~Taiuti$^{d,al}$,
C.~Tamburini$^s$,
Y.~Tayalati$^{am}$,
A.~Trovato$^v$,
B.~Vallage$^{ah}$,
C.~Vall\'ee$^g$,
V.~Van~Elewyck$^h$,
E.~Visser$^f$,
D.~Vivolo$^{t,u}$,
S.~Wagner$^e$,
J.~Wilms$^{ad}$,
E.~de~Wolf$^{f,ac}$,
K.~Yatkin$^g$,
H.~Yepes$^i$,
J.D.~Zornoza$^i$,
J.~Z\'u\~{n}iga$^i$}


\afiliations[{\noindent\scriptsize{$^a$ Institut d'Investigaci\'o per a la Gesti\'o Integrada de les Zones Costaneres (IGIC) - Universitat Polit\`ecnica de Val\`encia. C/  Paranimf 1 , 46730 Gandia, Spain\\}}
\afiliations[{\scriptsize{$^b$ GRPHE -Universit\'e de Haute Alsace \& Institut universitaire de technologie de Colmar, 34 rue du Grillenbreit BP 50568 - 68008 Colmar, France\\}}
\afiliations[{\scriptsize{$^c$ Technical University of Catalonia, Laboratory of Applied Bioacoustics, Rambla Exposici\'o,08800 Vilanova i la Geltr\'u,Barcelona, Spain\\}}
\afiliations[{\scriptsize{$^d$ INFN - Sezione di Genova, Via Dodecaneso 33, 16146 Genova, Italy\\}}
\afiliations[{\scriptsize{$^e$Friedrich-Alexander-Universit\"at Erlangen-N\"urnberg, Erlangen Centre for Astroparticle Physics, Erwin-Rommel-Str. 1, 91058 Erlangen, Germany\\}}
\afiliations[{\scriptsize{$^f$Nikhef, Science Park,  Amsterdam, The Netherlands\\}}
\afiliations[{\scriptsize{$^g$CPPM, Aix-Marseille Universit\'e, CNRS/IN2P3, Marseille, France\\}}
\afiliations[{\scriptsize{$^h$APC, Universit\'e Paris Diderot, CNRS/IN2P3, CEA/IRFU, Observatoire de Paris, Sorbonne Paris Cit\'e, 75205 Paris, France\\}}
\afiliations[{\scriptsize{$^i$IFIC - Instituto de F\'isica Corpuscular, Edificios Investigaci\'on de Paterna, CSIC - Universitat de Val\`encia, Apdo. de Correos 22085, 46071 Valencia, Spain\\}}
\afiliations[{\scriptsize{$^j$LAM - Laboratoire d'Astrophysique de Marseille, P\^ole de l'\'Etoile Site de Ch\^ateau-Gombert, rue Fr\'ed\'eric Joliot-Curie 38,  13388 Marseille Cedex 13, France\\}}
\afiliations[{\scriptsize{$^k$INFN - Sezione di Bologna, Viale Berti-Pichat 6/2, 40127 Bologna, Italy\\}}
\afiliations[{\scriptsize{$l$Dipartimento di Fisica dell'Universit\`a, Viale Berti Pichat 6/2, 40127 Bologna, Italy\\}}
\afiliations[{\scriptsize{$^m$INFN -Sezione di Roma, P.le Aldo Moro 2, 00185 Roma, Italy\\}}
\afiliations[{\scriptsize{$^n$Dipartimento di Fisica dell'Universit\`a La Sapienza, P.le Aldo Moro 2, 00185 Roma, Italy\\}}
\afiliations[{\scriptsize{$^o$Institute for Space Sciences, R-77125 Bucharest, M\u{a}gurele, Romania\\}}
\afiliations[{\scriptsize{$^p$Laboratoire de Physique Corpusculaire, Clermont Universit\'e, Universit\'e Blaise Pascal, CNRS/IN2P3, BP 10448, F-63000 Clermont-Ferrand, France\\}}
\afiliations[{\scriptsize{$^q$G\'eoazur, Universit\'e Nice Sophia-Antipolis, CNRS, IRD, Observatoire de la C\^ote d'Azur, Sophia Antipolis, France \\}}
\afiliations[{\scriptsize{$^r$INFN - Sezione di Bari, Via E. Orabona 4, 70126 Bari, Italy\\}}
\afiliations[{\scriptsize{$^s$Aix Marseille Universit\'e, CNRS/INSU, IRD, Mediterranean Institute of Oceanography (MIO), UM 110, Marseille, France ; Universit\'e de Toulon, CNRS, IRD, Mediterranean Institute of Oceanography (MIO), UM 110, La Garde, France\\}}
\afiliations[{\scriptsize{$^t$INFN -Sezione di Napoli, Via Cintia 80126 Napoli, Italy\\}}
\afiliations[{\scriptsize{$^u$Dipartimento di Fisica dell'Universit\`a Federico II di Napoli, Via Cintia 80126, Napoli, Italy\\}}
\afiliations[{\scriptsize{$^v$INFN - Laboratori Nazionali del Sud (LNS), Via S. Sofia 62, 95123 Catania, Italy\\}}
\afiliations[{\scriptsize{$^w$Univ. Paris-Sud , 91405 Orsay Cedex, France\\}}
\afiliations[{\scriptsize{$^x$Kernfysisch Versneller Instituut (KVI), University of Groningen, Zernikelaan 25, 9747 AA Groningen, The Netherlands\\}}
\afiliations[{\scriptsize{$^y$Institut f\"ur Theoretische Physik und Astrophysik, Universit\"at W\"urzburg, Emil-Fischer Str. 31, 97074 W\"urzburg, Germany\\}}
\afiliations[{\scriptsize{$^z$INFN - Sezione di Catania, Viale Andrea Doria 6, 95125 Catania, Italy\\}}
\afiliations[{\scriptsize{$^{aa}$Royal Netherlands Institute for Sea Research (NIOZ), Landsdiep 4,1797 SZ 't Horntje (Texel), The Netherlands\\}}
\afiliations[{\scriptsize{$^{ab}$Universiteit Utrecht, Faculteit Betawetenschappen, Princetonplein 5, 3584 CC Utrecht, The Netherlands\\}}
\afiliations[{\scriptsize{$^{ac}$Universiteit van Amsterdam, Instituut voor Hoge-Energie Fysica, Science Park 105, 1098 XG Amsterdam, The Netherlands\\}}
\afiliations[{\scriptsize{$^{ad}$Dr. Remeis-Sternwarte and ECAP, Universit\"at Erlangen-N\"urnberg,  Sternwartstr. 7, 96049 Bamberg, Germany\\}}
\afiliations[{\scriptsize{$^{ae}$Moscow State University,Skobeltsyn Institute of Nuclear Physics,Leninskie gory, 119991 Moscow, Russia\\}}
\afiliations[{\scriptsize{$^{af}$Dipartimento di Fisica ed Astronomia dell'Universit\`a, Viale Andrea Doria 6, 95125 Catania, Italy\\}}
\afiliations[{\scriptsize{$^{ah}$Direction des Sciences de la Mati\`ere - Institut de recherche sur les lois fondamentales de l'Univers - Service de Physique des Particules, CEA Saclay, 91191 Gif-sur-Yvette Cedex, France\\}}
\afiliations[{\scriptsize{$^{ai}$IPHC-Institut Pluridisciplinaire Hubert Curien - Universit\'e de Strasbourg et CNRS/IN2P3  23 rue du Loess, BP 28,  67037 Strasbourg Cedex 2, France\\}}
\afiliations[{\scriptsize{$^{aj}$ITEP - Institute for Theoretical and Experimental Physics, B. Cheremushkinskaya 25, 117218 Moscow, Russia\\}}
\afiliations[{\scriptsize{$^{ak}$Universiteit Leiden, Leids Instituut voor Onderzoek in Natuurkunde, 2333 CA Leiden, The Netherlands\\}}
\afiliations[{\scriptsize{$^{al}$Dipartimento di Fisica dell'Universit\`a, Via Dodecaneso 33, 16146 Genova, Italy\\}}
\afiliations[{\scriptsize{$^{am}$University Mohammed I, Laboratory of Physics of Matter and Radiations, B.P.717, Oujda 6000, Morocco\\}}
\afiliations[{\scriptsize{$^{\dagger}$Deceased\\}}



%
%

\newpage
\null
\newpage
\setcounter{tocdepth}{0}
\begin{onecolumn}
\vspace*{0.4cm}

\tableofcontents

\end{onecolumn}
\newpage
\null
\newpage



\id{id_ak}
\addcontentsline{toc}{part}{\arabic{IdContrib} - {\sl Antoine Kouchner - Highlight Talk} : Recent results from the {\sc Antares} neutrino telescope%
}

\def\Journal#1#2#3#4{{#1} {\bf #2} (#4) #3}
\def\NCA{Nuovo Cimento}
\def\NIM{ Nucl.~Instr~Methods}\def\NIMA{NIM~A}
\def\NPB{Nucl.~Phys.~B}
\def\PLB{Phys.~Lett.~B}
\def\PRL{Phys.~Rev.~Lett.}
\def\PRD{Phys.~Rev.~D}
\def\ZPC{Z.~Phys.~C}
%
%


\title{\arabic{IdContrib} - Recent results from the {\sc antares} neutrino telescope}

\shorttitle{\arabic{IdContrib} - {\sc antares} results}

\authors{
Antoine Kouchner$^{1}$
for the {\sc antares} Collaboration.
}

\afiliations{
$^1$ APC, Universit\'e Paris Diderot, CNRS/IN2P3, CEA/IRFU, Observatoire de Paris, Sorbonne Paris Cit\'e, 75205 Paris, France\\
}

\email{kouchner@apc.univ-paris7.fr}

\abstract{The {\sc antares} detector, located 40 km off the French coast, is the largest deep-sea
neutrino telescope in the world. It consists of an array of 885
photomultipliers detecting the Cherenkov light induced by charged leptons produced
by neutrino interactions in and around the detector.
The primary goal of {\sc antares} is to search for astrophysical neutrinos in the TeV-PeV
range. This comprises generic searches for any diffuse cosmic neutrino flux as
well as more specific searches for astrophysical sources such as active galactic
nuclei or Galactic sources. The search program also includes multi-messenger
analyses based on time and/or space coincidences with other cosmic probes.
The {\sc antares} observatory is sensitive to a wide-range of other phenomena, from
atmospheric neutrino oscillations to dark matter annihilation or potential exotics such
as nuclearites and magnetic monopoles.
The most recent results are reported.}


\maketitle

\section{Neutrino Astronomy} \label{ak:sec:physcase}

Neutrino astronomy has a key role to play towards a multi-messenger coverage of the high-energy (HE) sky, as HE neutrinos provide a unique tool to observe the non thermal Universe. Their main characteristic is that they can travel over cosmological distances without being absorbed or deflected.  \\
The detection of a HE astrophysical neutrino source would in particular unambiguously identify one of the so far unknown acceleration sites of HE cosmic rays.
Indeed, HE neutrinos ($\nu$) are produced in a beam dump
scenario via meson (pion $\pi$ and mainly kaon at high energy) decay, when the accelerated hadrons (protons $p$ or nuclei $A$)
interact with ambient matter or dense photon fields ($\gamma$): 
{\small
\[
  p/A + A/\gamma \longrightarrow
\begin{array}[t]{l}
 \pi^0 \\
 \downarrow \\
 \gamma + \gamma
\end{array} +
\begin{array}[t]{l}
 \pi^\pm \\
 \downarrow \\
 \mu^\pm + \nu_\mu ~(\overline{\nu}_\mu) \\
 \downarrow \\
 e^\pm +\nu_e ~(\overline{\nu}_e) + \overline{\nu}_\mu ~(\nu_\mu)
\end{array} + N +...
\]}
\noindent 
In this so-called ``bottom-up'' scenario, the production of HE neutrinos
is associated with the acceleration of nuclei
through Fermi-like mechanisms, and with the production of HE gamma-rays, through $\pi^0$ decays.
Following this association, various authors have inferred benchmark fluxes of cosmic neutrinos
based on the observed ultra HE cosmic rays (e.g the so-called WB bound~\cite{bibak:wb}).


HE neutrinos can also be produced by more exotic processes such as the decay of massive particles ("top down'' scenarios, currently not favoured) or the annihilation of dark matter (DM) gravitationally trapped inside massive objects like the Sun, the Earth or the Galactic centre~\cite{bibak:hooper}.

If the weak interaction of neutrinos with matter is an asset for astronomy, it also makes the detection challenging.
This requires the instrumentation of large volumes of water (or ice) with photomultipliers (PMTs) to detect the Cherenkov radiation induced by charged leptons (mainly muons, but also electron- or tau-induced showers) produced by cosmic neutrino interactions with the target transparent medium, inside or near the instrumented volume (see \S~\ref{ak:sec:antares} for a description of the {\sc antares} detector). PMT signals (timing and amplitude) are used to reconstruct the muon trajectory and the energy of the neutrino. In order to reduce the background due to the intense flux of downgoing atmospheric muons present at ground, such detectors are buried deep under the surface. Moreover, since the Earth acts as a shield against all particles but neutrinos, their design is optimized for the detection of upgoing muons produced by neutrinos which have traversed the Earth and interacted in the vicinity of the detector, with the consequence that one neutrino telescope can efficiently monitor one half of the sky in the TeV-
PeV range.\\
Atmospheric neutrinos (see \S~\ref{ak:sec:atmospheric}) produced in the atmospheric cascades can also travel through the Earth
 and interact close to the detector, producing an irreducible background with typical energy spectrum $\frac{dN}{dE} \propto E^{-3.7}$. 
To discover extraterrestrial neutrinos, one searches for an excess of events above a certain energy (diffuse flux, \S~\ref{ak:sec:diffuse}) and/or in a given direction (point sources,\S~\ref{ak:sec:point}). Another possible way to claim the discovery of cosmic neutrinos is to observe events in coincidence (in direction and/or time) with other 
messengers. This multi-messenger approach allows to strongly reduce the background by looking in a known direction for the reduced period 
of time, making the detection of a few events enough to claim a signal (see \S~\ref{ak:sec:too}).


\section{The {\sc antares} Neutrino Telescope}\label{ak:sec:antares}

The {\sc antares} (Astronomy with a Neutrino Telescope and Abyss environmental RESearch)
Collaboration started in March 2006 the deployment of a large scale detector $\sim40$ km off La-Seyne-sur-Mer  (French Riviera),
at a depth of 2475~m. The construction phase ended in May 2008. Since then, {\sc antares} is the largest neutrino telescope constructed in the Northern hemisphere, providing unprecedented sensitivity to the central region of our Galaxy. \\
The full detector~\cite{bibak:detector} comprises 885 photomultipliers distributed in a three dimension array on twelve 450~m high vertical detection lines (Fig.\ref{ak:fig:detector}). 
The lines are separated with a typical interline spacing of 60-70m and each line comprises 25~storeys, separated by
14.5~m. A storey hosts a triplet of Optical Modules (OM) orientated at $45$ degrees with respect to the vertical, in order to maximise the
sensitivity to Cherenkov light from upcoming neutrinos. The OMs contain a 10 inch PMT protected
in a 17 inch pressure resistant glass sphere. The lines are connected to a junction box, via interlink cables
on the seafloor. It provides electrical power and
gathers together the optical fibres from each line into
a single electro-mechanical fibre optic cable for
transmission of the data to and from the shore station.\\
The infrastructure also hosts a thirteenth line, the
instrumentation line (labelled as IL07 in Fig.~\ref{ak:fig:detector} ), which provides
measurement of environmental parameters such as
sea current, temperature and also hosts a part of the
AMADEUS system~\cite{bibak:amadeus}, a test bed for the acoustic
detection of ultra-high energy neutrinos. The line now also hosts a fully functional prototype OM designed for the next generation telescope KM3NeT, housing 31 3" PMTs. In December
2010, a secondary junction box, dedicated to host
sensors for various Earth and Marine science projects,
was also connected to the main junction box.

 \begin{figure}[ht]
  \centering
  \includegraphics[width=0.4\textwidth,clip]{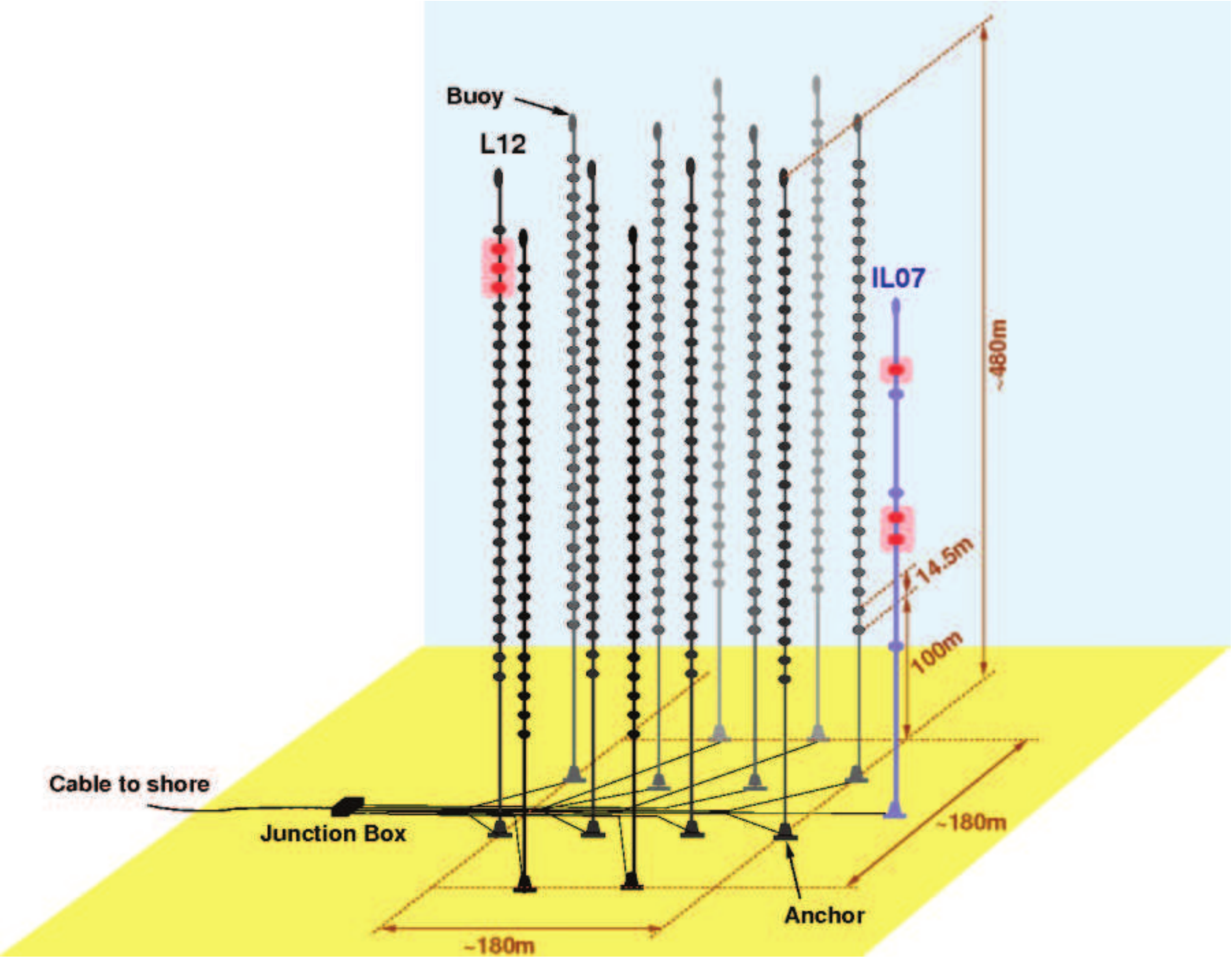}
  \caption{The {\sc antares} detector configuration.}
  \label{ak:fig:detector}
 \end{figure}


Timing calibration~\cite{bibak:calibration} is ensured by a dedicated network of laser and LED beacons, and line motions are monitored by acoustic devices and by inclinometers
regularly spread along the line, allowing redundancy. The angular resolution is within design expectations ($<0.5^{\circ}$) and allows to look for point sources with high sensitivity, as reported in section~\ref{ak:sec:point}. 

\section{Atmospheric Neutrino Studies} \label{ak:sec:atmospheric}
Atmospheric neutrinos represent a background in the search for astrophysical neutrinos. Nevertheless their study offers some intrinsic interest for particle physics, in particular at low energy ($O(GeV)$), where the mixing of flavors can be observed (see \S~\ref{ak:sec:oscillations}). At higher energies, the measurement of the spectrum is a demonstration of the good understanding of the detector behaviour (see \S~\ref{ak:sec:atmnuspectrum}).

\subsection{Neutrino Oscillations} \label{ak:sec:oscillations}

A measurement of the neutrino mixing parameters in the atmospheric sector has been performed with the data taken from 2007 to 2010~\cite{bibak:osc}, for an overall live time of 863 days. In the simplified framework of vacuum oscillations between two flavours ($\nu_\mu$ to $\nu_\tau$),  the $\nu_\mu$ survival probability can be written as
\begin{eqnarray}
P(\nu_\mu \rightarrow \nu_\mu)  = & 1- \sin^2 2\theta_{23} \sin^2\frac{1.27 \Delta m^2_{23} L}{E_\nu} \hspace{.6cm }\\
                                                          = &  1- \sin^2 2\theta_{23} \sin^2\frac{16200 \Delta m^2_{23} \cos \theta}{E_\nu} ,
\label{ak:eq:nuprob}
\end{eqnarray}

\noindent where $L$ and $\theta$ are respectively the neutrino path length and zenith angle, $\Delta m^2_{23}$ is the difference of the squares
of the mass eigenstates $m_2$ and $m_3$ and $\theta_{23}$ is the corresponding mixing angle. Considering values of  $\Delta m^2_{23}=~2.43~\times~10^{-3}{\rm eV^2}$
and $\sin^22\theta_{23} =1$, as reported by the MINOS experiment~\cite{bibak:minos}, the first oscillation maximum is found for vertical upgoing neutrinos at $E_\nu=24$~GeV, leading to a suppression of vertical upgoing muon neutrinos around this energy. \\ 
The oscillation parameters are thus inferred by fitting the event rate as a function of the ratio of the 
neutrino energy and the reconstructed incoming zenith angle (eq.~\ref{ak:eq:nuprob}), the energy being estimated from the observed muon
range in the detector. Assuming a maximal mixing, a mass difference of $\Delta m^2 = (3.1 \pm 0.9) \times 10^3 {\rm eV^2}$ is found. The corresponding contour plot for different confidence levels is shown in Fig.~\ref{ak:fig:contour}, together with measurements by other experiments for comparison. \\
The compatibility of the {\sc antares} measurement with the world data gives confidence in the understanding of the detector, even close to the detection threshold.
This, in addition, paves the way to additional measurements of the neutrino fundamental parameters by neutrino telescopes, such as the mass hierarchy, which has been recently put forward (e.g~\cite{bibak:ARS, bibak:MH}) thanks to the large measured value of the mixing angle $\theta_{13}$.  A feasibility study, dubbed ORCA, for such a measurement with a deep-sea detector is presently being performed in the framework of the KM3NeT Collaboration~\cite{bibak:ORCA}. \\

\begin{figure}[ht]
  \centering
  \includegraphics[width=0.4\textwidth,clip]{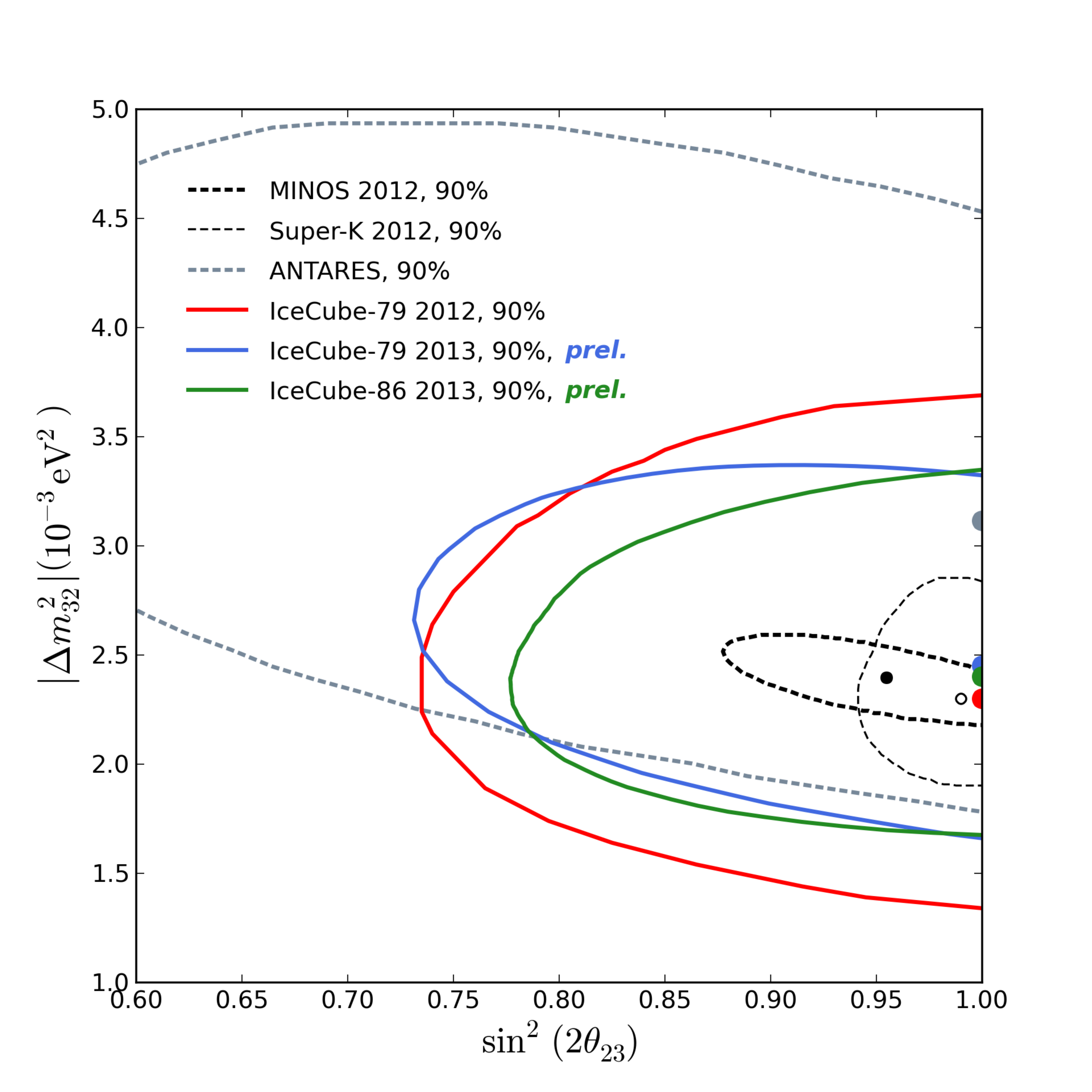}
  \caption{90\% C.L. contour of the neutrino oscillation parameters as derived from the fit of the $E_R/\cos\theta_R$ distribution.  Results from MINOS and Super-Kamiokande are shown for comparison, as well as recent results from the IceCube Neutrino Telescope~\cite{bibak:ICosc}}
  \label{ak:fig:contour}
 \end{figure}

\subsection{Measurement of the Atmospheric Neutrino Spectrum}  \label{ak:sec:atmnuspectrum}
 
At higher energy, where oscillations are suppressed, the neutrino energy cannot be estimated based on the muon path length, since the latter generally exceeds the dimension of the detector. The method used to reconstruct the muon energy relies instead on the total amount of detected light knowing the event geometry and the water properties. The event-by-event energy reconstruction is nevertheless not sufficient to derive the energy spectrum of the detected atmospheric neutrinos. The fact that only the induced muon is observed, the stochastic nature of muon energy losses, the limited energy resolution and the detection inefficiencies have to be accounted for in an unfolding procedure~\cite{bibak:antares_nu_spectrum}.\\
In this study, the analysed data sample covers acquisitions from December 2007 to December 2011, corresponding to an equivalent live time of 855 days.  For each data run a Monte Carlo counterpart is available, reproducing acquisition conditions of the detector. All the cuts to select a pure sample of atmospheric neutrinos were optimised with this Monte Carlo. A 10\% fraction of the data was initially used to check the agreement between the observed and expected quantities.\\
The inferred neutrino energy spectrum, averaged over zenith angles larger than $90^\circ$, is shown as a function of the energy in Fig.~\ref{ak:fig:nuspec}. The measured spectrum spans a range from 100 GeV to 200 TeV. It is 25\% higher than the prediction from the Bartol group (but still within uncertainties) and intermediate between that measured by the AMANDA-II~\cite{bibak:amandanu} and IC40~\cite{bibak:icecubenu} Antarctic neutrino telescopes. This good overall agreement demonstrates the proper understanding of the energy scale of the detector.

\begin{figure}[ht]
  \centering
  \includegraphics[width=0.5\textwidth,clip]{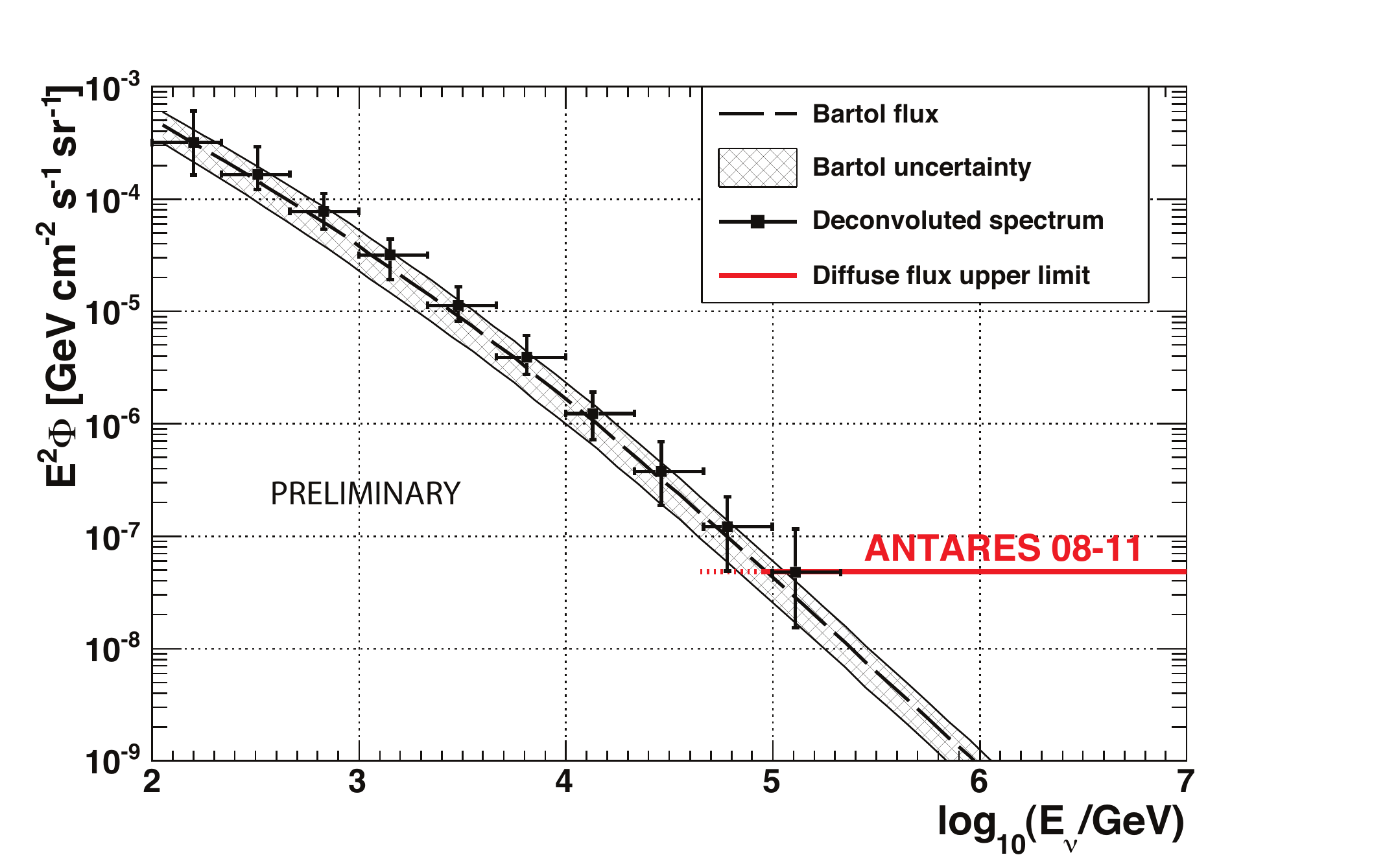}
  \caption{Atmospheric neutrino energy spectrum as measured by {\sc antares}. The flux is compared with the expectations from~\cite{bibak:bartol1} while the grey band corresponds to the uncertainty in the flux calculation~\cite{bibak:bartol2}.}
  \label{ak:fig:nuspec}
 \end{figure}

\section{Searches for Diffuse Fluxes} \label{ak:sec:diffuse}


\subsection{Search for a Cumulative Flux from Unresolved Astrophysical Sources}

The measurement in the highest energy region of the atmospheric neutrino spectrum presented in the previous section has been used to put a constraint on
the diffuse flux of cosmic neutrinos. Indeed, a harder spectrum (typically $\frac{dN}{dE}\propto E^{-\alpha}$ with $\alpha=1-2$), is expected for neutrinos of astrophysical origin, as mentioned in section~\ref{ak:sec:physcase}. After a critical value of the energy, which depends on the absolute normalization of the cosmic neutrino flux, an excess of events over the atmospheric background is expected.\\
The same statistical method as the one described in reference~\cite{bibak:antaresdiffuse}, which presents the first {\sc antares} limit based on the 2008-2009 data set, was used.  The main difference arises from the new energy estimator employed. Here the muon energy deposit per unit path length is approximated by an estimator
which can be derived from measurable quantities such as the amplitude of the hit PMTs. To remove the contribution from background light, only the hits used by the final tracking algorithms (filtered out following strict causality criteria) are used. The energy estimator was then used to determine the cut yielding the best sensitivity, applying the Model Rejection Factor (MRF) procedure~\cite{bibak:mrf}.
In the considered live time and with the track quality cuts applied, the conventional atmospheric neutrinos revealed a 28\% deficit, well within uncertainties,  with respect to the data. Therefore the predicted background from the simulation is normalised to the data. After normalization, 8.4 atmospheric events were expected and 8 events were observed in the high energy region (2.3 signal events were expected from a test flux $E^2 \Phi_{\rm test} = 2 \times 10^{-8} {\rm GeV cm^{-2} s^{-1} sr^{-1}}$). This translates into a 90\% C.L. upper limit of $$E^2\frac{dN}{dE} = 4.8 \times 10^{-8} {\rm GeV cm^{-2} s^{-1} sr^{-1} }$$ in the energy range 45 TeV - 10 PeV, as indicated in Fig.~\ref{ak:fig:nuspec}. As can be seen in Fig.~\ref{ak:fig:diffuse}, this updated result of {\sc antares} is close to the WB predictions. It is only a factor $\sim4$ above the limit obtained by IceCube in its 59 string configuration and applies complementarily to the southern hemisphere.

\begin{figure}[ht]
  \centering
  \includegraphics[width=0.5\textwidth,clip]{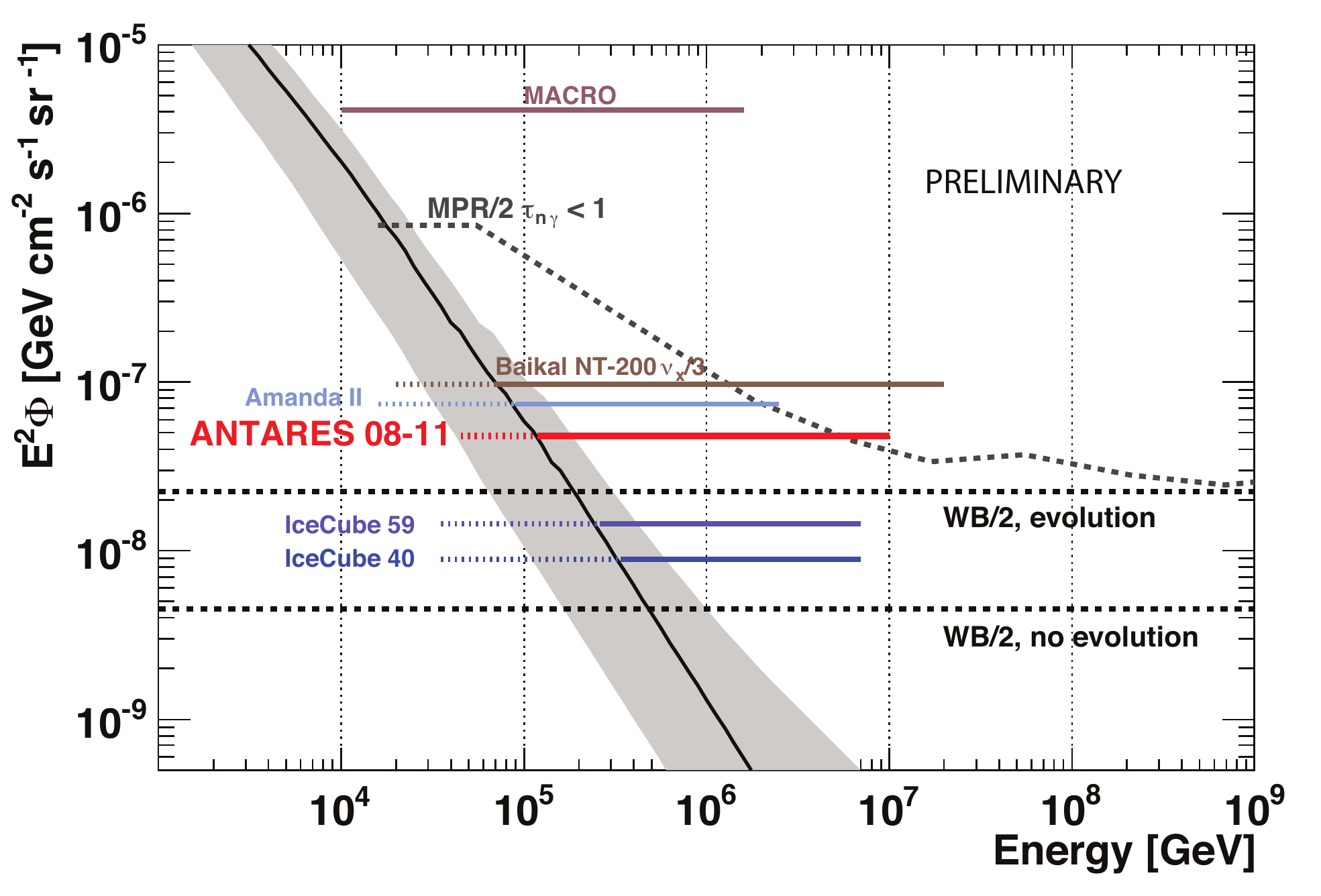}
  \caption{The updated {\sc antares} upper limit (90\% C. L.) for an $E^{-2}$ spectrum compared to 90\% C.L. upper limits from other experiments, including IceCube limits obtained with 40~\cite{bibak:IC40} and 59 strings~\cite{bibak:IC59}, all for single neutrino flavour.}
  \label{ak:fig:diffuse}
 \end{figure}

\subsection{Fermi Bubbles}

The Fermi-LAT data has revealed a large ($\sim$10kpc) bilateral structure
originating from the Galactic Centre and perpendicular to the Galactic plane (and therefore, visible mostly in the southern hemisphere). This structure is generally referred to as "Fermi Bubbles" (FBs).
A possible explanation of the observed gamma-rays (which have an almost uniform $E^{-2}$ spectrum, with a normalisation uncertainty of a factor of $\sim$2~\cite{bibak:FBflux}) is the presence of a Galactic wind in which accelerated cosmic rays interact with an interstellar medium producing pions~\cite{bibak:fb}. In such a scenario, neutrinos would be naturally produced. A dedicated search with the {\sc antares} data from May 2008 to December 2011 (806 days) has thus been performed by comparing the rate of HE events observed in the region of the FBs (ON zone) to that observed in equivalent areas of the Galaxy excluding the FBs
(OFF zones). A small excess of events (1.2$\sigma$) was observed in the ON region above the chosen energy cut. Limits were placed on possible fluxes of neutrinos for various assumptions on the energy cutoff at the source (Fig.~\ref{ak:fig:fb}). These limits are close to the theoretical expectations, meaning that the hadronic origin of the observed gamma-rays from the FBs could be probed with the full {\sc antares} data set, or after one year of data taking with the next generation neutrino telescope to be built in the Mediterranean~\cite{bibak:fbkm3net}. 

\begin{figure}[ht]
  \centering
  \includegraphics[width=0.5\textwidth,clip]{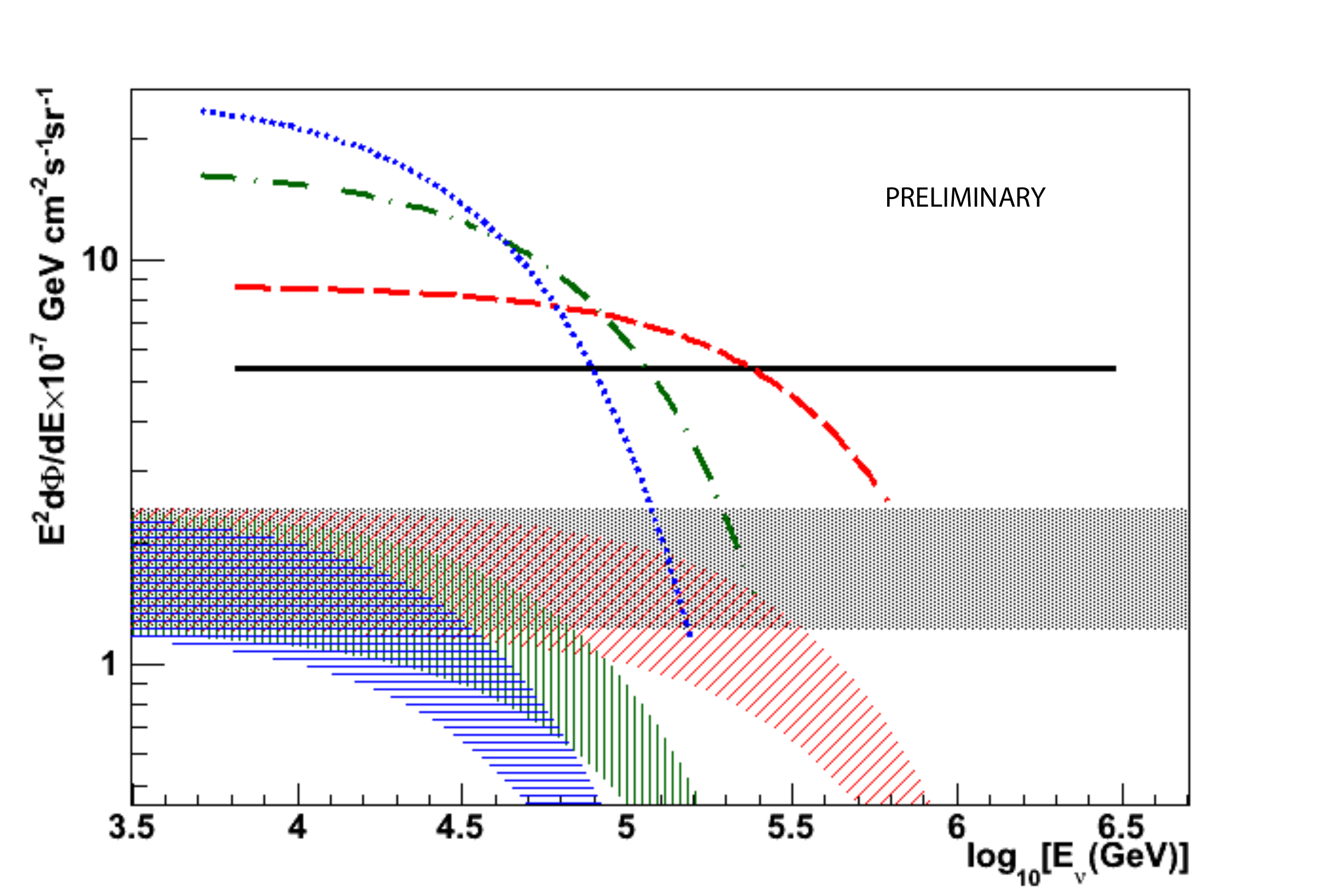}
  \caption{Upper limits on the neutrino flux from the Fermi Bubbles, estimated for different source cutoffs: no cutoff (black solid), 500 TeV (red dashed), 100 TeV
(green dot-dashed), 50 TeV (blue dotted) together with the theoretical predictions for the case of a pure hadronic model (the same colours,
areas filled with dots, inclined lines, vertical lines and horizontal lines
respectively). The limits are drawn for the energy range where 90\% of
the signal is expected.}
  \label{ak:fig:fb}
 \end{figure}

\section{Searches for Sources of Cosmic Neutrinos } \label{ak:sec:point}

For the search of resolved sources of cosmic neutrinos, several approaches were followed. The studies presented below have been first applied to the 2007-2010 data 
(for a total live time of 813 days), corresponding to 3058 events passing the optimised selection criteria~\cite{bibak:antaresps}. A recent update, based on data from the 31st of January
2007 until the 31st of December 2012 is also presented~\cite{bibak:psicrc}, corresponding to a total live time of $\sim$1334 days
of which 183 days correspond to the period when the detector was working with the first 5 deployed lines. After the quality cuts, a total of 5516 neutrino candidates
are selected, of which 90 \% should be atmospheric neutrinos and the rest is composed of misreconstructed
muons. The median uncertainty on the reconstructed neutrino direction, assuming an $E^{-2}$ neutrino energy spectrum, is $0.4^\circ\pm0.1^\circ$. Almost 90\% of the events are reconstructed within $1^\circ$ from their true direction. With the selected events, a time-integrated full-sky search was performed, which is presented in section~\ref{ak:sec:fullsky}. Another search focused on $\sim$50 neutrino source candidates of various types (section~\ref{ak:sec:list}). Special attention was also paid to the case of some extended sources (section~\ref{ak:sec:extended}). It is also worth mentioning that a two-point autocorrelation method was also applied to the 2007-2010 data sample (section\ref{ak:sec:2pt}). Such method offers the advantage of being sensitive to a large variety of cluster morphologies and does not depend on Monte Carlo simulations. 

\subsection{Full Sky Search} \label{ak:sec:fullsky}

In the full-sky search an excess of events over the atmospheric muon and
neutrino backgrounds is looked for in the declination
range [$-90^{\circ}; +48^{\circ}$]. The algorithm used
is based on the likelihood of the observed events, where
the knowledge of the point spread function, of the expected
background rate as a function of the declination and of the number of hits used in the reconstruction are
included. The error estimate on the track direction, as given by the fit procedure, is also included in the likelihood on an event-by-event basis. A
skymap with the position of every selected point in the sky is shown in equatorial coordinates in Fig.~\ref{ak:fig:skymap}. 
The most significant cluster is located at (R.A=-47.8$^{\circ}$; $\delta$=-64.9${\circ}$), containing 14 events. This
corresponds to a fitted number of signal events of 6.3
on top of the expected background. The post-trial p-value is 2.1\%, equivalent to $2.3\sigma$. It is worth noticing that this is
the same most significant cluster as found in the 2007-2010 data set, plus 6 additional events. Nonetheless, this "warm spot" was searched for counterparts in the electromagnetic 
band with data from other instruments, without success~\cite{bibak:hotspot}.

\begin{figure}[ht]
  \centering
  \includegraphics[width=0.5\textwidth,clip]{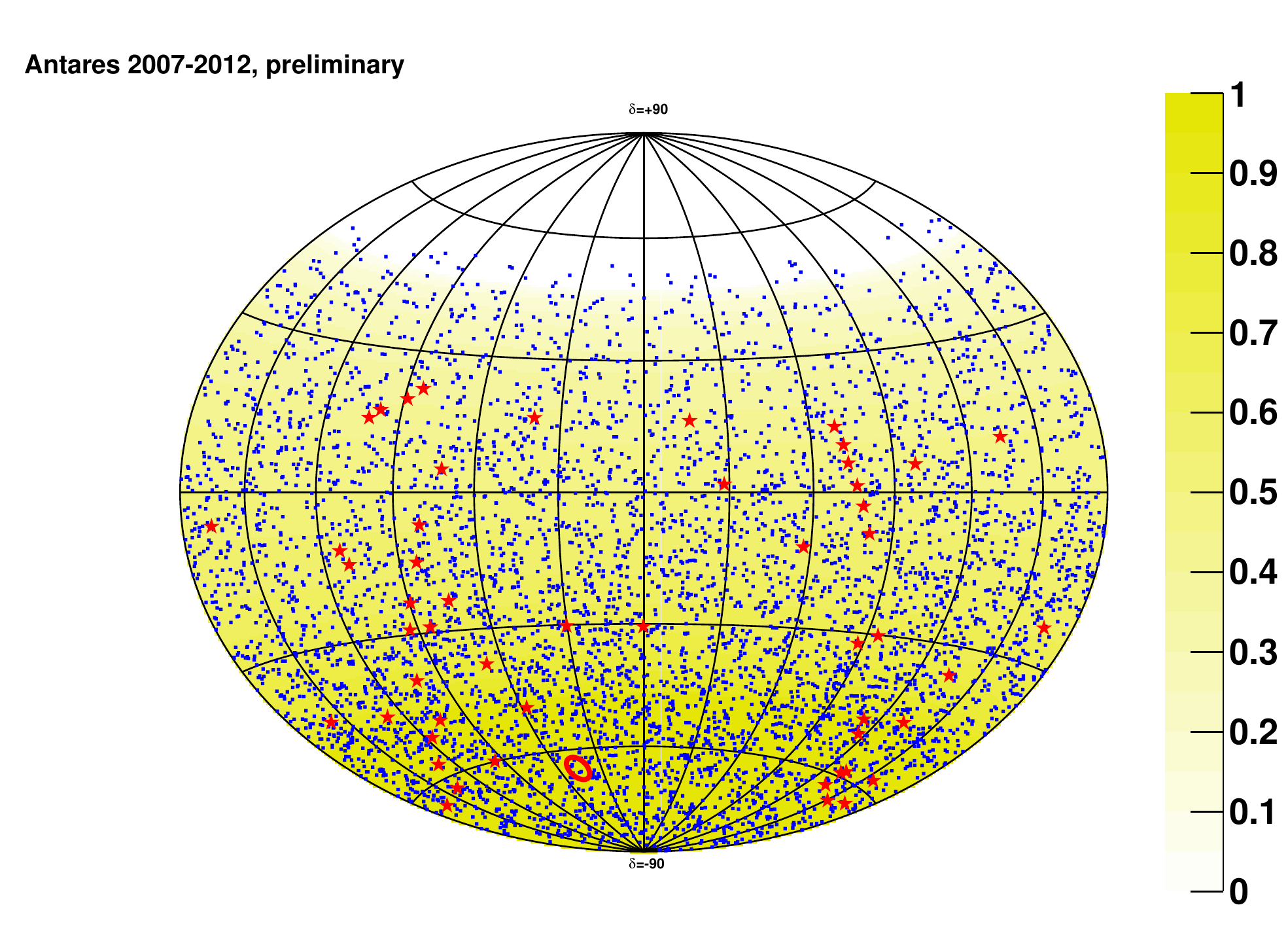}
  \caption{Sky map in equatorial coordinates showing the 5516 selected data events.
The position of the most signal-like cluster is indicated by
the red circle. The red stars denote the position of the candidate
sources.}
  \label{ak:fig:skymap}
 \end{figure}

\subsubsection{Correlation Searches} \label{ak:sec:2pt}

As a cross-check approach, the 2007-2010 point source data set was also studied with a two-point autocorrelation method. The main interest of the method is that it is sensitive to a
large variety of cluster morphologies and, on the other side does not rely on Monte Carlo simulations. As such, it offers complementary information to the all-sky search presented above.
The standard 2pt-correlation function, used to search for clusters in a set of $N$ event, is defined as the differential distribution of the number of observed event pairs $\mathcal{N}_\mathrm{p}$  in the dataset as a function of their mutual angular distance $\Delta \Omega$.  The correlation function has been modified to incorporate information from an estimator of the event energy. The cumulative autocorrelation distribution used is defined as \begin{equation}
\mathcal{N}_\mathrm{p} (\Delta \Omega) = \sum\limits_{i=1}^{N} \sum\limits_{j=i+1}^{N} w_{ij} \times H(\Delta \Omega_{ij} - \Delta \Omega), \label{equ:autocor}
\end{equation}
where $H$ is the Heaviside step function. The events are weighted by the factor $w_{ij}=w_i \times w_j$ where the individual  weights $w_i$ represents the p-value related to their energy estimate.
Such a modification of the standard autocorrelation function leads to a significant increase of sensitivity to detect clustering of events of astrophysical origin, as quantified in~\cite{bibak:2pt_ps}. 
The cumulative distributions, both for the {\sc antares} data set and for an isotropic data set, are given in Fig.\ref{ak:fig:2pt_ps}. No significant deviation from isotropy was observed.\\
To improve the sensitivity of the search for astrophysical neutrinos, equation~\ref{equ:autocor} was further modified to allow for cross-correlations with external source catalogues. 
Two such catalogues were tried: the 2FGL point source Fermi catalogue and the GWGC catalogue hosting 53295 galaxies within a distance of D $< 100$Mpc. In both cases, no correlation was observed.\\
The above described studies are currently being reproduced with the full 2007-2012 data set.

\begin{figure}[ht]
  \centering
  \includegraphics[width=0.5\textwidth,clip]{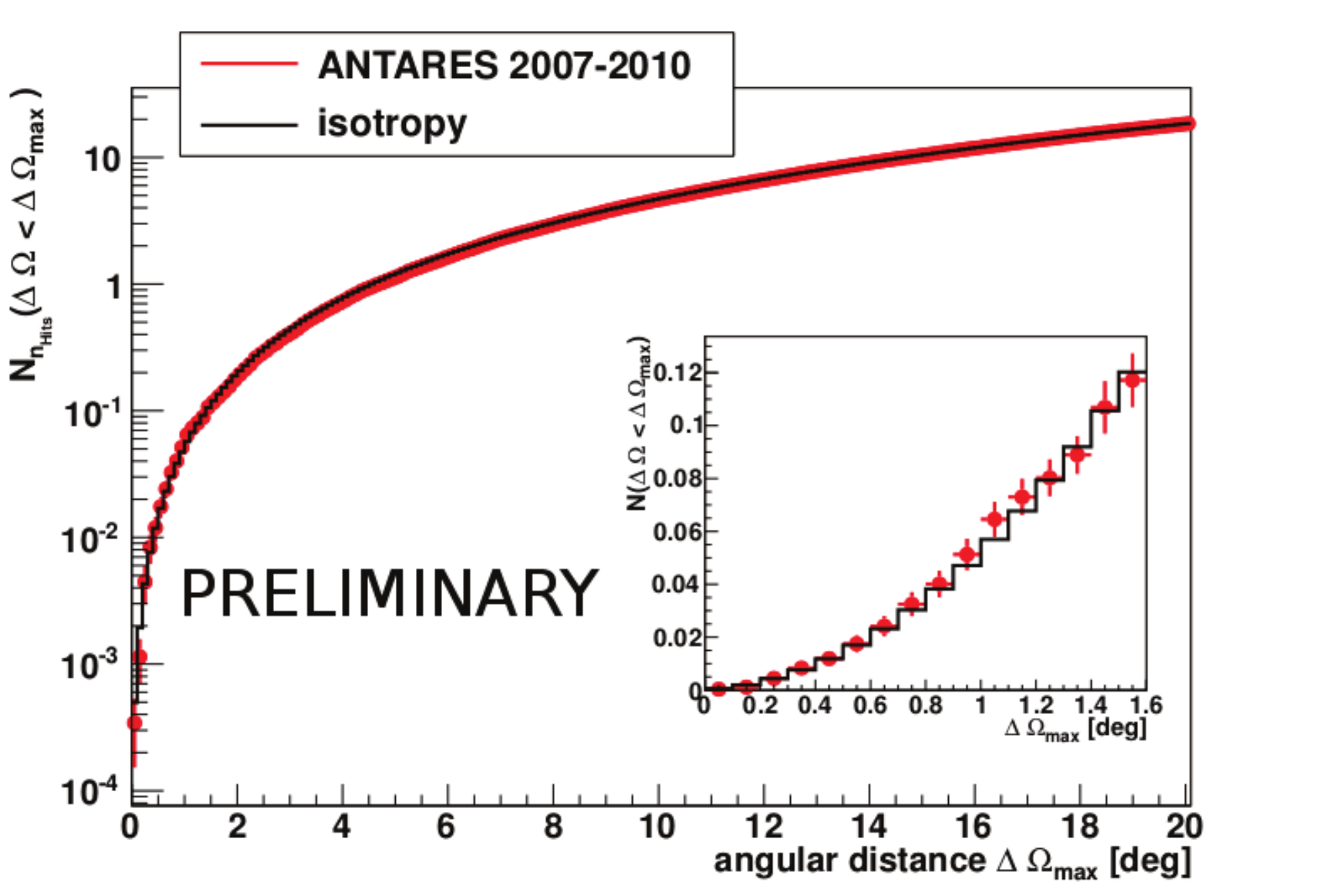}
  \caption{Cumulative autocorrelation function of  the 2007-2010 data set. The red markers denote the
{\sc antares} data and the black histogram represents the reference distribution expected for an isotropic dataset. The inset
shows is a zoom on the smallest angular distances.}
  \label{ak:fig:2pt_ps}
 \end{figure}

\subsection{Candidate-list Search}\label{ak:sec:list}

Good candidates for HE $\nu$ production are active galactic
nuclei (AGN) where the accretion of matter by a supermassive 
black hole may lead to relativistic ejecta~\cite{agn_jet1,agn_jet2}. 
TeV activity from our Galaxy has also been reported by ground based gamma-ray telescopes. 
Many of the observed Galactic sources~\cite{galactic_sources} are candidates of hadron acceleration and subsequent neutrino production. 
This includes, among others, supernovae remnants, pulsar wind nebulae or microquasars.
A list of $\sim$ 50 sources of interest has been established for a search in the corresponding specific directions.
No significant excess was found. Upper limits on the neutrino flux from these sources were derived as
shown in Fig.\ref{ak:fig:ps_limits}. In addition to the above mentioned sources, cosmic neutrinos were also looked for in the direction of a selected sample 
of gravitational lenses, with the same method. The results of this specific study, applied to the 2007-2010 data set, are presented in~\cite{bibak:antareslensing}.\\
For TeV-PeV sources, these limits are the best for the Southern hemisphere 
since, for this region of the sky, the IceCube limits are relevant above  $\sim$~1~PeV, an energy threshold applied in order to
mitigate the downgoing muon background. 

\vskip 0.5cm
\subsection{Extended Sources}\label{ak:sec:extended}

Several of the sources included in the candidate source list are not point-like, but have an intrinsic spatial structure that can be resolved by {\sc antares}.  This applies to sources such as the shell-type
supernova remnant RX J1713.7-3946 and the pulsar wind nebula Vela X. For these sources, the assumption of point-like sources is not optimal and may be improved.
Assuming the neutrino flux to be
related to the gamma-ray flux and taking into account the measured extensions, special studies were made for these two sources.  Upper limits (90\% C.L.) on the flux normalisation were
computed, as well as the corresponding
model rejection factor (MRF). For RX J1713.7-3946 the upper limit is a factor
6.4 higher than the theoretical prediction.
For Vela X the upper limit is a factor 9.7 higher than the model, which is slightly worse than previously obtained with the 2007-2010 data set, because of the presence of 4 additional events.
In both cases {\sc antares} limits are the most restrictive ones for the emission models considered (see~\cite{bibak:psicrc} for more details).

\begin{figure}[ht]
  \centering
  \includegraphics[width=0.5\textwidth,clip]{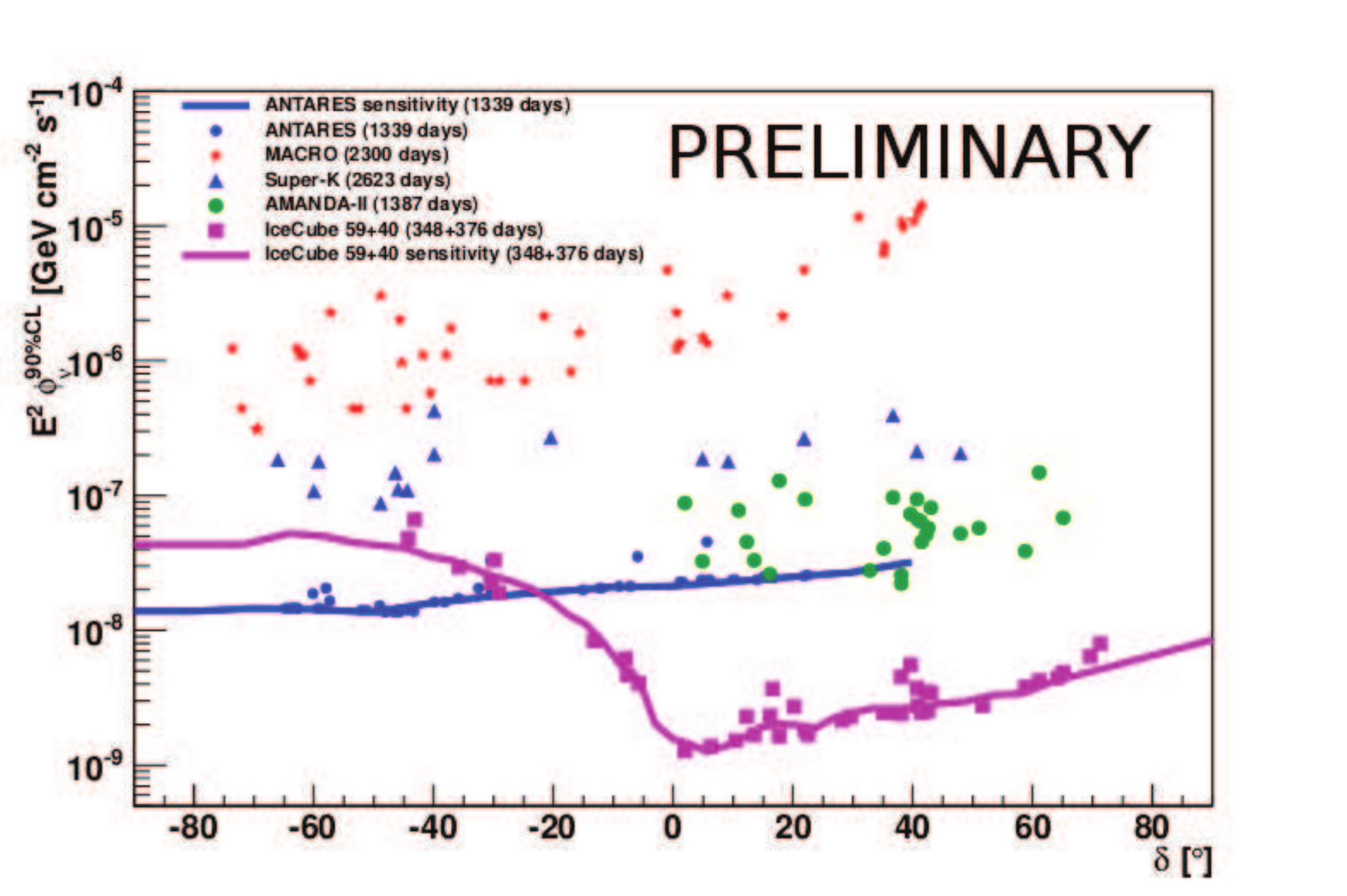}
  \caption{Upper limits (at 90\% C.L.) on the $E^{-2}$ neutrino flux from
selected candidate sources as function of their declination. Upper
limits from other experiments are also indicated. The lines
show the expected median sensitivities.}
  \label{ak:fig:ps_limits}
 \end{figure}

\section{Searches for Transient Sources} \label{ak:sec:too}

Other potential sources of extra-galactic 
HE neutrinos are transient sources like
gamma ray bursts (GRBs, section~\ref{ak:sec:grb}). The flux of HE neutrinos from GRBs~\cite{nu_grb} is lower than the one expected from 
steady sources, but the background can be dramatically reduced by 
requiring a directional and temporal coincidence 
with the direction and time of a GRB detected by a satellite. 
Similar studies can also be applied to AGNs or microquasars while observed in high flaring periods (section~\ref{ak:sec:flares}).

\subsection{GRB} \label{ak:sec:grb}

\begin{figure}[ht]
  \centering
  \includegraphics[width=0.43\textwidth,clip]{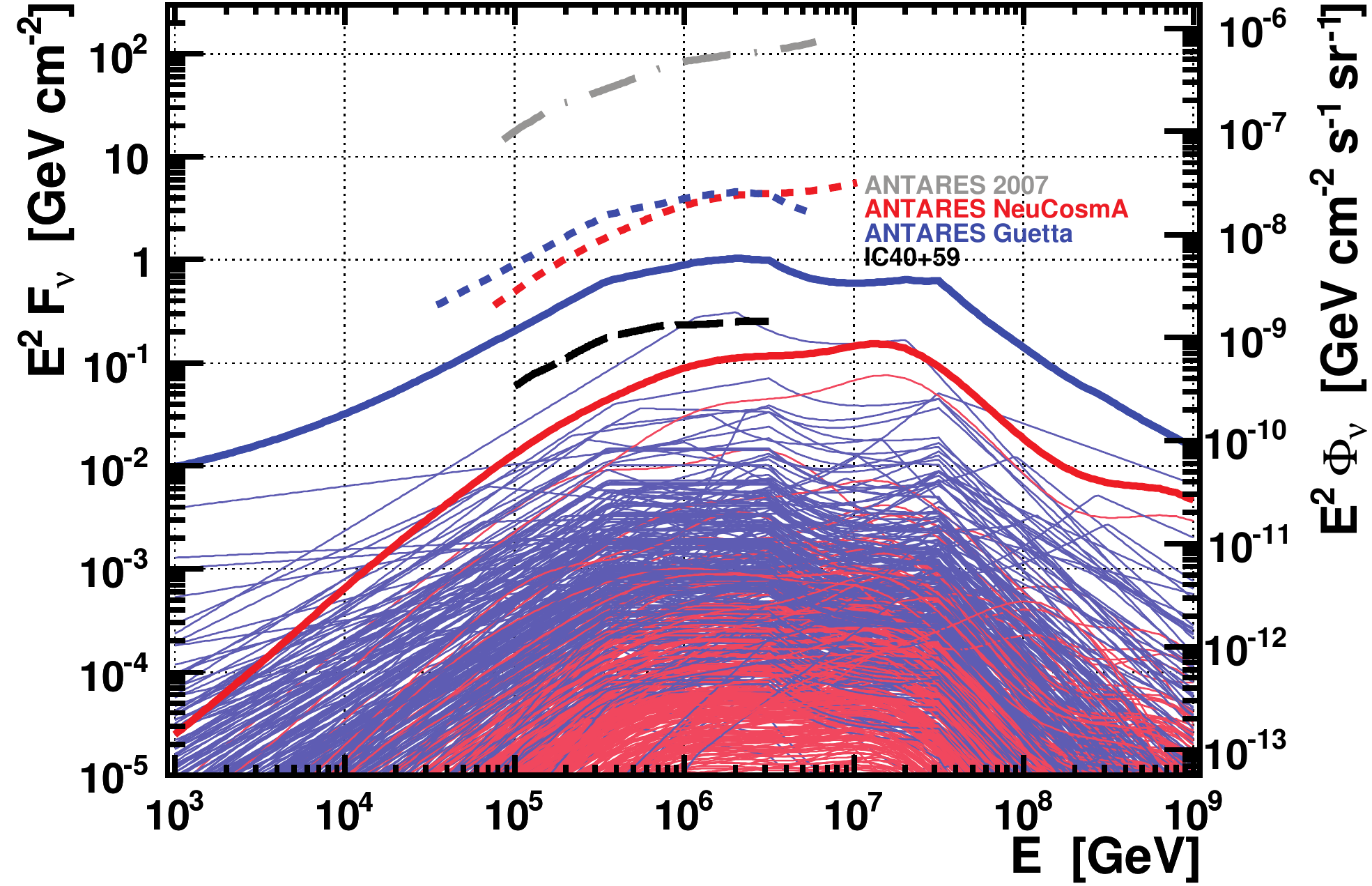}
  \caption{Individual spectra of all analysed bursts in the second study (thin solid lines) and their sum (thick solid lines) for an analytical model (blue) and a more recent numerical model (red). Experimental limits on the total fluence
are shown in dashed lines. The righthand axis gives the inferred quasi-diffuse flux limit $E^2\Phi_\nu$.}
\label{ak:fig:grblimits}
 \end{figure}
 
A first search for neutrino-induced muons in correlation with a selection of 40 GRBs that occurred in 2007, while the detector was in its 5-line configuration, is reported in~\cite{antares_grb}. A second search, based on the 2008 - 2011 data sample, was recently performed, corresponding to an analysed set of 296 GRBs, representing a total equivalent live time of 6.6 hours.  In contrast to the previous searches, this study has been optimised for a fully numerical neutrino-emission model, based on~\cite{bibak:cosmA}. The optimisation relied on an Extended Maximum Likelihood Ratio built with an a priori knowledge of the background estimated from real data from the whole data sample period, in the GRB direction, adjusted to match the detector efficiency at the respective trigger time. \\
In both cases, no significant number of events was found in correlation with the prompt photon emission of the GRBs and upper limits were placed on the total fluence from the whole sample as well as on the inferred quasi-diffuse flux of neutrinos for different models, as shown in Fig.~\ref{ak:fig:grblimits}. Although not competitive with the IceCube limits, the {\sc antares} results are obtained on a $~90\%$ different sample of GRBs, thus offering some complementarity.\\
On April $\rm 27^{th}$, 2013, the brightest GRB of the last 29 years has been detected by five $\gamma$-ray telescopes simultaneously, with the highest-energy photon detected by Fermi-LAT reaching a record-holding energy of 94 GeV. Consequently, a dedicated search for neutrino emission from this exceptionally nearby GRB130427A has been conducted. No excess over background has been found in coincidence with the electromagnetic signal, thus the first limits have been placed on the neutrino emission from this burst\cite{bibak:GRB130427A}, as shown in Fig.~\ref{ak:fig:GRB130427A}\\

The above mentioned studies are triggered searches based on external alerts. Reversely, {\sc antares} can also send alerts thanks to the ability of its data acquisition system to rapidly filter and reconstruct events in real-time~\cite{bibak:alerts}. In this context, the $2\pi$ instantaneous sky coverage and the high duty cycle of the detector are relevant assets.
Alerts consist either of doublets occurring within 15 minutes and separated by $3^\circ$ or of HE events (typically above 5 TeV). Recenty, a third criterion has been implemented: events closer than $0.3^\circ$ from a local galaxy (within 20~Mpc) generate an alert as well. Since 2009, alerts have been sent on a regular basis to a network of fast-response, wide field of view ($1.9^\circ\times1.9^\circ$) robotic telescopes (TAROT, ROTSE, ZADKO) and more recently also to the SWIFT/XRT telescope. The rapid response of the system ($\sim20$s) is particularly well suited for the detection of optical afterglows from GRBs. Up to now, no optical counterpart associated with a neutrino alert was observed and limits on the magnitude of a possible GRB afterglow were derived~\cite{bibak:icrtatoo}.

\subsection{Flaring Sources} \label{ak:sec:flares}

To further improve the sensitivity, specific searches have been performed during reported period of intense activity of some sources. This approach enhances the
sensitivity by a factor 2-3 with respect to a standard time-integrated  search, for a flare duration of 1-100 days.
The first analysis of this kind was made with a total of ten blazars selected based on their
Fermi-LAT light curve profiles in the last four months of 2008. In this method, the assumed neutrino time distribution is directly extracted from the gamma-ray light curve. The search for a signal is performed using
an unbinned likelihood ratio maximisation, where the data are parameterised as a two components
mixture of signal and background. For one of these AGNs (3C279), one neutrino candidate was
found to be in spatial ($0.56^\circ$) and time coincidence
with the flare. The post-trial probability for this to
occur randomly in the only-background hypothesis was evaluated as 10\%. The 90\% C.L. upper limits derived on the neutrino fluence for these
AGNs are presented in~\cite{bibak:blazars}.  An extension of the search was recently done with data till December 31st, 2011 (750 live time days). A total of 86 flaring periods from 41 blazars was studied with an improved likelihood incorporating the number of hits as an energy proxy, yielding an additional
improvement of about 25 \%. Again, the most significant flare, with a post trial p-value of 12\%, is found for 3C279 with a second event passing the selection. More details are given in~\cite{bibak:icrc_blazars}.\\
Another search of this kind has been conducted with a list of 6 microquasars with x-ray or gamma-ray outbursts in the
2007-2010 satellite data (RXTE/ASM, Swift/BAT and Fermi/LAT). No significant excess of neutrino events in spatial and time coincidence with the flares
was found. The inferred limits on the neutrino flux are close to theoretical predictions\cite{bibak:microquasars} which may be reached by {\sc antares} in the following years, in particular for what concerns GX339-4 and CygX-3.

\begin{figure}[ht]
  \centering
  \includegraphics[width=0.5\textwidth,clip]{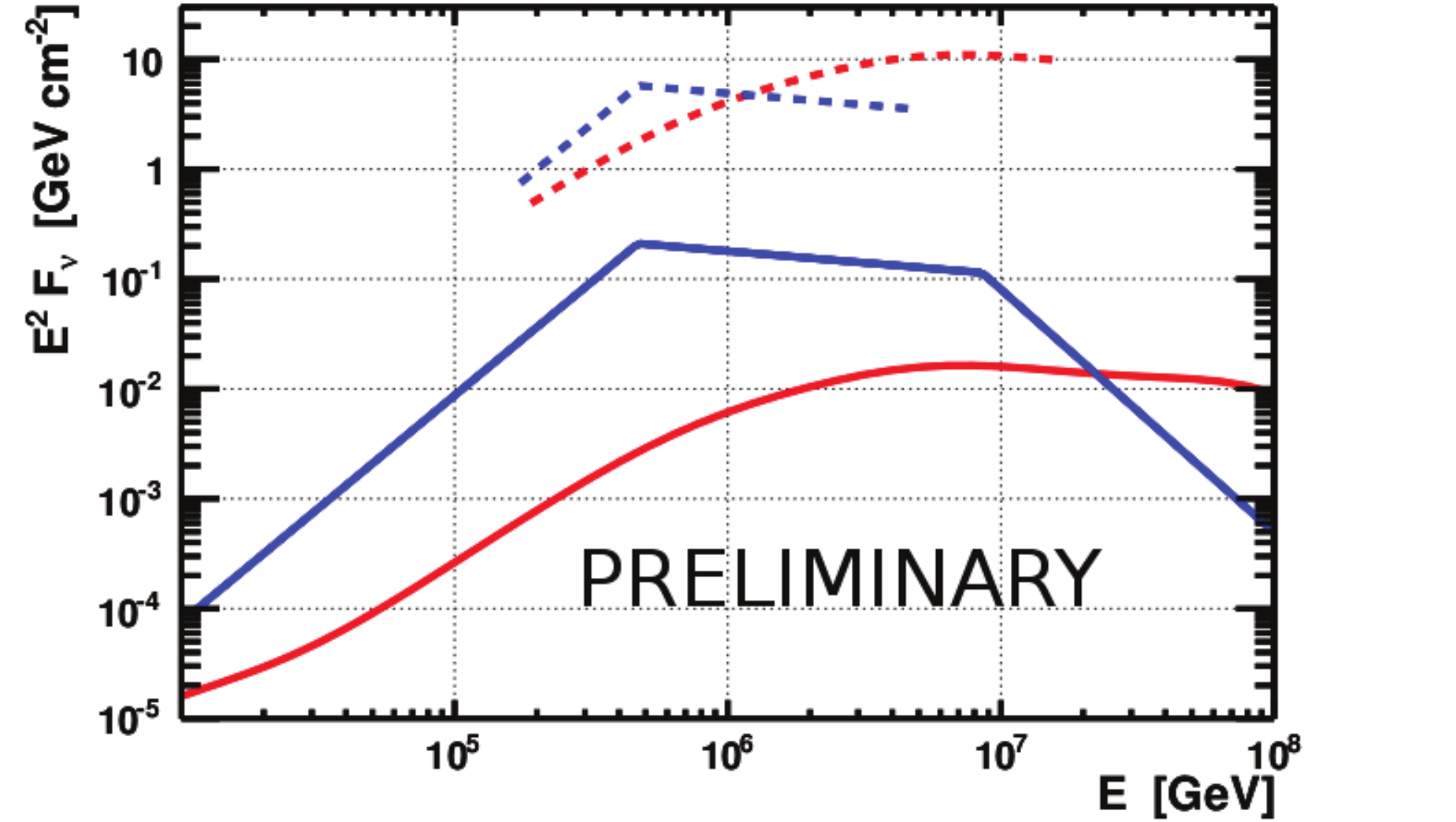}
  \caption{Upper limits on the neutrino fluence from GRB130427A (dashed lines), both for the analytical model (blue) and the more recent numerical model (red). The predicted fluences are shown as a solid line (same color code).}
\label{ak:fig:GRB130427A}
 \end{figure}

\subsection{Sources emitting Gravitational Waves}

Another example of a multi-messenger search with {\sc antares} candidate events used as a trigger, is the search for joint sources of HE neutrinos and Gravitational Waves (GWs) with the VIRGO and LIGO interferometers. In addition to already know type of sources, the association of HE neutrino candidates with GWs could reveal new, hidden sources that are not observed by conventional photon astronomy (e.g. failed GRBs). This has motivated the signature of an agreement for data exchange between the {\sc antares} Collaboration, the LIGO Scientific Collaboration and the Virgo Collaboration.
Two common data set exists. The first one covers the period from January to September 2007 which coincides with the fifth and first science runs of LIGO and Virgo, respectively, and with data from the 5 first {\sc antares} lines. This common data set was jointly analysed. No significant number of coincident event was observed and limits on the density of joint HE neutrino - GW emission events in the local universe were placed~\cite{bibak:GWHEN}.\\
An additional data set corresponding to the second VIRGO-LIGO common science data sample (from July  2009 to October 2010) and to the full {\sc antares} configuration is currently being analysed with improved methods on both sides.

\section{Beyond Astrophysics}

The {\sc antares} detector offers the possibility to study a broad field of physics, beyond astrophysics.
Here we briefly report on searches for DM in the form of Weakly Interacting Massive Particles (WIMPs, section~\ref{ak:sec:wimps}) and further exotic physics such as magnetic monopoles and nuclearites (section~\ref{ak:sec:exotica}). 

\subsection{Dark Matter Searches} \label{ak:sec:wimps}

\begin{figure}[ht]
  \centering
  \includegraphics[width=0.5\textwidth,clip]{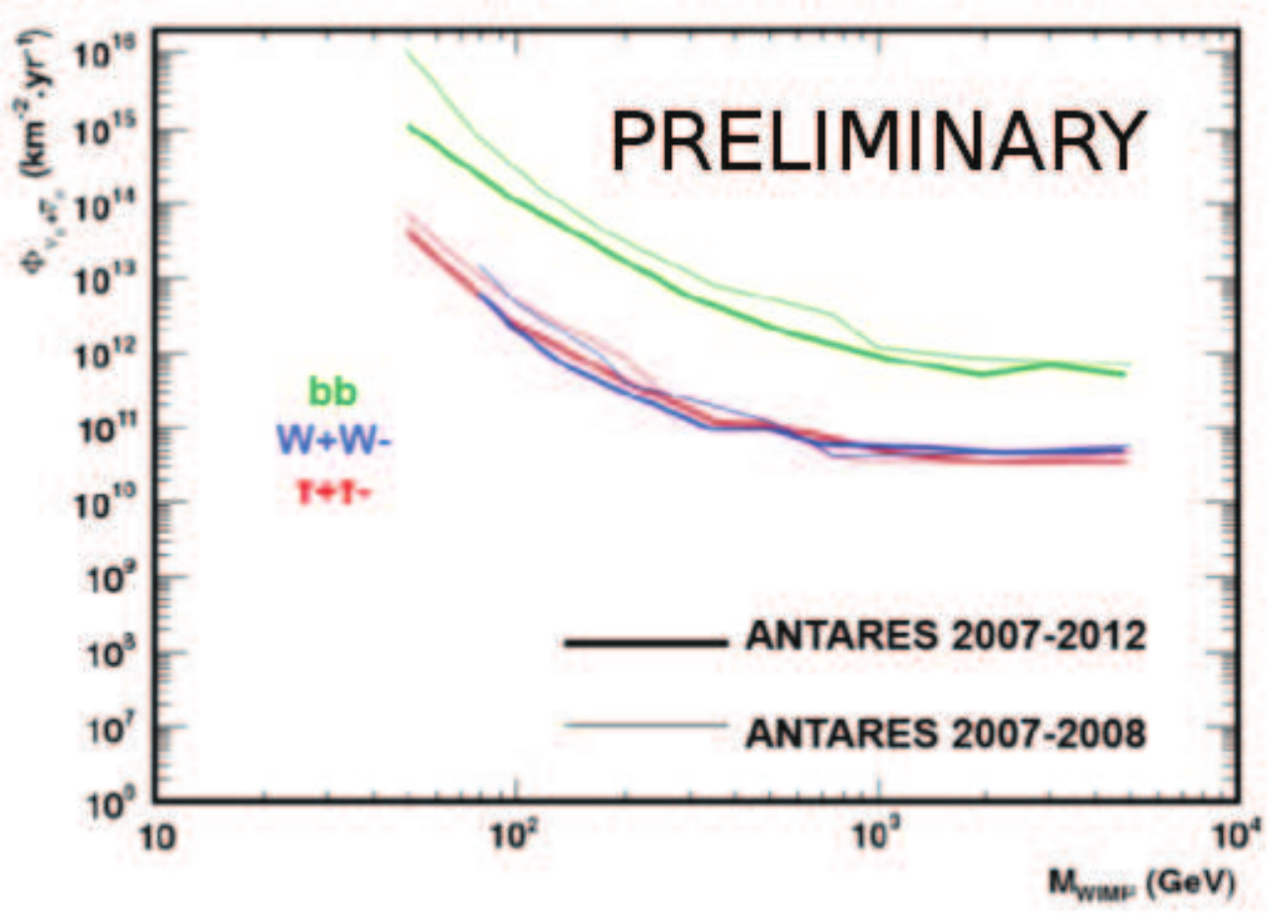}
  \caption{90\% C.L. upper limit on the muon  neutrino flux as a function of the WIMP mass for the three self-annihilation channels $b\bar{b}$ (green),
$W^+W^-$ (blue) and $\tau^+\tau^-$ (red). }
  \label{ak:fig:DM1}
 \end{figure}

WIMPs are advocated in many theoretical models to explain the missing matter in the Universe. As massive particles, they are gravitationally trapped in dense bodies such as the Sun, the Earth or the Galactic centre, where they can subsequently self-annihilate. As by-product of the annihilation process, neutrinos can be produced, which may be the only particles able to escape, with an energy of the order of the mass of the WIMP. A signature for DM would therefore be an excess of neutrinos in the direction of such bodies in the $\sim$10~GeV ~-~1~TeV energy domain.\\
A first search in the direction of the Sun has been performed using the data recorded by  {\sc antares}  during 2007 and 2008, corresponding to a total live time of 294.6
days. For each WIMP mass and each annihilation channel envisaged, the quality cuts were chosen to minimise the
average 90\% C.L. upper limit on the DM induced neutrino flux. No excess above atmospheric background was found. The results are reported in~\cite{bibak:icrcdm}.\\
An improved analysis has been recently unblinded, relying on 1321 effective days from the 2007-2012 data set. Improvements arise at low energy from the addition of events reconstructed with only one line (offering poor azimuth accuracy but valuable zenith information). Another source of improvement comes from the use of 2 independent track reconstructions, the best one being kept for each tested WIMP mass. Again, no significant number of event was found over background and upper limits were derived on the flux of neutrinos from the Sun (see Fig.~\ref{ak:fig:DM1} ). Additionally, in Fig.~\ref{ak:fig:DM2}, the 90\% C.L. upper limits in terms of spin-dependent  WIMP-proton cross-sections are derived and compared to predictions of the MSSM-7~\cite{bibak:mssm-7} model, a simplified version of the Minimal SuperSymmetric Model containing a neutralino as lightest stable particle.

\begin{figure}[ht]
  \centering
  \includegraphics[width=0.5\textwidth,clip]{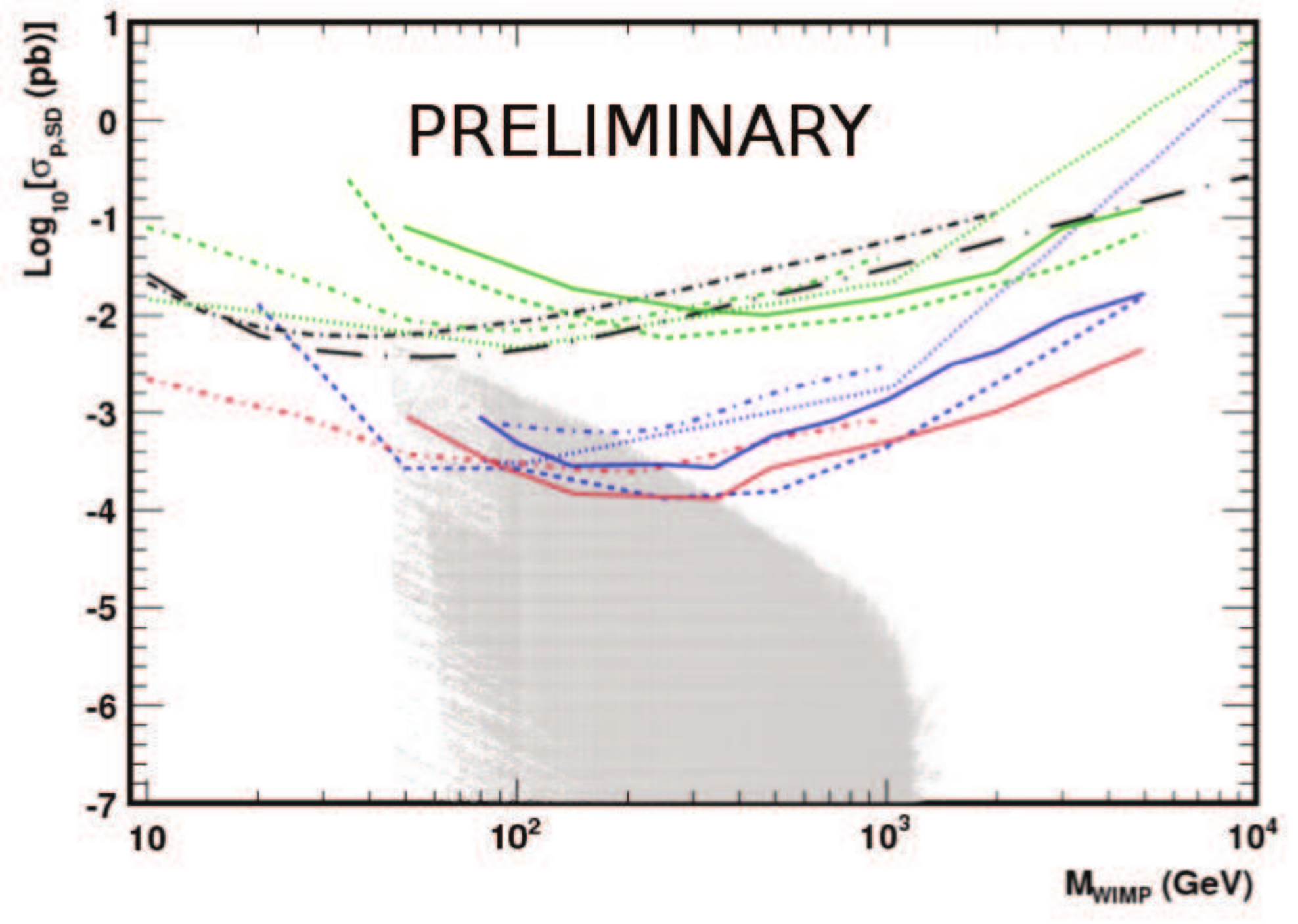}
  \caption{ {\sc antares} limits on the spin-dependent cross-section of WIMPS on
protons inside the Sun. Non-excluded MSSM-7 models from a scan
over its parameter space are indicated. The results from Baksan 1978-2009 (dash-dotted lines), SuperKamiokande 1996-2008 (dotted lines) and IceCube-79 2010-2011(dashed lines) are also shown, with
the addition of the direct detection limits from KIMS 2007
(dashed-dotted black line) and COUP 2011 (dashed black line).}
  \label{ak:fig:DM2}
 \end{figure}

\subsection{Exotic Searches} \label{ak:sec:exotica}

\subsubsection{Magnetic Monopoles}

Magnetic monopoles are predicted by various gauge theories to have been produced in the early universe. Their predicted signature in a neutrino telescope is quite visible, as the intensity of the light they emit in the detector is $O(10^4)$ times greater than the Cherenkov light, depending on the velocity $\beta$.
A search for relativistic upgoing magnetic monopoles was performed with 116 days live time of {\sc antares} 2007-2008 data~\cite{bibak:antaresmonopoles}. In order to improve the sensitivity for low velocity monopoles, the tracking algorithm has been modified so as to leave
the velocity as a free parameter to be fitted. One event was observed, compatible with the background only hypothesis.
The derived limits on the upgoing magnetic monopole flux above the Cherenkov
threshold for $0.625  \leq \beta \leq  0.995$ are shown in Fig.~\ref{ak:fig:monopoles}. 

\subsubsection{Nuclearites}

Nuclearites are hypothetical massive stable particles made of lumps of up, down and strange quarks~\cite{bibak:nuclearites}. They could be present in the cosmic radiation, either as relics of the early Universe or as debris of supernovae or strange star collisions. They could be detected in a  neutrino telescope through the blackbody radiation
emitted by the expanding thermal shock wave along their path. \\
A dedicated search was performed with data collected in 2007 and 2008. The study was optimized for non relativistic nuclearites. Assuming a velocity $\beta=10^{-3}$ outside the atmosphere, the nuclearite should have a mass larger than $\sim10^{22}$ GeV in order to cross the Earth. Since the nuclearite flux is expected to decrease with
the nuclearite mass, only downgoing nuclearites were considered in the analysis. As for monopoles, the light yield in the detector depends on the nuclearite mass. It is expected to be much greater than for relativistic muons for nuclearite masses larger than few $10^{13}$~GeV. After unblinding, no significant excess of nuclearite-like events was found. Upper limits on the flux of downgoing nuclearites were established between $7.1 \times 10^{-17}$ and $6.7 \times 10^{-18} {\rm cm^{-2} s^{-1} sr^{-1}}$,
for the mass range $10^{14} \leq M_N \leq 10^{17}$~GeV.
\begin{figure}[ht]
  \centering
  \includegraphics[width=0.5\textwidth,clip]{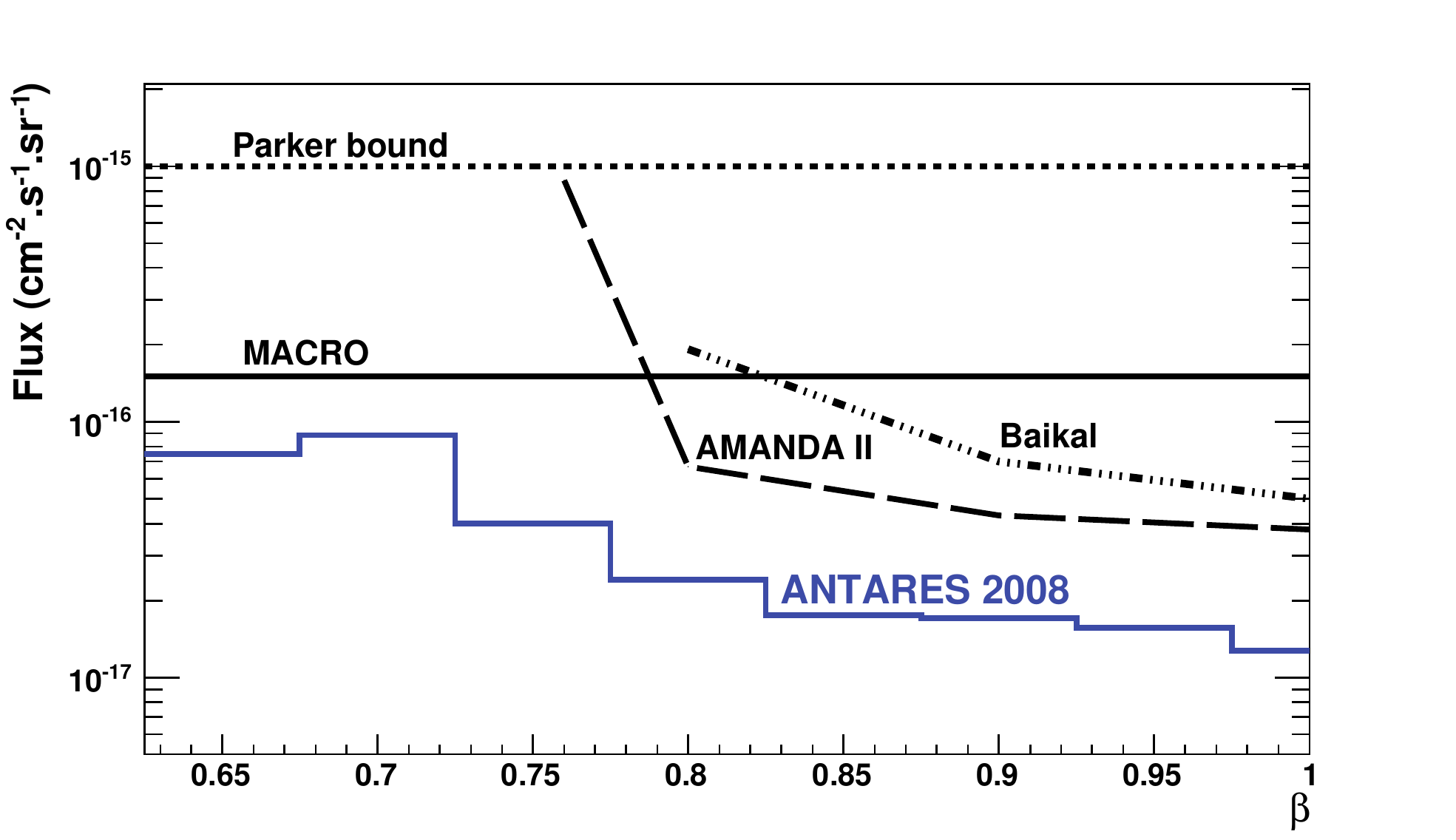}
  \caption{The {\sc antares} 90\% C.L. upper limit on an upgoing magnetic monopole flux for relativistic velocities 0.625  $\leq \beta \leq$  0.995 obtained in this analysis, compared to the theoretical Parker bound~\cite{bibak:parker} and previous results from other experiments.}
  \label{ak:fig:monopoles}
 \end{figure}

\section{Conclusions}

The {\sc antares} neutrino telescope is the first and only deep sea neutrino telescope currently in operation. It has been continuously monitoring the Southern Sky with unprecedented sensitivity (in particular to the TeV sources of the central region of our Galaxy) since the deployment of the first detection lines in 2007. \\
Besides atmospheric neutrino studies, several searches for neutrinos of astrophysical origin have been performed, including a rich multi-messenger program. \\
Beyond astrophysics, competitive searches for dark matter and even more exotic particles have been carried out. All the studies presented in the proceedings will be updated with the data remaining to be analysed and/or recorded. The data acquisition is foreseen at least until the end of 2016, when the detector gets eventually superseded by the next generation {\sc km3net} detector~\cite{bibak:km3net}, 
currently being built in the Mediterranean Sea.

\setcounter{figure}{0}
\setcounter{table}{0}
\setcounter{footnote}{0}
\setcounter{section}{0}
\setcounter{equation}{0}

\newpage
\addcontentsline{toc}{part}{{\sc Search for Point Sources and Diffuse Fluxes}%
\vspace{-0.5cm}
}
\id{id_schulte}
\addcontentsline{toc}{part}{\arabic{IdContrib} - {\sl Stephan Schulte} : Update on the {\sc Antares} full-sky neutrino point source search%
\vspace{-0.5cm}
}

%
\lefthyphenmin=1 
\righthyphenmin=1

\title{\arabic{IdContrib} - Update on the ANTARES full-sky neutrino point source search}

\shorttitle{\arabic{IdContrib} - ANTARES full-sky neutrino point source search}

\authors{
Stephan Schulte$^{1}$ for the ANTARES Collaboration.
}

\afiliations{
$^1$ FOM Instituut voor Subatomaire Fysica Nikhef, Science Park 105, 1098 XG Amsterdam, The Netherlands \\
}

\email{stephans@nikhef.nl}

\abstract{
Identifying the sources
of galactic and intergalactic high energy neutrinos is
one of the main goals of the ANTARES experiment.
For several galactic supernova remnants, which have been confirmed by the Fermi satellite as sources of cosmic rays,
ANTARES is the most sensitive detector currently probing neutrino emission
in the relevant sub-PeV energy range.
The most recent results of a full-sky search will be presented as
well as results for a number of preselected gamma-ray sources
of interest. In addition, results for a specific model describing two extended sources, RX 1713 and Vela X, are discussed.}


\maketitle

\section{Introduction}

Cosmic rays (CRs) are known for more than 100 years, since their discovery by Victor Hess.  
Although progress has been made recently in measuring the flux with very high precision up to the highest energies as well as determining the composition of the CRs, their origin where these particles are accelerated up to energies of $10^{18}$ eV or even higher, remains unknown.  
Among existing models of possible acceleration sites one can find galactic and extra-galactic objects, e.g. supernova remnants or active galactic nuclei \cite{bibschulte:becker}. 
However, the identification of the true origin becomes challenging due to magnetic fields deflecting the CRs on their path from the source to Earth causing the information about their origin to be lost.
In this context, high energy neutrinos as neutral particles offer a unique opportunity to identify and study the most violent objects in the universe due to the fact that they can  be produced through hadronic interactions at the same acceleration site.

\subsection{The ANTARES detector}

The ANTARES detector \cite{bibschulte:Antoine} is an underwater neutrino telescope in the Mediterranean Sea ($42^{\circ} 48 N, 6^{\circ} 10E$) based 40 km south of Toulon at a depth of 2475 m. 
Twelve vertical lines are forming the basic detector structure.
They are kept taut by a buoy attached to their tops. 
Each line consists of 25 detection stories, which are 14.5 m apart, equipped with three downward looking 10-inch photomultiplier tubes (PMTs) with an angle of $45^{\circ}$ towards the axis of the line.
The average distance among the lines is between 60 and 70~m. 
Neutrinos are detected through their interaction with the detector surrounding sea water and rock creating charged particles which then induce Cherenkov radiation.
These photons are detected by the PMTs with the corresponding time stamps and charges.  
This information is then digitized into 'hits' \cite{bibschulte:Electronics} and send to the shore station. 

\section{Data Selection}

The data set covers the period from the 31st of January 2007 until the 31st of December 2012.
During the first two years the detector was still being constructed, i.e. in 2007 only 5 lines were in operation and in 2008 9, 10 and finally 12 were taking data.   
After applying data quality cuts, we end up with a total lifetime of the detector of 1338.98 days whereupon 183 days correspond to the 5 line period. \\
All recorded events are reconstructed using the time and position information of the corresponding hits by means of a modified maximum likelihood method (MLL) \cite{bibschulte:Aart}. 
This algorithm consists of a multi-step fitting procedure to optimize the direction of the reconstructed muon by maximizing the MLL-parameter $\Lambda$.
The corresponding distributions for only upward-going reconstructed tracks are shown in Fig.\ref{schulte:fig:quality_reco} including the Monte Carlo predicted contributions of atmospheric neutrinos according to the Bartol flux \cite{bibschulte:Bartol}, of misreconstructed muons and of the data.  
The neutrinos and muons are simulated using the GENHEN and the MUPAGE-package \cite{bibschulte:MUPAGE}, respectively. \\
We select neutrino candidates by applying three cuts after the reconstruction, namely $\Lambda > -5.2$ which is a measure of the reconstruction quality, $cos\theta < 0.1$, the zenith angle, to select only upward going tracks and $\beta < 1^{\circ}$, the estimated angular resolution. 
The last cut reduces significantly the amount of misreconstructed atmospheric muons in the data sample. 
In total, we end up with 5516 neutrino candidates of which ~90 \% should be atmospheric neutrinos according to the MC estimation and the rest is composed of misreconstructed muons. 

 \begin{figure}[h]
 \centering
 \includegraphics[width=8.3cm]{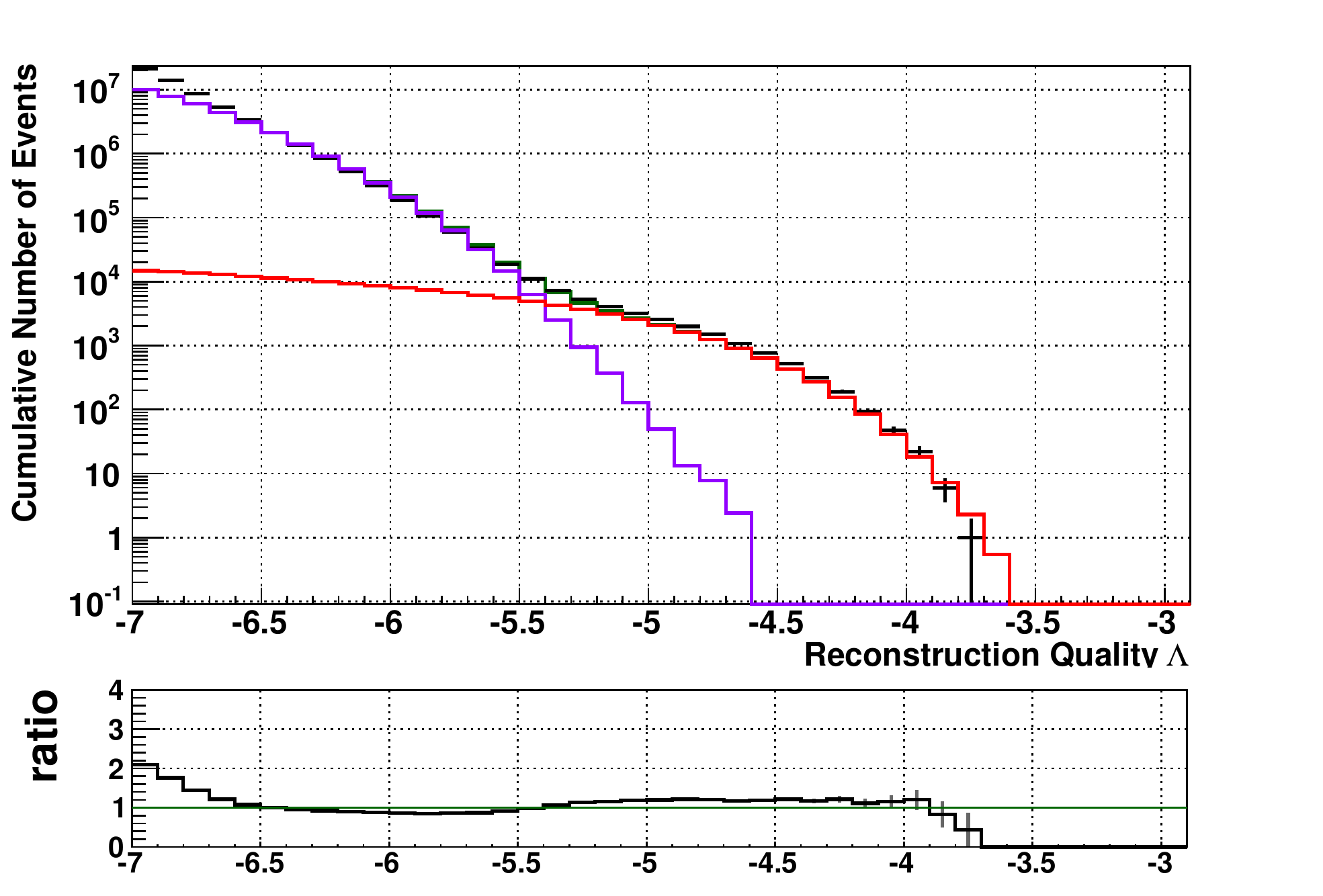}
 \caption{Cumulative distribution of the number of events depending on the quality parameter $\Lambda$. The purple line shows the contribution of the muon MC, while the red one corresponds to the atmospheric neutrino MC. The bottom plot gives the ratio between the data and the sum of both MCs.}
 \label{schulte:fig:quality_reco}
 \end{figure}

\section{Detector Performance}

By applying cuts mentioned above to the MC data sets, we calculated the corresponding angular resolution and acceptance of the detector for a typical $E^{-2}$ signal flux. 

\subsection{Angular Resolution}
\label{schulte:sec:ang_res}
Previously \cite{bibschulte:ICRC2011}, it has been reported that an ad hoc 2 ns smearing was necessary to account for the unknown time transit spread (TTS) within the PMTs.
Recently, the timing accuracy of the PMTs has been revised and a more accurate description of the TTS has been obtained which has a much sharper peak but also some long non-Gaussian tails.
This leads to an improvement of the median from 0.46 to 0.40 degrees. 
The corresponding cumulative distribution of the angular resolution can be found in Fig. \ref{schulte:fig:reco_angle}.

 \begin{figure}[h]
 \centering
 \includegraphics[width=7.5cm]{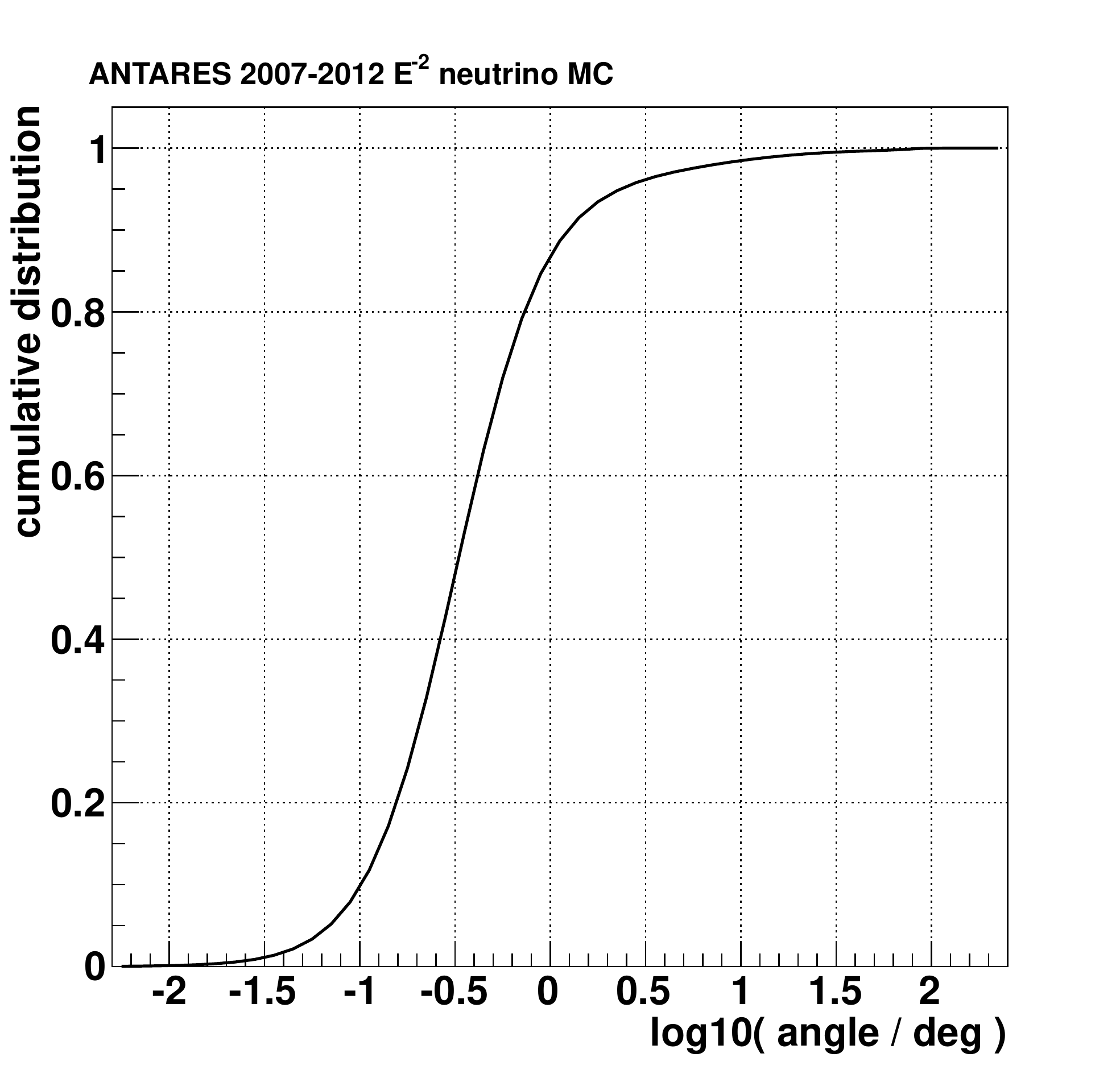}
 \caption{Cumulative distribution of the angle between the reconstructed muon direction and the true neutrino direction for simulated upward going neutrinos that pass the cuts described in Section \ref{schulte:sec:ang_res} assuming a $E_{\nu}^{-2}$ neutrino spectrum.}
 \label{schulte:fig:reco_angle}
 \end{figure}

\subsection{Acceptance}
The MC data sets were also used to obtain an estimate for the acceptance. 
The discussed flux has the form
\begin{eqnarray}
\frac{dN}{dE} & = & \Phi \left(\frac{E_{\nu}}{GeV}\right)^{-2} GeV^{-2} s^{-1} cm^{-2},
\end{eqnarray}
with $\Phi$ the flux normalization.
The acceptance is the proportionality factor between a given flux and corresponding number of signal events expected in the detector. 
For the whole data period this comes down to 1.49 (0.87) $\times$ 10$^8$ GeV cm$^2$ s for a declination $\delta$ = -90$^{\circ}$ (0$^{\circ}$)(see Fig.\ref{schulte:fig:acceptance} for an E$^{-2}$ signal neutrino flux).

\begin{figure}[h]
 \centering
 \includegraphics[width=7.cm]{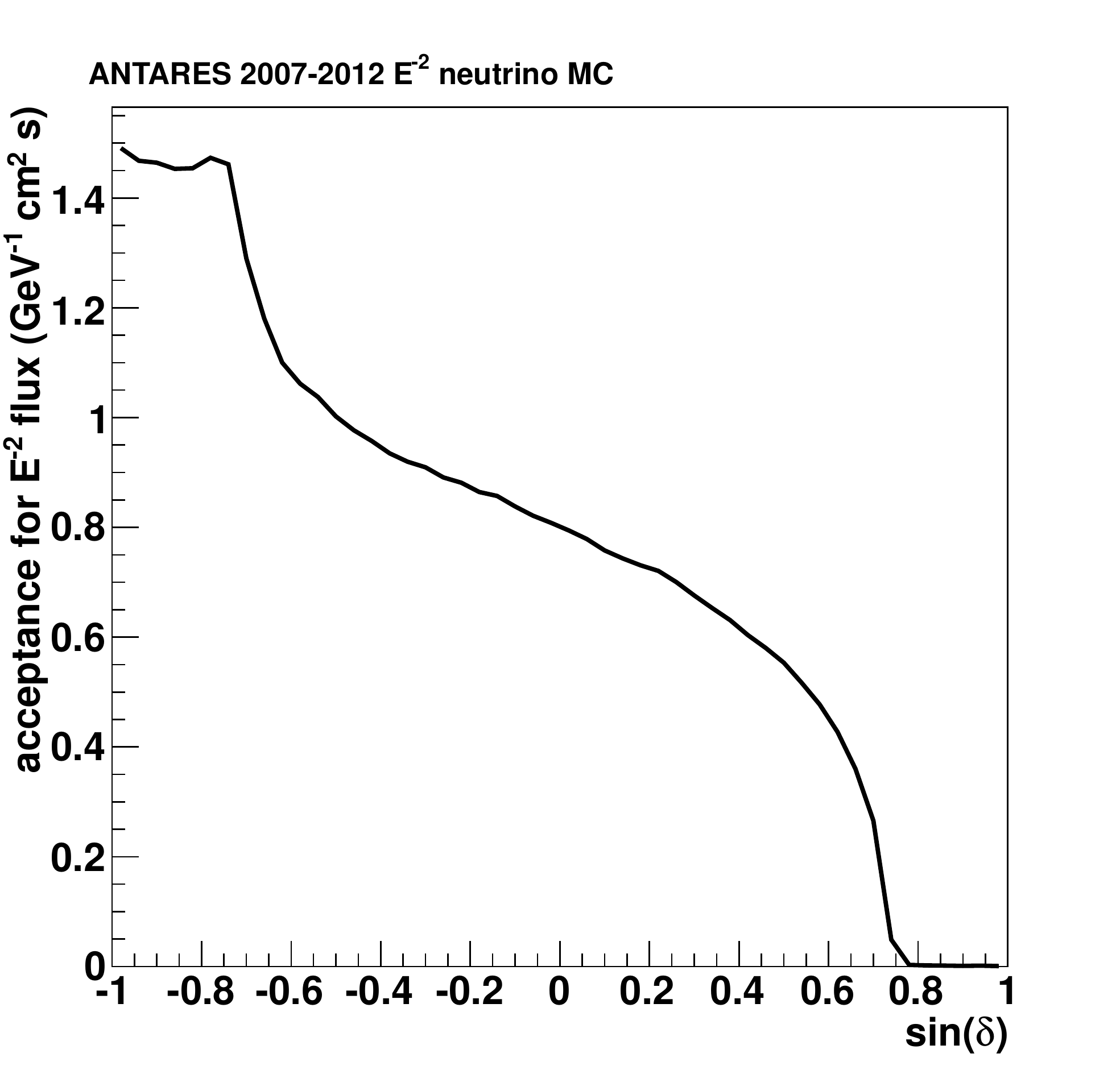}
 \caption{Acceptance, i.e. the constant of proportionality between the normalisation factor for an $E^{-2}$ flux and the selected number of events. }
 \label{schulte:fig:acceptance}
 \end{figure}

\section{Search Method}
The algorithm used on the analysis is based on the likelihood of the observed events which is defined as:
\begin{eqnarray*}
  \log \cal L_{\rm s+b} &=& \sum_i \log (\frac{n_s}{N} \frac{1}{2\pi \beta_i} e^{-\frac{\left| x_i - x_s \right|^2}{2\beta_i^2}}  {\cal N}(N_{hits}^{i,sig})\\ 
  &+& \left( 1 - \frac{n_s}{N} \right) 
  \frac{F(\delta_i)}{2\pi}  {\cal N}(N_{hits}^{i,bkg}))
\end{eqnarray*}
\noindent where the sum is over the events,  $\vec{x_i} = (\alpha, \delta)$ is the corresponding direction of event i which is weighted with a gaussian distribution around the possible source direction $\vec{x_s}$ with $\beta_i$ being the corresponding angular error.
The parameter $n_s$ represents the number of signal events for a particular source over the total number of events $N$ in the sample;$F(\delta_i)$ is a parametrization of the background rate, obtained from the observed declination distribution of the events and ${\cal N}(N_{hits}^{i})$ is the probability for an event i to be reconstructed with $N_{hits}$ number of hits.
In the case of the full sky search, the $\vec{x_s}$ as well as the number of signal events are varied to find the optimal combination which maximizes $\log {\cal L}_{\rm s+b} $.
During the second approach, a candidate list of possible sources is composed and fed to the algorithm in terms of the known position of the objects.
Thus, only the number of signal events needs to be optimized.
The test statistic (TS) of our analysis is defined as 
\begin{eqnarray*}
TS = \log \cal L_{\rm s+b} - \cal L_{\rm b} 
\end{eqnarray*}
where $ \log {\cal L}_{\rm b}$ represents the background only hypothesis ($n_s$ = 0). 
With increasing difference between the two functions, the dataset becomes more signal-like.

\section{Results}
As mentioned above, two different kinds of analyses have been performed, on the one hand a full-sky search for a signal-like excess anywhere in the field of view of the ANTARES telescope and on the other hand a fixed search, considering an \emph{a priori} defined list of promising cosmic neutrino sources.  
In the following the findings of both analyses are reported.
\subsection{Full sky search}
In the whole data set, the most significant cluster has been found at $\alpha_i, \delta_i$ = (-47.8$^{\circ}$, -64.9$^{\circ}$) containing 14 events.
This corresponds to a fitted number of signal events of 6.3 in addition to the expected background in that direction. 
Inserting the obtained values for the fitting parameters into the test statistics yields a value for $TS$ of 14 which translates into a probability of 2.1 \%. 
It is worth noticing that this is the same cluster as two years ago plus 6 more events.
A skymap with the position of this cluster and all assigned events is shown in Fig. \ref{schulte:fig:skymap}.

\subsection{Candidate list search}
In total 50 possible neutrino sources have been selected which can be found in Table \ref{schulte:table:lims} including the results of the fixed search. 
None of these shows an excess of clustered events exceeding significantly the number of expected background events . 
The smallest post-trial p-value belongs to the astrophysical object HESS J0632+057X, namely 7.3 \%.
Based on these results, Fig. \ref{schulte:fig:lims} shows the limit on the corresponding cosmic neutrino flux for all objects in the candidate list depending on the declination.
In addition, the sensitivity of the ANTARES detector is presented for this time period which is 25 \% below the results reported in \cite{bibschulte:ICRC2011} as well as the results from other experiments.
To put the different limits in the correct perspective, it is important to mention that under the assumption of an E$^{-2}$-flux the IceCube experiment is mostly sensitive in the PeV energy range in the southern hemisphere \cite{bibschulte:Icecube} while ANTARES is in fact in the TeV regime.

 \begin{figure}[h]
 \centering
 \includegraphics[width=8.0cm]{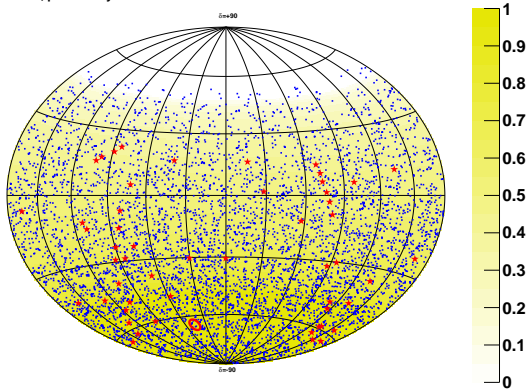}
 \caption{Equatorial skymap showing the 5516 data events. The position of the most signal-like cluster is indicated by the circle. The stars denote the position of the 50 candidate sources.}
 \label{schulte:fig:skymap}
 \end{figure}
 
 \begin{figure}[h]
 \centering
 \includegraphics[width=8cm]{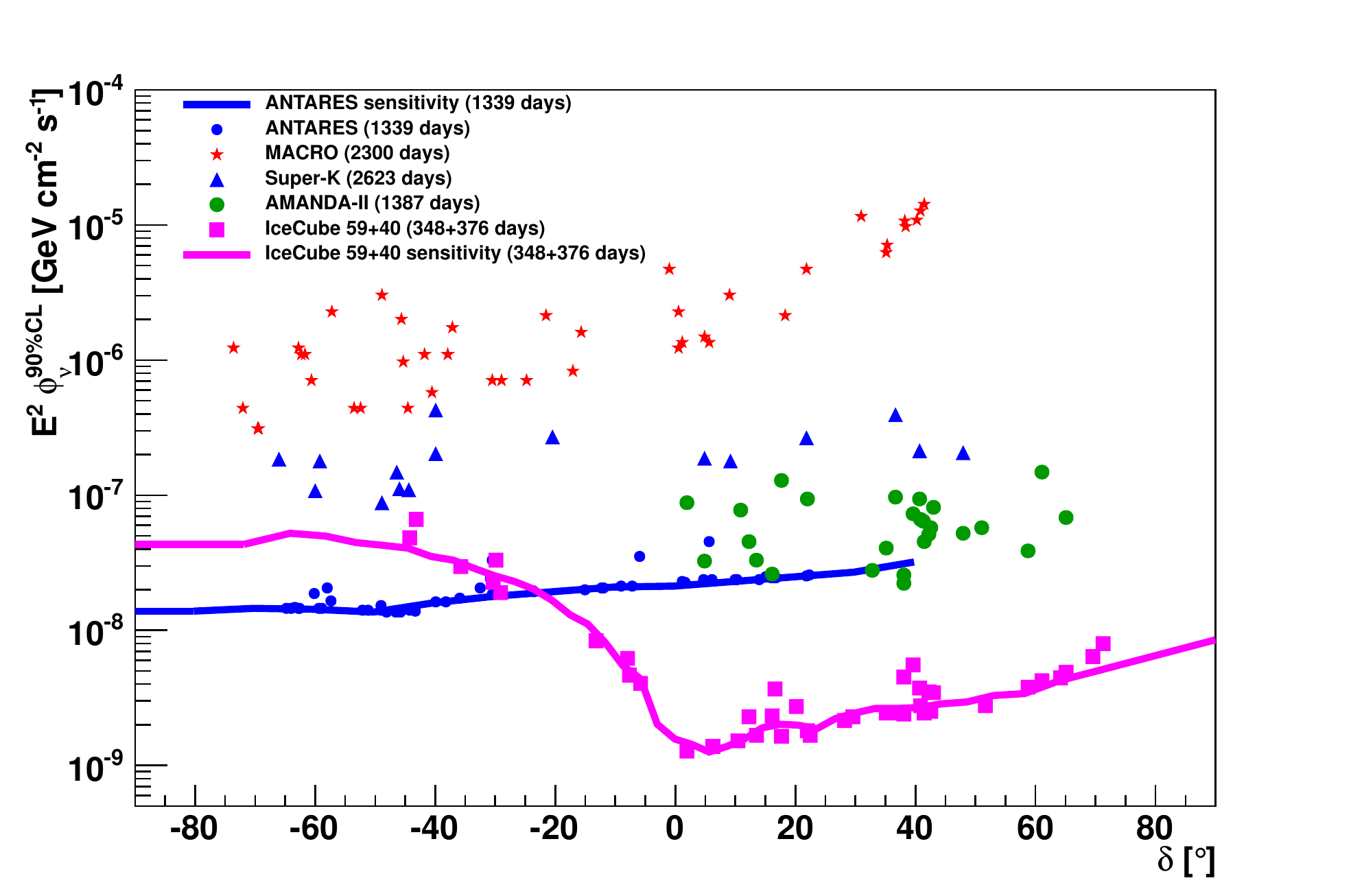}
 \caption{Limits set on the normalisation $\phi$ of an $E^{-2}_{\nu}$ spectrum of high energy
         neutrinos from selected candidates (see Table \ref{schulte:table:lims}). Also shown is the
         sensitivity, which is defined as the median expected limit.
         In addition to the present result, several previously published limits on sources in both
         the Southern and Northern sky are also shown.}
 \label{schulte:fig:lims}
 \end{figure}

\section{Source Morphology}

Until now, all sources were assumed to be point like.
However, several of those included in the candidate source list are in fact not point like but do have an extended structure which can be resolved by ANTARES.
Two of them are the SNR RXJ1713.7-3946 (RXJ) and the pulsar wind nebula VelaX.
Both have been proposed to in part hadronic acceleration processes.
Based on the gamma ray spectrum measured by H.E.S.S., Kappes et al. \cite{bibschulte:Kappes_2006fg} estimated the expected neutrino flux by assuming a gaussian distribution to model the extended sources. 
The general description of the flux is as follows
\begin{eqnarray*}
\frac{dN}{dE} = \Phi \times 10^{-15} \left[ \frac{E}{TeV}\right]^{-\gamma_{\nu}} exp( \sqrt{E/E_{cut}}) GeV^{-1} s^{-1} cm^{-2}
\end{eqnarray*}
with $\Phi$ the flux normalization being 16.8 (11.75) for RXJ (VelaX), $\gamma_{\nu}$ the spectral index of 1.72 (0.98) and $E_{cut}$ the cut-off energy of 2.1 (0.84) TeV.
Assuming these models, 90 \% CL limits on the flux normalization and the corresponding model rejection factor (MRF) were computed for both sources which are 6.4 and 9.7, respectively.
The results of point source and the morphology study are presented in Fig. \ref{schulte:fig:extended}.
Compared the latest ANTARES publication \cite{bibschulte:AdrianMartinez_2012rp}, the MRF for the RXJ studied decreased from 8.8. 
However, the MRF for Vela X increased from 9.1. 
This is due to the fact, that during our last analysis only one event close by was found allowing us to set a more stringent limit.
With the 2007 - 2012 data set we obtained 4 additional events, although only 60 \% more data were added.
This led to a slightly worse limit.

 \begin{figure}[h]
 \centering
 \includegraphics[width=8.5cm]{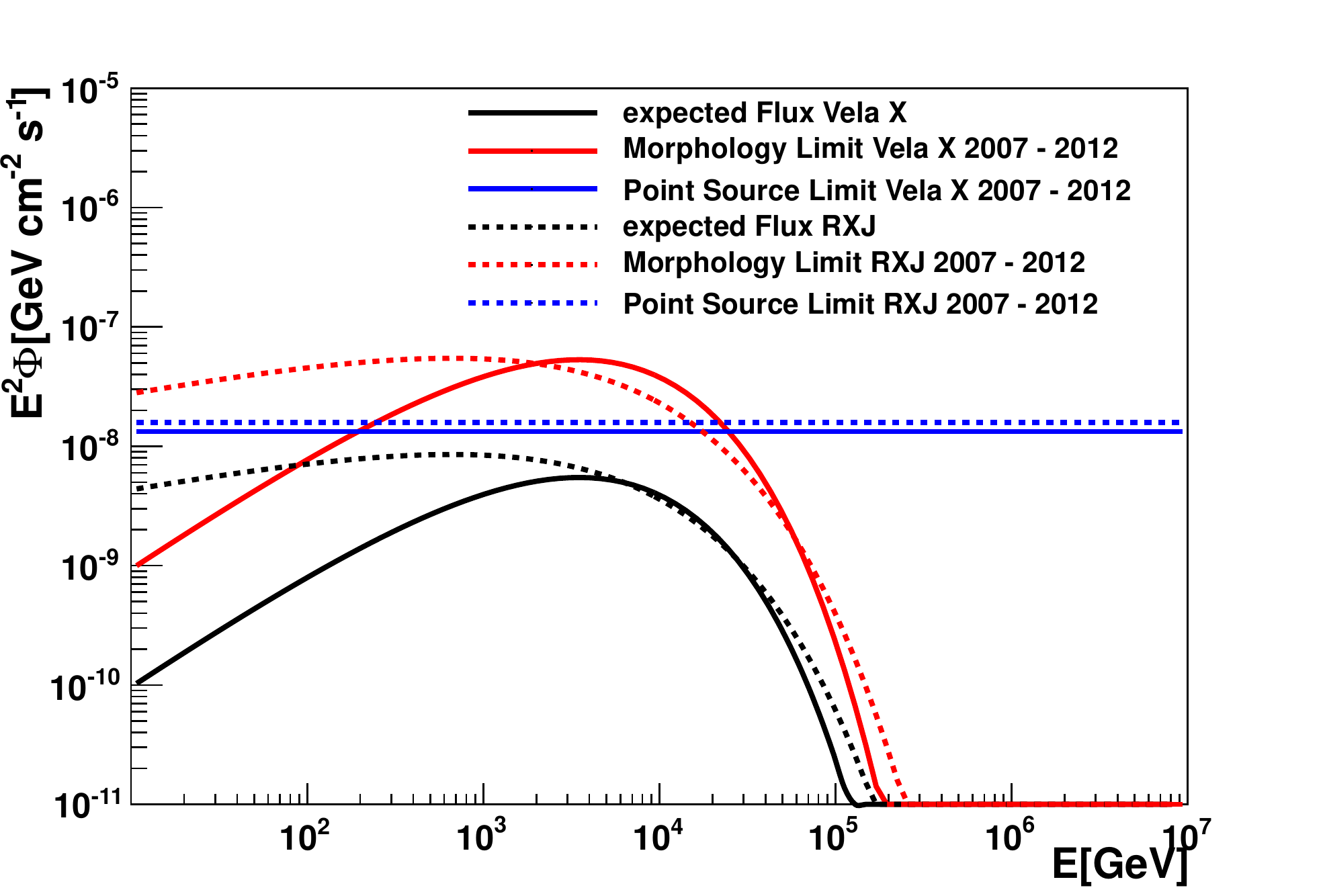}
 \caption{Here we present the results of the Morphology Study for RXJ (dashed) and Vela X (straight). The neutrino flux models are shown in black, the 90 \% CL limits for point like sources in red and the $E^{-2}$ point source limit presented in Tab. \ref{schulte:table:lims}.}
 \label{schulte:fig:extended}
 \end{figure}

\section{Conclusions}

The ANTARES detector has been used to search for high energy cosmic neutrinos.
For this purpose, 5 years of data taking were available whereupon during most of the first year only five lines were deployed and afterwards 9, 10 and 12 lines for the rest of the period. 
According to the corresponding MonteCarlo production, an angular resolution of 0.40 degrees is achieved. 
In both, the full-sky as well as the candidate list search, no excess above the expected background has been found. 
Thus, limits have been calculated on the cosmic neutrino flux. 
Finally, specific models of the two extended sources RXJ1713 and VelaX were discussed and the corresponding MRFs have been presented.

\begin{table}
\begin{tabular}{lr@{.}lr@{.}l@{\hspace{0.18cm}}ll}
\hline
         source
         & \multicolumn{2}{c}{$\alpha_s(^{\circ})$ }
         & \multicolumn{2}{c}{$\delta_s(^{\circ})$ }
         & \multicolumn{1}{c}{$p$}
         & \multicolumn{1}{c}{$\phi^{90\% \rm CL}$} \\
\hline
HESSJ0632+057   	& 98&24         & 5&81  & 0.07  & 4.40 \\
HESSJ1741-302   	& 265&25        & -30&20        & 0.14  & 3.23 \\
3C279   			& 194&05        & -5&79         & 0.39  & 3.45 \\
HESSJ1023-575   & 155&83        & -57&76        & 0.82  & 2.01 \\
ESO139-G12      & 264&41        & -59&94        & 0.95  & 1.82 \\
CirX-1          & 230&17        & -57&17        & 1.00  & 1.62 \\
PKS0548-322     & 87&67         & -32&27        & 1.00  & 2.00 \\
GX339-4         & 255&70        & -48&79        & 1.00  & 1.50 \\
VERJ0648+152    & 102&20        & 15&27         & 1.00  & 2.45 \\
PKS0537-441     & 84&71         & -44&08        & 1.00  & 1.37 \\
MGROJ1908+06    & 286&99        & 6&27  & 1.00  & 2.32 \\
Crab    & 83&63         & 22&01         & 1.00  & 2.46 \\
HESSJ1614-518   & 243&58        & -51&82        & 1.00  & 1.39 \\
HESSJ1837-069   & 279&41        & -6&95         & 1.00  & 2.09 \\
PKS0235+164     & 39&66         & 16&61         & 1.00  & 2.39 \\
Geminga         & 98&31         & 17&01         & 1.00  & 2.39 \\
PKS0727-11      & 112&58        & -11&70        & 1.00  & 2.01 \\
PKS2005-489     & 302&37        & -48&82        & 1.00  & 1.39 \\
PSRB1259-63     & 195&70        & -63&83        & 1.00  & 1.41 \\
HESSJ1503-582   & 226&46        & -58&74        & 1.00  & 1.41 \\
PKS0454-234     & 74&27         & -23&43        & 1.00  & 1.92 \\
PKS1454-354     & 224&36        & -35&67        & 1.00  & 1.70 \\
HESSJ1834-087   & 278&69        & -8&76         & 1.00  & 2.06 \\
HESSJ1616-508   & 243&97        & -50&97        & 1.00  & 1.39 \\
H2356-309       & 359&78        & -30&63        & 1.00  & 2.35 \\
HESSJ1912+101   & 288&21        & 10&15         & 1.00  & 2.31 \\
PKS0426-380     & 67&17         & -37&93        & 1.00  & 1.59 \\
W28     & 270&43        & -23&34        & 1.00  & 1.89 \\
MSH15-52        & 228&53        & -59&16        & 1.00  & 1.41 \\
RGBJ0152+017    & 28&17         & 1&79  & 1.00  & 2.19 \\
W51C    & 290&75        & 14&19         & 1.00  & 2.32 \\
PKS1502+106     & 226&10        & 10&52         & 1.00  & 2.31 \\
HESSJ1632-478   & 248&04        & -47&82        & 1.00  & 1.33 \\
HESSJ1356-645   & 209&00        & -64&50        & 1.00  & 1.42 \\
1ES1101-232     & 165&91        & -23&49        & 1.00  & 1.92 \\
HESSJ1507-622   & 226&72        & -62&34        & 1.00  & 1.41 \\
RXJ0852.0-4622          & 133&00        & -46&37        & 1.00  & 1.33 \\
RCW86   & 220&68        & -62&48        & 1.00  & 1.41 \\
RXJ1713.7-3946          & 258&25        & -39&75        & 1.00  & 1.59 \\
SS433   & 287&96        & 4&98  & 1.00  & 2.32 \\
1ES0347-121     & 57&35         & -11&99        & 1.00  & 2.01 \\
VelaX   & 128&75        & -45&60        & 1.00  & 1.33 \\
HESSJ1303-631   & 195&77        & -63&20        & 1.00  & 1.43 \\
LS5039          & 276&56        & -14&83        & 1.00  & 1.96 \\
PKS2155-304     & 329&72        & -30&22        & 1.00  & 1.79 \\
GalacticCenter          & 266&42        & -29&01        & 1.00  & 1.85 \\
CentaurusA      & 201&36        & -43&02        & 1.00  & 1.36 \\
W44     & 284&04        & 1&38  & 1.00  & 2.23 \\
IC443   & 94&21         & 22&51         & 1.00  & 2.50 \\
3C454.3         & 343&50        & 16&15         & 1.00  & 2.39 \\
\hline
\end{tabular}
\caption{Results of the candidate source search. The source coordinates and the p-values ($p$) are shown as well as the
         limits on the flux intensity $\phi^{90\% \rm CL}$; the latter has units $10^{-8} \rm GeV^{-1} cm^{-2} s^{-1}$.}
\label{schulte:table:lims}
\end{table}
\setcounter{figure}{0}
\setcounter{table}{0}
\setcounter{footnote}{0}
\setcounter{section}{0}
\setcounter{equation}{0}

\newpage
\id{id_schussler1}

%


\title{\arabic{IdContrib} - Energy reconstruction in neutrino telescopes}
\addcontentsline{toc}{part}{\arabic{IdContrib} - {\sl Fabian Sch\"ussler} : Energy reconstruction in neutrino telescopes%
\vspace{-0.5cm}
}

\shorttitle{\arabic{IdContrib} - Energy reconstruction in neutrino telescopes}

\authors{
Fabian Sch\"ussler$^{1}$ for the ANTARES Collaboration.
}

\afiliations{
$^1$ Commissariat \`a l'\'energie atomique et aux \'energies alternatives / Institut de recherche sur les lois fondamentales de l'Univers \\
}

\email{fabian.schussler@cea.fr}

\abstract{The energy is one of the most important parameters to discriminate between atmospheric and astrophysical events recorded by neutrino telescopes like ANTARES and IceCube. Here we introduce and describe a method to reconstruct the energy deposited by muons along their track, $\mathrm{d}E/\mathrm{d}X$, while crossing the fiducial volume of such a detector. Exploiting the close correlation between the energy deposit and the energy of charged particles above a few hundred GeV we use the reconstructed $\mathrm{d}E/\mathrm{d}X$ to derive the energies of the incident muon and the primary neutrino. We describe the basic ideas behind the algorithm and, applied to the ANTARES neutrino telescope, quantify its performance and discuss systematic uncertainties using both data and detailed Monte Carlo simulations.}


\maketitle

\section{Introduction}
The detection of astrophysical neutrinos and the identification of their sources is one of the main aims of large neutrino telescopes operating at the South Pole (IceCube), in Lake Baikal and in the Mediterranean Sea (ANTARES). The vast majority of the neutrino candidates recorded by these experiments are of atmospheric origin. To discriminate and select events of potential astrophysical origin, the energy of the events is the prime parameter. It is expected that the astrophysical neutrino flux follows a harder spectrum (typically described by an $E^{-2}$ energy dependence), whereas the atmospheric flux is falling more rapidly with increasing energy ($E^{-3.7}$ in the energy range typically accessible with current neutrino telescopes~\cite{bibschussler1:Bartol}).

Here we introduce and describe an algorithm to reconstruct the energy of both the muon traversing the detector and the primary neutrino. The algorithm has been developed within the ANTARES collaboration~\cite{bibschussler1:Antares_DetectorPaper} and its use in several analyses lead to significant increase of their sensitivities~\cite{bibschussler1:ICRC2013_Spectrum, bibschussler1:Moriond2013_DiffuseFlux}. The underlying principles are nevertheless valid for all neutrino telescopes and similiar algorithms are being developed for example within the IceCube Collaboration~\cite{bibschussler1:IceCube_dEdX}.

The different neutrino interaction modes lead to different experimental signatures in neutrino telescopes. The signature of neutral current interactions are particle showers, i.e. very localized energy deposits and light emission. The rate of these events is limited by the available instrumented volume which acts as interaction volume. On the other hand their energy reconstruction is possible with rather good precision as they are usually fully contained. Muons emerging from charged current interactions of muon neutrinos provide the bulk of the neutrino induced data of ANTARES and other neutrino telescopes due to the extension of the fiducial volume beyond the instrumented volume. Whereas the direction of the muon track can be reconstructed with good precision, the reconstruction of its energy however is, due to the intrinsic fluctuations of the energy deposited within the detector volume, less obvious and the subject of this paper.

The fundamental idea behind the presented algorithm is to exploit the correlation between the energy of a charged particle in a medium and its energy loss. The latter is deposited along the muon track and can be denoted as energy deposit $\mathrm{d}E$ per tracklength $\mathrm{d}X$. At energies above the critical energy of a few hundred GeV, energy losses due to Bremsstrahlung become more important with respect to ionisation losses and a clear correlation between $\mathrm{d}E/\mathrm{d}X$ and the particle energy can be expected. If a significant amount of this energy deposit happens within or close to the instrumented volume of a neutrino telescope it can be detected via the recording of the emitted light along the muon track. One will then be able to reconstruct a measure of the (local) energy loss by dividing the measured amount of energy deposit by the reconstructed length of the track within the fiducial volume. Detailed Monte Carlo simulations are then used to exploit the discussed correlations to 
estimate the energy of the muon and the incident neutrino.\\

\section{The $\mathrm{d}E/\mathrm{d}X$ energy estimator}
\begin{figure*}[!t]
\centerline{
  \includegraphics[width=0.41\textwidth]{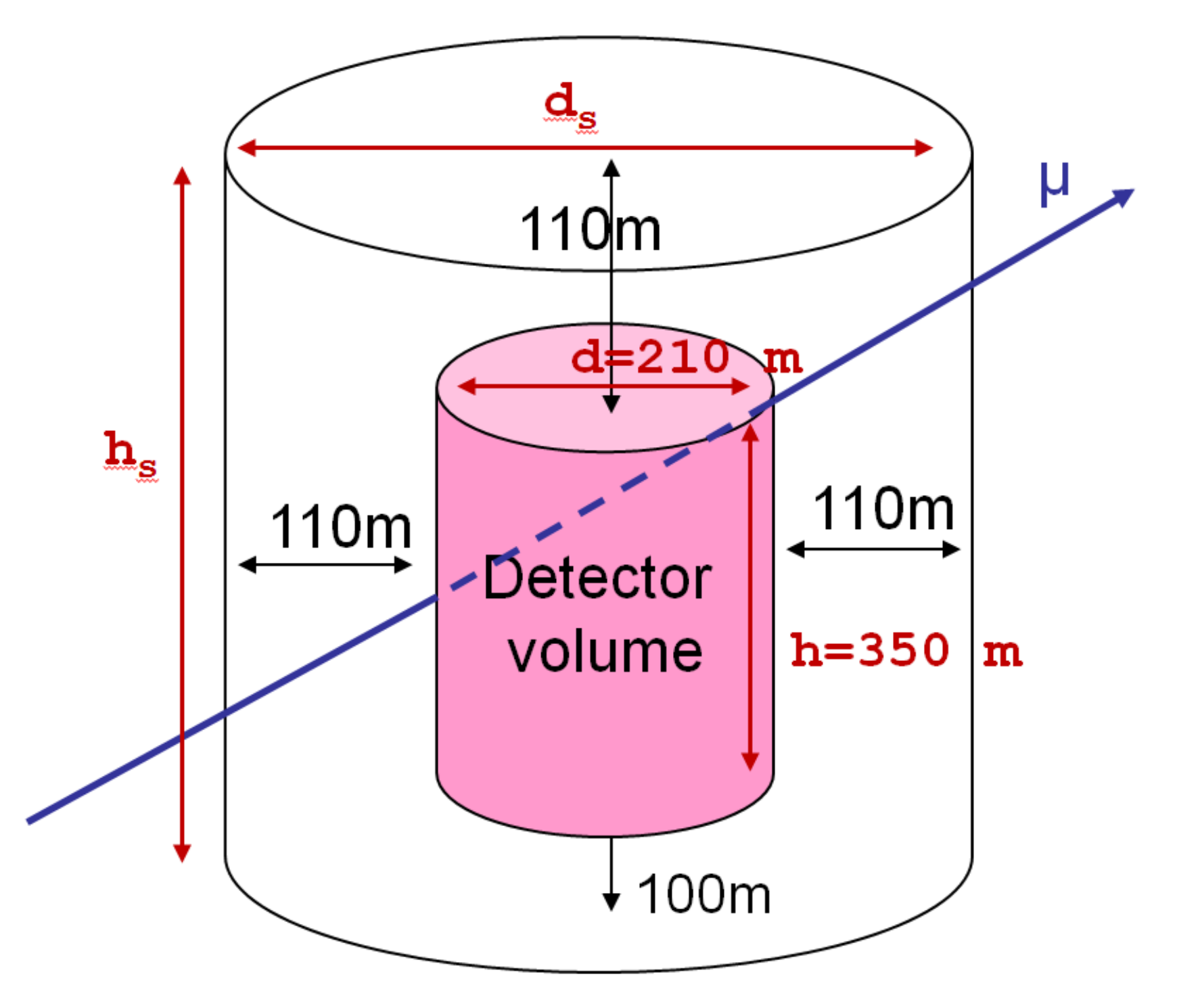}
\hfill
   \includegraphics[width=0.48\textwidth]{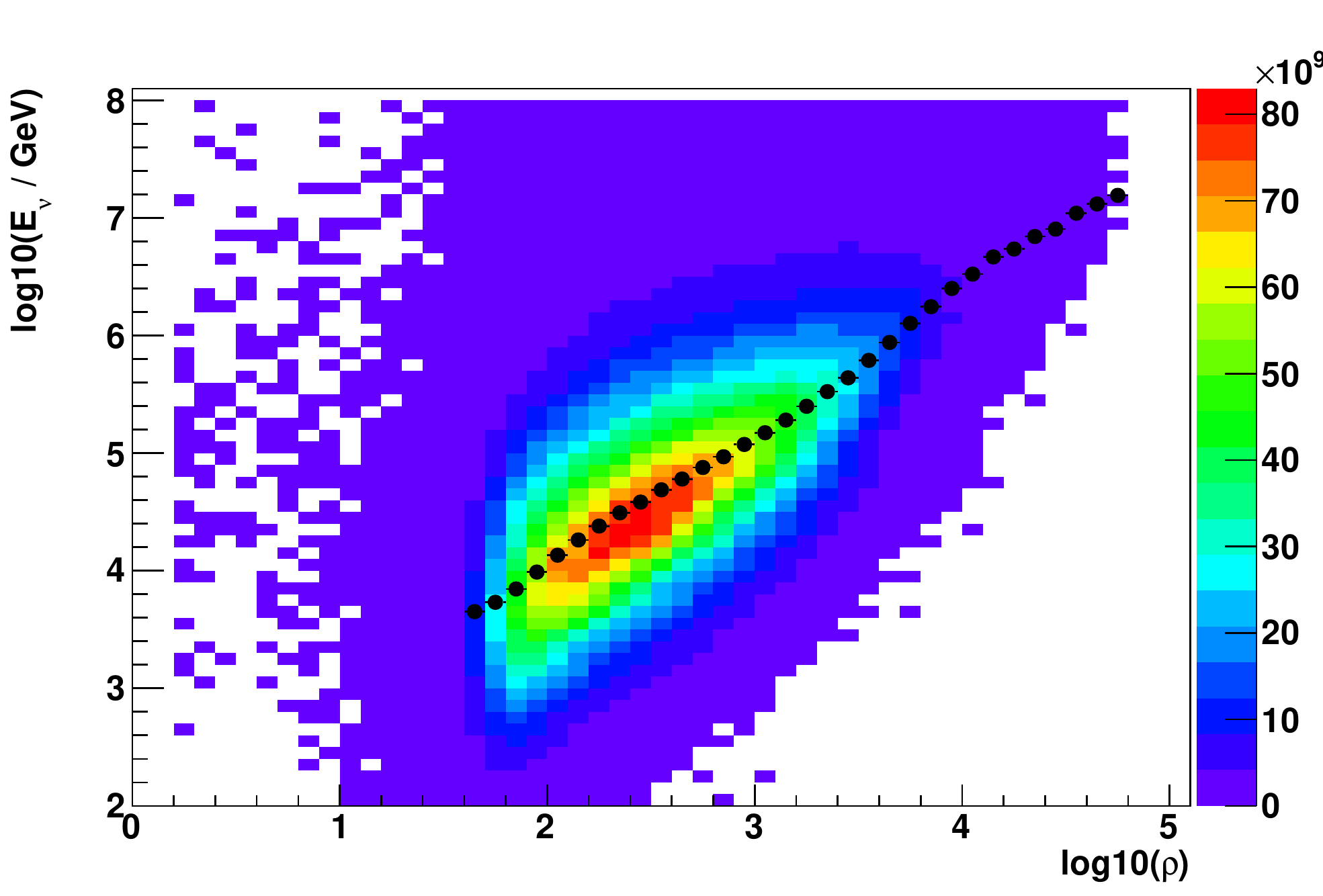}
   }
  \caption{Left plot: Definition of the fiducial volume used to calculate the length of the muon track $L_\mu$. The total size of the volume is given by $d_s = 430~\mathrm{m}$ and $h_s = 560~\mathrm{m}$. Right plot: The correlation between the reconstructed $\mathrm{d}E/\mathrm{d}X$ and the true energy is used to calibrate the energy estimator. The black markers denotes the derived calibration table, i.e. the average true energy per $\mathrm{d}E/\mathrm{d}X$ bin.}
     \label{figschussler1:volumecalibration}
 \end{figure*}
\noindent
   
\subsection{$\mathrm{d}E/\mathrm{d}X$ estimation}
We approximate the total muon energy deposit $\mathrm{d}E/\mathrm{d}X$ by an estimator $\rho$ which can be derived on an event-by-event basis from quantities measured by the ANTARES detector:
\begin{equation}\label{eqschussler1:rho}
\mathrm{d}E/
\mathrm{d}X \approx \rho =  \frac{\sum\limits^{\mathrm{nHits}} Q_i}{\epsilon(\vec{x})} \cdot \frac{1}{L_\mu(\vec{x})}
\end{equation}
$\epsilon(\vec{x})$ is the light detection efficiency and will be described in detailed below. $Q_i$ denotes the charge recorded by a given photomultiplier tube $i$ of the ANTARES detector. To suppress the influence of background light, we only consider the hits that remain after a hit selection based on the causality criterion assuming a Cherenkov light cone and that have been selected for the final step of the track reconstruction. The track length $L_\mu$ is taken as the length of the reconstructed muon path within a sensitive volume. This volume has been defined as the cylinder of the ANTARES instrumented volume extended by twice the approximate light attenuation length ($L_\mathrm{att}=55~m$) to take into account the possibility of light entering the instrumented volume from the outside. It is depicted in Fig.~\ref{figschussler1:volumecalibration}, left plot.
 
The ANTARES light detection efficiency is depending on the geometrical position and direction of the muon track $\vec{x}$. This efficiency $\epsilon$ can be derived on an event-by-event basis as:

\begin{equation}\label{eqschussler1:epsilon}
\epsilon(\vec{x}) = \sum\limits^\mathrm{nOMs} \exp\left(- \frac{r_i} {L_\mathrm{abs}} \right) \cdot \frac{\alpha_i(\theta_i)} {r_i} 
\end{equation}
Here, the sum runs over all optical modules (OMs) that were active at the time the event was recorded. Modules become inactive for short periods of time, due to localized bioluminescence bursts which cause the data acquisition for modules close by to be stopped, or permanently, due to mechanical or electronical failures. The distance to the muon track $r$ and the angle of incidence $\theta$ of the Cherenkov light is calculated for all $\mathrm{nOM}$ active modules. The latter is used to derive the angular acceptance $\alpha(\theta)$ of the optical modules. $r$ is used to correct for light absorption in the water, with $L_\mathrm{abs}$ being the light absorption length. Finally a factor $1/r$ is applied to take into account the light distribution within the Cherenkov cone.

\subsection{Energy estimation}\label{secschussler1:calibration}
Charged current muon neutrino simulations in combination with a time dependent detector simulation reproducing the actual data taking conditions of the ANTARES detector have been used to correlate  the $\mathrm{d}E/\mathrm{d}X$ values calculated following Eq.~\ref{eqschussler1:rho} with the true energy of the incident neutrino or of the muon passing through the detector. These correlations are shown in the right plot of Fig.~\ref{figschussler1:volumecalibration}. Averaging the result in small $\mathrm{d}E/\mathrm{d}X$ bins ($\Delta(\log(\mathrm{d}E/\mathrm{d}X) = 0.1$), the distributions have been condensed into the final calibration tables. Given a $\mathrm{d}E/\mathrm{d}X$ value, these tables can be used easily to derive the corresponding estimated energy. Linear interpolations in log-log scale are used between the discrete bins of the tables. As baseline, this calibration step is performed using neutrino simulations fulfilling the quality cuts described in~\cite{bibschussler1:Antares_PointSources2010}. It should be noted that, depending 
on the intended application of the energy estimator, a dedicated calibration might become necessary (e.g. energy reconstruction of atmospheric muons, etc.).

\section{Data vs. Monte Carlo comparison}\label{secschussler1:DataMC}
To make sure that the energy estimation will be as reliable for real data as it is for simulated events (see Sec.~\ref{secschussler1:Performance} below), a detailed data vs. Monte Carlo comparison has been performed. This comparison has been conducted at several levels, ranging from the input parameters that are used for the energy estimation as given in Eq.~\ref{eqschussler1:rho} and \ref{eqschussler1:epsilon} to the distribution of the final reconstructed energies and for the main event signatures available with sufficient statistics: atmospheric muons and muon neutrinos. Several event selection criteria have been tested and all distributions show a very satisfactory agreement between data and simulations. Examples are shown in Fig.~\ref{figschussler1:DataMC}. It can therefore be expected to obtain results similar to those for Monte Carlo simulations when the estimator is applied to real data. As final example, the distribution of the $\rho$ estimator (see Eq.~\ref{eqschussler1:rho}) is shown in the left plot of Fig.~\ref{figschussler1:Rho+Resolution} for events 
fulfilling the high quality event selection criteria used for the determination of the atmospheric neutrino spectrum~\cite{bibschussler1:ICRC2013_Spectrum}. It should be noted that the total number of events selected from data is commonly about $25~\%$ higher with respect to the expectations from flux parametrizations (see for example~\cite{bibschussler1:Moriond2013_DiffuseFlux, bibschussler1:ICRC2011_DiffuseFlux}). Here we are only interested in the agreement of the shape, the distributions have therefore been normalized to unity.

\begin{figure*}[!t]
   \centerline{
   			  \includegraphics[width=0.32\textwidth]{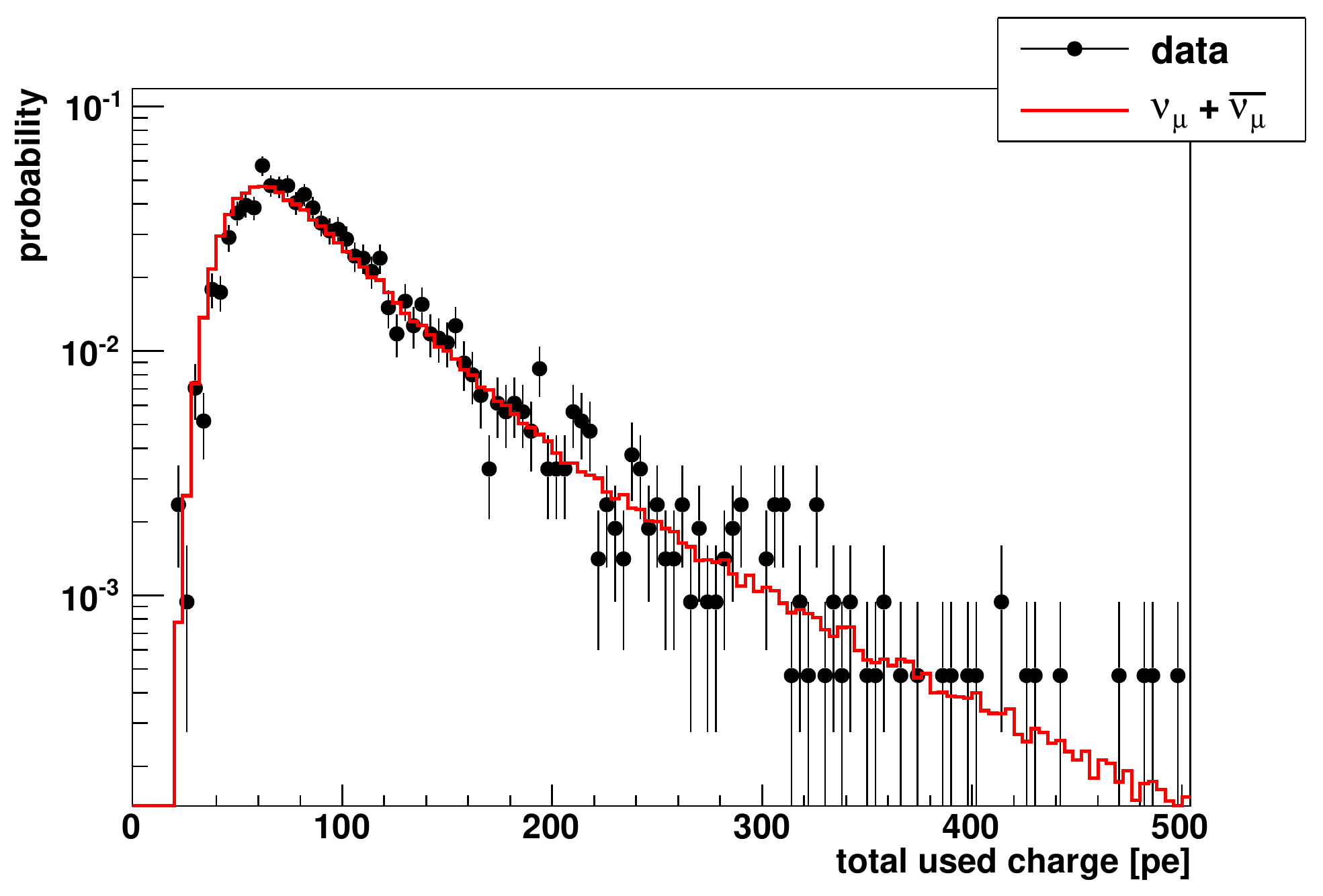}
              \hfill
              \includegraphics[width=0.32\textwidth]{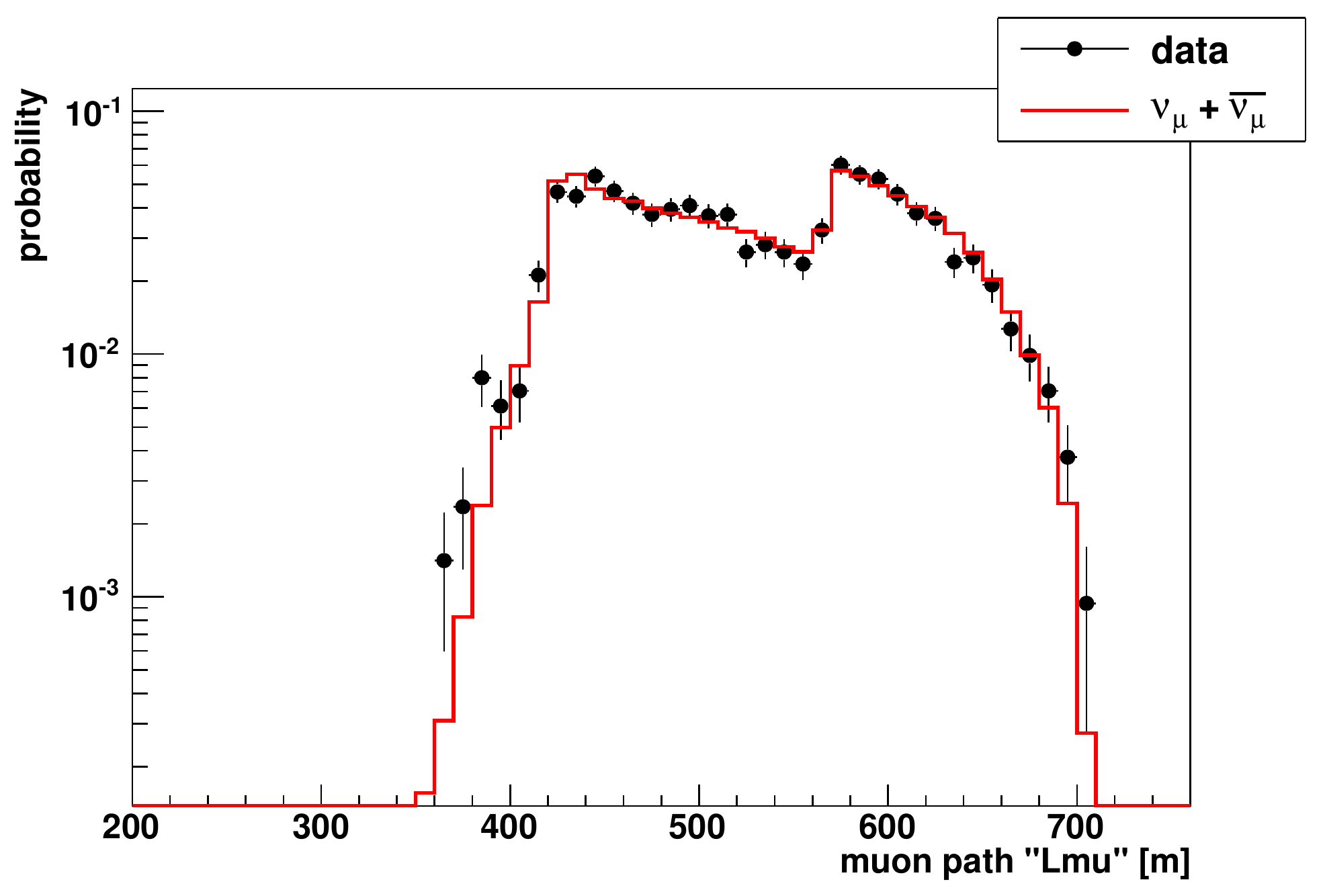}
              \hfill
              \includegraphics[width=0.32\textwidth]{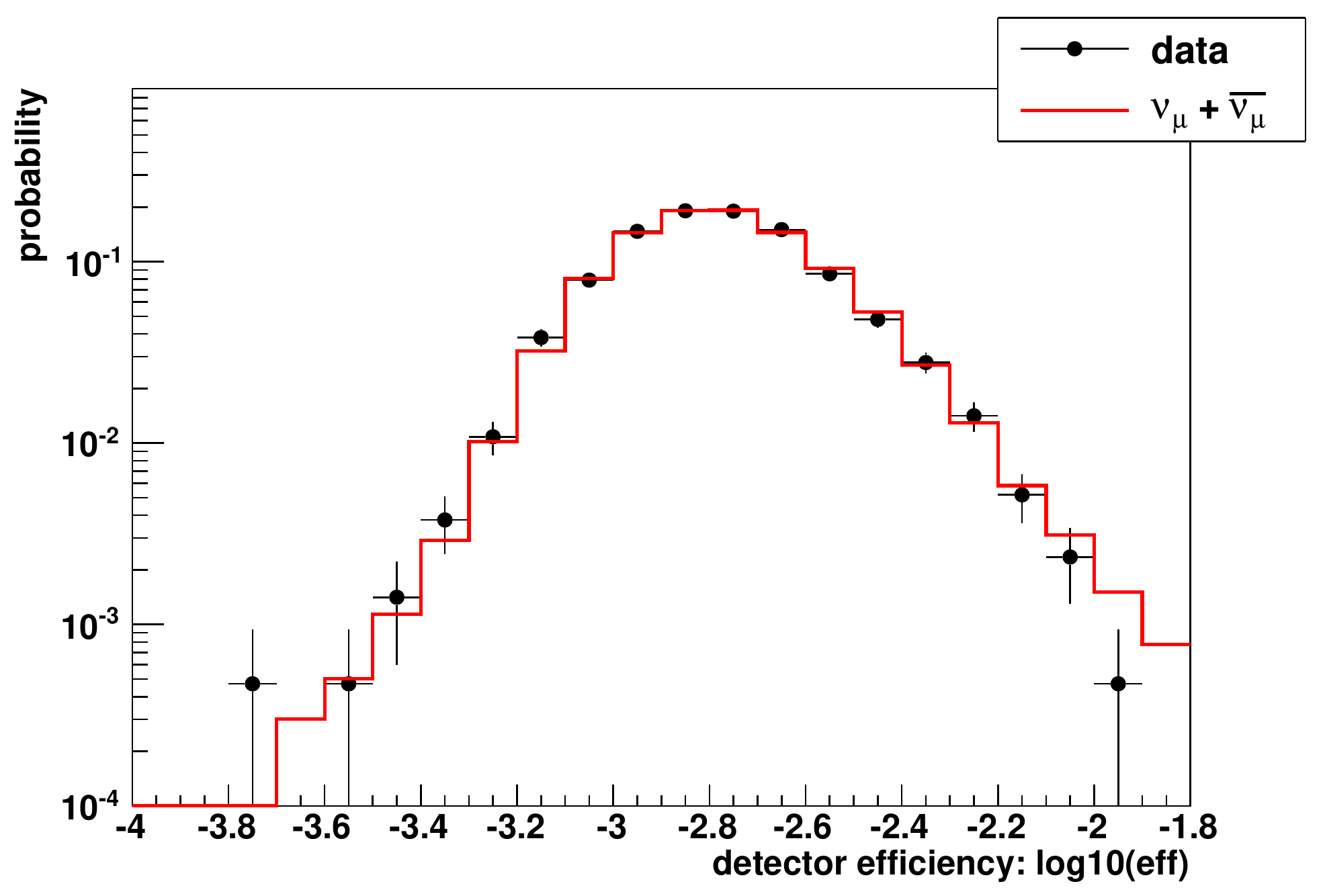} 
             }
   \caption{Comparison between data (black markers) and Monte Carlo (red histogram) for events selected for the determination of the atmospheric neutrino spectrum~\cite{bibschussler1:ICRC2013_Spectrum} for the main input variables to the energy estimator (cf. Eq.~\ref{eqschussler1:rho}). Left plot: The total charge of all used hits $\sum^{\mathrm{nHits}} Q_i$. Middle plot: The tracklength within the fiducial volume $L_\mu(\vec{x})$. Right plot: The detection efficiency $\epsilon(\vec{x})$. }
   \label{figschussler1:DataMC}
 \end{figure*}
\noindent

\section{Performance}\label{secschussler1:Performance}
\subsection{Event selection}
After the verification of agreement between data and Monte Carlo, the performance of the energy estimator can be derived from Monte Carlo simulations. The described method has therefore been applied to charge current neutrino simulations reproducing the ANTARES data taken in the period 01/2008-12/2011. As an example, the event selection criteria follow the ones developed during the search for point like sources~\cite{bibschussler1:Antares_PointSources2010}. They contain a cut on the reconstructed zenith angle $\theta > 90^\circ$, a requirement on the reconstruction quality parameter $\Lambda>-5.2$ as well as a cut on the estimated angular uncertainty of the track reconstruction $\beta < 1^\circ$. 

To improve the energy reconstruction quality, two additional criteria based on internal parameters of the energy estimator have been developed:
\begin{itemize}
\item $\log(\rho) > 1.6$
\item $L_\mu > 380~\mathrm{m}$
\end{itemize}

Events with path lengths within the fiducial volume $L_\mu$ shorter than $380~\mathrm{m}$ are dominated by events passing outside the instrumented volume which leads to an overestimation of the energy. The cut at the limit of the $\rho - E_\mathrm{MC}$ table ($\log(\rho) > 1.6$) is necessary to define the validity of the energy estimator: the correlation between energy and energy deposit practically disappears at low energies and, in addition, the current version of the calibration table is only valid above $\log(\rho)>1.6$ (see Fig.~\ref{figschussler1:volumecalibration}).

\subsection{Efficiency}
The efficiency of the algorithm has been estimated with the help of the above mentioned Monte Carlo simulations. An efficiency of 1 is found over a wide range of energies for events fulfilling the reconstruction quality cuts. The two additional selection criteria related to the energy estimator naturally degrade the efficiency. Whereas the cut on $\log(\rho)$ simply reflects the validity range of the estimator, the minimal track length requirement of $L_\mu > 380~m$ removes high energy tracks that pass outside the instrumented volume. Nevertheless this criterion is necessary to avoid a bias introduced by these external events and starts to degrade the efficiency of the algorithm for events above roughly $100~\mathrm{TeV}$, therefore affecting only a marginal amount of ANTARES data.

\subsection{Resolution}
Applying all selection criteria and weighting the neutrino simulations to follow an astrophysical $E^{-2}$ energy spectrum, the performance of the energy estimator has been derived. As can be seen in Fig.~\ref{figschussler1:Rho+Resolution}, an average resolution of $\log(E) \approx 0.45$ ($\log(E) \approx 0.7$) has been achieved for the reconstruction of the muon (neutrino) energy. 

The main limitation to the energy resolution is the limited size of the detector, which, combined with the statistical nature of the energy loss processes, leads to an insufficient sampling of the energy losses along the muon track. The reconstruction of the neutrino energy suffers in addition from the fluctuations induced in the charged current interaction. Minor additional contributions are related to the uncertainties of the directional reconstruction and the selection of the hits used as input for the energy estimation.

\begin{figure*}[!t]
  \centerline{
  \includegraphics[width=0.46\textwidth]{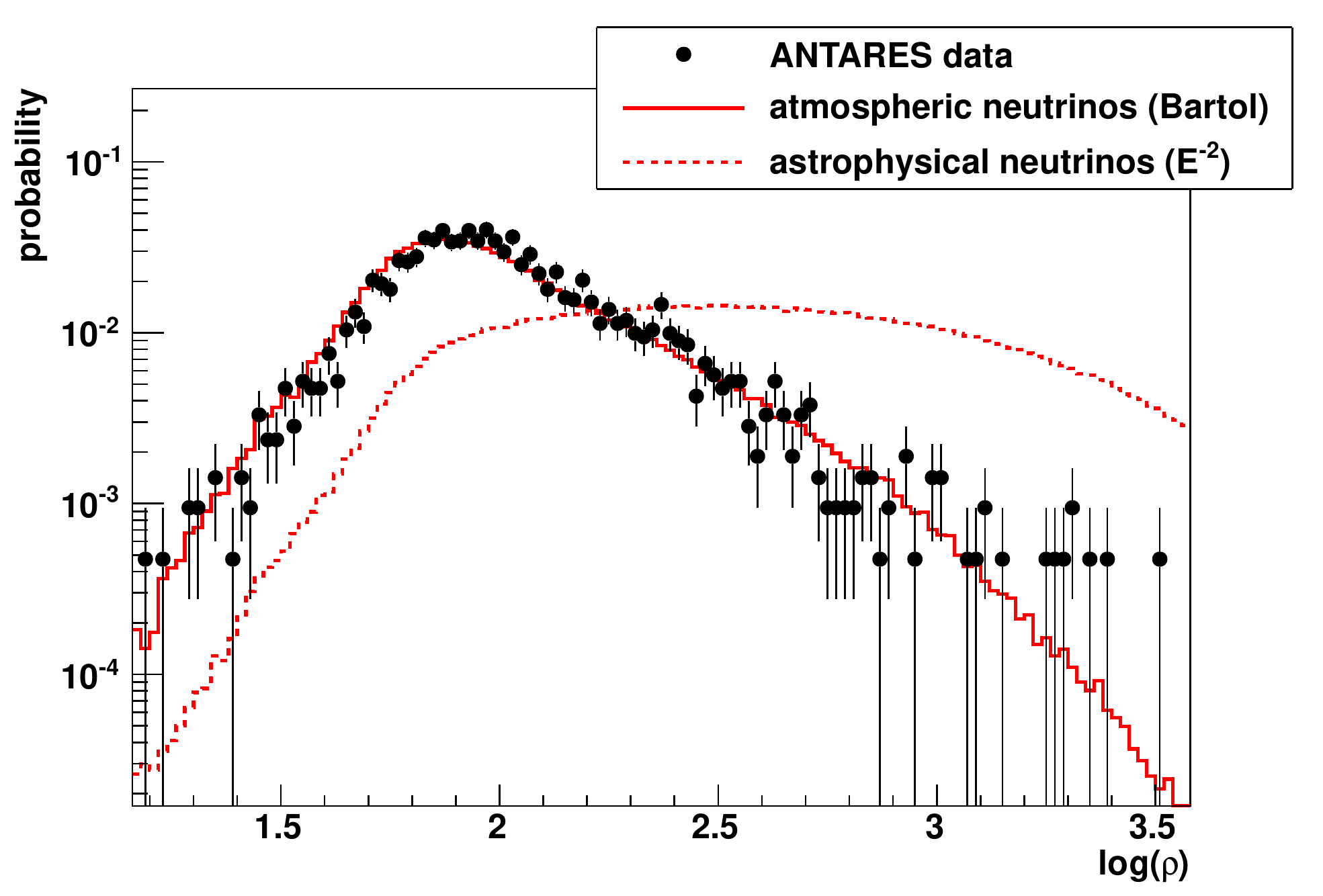}
  \hfill
  \includegraphics[width=0.46\textwidth]{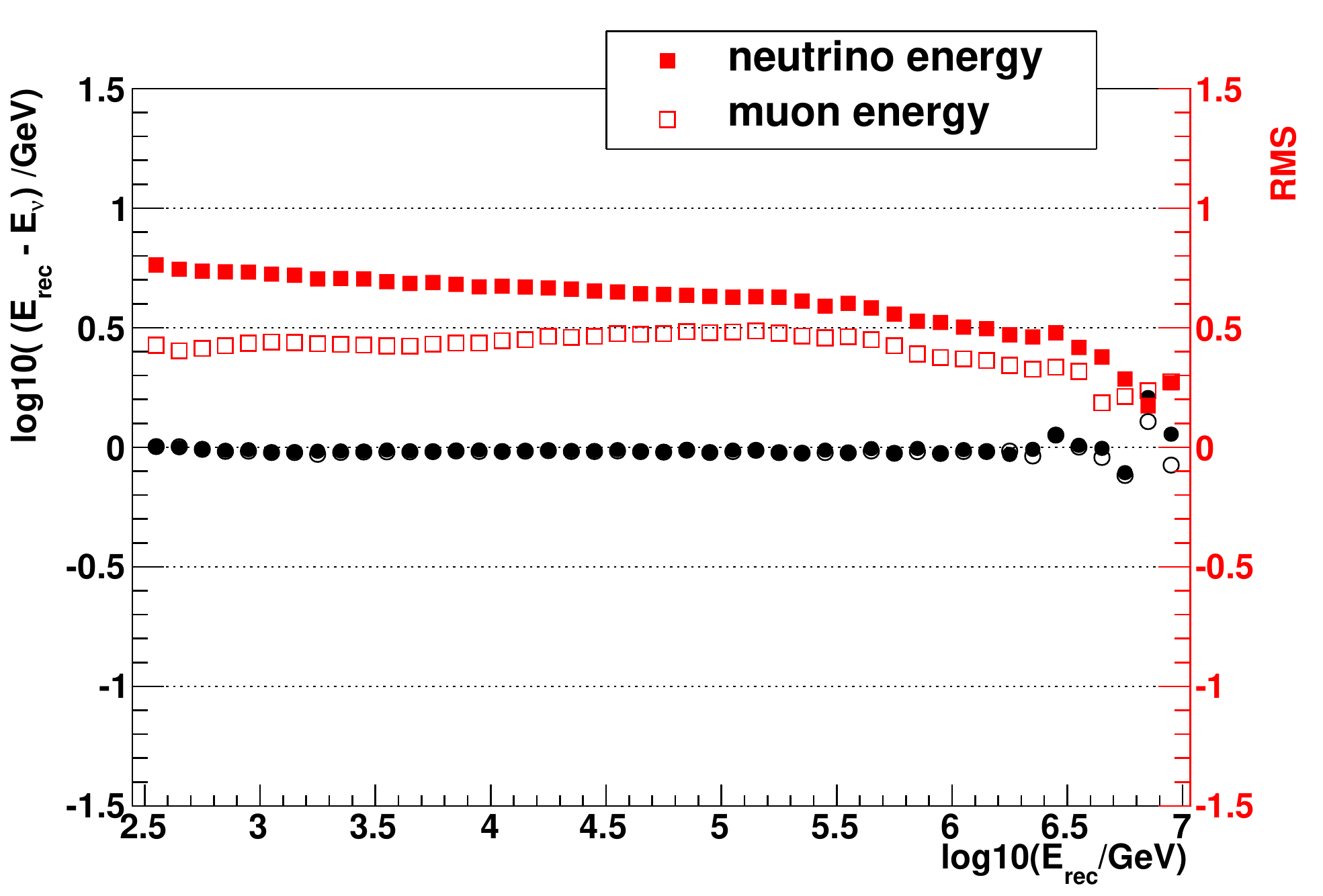}
          }	
  \caption{Left plot: Distributions of the $\rho$ parameter showing the good agreement between data (black markers) and Monte Carlo simulations (solid, red line). For illustration the expectation for an astrophysical neutrino flux is also shown (dotted, red line). Right plot: The stability of the mean (black markers) and the evolution of the RMS (red markers) of the difference between reconstructed and true energy as function of the reconstructed energy. Filled (open) markers denote the reconstruction of the neutrino (muon) energy.}
  \label{figschussler1:Rho+Resolution}
 \end{figure*}

\section{Systematic uncertainties}
\subsection{Energy estimator calibration}\label{secschussler1:syscalibration}
The derived energy estimator relies on Monte Carlo simulations for the correspondence between the estimator, i.e. the $\rho \approx \mathrm{d}E/\mathrm{d}X$ values, and the muon (neutrino) energy. If the used simulations do not perfectly describe the real data, systematic biases might be introduced. No significant difference between data and MC in the distributions of the parameters used as input for the energy reconstruction have been found (cf. Sec.~\ref{secschussler1:DataMC}). The related systematic uncertainty is therefore expected to be reasonably small. In order to quantify the remaining uncertainty, we studied the influence of changes in the Monte Carlo simulations on the energy reconstruction. In an end-to-end approach, a dedicated $\rho \rightarrow E$ calibration (cf. Sec.~\ref{secschussler1:calibration}) using only a subset (corresponding to data taken in 2008) of a modified Monte Carlo simulation set has been used. Applying this calibration to the simulations described above covering the full period (2007-2011) 
several potential effects are included. Among them are uncertainties in the charge and time calibration of the detector, different background noise levels, as well as different detector layouts induced by maintenance and (to a lesser extend) hardware failures. From this study, the overall systematic uncertainty due to imperfections of the Monte Carlo simulations has been estimated to be less than $0.1$ in $\log(E)$.\\[1.5ex]

\subsection{Detector and time evolution}
The ANTARES data taking conditions are not stable in time mainly due to changes in the background rate (induced in majority by bioluminescence) and changes in the detector configuration (construction, maintenance, etc.). The reliability of the energy reconstruction algorithm has therefore been studied as function of both contributions. Thanks to the very robust hit selection, which is able to remove the majority of noise induced hits, no strong dependence of the energy reconstruction quality as function of the background rate has been found. On the other hand, due to the very different detector configuration of only 5 active detection lines in 2007, a significant bias of almost $0.5$ in $\log(E)$ has been found for that period. It should be noted that the $\rho$ estimator itself is not influenced by this bias and that a dedicated calibration table for that period removes it completely.\\

\subsection{Dependence on quality criteria}\label{secschussler1:sysQual}
Other potential systematic effects might arise when events are selected based on different quality criteria. To study this behaviour a cut variation analysis has been performed. Events have been selected based on the criteria given in Sec.~\ref{secschussler1:Performance} removing only the cut under study one by one. No strong dependence on the reconstructed muon direction nor on the exact value of the quality selection criteria (here: angular uncertainty $\beta$ and reconstruction quality $\Lambda$) is found. The dependence on the muon path length within the fiducial volume (cf. Fig.~\ref{figschussler1:volumecalibration}) can be removed by the requirement to have path length $L_\mu > 380~\mathrm{m}$ as discussed above.

Several dependencies of the energy reconstruction stability and performance on intrinsic parameters have been studied in addition. The quality of the energy reconstruction depends for example on the value of the detection efficiency defined in Eq.~\ref{eqschussler1:epsilon}. No clear evidence for its origin could be determined. The dependency is neither correlated to detector effects nor to the underlying event geometry. As some events are affected by a significant underestimation of the energy, an a-posteriori correction has been developed. This correction is based on a fit to the mean energy bias as a function of the detection efficiency. Applying this correction to the full data set on an event-by-event basis leads to an improvement of the energy resolution by $\log(E)=0.01$ with respect to the default values obtained in Sec.~\ref{secschussler1:Performance}.\\

\subsection{Energy spectrum}\label{secschussler1:spectrum}
The selection of simulated events used for the calibration of the energy estimator is subject to individual choices. The estimator can for example be {\it trained} on atmospheric muons, atmospheric neutrinos or (as used throughout this paper) on astrophysical neutrinos following an $E^{-2}$ spectrum. Although a dedicated calibration of the energy estimator for each application is highly advised, an uncertainty on the expected energy spectrum of the analysed events will probably remain. To quantify this uncertainty, the energy of neutrino events were reconstructed using the default $E^{-2}$ calibration. They were then weighted to follow a power-law with index $\alpha = -2.2$. The obtained resolution and its energy dependence show only a small systematic bias of about $0.05$ and an degradation of the resolution by $0.03$ in $\log(E)$. These uncertainties are comparable with the ones introduced by modifications of the Monte Carlo simulations and extending the validity of the calibration in time (cf. Sec.\ref{
secschussler1:syscalibration}).\\[0.5ex]


\section{Summary}
We presented an algorithm able to use the deposited energy $\mathrm{d}E/\mathrm{d}X$ to estimate the energy of muons and neutrinos detected by large scale neutrino telescopes. Applied to data and Monte Carlo simulations of the ANTARES detector the method has been validated and a resolution of $\log(\Delta E) \approx 0.45$ for muons  and $\log(\Delta E) \approx 0.7$ for neutrinos has been obtained. The systematic uncertainty has been conservatively estimated to be $0.1$ in $\log(E)$ with the main contribution due to the uncertainty in the energy spectrum.

The energy estimation method is mainly limited by the available detection volume. A significance performance increase can therefore be expected by the next generation of neutrino telescopes like KM3NeT~\cite{bibschussler1:KM3NeT}.


\setcounter{figure}{0}
\setcounter{table}{0}
\setcounter{footnote}{0}
\setcounter{section}{0}
\setcounter{equation}{0}

\newpage
\id{id_fusco}

%
\title{\arabic{IdContrib} - Measurement of the atmospheric $\nu_\mu$ energy spectrum with the ANTARES neutrino telescope}
\addcontentsline{toc}{part}{\arabic{IdContrib} - {\sl Luigi Antonio Fusco} : Measurement of the atmospheric $\nu_\mu$ energy spectrum%
\vspace{-0.5cm}
}

\shorttitle{\arabic{IdContrib} - Atmospheric neutrinos with {\sc Antares}}

\authors{
  Luigi Antonio Fusco$^{1,2}$, Vladimir Kulikovskiy$^{3,4}$, on behalf of the ANTARES Collaboration.
}

\afiliations{
  $^1$INFN - Sezione di Bologna, Viale Berti-Pichat 6/2, 40127 Bologna, Italy  \\
  $^2$Dipartimento di Fisica dell'Universit\`a, Viale Berti-Pichat 6/2, 40127 Bologna, Italy\\
  $^3$INFN - Sezione di Genova, Via Dodecaneso 33, 16146 Genova, Italy\\
  $^4$Moscow State University, Skobeltsyn Institute of Nuclear Physics, Leninskie gory, 119991 Moscow, Russia
  \scriptsize{
  }
}

\email{lfusco@bo.infn.it}

\abstract{
  Atmospheric neutrinos are produced by the decays of unstable particles in air showers. The main contribution to the atmospheric neutrino flux is given by the decays of pions and kaons, resulting in a very steep energy spectrum. Additional contributions from short-lived charmed hadrons are expected to give a harder energy spectrum above several tens of TeV.
  
   We present a measurement of the atmospheric $\nu_\mu + \bar{\nu}_\mu$ energy spectrum in the energy range 100 GeV - 200 TeV, using data collected by the ANTARES neutrino telescope from 2008 to 2011. The measured flux is consistent with the expectations from theoretical models as well as with the spectra measured by the AMANDA-II and IC40 Antarctic neutrino telescopes.
}


\maketitle

\section{Introduction} \label{secfusco:intro}

The ANTARES detector \cite{bibfusco:antares} is the largest operating undersea neutrino telescope, located at a depth of 2475 m in the Mediterranean Sea, 40 km offshore from Toulon, France. The detector consists of a three dimensional array of 885 photomultiplier tubes (PMTs), distributed along 12 vertical flexible lines and located in pressure resistant spheres (optical modules, OMs), designed to detect the Cherenkov light produced by charged particles, mostly muons, crossing the instrumented volume. All the detected signals are transmitted via an optical cable to a shore station, where filtering and triggering of events takes place. The muon direction is determined by maximising a likelihood which compares the times of the detected signals (\textit{hits}) with the expectations from the Cherenkov signal from a muon track \cite{bibfusco:2012PS}. A more complete description of the detector and recent achievements can be found in these procedings \cite{bibfusco:antoine}.

The main scientific goal of the ANTARES neutrino telescope is to detect muons induced by high energy neutrinos of cosmic origin. Atmospheric neutrinos constitute an irreducible background for these searches. The main contribution to the atmospheric neutrino flux is given by the ``conventional'' component, deriving from the decays of charged pions and kaons. Many theoretical predictions are available for this component of the flux, such as the ones by Barr et al. \cite{bibfusco:bartol} and Honda et al. \cite{bibfusco:honda}. The competition between interaction and decay of pions and kaons generates a very steep energy spectrum for the resulting neutrinos, asymptotically proportional to $E_\nu^{-3.7}$. Above several tens of TeV the so-called ``prompt'' contributions are expected to arise from the decays of charmed hadrons. These hadrons being much more short-lived than charged $\pi$ and $K$, only decays can occur, giving a harder energy spectrum. Many predictions are available for this contribution (e.g. \cite{bibfusco:martin, bibfusco:enberg}), but none of them can be confirmed at the moment.

The atmospheric neutrino energy spectrum has been measured above the TeV region only by AMANDA-II \cite{bibfusco:amanda} and IceCube40 \cite{bibfusco:ic40} using Antarctic ice has converting medium; these measurements differ by $\sim$50\% and are dominated by systematic uncertainties. In these proceedings the first measurement of the atmospheric neutrino energy spectrum performed under seawater, with completely different systematic effects, will be presented.

\section{Energy reconstruction} \label{secfusco:eest}

Relativistic muons passing through the detector lose energy by means of ionization or radiative processes. The former are dominant for $E_\mu<500$ GeV in water, giving energy losses which are only slightly energy dependent; the latter are largely dominant at higher energies and the consequent energy loss along the muon path is linearly dependent on the muon energy. Pair production, bremsstrahlung radiation and photonuclear interactions are responsible for the increase of energy losses at high energies. The energy loss per unit length can be described by:
\begin{equation}
  \frac{\textrm{d}E_\mu}{\textrm{d}X} = -\alpha(E_\mu)-\beta(E_\mu) E_\mu
  \label{eqfusco:Eloss}
\end{equation}
where the first term takes into account ionization losses while the second term describes radiative processes.

Radiative processes generate electromagnetic and hadronic showers along the muon track, whose particles are above the Cherenkov threshold and produce detectable light. These phenomena being linearly dependent on the muon energy, a more energetic muon will produce more detectable light. From this assumption, it is possible to develop muon energy estimation methods, such as the ones described here. 

\subsection{Maximum likelihood approach} \label{secfusco:eestML}

A first approach tries to maximise the agreement between the expected and the observed amount of light on each OM. Starting from the direction information of the reconstructed track and keeping the energy of the muon $E_\mu$ as a free parameter, a likelihood function is built. The likelihood function is defined as the product of individual likelihood function $\mathcal{L}_i(E_\mu)$:
\begin{equation}
  \mathcal{L}(E_\mu) = \frac{1}{N_{OM}} \prod_{i}^{N_{OM}}\mathcal{L}_i(E_\mu).
  \label{eqfusco:Likelihood_def}
\end{equation}
The product is taken over all the $N_{OM}$ optical modules positioned up to 300 m from the reconstructed track, regardless of whether a hit was recorded or not. At further distances, no hit is expected to be recorded. Each individual likelihood function $\mathcal{L}_i(E_\mu)$ depends on the probability of observing a pulse of measured amplitude $Q_i$ given a certain number of photoelectrons induced on the i$^{th}$ OM. These individual likelihood functions $\mathcal{L}_i(E_\mu)$ are constructed as:
\begin{equation}
  \mathcal{L}_i(E_\mu) \equiv P(Q_i; \langle n_{pe} \rangle) = \sum_{n_{pe}=1}^{n_{pe}^{max}} P(n_{pe}; \langle n_{pe} \rangle) \cdot G(Q_i; n_{pe}),
  \label{eqfusco:Likelihood_ind}
\end{equation}
when a hit is recorded on the i$^{th}$ OM and
\begin{equation}
  \mathcal{L}_i(E_\mu) \equiv P(0; \langle n_{pe} \rangle) = e^{-\langle n_{pe}\rangle}+P_{th}(\langle n_{pe}\rangle),
  \label{eqfusco:Likelihood_ind_0}
\end{equation}
when there is no hit on the optical module. Equation (\ref{eqfusco:Likelihood_ind}) takes into account the Poisson probability $P(n_{pe}; \langle n_{pe} \rangle)$ of having $n_{pe}$ photoelectrons given that the expectation is $\langle n_{pe} \rangle$, together with a Gaussian term $G(Q_i; n_{pe})$ which expresses the probability that $n_{pe}$ photoelectrons on the photocathode will yield the measured amplitude $Q_{i}$. Equation (\ref{eqfusco:Likelihood_ind_0}) consists of a term describing the Poisson probability of observing zero photoelectrons when the expected value is $\langle n_{pe} \rangle$, and a term, $P_{th}(\langle n_{pe}\rangle)$, describing the probability that a photon conversion in the optical module will give an amplitude below the threshold level of 0.3 photoelectrons. The performances of this energy reconstruction method are discussed in \cite{bibfusco:dimitris}.

\subsection{Energy loss approach} \label{secfusco:eestdEdX}

The second muon energy estimation method relies on the muon energy loss along its trajectory. The muon energy deposit per unit path length (eq. \ref{eqfusco:Eloss}) is approximated by an estimator $\rho$ which can be derived from measurable quantities:
\begin{equation}
  \frac{dE}{dX} \propto \rho = \frac{\sum^{nHits}Q_i}{\epsilon}\cdot\frac{1}{L_\mu} .
  \label{eqfusco:dEdX_def}
\end{equation}
The quantity $L_\mu$ represents the muon track length within a sensitive volume defined by extending the radius and height of the cylinder surrounding the instrumented detector volume by twice the light attenuation length. $Q_i$ is (as before) the measured amplitude on the i$^{th}$ OM. To remove the contribution from background light a causality criterion embedded in the reconstruction algorithm is used. Finally, the quantity $\epsilon$ represents the overall ANTARES light detection efficiency. This quantity depends on the geometrical position and direction of the muon track. It can be derived on an event-by-event basis as:
\begin{equation}
  \epsilon = \sum_{i=1}^{nOMs}\exp\left(-\frac{r_i}{L_{abs}}\right) \cdot \frac{\alpha_i(\theta_i)}{r_i}.
  \label{eqfusco:dEdX_epsilon}
\end{equation}
Here the sum runs over all the active optical modules. The distance from the muon track $r_i$ and the photon angle of incidence $\theta_i$ are calculated for each event; $\theta_i$ is used to obtain the corresponding angular acceptance $\alpha_i(\theta_i)$ of the involved OM. The distance $r_i$ is used to correct for the light absorption ($L_{abs}$, absorption length) in water taking into account the light distribution within the Cherenkov cone. A complete description of this energy estimation method and its performances can be found in \cite{bibfusco:fabian}.

\section{Energy spectrum measurement} \label{spectrum}

\subsection{Unfolding techniques} \label{unfolding}

The atmospheric neutrino energy spectrum cannot be reconstructed on an event-by-event basis. This is mainly due to the limited resolution of the energy reconstruction: the atmospheric neutrino energy spectrum is extremely steep and the overestimation of the event energy would introduce a large distortion of the spectrum at high energy. This can be overcome by the use of unfolding techniques. 

The problem to be solved can be modeled as a set of linear equations of the form:
\begin{equation}  \label{eqfusco:Axa}
  A\mathbf{E}=\mathbf{X} \ .
\end{equation}
The vector $\mathbf{E}$ represents the true unknown energy distribution in a discrete number of intervals, the vector $\mathbf{X}$ is the measured distribution of the observable and the matrix $A$, called the response matrix, is the transformation matrix between these two. The response matrix is built using Monte Carlo simulations. A simple direct inversion of the response matrix leads in most case to a rapidly oscillating solution and large uncertainties since the matrix $A$ is ill-conditioned \cite{bibfusco:regul}: minor fluctuations on the data vector $\mathbf{X}$ can be catastrophically propagated to the solution $\mathbf{E}$.

A \textit{singular value decomposition} (SVD) approach to unfold the energy spectrum \cite{bibfusco:svd} solves this problem by the decomposition of the response matrix as $A=USV^T$ where $S$ is a diagonal matrix and $U$ and $V$ are orthogonal matrices. This is equivalent to expressing the solution vector $\mathbf{E}$ as a sum of terms weighted with the inverse of the singular values of the matrix $S$. Small singular values can enhance the statistically insignificant coefficients in the solution expansion: this can be overcome by imposing an external constraint on how the solution is expected to behave. The process of imposing such a constraint is called regularization.

An unfolding method not relying on the regularization procedure is the iterative method based on Bayes' theorem described in \cite{bibfusco:bayes}. Considering the case of an energy spectrum measurement, Bayes' theorem states that the probability $\mathcal{P}(E_i|X_j)$ that $E_i$ is the content of the i$^{th}$ bin of the true energy distribution, given the measurement of a value $X_j$ for the j$^{th}$bin of the energy estimator distribution is equal to:
\begin{equation}
  \mathcal{P}(E_i|X_j) = \frac{A(X_j|E_i)p_0(E_i)}{\sum_{l=1}^{n_E}A(X_j|E_l)p_0(E_l)} 
  \label{eqfusco:BayesTh}
\end{equation}
where $A(X_j|E_i)$ is the probability (calculated from Monte Carlo simulations) of measuring an estimator value equal to $X_j$ when the true energy is $E_i$. This quantity corresponds to the element $A_{ij}$ of the response matrix. The \textit{a priori} probability $p_0(E_j)$ is the expected energy distribution at the detector from theoretical expectations and Monte Carlo simulations. The sum in the denominator runs over all the $n_E$ bins of the true energy distribution. Given the observed estimator distribution, the energy distribution at the detector can be obtained applying iteratively eq. \ref{eqfusco:BayesTh}. At the $n^{th}$ iteration, the energy distribution at the detector $P_n(E_j)$ is calculated taking into account the observed number of events in the estimator distribution and the expectations from $P_{n-1}(E_j)$. The result rapidly converges to a stable solution.

\subsection{Uncertainties} \label{uncertainties}

The unfolding process is dependent on Monte Carlo simulations via the construction of the response matrix since simulations use a certain number of parameters whose uncertainties systematically influence the unfolding result. The largest variations in the unfolded spectrum derive from uncertainties on the optical modules sensitivity (efficiency and acceptance as a function of the photon incident angle) and on water properties. The impact of the variations of these parameters was estimated using different specialized neutrino simulations datasets, varying only one parameter each time. The overall sensitivity of the optical modules has been modified by enhancing/decreasing by $10\%$ the probability that a photon reaching the optical module converts into a photoelectron on the PMT photocathode. A second uncertainty related to the optical modules is associated with the angular acceptance, i.e. on the angular dependence of the light collecting efficiency of each OM. Two different response curves, centred on the 
nominal value and departing from it in opposite directions, were used as input of the dedicated Monte Carlo simulation. The uncertainties on water properties are taken into account by applying a rescaling factor to the curve describing the absorption length of light in water as a function of the wavelength, in order to modify by $\pm 10\%$ a standard value of 55 m for a light wavelength of 470 nm.

The simulation set obtained with the standard parameters, corresponding to our best estimate of the considered parameters, is used to construct the default response matrix. Each modified Monte Carlo sample has then been used as pseudo-data and unfolded. The deviation in each energy bin from the spectrum obtained with the default value of the parameter corresponds to the systematic uncertainty associated with the parameter variation. The outcomes of each of these independent uncertainty calculations are added quadratically, separately for positive and negative deviations. By construction of the energy estimators and of the two unfolding methods the systematic uncertainties as a function of the neutrino energy are different for the two methods. 

An additional uncertainty arises from the choice of the underlying spectrum in the construction of the response matrix; slight changes in the spectrum have been introduced and the deriving systematic effects have been estimated.

\begin{figure}
 \centering
  \includegraphics[width=0.5\textwidth]{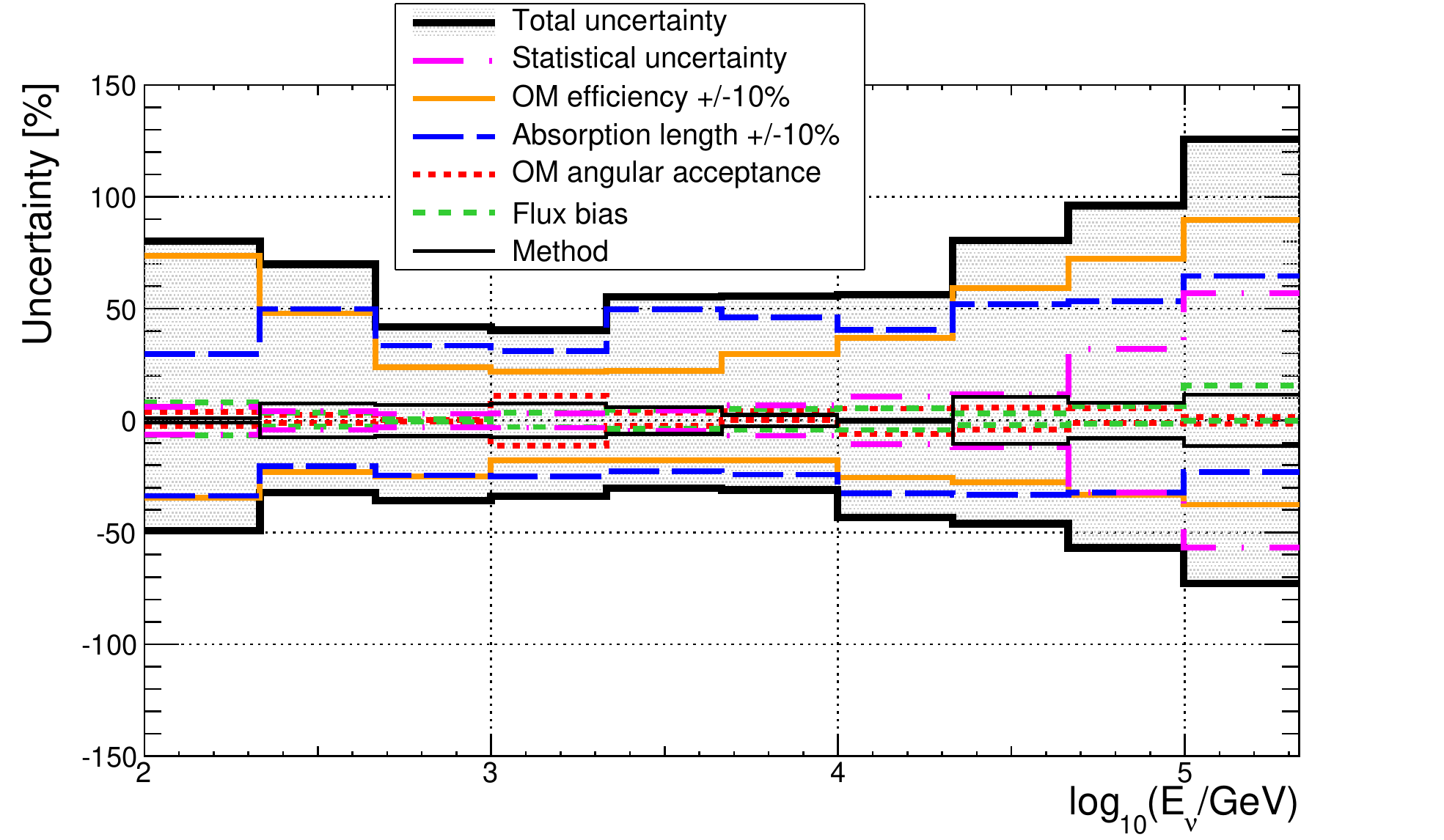}
  \caption{Systematic uncertainties calculated as a function of the neutrino energy. Orange, continuous line represents the effects given by changing the OM efficiency; blue, dashed line is for a change in the water properties; red, dotted line is for the OM angular acceptance variation; green, short dashed line shows the effects given by a change in the underlying model in the response matrix; magenta, dash dot line is the statistical uncertainties given by the unfolding method; black, thin line represent the differences in the unfolding output by the two methods. The shaded area represent the total estimated uncertainty (see text).}
  \label{figfusco:Nu_unc}
\end{figure}

\subsection{Data analysis} \label{secfusco:analysis}

The two described unfolding procedures implemented in the \textit{RooUnfold} package \cite{bibfusco:roounfold} have been applied in two independent analyses (namely $a$ and $\alpha$) on data collected by the ANTARES neutrino telescope from Dec. 2007 to Dec. 2011 to obtain the atmospheric neutrino energy spectrum. The corresponding equivalent livetime is 855 days. Analysis $a$ uses the maximum likelihood energy estimator (\S\ref{secfusco:eestML}) and the SVD unfolding technique (\S\ref{unfolding}); analysis $\alpha$ applies the Bayesian unfolding technique (\S\ref{unfolding}) on the energy loss estimator (\S\ref{secfusco:eestdEdX}). The final result is given by the averaging of the two results, together with an estimation of the uncertainties coming from both analyses.

The main background for the analysis of atmospheric neutrinos is given by downgoing atmospheric muons, misreconstructed as upgoing tracks. The rejection of these tracks is obtained by appropriate cuts on the quality variables of the tracking algorithm $\Lambda$ - related to the maximum likelihood of the tracking algorithm - and $\beta$ - an estimation of the angular error on the reconstructed track \cite{bibfusco:2012PS} - and on the reconstructed zenith angle. Two different sets of selection cuts have been obtained by the two analyses from Monte Carlo simulations, leading to analogous muon contamination (below 0.3\% of the final sample) and similar neutrino candidate rates - $\nu_a \sim$ 1.7 ev/day and $\nu_\alpha \sim$ 1.8 ev/day respectively for $a$ and $\alpha$.

The two unfolding methods described in \S\ref{unfolding} have been applied on the observed energy estimator distributions. The result of the unfolding procedure is the neutrino energy distribution at the detector. In order to obtain the neutrino energy spectrum at the surface of the Earth in the common units of GeV$^{-1}$ s$^{-1}$ sr$^{-1}$ cm$^{-2}$ the ANTARES neutrino effective area is calculated. The correction factor to be applied on the unfolded energy distributions takes into account the neutrino detection and selection efficiency, together with the neutrino propagation through the Earth. Using the neutrino effective area calculation, the atmospheric neutrino energy spectrum at the surface of the Earth, averaged over the zenith angle from 90$^\circ$ to 180$^\circ$ is obtained for the two different strategies. The atmospheric neutrino energy spectrum presented in table \ref{tab:Nu_spectrum} is given by the averaging of the result of the two. The reported uncertainty is calculated taking the quadratic 
sum of the largest uncertainty for each considered effect from the two different analysis strategies (figure \ref{figfusco:Nu_unc}).

\begin{table} 
  \[
  \begin{array}{lcr}
    \hline
    \textrm{Energy range}               & {E_\nu}^2\cdot {d\Phi_\nu/dE_\nu}  &\textrm{ Uncertainty}  \\
    \log_{10}({E}_{\nu}/\textrm{GeV})     &\textrm{[GeV s$^{-1}$ sr$^{-1}$ cm$^{-2}$]}& [\%]   \\
    \hline\hline
    2.00 - 2.33                         &    3.2 \times 10^{-4}           &  +80, -49       \\
    2.33 - 2.66                         &    1.7 \times 10^{-4}           &  +69, -32       \\
    2.66 - 3.00                         &    7.8 \times 10^{-5}           &  +41, -36       \\ 
    3.00 - 3.33                         &    3.2 \times 10^{-5}           &  +40, -34       \\
    3.33 - 3.66                         &    1.1 \times 10^{-5}           &  +55, -30       \\
    3.66 - 4.00                         &    3.9 \times 10^{-6}           &  +56, -31       \\
    4.00 - 4.33                         &    1.2 \times 10^{-6}           &  +56, -43       \\
    4.33 - 4.66                         &    3.8 \times 10^{-7}           &  +80, -46       \\
    4.66 - 5.00                         &    1.2 \times 10^{-7}           &  +96, -57       \\
    5.00 - 5.33                         &    4.8 \times 10^{-8}           &  +125, -73      \\
    \hline
  \end{array}
  \]
  \caption{The unfolded atmospheric neutrino energy spectrum measured with the ANTARES neutrino telescope multiplied by E$_\nu^2$. Each row shows the energy interval of the bin, the measured flux multiplied by $E_\nu^2$ and the percentage uncertainty on the flux.}
  \label{tab:Nu_spectrum}
\end{table}

Figure \ref{figfusco:Nu_spectrum} shows the results of this analysis together with the published results from AMANDA-II \cite{bibfusco:amanda} and IceCube40 \cite{bibfusco:ic40}. Along with the experimental results from Antarctic neutrino telescope, the ANTARES result is also compared with the expectations from Barr et al. \cite{bibfusco:bartol} conventional flux model plus the prompt contribution from Martin et al. \cite{bibfusco:martin} and Enberg et al. \cite{bibfusco:enberg}.

The result of ANTARES is averaged over the zenith angle from 90$^\circ$ to 180$^\circ$ while AMANDA-II and IceCube40 start respectively from 100$^\circ$ and 97$^\circ$; since the atmospheric neutrino energy spectrum is zenith dependent, a correction factor (from $\sim 3\%$ at 100 GeV up to $\sim 40\%$ at 100 TeV) is to be applied to the fluxes reported by the Antarctic neutrino telescopes in order to be directly comparable with the ANTARES result. Considering this correction to be applied to the reported AMANDA-II and IceCube40 spectra, the ANTARES result is completely compatible with the two measurements within the reported errors. Moreover the ANTARES result is above the neutrino flux expectations from Barr et al. by $\sim$25\%, well within the reported uncertainties on the theoretical flux normalization. 

\begin{figure}
 \centering
  \includegraphics[width=0.5\textwidth]{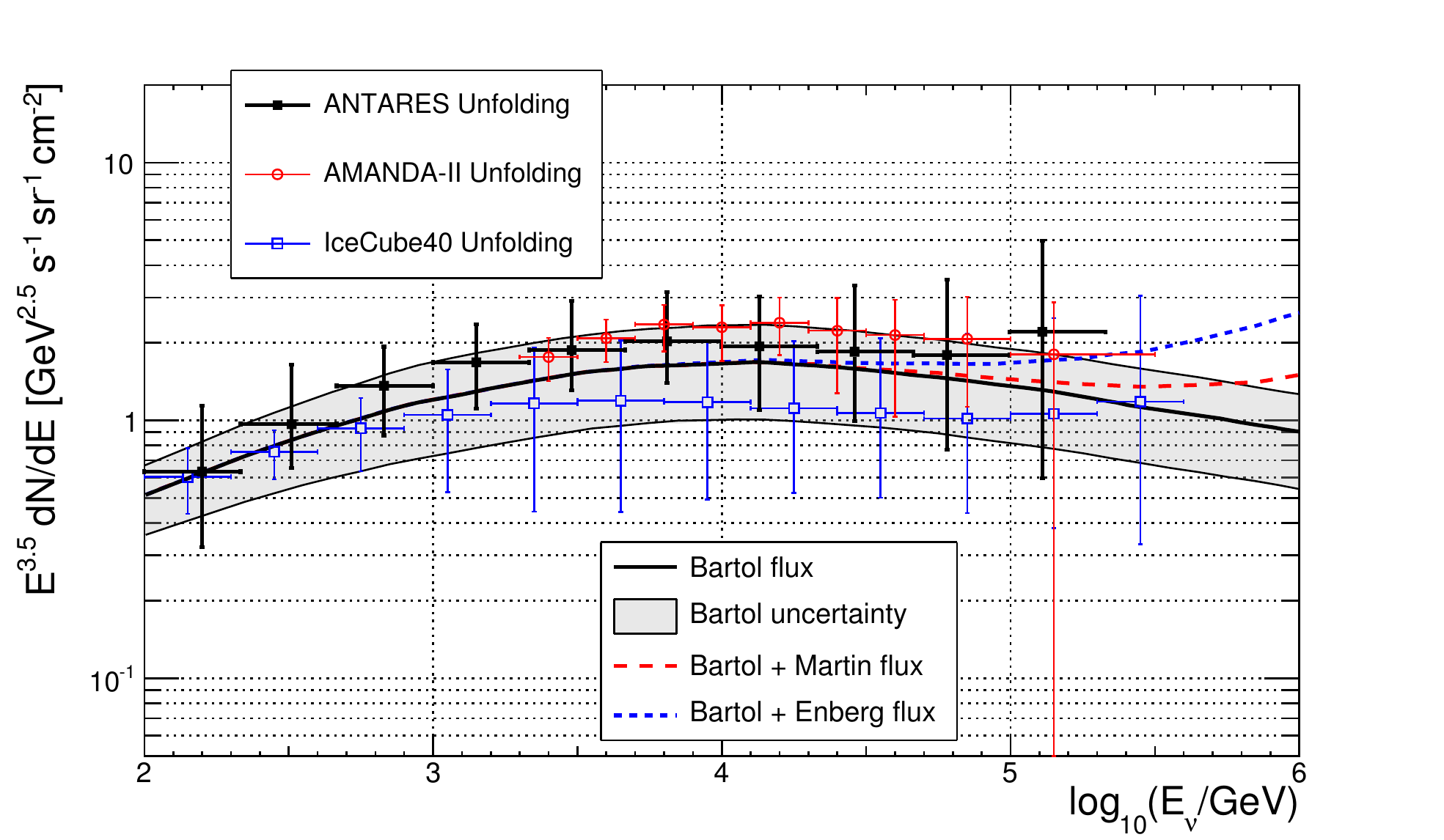}
  \caption{Atmospheric neutrino energy spectrum multiplied by E$_\nu^{3.5}$. The ANTARES result (black - full squares) is shown together with the results from the AMANDA-II (red - empty circles) \cite{bibfusco:amanda} and IceCube40 (blue - empty squares) \cite{bibfusco:ic40}. The ANTARES spectrum is zenith-averaged from 90$^\circ$ to 180$^\circ$, while IceCube is from 97$^\circ$ to 180$^\circ$ and AMANDA-II  from 100$^\circ$ to 180$^\circ$ (see text). The experimental results are shown together with the conventional flux model from \cite{bibfusco:bartol} (black - solid line) to which the prompt expectations from \cite{bibfusco:martin} (red - dashed line) and \cite{bibfusco:enberg} (blue - dashed line) are added. The shaded area represents the reported uncertainty on the conventional flux from \cite{bibfusco:bartol}. Theoretical expectations are averaged from 90$^\circ$ to 180$^\circ$.}
  \label{figfusco:Nu_spectrum}
\end{figure}

\section{Conclusions}

The atmospheric neutrino energy spectrum from 100 GeV to 200 TeV measured using data collected with the ANTARES neutrino telescope from 2008 to 2011 has been presented in these proceedings. Two separate analyses have been conducted, using different energy estimators and unfolding methods; the results of the two methods have been merged to obtain the atmospheric neutrino spectrum. The final result is compatible within the reported errors with the theoretical expectations from conventional models. At the moment it is not possible to confirm or reject the presence of any prompt contribution to the atmospheric neutrino flux beacuse of the large systematic uncertainties. The presented result is compatible within the reported errors with the measurement under the Antarctic ice by AMANDA-II and IceCube40.  
\setcounter{figure}{0}
\setcounter{table}{0}
\setcounter{footnote}{0}
\setcounter{section}{0}
\setcounter{equation}{0}

\newpage
\id{id_vladimir}
%

\title{\arabic{IdContrib} - A search for Neutrino Emission from the Fermi  Bubbles with the ANTARES Telescope}
\addcontentsline{toc}{part}{\arabic{IdContrib} - {\sl Vladimir Kulikovskiy} : A search for Neutrino Emission from the Fermi Bubbles%
 \vspace{-0.5cm}
 }

\shorttitle{\arabic{IdContrib} - A search for Neutrino Emission from the Fermi Bubbles with the {\sc Antares} Telescope}

\authors{
V. Kulikovskiy$^{1,2,3}$ for the ANTARES Collaboration.
}

\afiliations{
$^1$ INFN Sezione di Genova, via Dodecaneso 33, Genoa, Italy \\
$^2$ APC, Universit\'e Paris Diderot, CNRS/IN2P3, CEA/IRFU, Observatoire de Paris, \\Sorbonne Paris Cit\'e, 75205 Paris, France \\
$^3$ SINP MSU, Vorobyevi gory, 1, Moscow, Russia \\
}

\email{vladimir.kulikovskiy@ge.infn.it}

\abstract{Analysis of the Fermi-LAT data has revealed two extended structures above and below the Galactic Centre emitting gamma rays with a hard spectrum, the so-called Fermi bubbles. Some of the promising explanations of this phenomenon assume that accelerated cosmic rays interact with an interstellar medium in the Fermi bubble regions producing pions. Gamma rays and high-energy neutrinos are expected with similar flux from the pion decay. The ANTARES detector is a neutrino telescope located in the Mediterranean Sea, a geographical position which enables good visibility to the Fermi bubble regions. Using ANTARES data from 2008 to 2011 upper limits on the neutrino flux for $E_{\nu}>5$ TeV from the Fermi bubbles were derived for various assumed energy cutoffs of the source. No statistically significant excess of events was observed using data corresponding to 3.5 years of lifetime.}


\maketitle

\section{Fermi bubbles}
Analysis of data collected with the Fermi-LAT experiment has revealed two large spherical structures centred around our Galactic Centre and perpendicular to the galactic plane~--- the so-called Fermi bubbles~\cite{bibvladimir:Su}. These structures are characterised by gamma-ray emission with a hard $E^{-2}$ spectrum and a relatively constant intensity over the full emission region. The approximate edges of the Fermi bubble regions seen in gamma rays using the Fermi-LAT data are shown in figure~\ref{figvladimir:fb_shape}. The size of the simplified shape shown in the same figure is 0.66 sr.
\begin{figure}[h]
\centering
\includegraphics[width=0.4\linewidth]{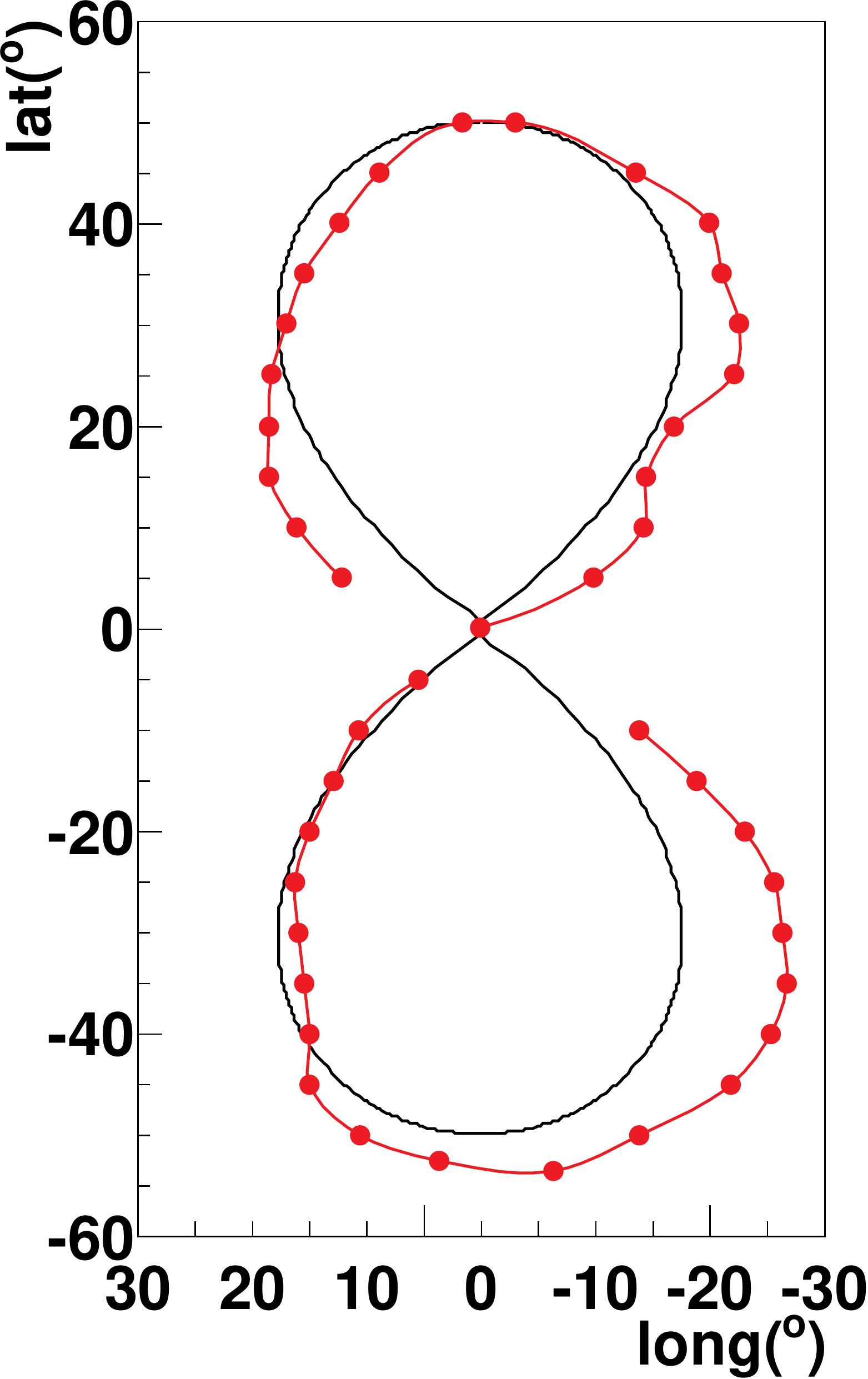}
\caption{Approximate edges (red line, circles) of the north and south Fermi bubbles respectively in galactic coordinates identified from the 1--5 GeV maps built from the Fermi-LAT data~\cite{bibvladimir:Su}. The contour line is discontinuous at the region of the Galactic Centre as the maps are severely compromised by the poor subtraction and interpolation over a large number of point sources in this region.  Simplified shape of the Fermi bubles used in this analysis (black line).}
\label{figvladimir:fb_shape}
\end{figure}

Signal from the Fermi bubble regions was also observed in the microwave band by WMAP~\cite{bibvladimir:Dobbler}, in the X-rays by ROSAT~\cite{bibvladimir:ROSAT} and, recently, in the radio-wave band~\cite{bibvladimir:radio}. Several proposed models explaining the emission include hadronic mechanisms, in which gamma rays together with neutrinos are produced by the collisions of cosmic-ray protons with very underdense interstellar matter~\cite{bibvladimir:CrockerAharonian,bibvladimir:Lacki,bibvladimir:Thoudam}.  Other types of the models exclude the neutrino emission or assume lower fluxes. The observation of a neutrino signal from the Fermi bubble regions may play an unique role in discriminating between models.

The estimated photon flux in the energy range 1--100~GeV covered by the Fermi-LAT detector from the Fermi bubble regions is~\cite{bibvladimir:Su}:
\begin{equation}
E^{2}\mathrm{d}\Phi_\gamma/\mathrm{d}E \approx 3-6\times10^{-7}\mathrm{\,GeV\,cm^{-2}\,s^{-1} sr^{-1}}
\label{eqvladimir:gamma}
\end{equation}
Assuming a hadronic model in which the gamma-ray and neutrino fluxes arise from the decay of neutral and charged pions respectively, the $\nu_\mu$ and $\overline\nu_\mu$ fluxes are proportional to the gamma-ray flux and their proportionality coefficients are about 0.211 and 0.195 correspondingly~\cite{bibvladimir:VillanteVissani}.
With this assumption and using Equation~(\ref{eqvladimir:gamma}):
\begin{equation}
E^{2}\mathrm{d}\Phi_{\nu_\mu+\overline\nu_\mu}/\mathrm{d}E\equiv A_\mathrm{theo},
\label{eqvladimir:fb_flux}
\end{equation}
\begin{equation}
A_\mathrm{theo}\approx 1.2-2.4\times10^{-7}\mathrm{\,GeV\,cm^{-2}\,s^{-1} sr^{-1}},
\end{equation}
the extrapolation of Equation~(\ref{eqvladimir:fb_flux}) towards higher energies can be represented by: 
\begin{equation}
E^{2}\mathrm{d}\Phi_{\nu_\mu+\overline\nu_\mu}/\mathrm{d}E \approx A_\mathrm{theo}\mathrm{e}^{-E/E^\mathrm{cutoff}_\nu}.
\label{eqvladimir:fb_flux2}
\end{equation}
The neutrino flux, as well as the gamma-ray flux, is expected to have an exponential energy cutoff represented in equation~\ref{eqvladimir:fb_flux2} by $E_\nu^\mathrm{cutoff}$.
The cutoff is determined by the primary protons which have a suggested cutoff $E_p^\mathrm{cutoff}$ in the range 1--10~PeV~\cite{bibvladimir:CrockerAharonian}. The corresponding neutrino-energy cutoff may be estimated by assuming that the energy transferred from $p$ to $\nu$ derives from the fraction of energy going into charged pions ($\sim20\%$) which is then distributed over four leptons in the pion decay. Thus: 
\begin{equation}
	E_\nu^\mathrm{cutoff} \approx E_p^\mathrm{cutoff}/20,
\end{equation}
which gives a range 50--500 TeV for $E_\nu^\mathrm{cutoff}$.
\section{The ANTARES neutrino telescope}
The ANTARES telescope is a deep-sea Cherenkov detector which is located 40 km from Toulon at a mooring depth of 2475~m. The energy and direction of incident neutrinos are measured by detecting the Cherenkov light produced in water from muons originating in the charged-current interactions of $\nu_{\mu}$ and $\bar{\nu}_{\mu}$. The light is detected with a three-dimensional array of 885 optical modules, each containing a 10 inch PMT. More details on the detector construction, its positioning system and the time calibration can be found in~\cite{bibvladimir:ANTARES,bibvladimir:Positioning,bibvladimir:timing}.

The ANTARES detector started data-taking with the first 5 lines installed in 2007. The full detector was completed in May 2008 and has been operating continuously ever since.

Not the only neutrinos produced by the cosmic sources present the events in the detector. Cosmic rays produce particle showers when interacting with the atmosphere. Muons and neutrinos created in these atmospheric showers provide two main background components for the search for cosmic neutrinos. The more than 2 km of water above the detector act as a partial shield against the atmospheric muons. Below the detector is protected by the Earth. As the downgoing atmospheric muon background at these depths is still bigger than the expected signal, only upgoing events can be used for the cosmic signal search. The ANTARES neutrino telescope, located in the Northern Hemisphere, has an excellent visibility by means of the upgoing neutrinos to the Fermi bubbles. Atmospheric neutrinos may traverse the Earth and lead to upward-going tracks in the detector, presenting an irreducible background. The signal-to-noise ratio can be improved by rejecting low-energy neutrino events as the spectrum of the atmospheric neutrinos is expected to be steeper than the expected source spectrum.

Tracks are reconstructed using the arrival time of the photons together with the positions and directions of the photomultipliers. Details of the tracking algorithm are given in~\cite{bibvladimir:pointsearch}. Only events  reconstructed as upgoing have been selected for this analysis. In addition, cuts on the reconstruction quality parameters have been applied in order to reject downgoing atmospheric muon events that are incorrectly reconstructed as upgoing tracks. These parameters are the quality $\Lambda$ of the track fit which is derived from the track fit likelihood and the uncertainty $\beta$ of the reconstructed track direction. Simulations for an $E^{-2}$ neutrino-energy spectrum yield a median angular resolution on the neutrino direction of less than $0.6^\circ$ for the events with $\Lambda > -5.2$ and $\beta<1^\mathrm{o}$.

A shower-like events can be identified by using the second tracking algorithm with a two $\chi^2$-like fits of each event, assuming the hypothesis of a relativistic muon ($\chi^2_\mathrm{track}$) and that of a shower-like event ($\chi^2_\mathrm{point}$)~\cite{bibvladimir:bbfit}. Events with $\chi^2_\mathrm{track}>\chi^2_\mathrm{point}$ were excluded from the analysis.

In this analysis the energy was estimated using Artificial Neural Networks~\cite{bibvladimir:Jutta}. The used parameters include the number of detected photons and the total deposited charge. The median energy resolution is roughly 50\% for neutrinos with an energy of 10~TeV.

\section{Analysis}
Data in the period from May 2008, when the detector started to operate in its complete configuration, till December 2011 were used. In total 806 days were selected for the analysis. A signal from the Fermi bubbles was searched for by comparing the number of selected events from this area (on-zone) with this in comparable regions with no expected signal (off-zones). The simplified shape of the Fermi bubbles used in this analysis is shown in Figure~\ref{figvladimir:fb_shape}. The events selection was based on two parameters, namely the track quality parameter $\Lambda^\mathrm{cut}$ and the reconstructed energy $E_\mathrm{Rec}^\mathrm{cut}$. The reconstructed energy was used to decrease the atmospheric neutrino background while $\Lambda$ was used mostly to remove the atmospheric muons. The analysis adopted a blinding strategy in which the cut optimisation was performed using simulated data for the signal and the events arriving from the off-zones for the background. 

Off-zones with the same size and shape as the on-zone were defined using events coming from the same solid angle in local coordinates as the on-zone events, but shifted with some fixed delay in time. This ensures the same expected number of background events, as their number is proportional to the efficiency of the detector, which is a function of the local coordinates only. The off-zones defined in this way are fixed in the sky. The size of the Fermi bubbles allows to select at maximum three non overlapping off-zones. The Fermi bubble regions and the three off-zones are shown in figure~\ref{figvladimir:fb_offzones} together with the sky visibility. The visibility to each point on the sky was calculated as a part of the sidereal day during which it is below the horizon (in order to produce upgoing events in the detector). The visibility to each zone is 0.68 (0.57 for the northern part and 0.80 for the southern part of the Fermi bubble regions).
\begin{figure}
\centering
\includegraphics[width=\linewidth]{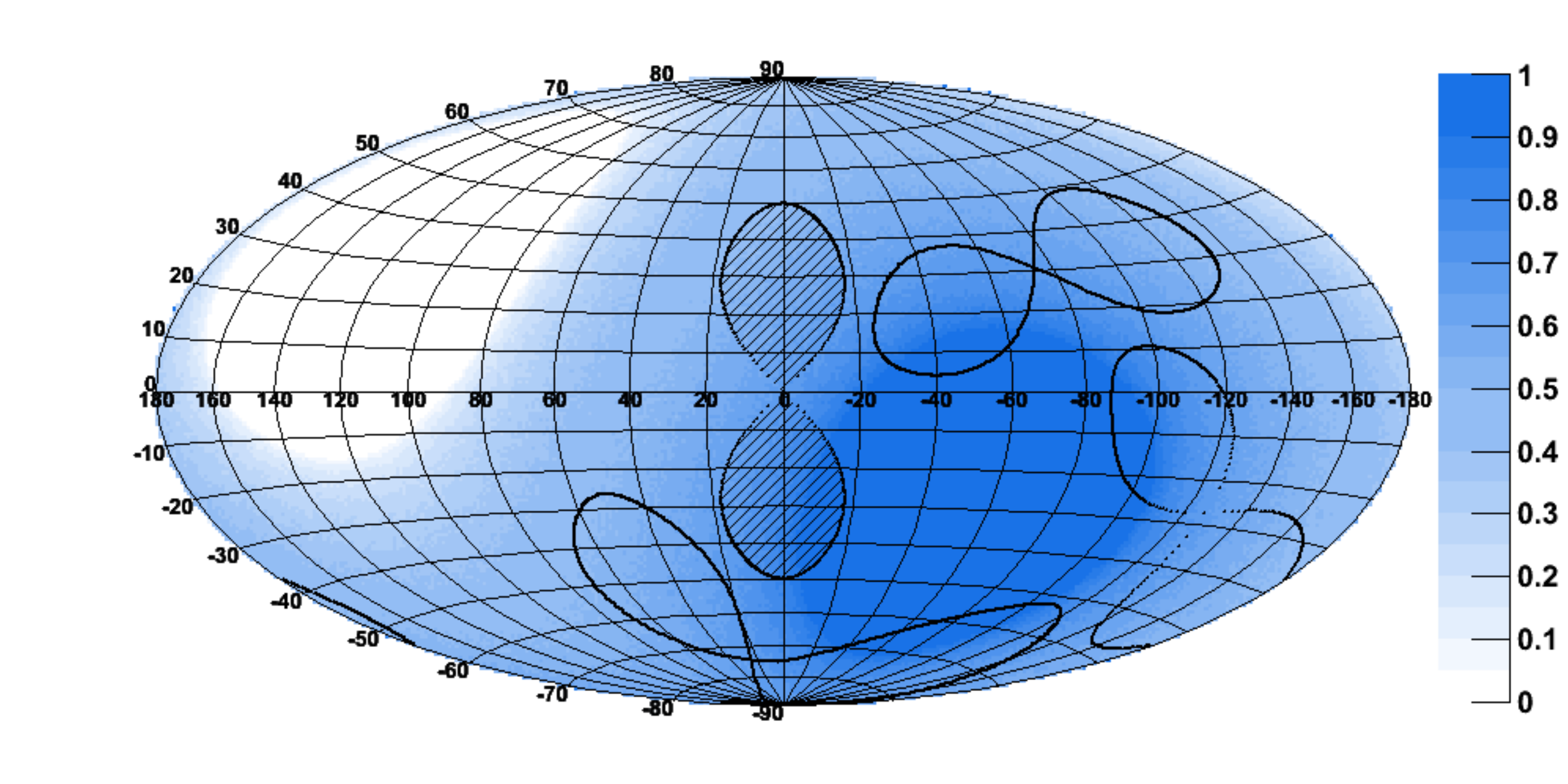}
\caption{Visibility of the sky at the ANTARES site in galactic coordinates. Maximum corresponds to a 24 h per day visibility. The area of the Fermi bubbles (on-zone) is in the centre. The regions corresponding to the three off-zones around the maximum of the visibility are also depicted.}
\label{figvladimir:fb_offzones}
\end{figure}

The difference in the number of background events between the on-zone and the three off-zones was tested. Firstly, the number of events in the off-zones was extracted from the data for various cuts ($\Lambda$, $E_\mathrm{Rec}$) and the difference in the event numbers between each pair of off-zones was calculated. This difference was compared with the statistical uncertainty and no deviation was seen beyond the expected statistical fluctuations. Secondly, the number of events in the on-zone together with the average number of events in the three off-zones was tested using the simulated atmospheric background and the difference was found to be within the expectation from the statistical uncertainty.

The simulation chain for ANTARES is described in~\cite{bibvladimir:Brunner}. For the expected signal from the Fermi bubbles the flux according to equation~\ref{eqvladimir:fb_flux2} was assumed. Four different cutoffs $E^{\nu}_\mathrm{cut}$ were considered in this analysis: no \mbox{cutoff} ($E_{\nu}^\mathrm{cutoff}=\infty$), 500~TeV, 100~TeV and 50~TeV which correspond to the suggested cutoff of the proton spectrum. Atmospheric neutrinos were simulated using the model from the Bartol group~\cite{bibvladimir:Bartol} which does not include the decay of charmed particles. At energies above 100~TeV the semi-leptonic decay of short-lived charmed particles might become a significant source of atmospheric neutrino background. The uncertainty introduced by the estimation of this flux contribution ranges over several orders of magnitude. Due to the comparison of on and off zones and the final cut $\sim10$~TeV (defined in the end of this section) the flux from the charmed particle decay does not have a significant impact on the analysis nor alter the final result on upper limits. 

Figure~\ref{figvladimir:fb_lambda} shows the distribution of data and simulated events as a function of the parameter $\Lambda$ for events arriving from the three off-zones. Here the events with at least 10 detected photons associated with the reconstructed track were selected with the requirement of angular error estimate $\beta<1^\circ$. The latter condition is necessary in order to ensure a high angular resolution to avoid events originating from the off-zone region being associated with the signal region and vice versa. The requirement on the number of photons removes most of the low-energy background events.
\begin{figure}
\center
\includegraphics[width=\linewidth]{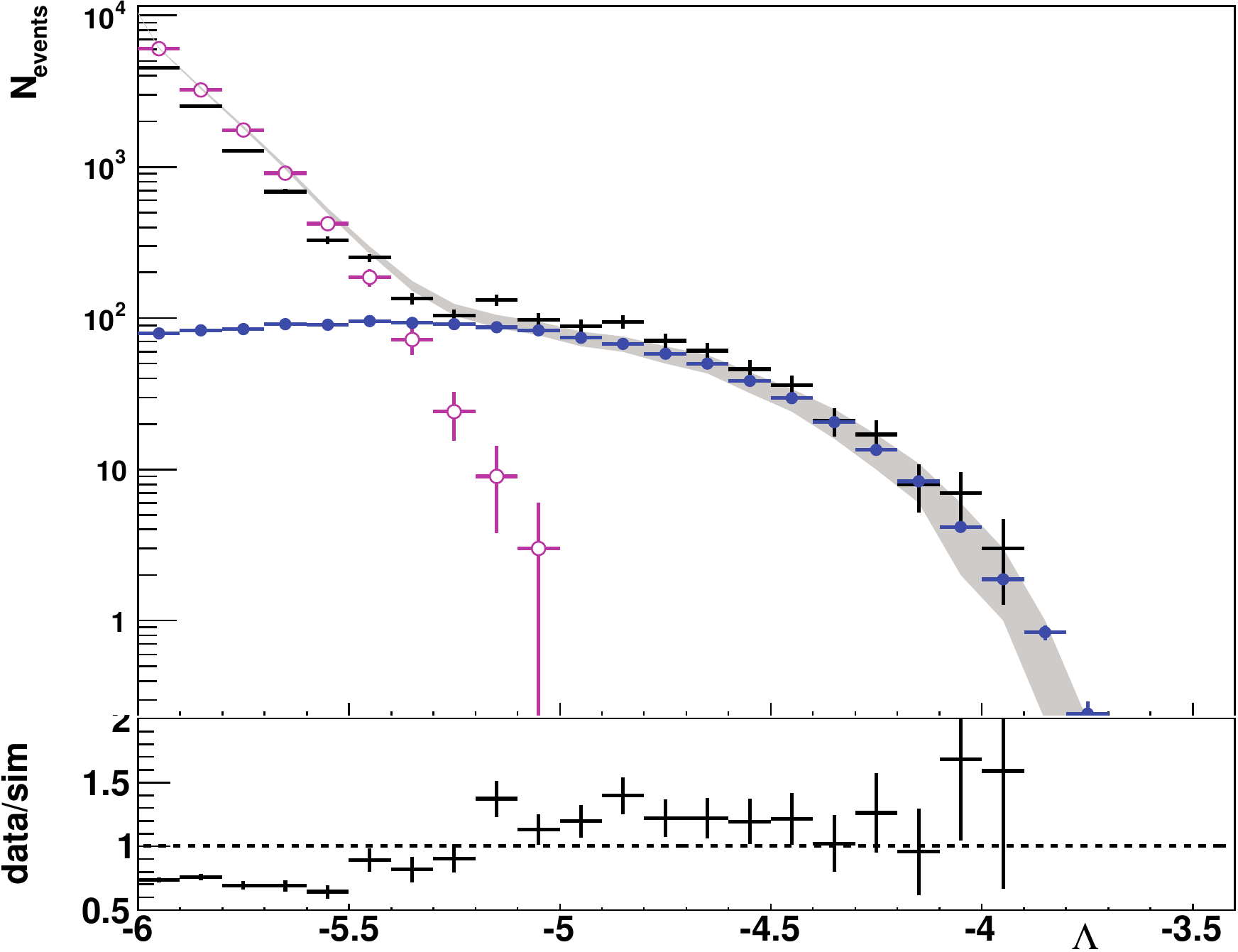}
\caption{Distribution of the fit-quality parameter $\Lambda$ for the upgoing events arriving from the three off-zones: data (black crosses), 64\% confidence area given by the total background simulation (grey area), $\nu^{\mathrm{sim}}_{\mathrm{atm}}$ (blue full cirles), $\mu^{\mathrm{sim}}_{\mathrm{atm}}$ (pink empty circles); bin-ratio of the data to the total background simulation (bottom).}
\label{figvladimir:fb_lambda}
\end{figure}

Figure~\ref{figvladimir:fb_lambda} shows a change of the main background component at $\Lambda\sim-5.35$: for $\Lambda<-5.35$ most of the events are misreconstructed atmospheric muons while for $\Lambda>-5.35$ the upgoing neutrino events are dominant. The flux of atmospheric neutrinos in the simulation is 23\% lower than observed in the data. This is well within the systematic uncertainty on the atmospheric neutrino flux and the simulation was scaled accordinly in the further analysis.
 
Table~\ref{tablevladimir:MRF} reports the optimal cuts ($\Lambda^\mathrm{cut}$, $E_\mathrm{Rec}^\mathrm{cut}$) obtained for the four chosen cutoff energies (50, 100, 500 TeV and $\infty$) of the neutrino source spectrum and the corresponding value of the average upper limit on the flux coefficient $\overline  A_{90\%}$. Additionally, the optimal cuts for $E^\mathrm{cutoff}_\nu = 100$~TeV were applied for the other neutrino-energy cutoffs. The values $\overline A_{90\%}^{100}$ are reported for comparison. As the obtained values $\overline A_{90\%}$ and $\overline A_{90\%}^{100}$ for each cutoff are similar, the 100~TeV cut was chosen for the final event selection.
\begin{table}
\caption{Optimisation results for each cutoff of the neutrino energy spectrum. Average upper limits on the flux coefficient $\overline A_{90\%}$ are presented in units of $10^{-7} \mathrm{\,GeV\,cm^{-2}\,s^{-1} sr^{-1}}$. Bold numbers highlight the cut used for the $\overline A_{90\%}$ calculation presented in the last row of the table.
}
\centering
\label{tablevladimir:MRF}
\begin{tabular}{l c c c c}
$E_\nu^\mathrm{cutoff}$ (TeV)&$\infty$&500&100&50\\
\hline 
$\Lambda^\mathrm{cut}$&-5.16&-5.14&{\bf--5.14}&--5.14\\
$\log_{10}(E_\mathrm{Rec}^\mathrm{cut}[\mathrm{GeV}])$&4.57&4.27&{\bf4.03}&3.87\\
$\overline A_{90\%}$&2.67&4.47&8.44&12.43\\
$\overline A_{90\%}^{100}$ (100 TeV cuts)&3.07&4.68&8.44&12.75\\
\end{tabular}
\end{table}
\section{Results}
The final event selection with the cut $\Lambda>-5.14$, $\log_{10}(E_\mathrm{Rec}[\mathrm{GeV}])>4.03$  was applied to the unblinded data. In the three off-zones the average number of background events $\overline n_\mathrm{bg}=(9+12+12)/3=11$ was observed. $N_\mathrm{obs}=16$ events were measured in the Fermi bubble regions. A significance 1.2 $\sigma$ as a standard deviation of the no-signal hypothesis was obtained using the method by Li \& Ma~\cite{bibvladimir:LiMa}.

The distribution of the energy estimator for both the on-zone and the average of the off-zones is presented in figure~\ref{figvladimir:unblinded_events}. A small excess of high energy events in the on-zone is seen with respect to both the average from the off-zones and atmospheric neutrino simulation. In addition, no evident clusters were observed by plotting events seen in the Fermi bubble regions.
\begin{figure}
\center
\includegraphics[width=\linewidth]{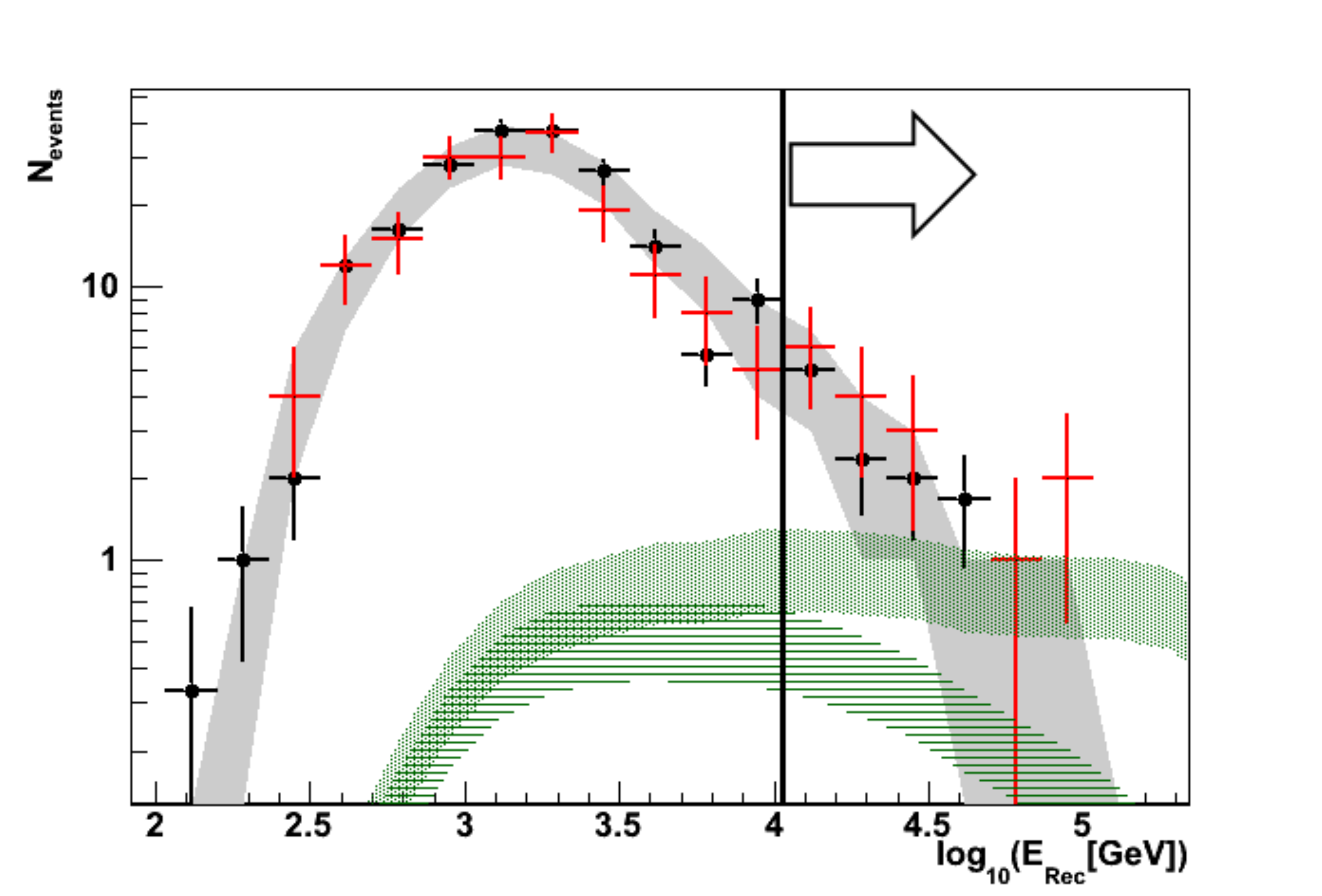}
\caption{Distribution of the reconstructed energy of the events after the final cut on $\Lambda$: events in on-zone (red crosses), average over off-zones (black circles), 64\% confidence area given by the total background simulation (grey area), expected signal from the Fermi bubbles without neutrino-energy cutoff (green area filled with dots) and 50 TeV cutoff (green area filled with horizontal lines). Black line with an arrow represents the final event selection.}
\label{figvladimir:unblinded_events}
\end{figure}

Upper limits on the number of signal events were calculated applying a Bayesian approach at 90\% C.L. using the probability distrubution with two Poisson distributions for the measurements in the on-zone and in the three off-zones. In order to account for systematic uncertainties in simulations of the signal a special study was performed in which the assumed absorption length in seawater was varied by $\pm10\%$ and the assumed optical module efficiency varied by $\pm10\%$. For each variation the number of events was calculated for each cutoff and compared with the value $s_\mathrm{sim}$ obtained using the standard simulation. The differences were calculated and summed in quadrature to obtain $\sigma_\mathrm{sim}$. A Gaussian distribution of the efficiency coefficient for the signal with mean $s_\mathrm{sim}$ and sigma $\sigma_\mathrm{sim}$ was added to the probability distribution.

The frequentist approach from~\cite{bibvladimir:Conrad} was used for a cross-check. The statistical uncertainty in the background and the efficiency uncertainty given by signal simulations were used. The results are summarised in Table~\ref{tablevladimir:flux} and are shown in figure~\ref{figvladimir:upper_limit_fluxes} . 
\begin{table}
\center
\caption{90\% C.L. upper limits on the neutrino flux coefficient $A$ for the Fermi bubbles presented in units of $10^{-7} \mathrm{\,GeV\,cm^{-2}\,s^{-1} sr^{-1}}$.}
\label{tablevladimir:flux}
\begin{tabular}{r c c c c}
$E_\nu^\mathrm{cutoff}$ (TeV)  & $\infty$  & 500 & 100 & 50 \\
\hline
$s_\mathrm{sim}$ & 2.9 & 1.9 & 1.1 & 0.7 \\
$\sigma_\mathrm{sim}$, \% & 14 & 19 & 24 & 27 \\
$A_{90\%}^\mathrm{upper}$& 5.4 & 8.7 & 17.0 & 25.9 \\
$A_{90\%}^\mathrm{upper}$ from~\cite{bibvladimir:Conrad} & 5.9 & 9.1 & 16.7 & 26.4 \\
\end{tabular}
\end{table}
\begin{figure}
\center
\includegraphics[width=\linewidth]{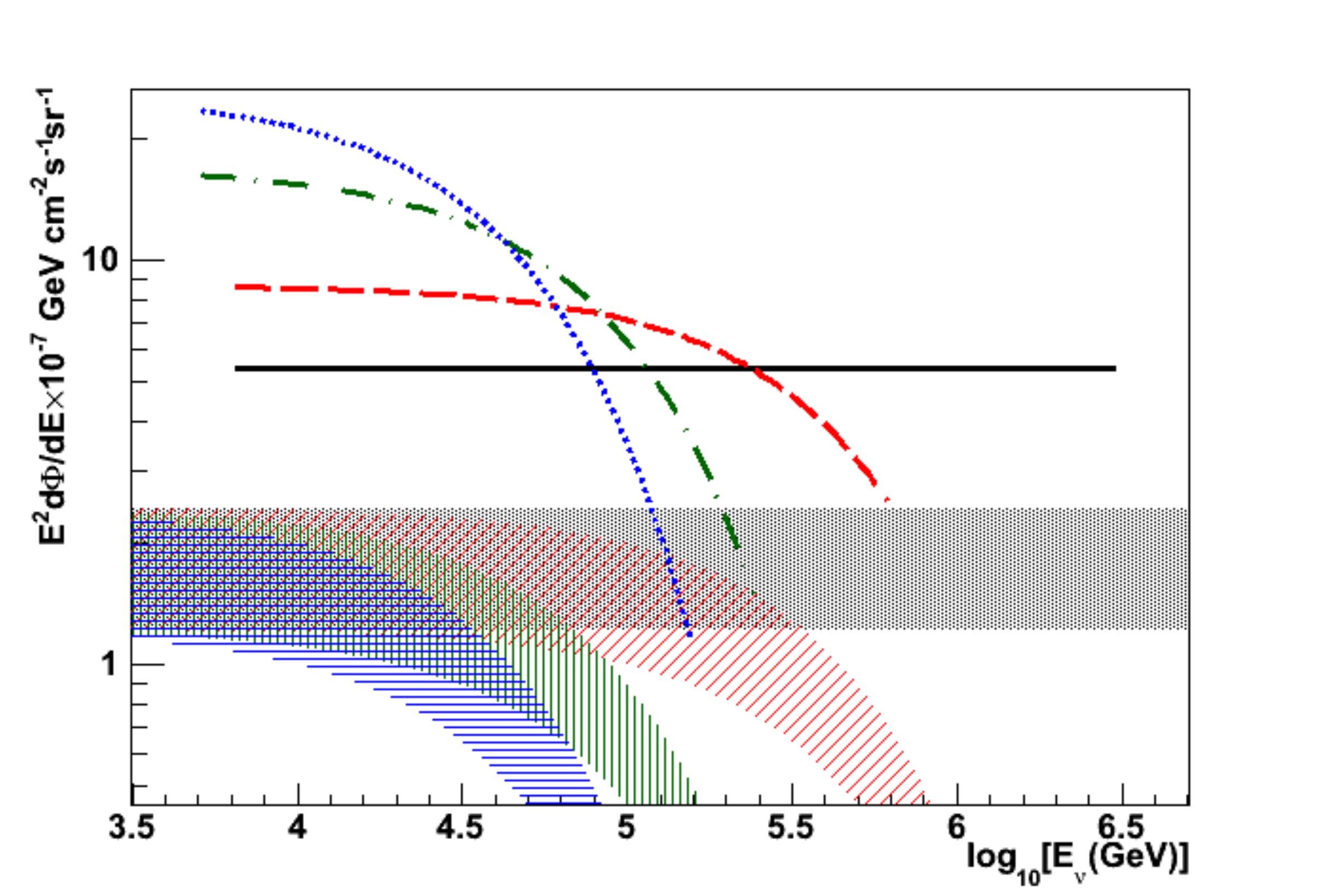}
\caption{Upper limits on the neutrino flux from the Fermi bubbles, estimated with Bayes' method for different cutoffs: no cutoff (black solid), 500~TeV (red dashed), 100~TeV (green dot-dashed), 50~TeV (blue dotted) together with the theoretical predictions for the case of a pure hadronic model (the same colours, areas filled with dots, inclined lines, vertical lines and horizontal lines correspondingly).}
\label{figvladimir:upper_limit_fluxes}
\end{figure}
\section{Conclusions}
High-energy neutrino emission from the region of the Fermi bubbles has been searched for using data from the ANTARES detector. 

An analysis of the 2008--2011 ANTARES data yielded a 1.2 $\sigma$ excess of events in the Fermi bubble regions, compatible with the no-signal hypothesis. For the optimistic case of no energy cutoff in the flux, the limits are within a factor of three of a pure hadronic model. The sensitivity will improve as more data is accumulated (more than 65\% gain in the sensitivity is expected once 2012--2016 data is added to the analysis). The next generation KM3NeT neutrino telescope~\cite{bibvladimir:km3} will provide more than an order of magnitude improvement in sensitivity.


\setcounter{figure}{0}
\setcounter{table}{0}
\setcounter{footnote}{0}
\setcounter{section}{0}
\setcounter{equation}{0}

\newpage
\id{id_mangano1}

%

\title{\arabic{IdContrib} - Gravitational Lensing and Neutrinos with the {\sc Antares} Deep-Sea Telescope}
\addcontentsline{toc}{part}{\arabic{IdContrib} - {\sl Salvatore Mangano} : Gravitational Lensing and Neutrinos%
  \vspace{-0.5cm}
 }

\shorttitle{\arabic{IdContrib} - Gravitational Lensing with Neutrinos}

\authors{Salvatore Mangano$^{1}$ and Juan Jos\'e Hern\'andez-Rey$^{1}$ for the ANTARES Collaboration.
}

\afiliations{
$^1$ IFIC - Instituto de F\'{\i}sica Corpuscular (CSIC- Universitat de
Val\`encia), E-46980 Paterna, Spain}

\email{manganos@ific.uv.es}

\abstract{
Galaxy clusters can enhance the photon
flux observed from sources behind them through the gravitational
lensing. Neutrino fluxes would also be enhanced by this
effect, and this could allow to observe sources otherwise below the detection
threshold. Neutrinos, contrary to photons, would not be  absorbed by the gravitational lens. 
Therefore, sources with a moderate observed gamma-ray flux could be interesting candidates for neutrino telescopes. 
This paper presents the outcome of a search for cosmic neutrinos 
in the direction of a selected samples of eleven of the most promising
gravitational lenses
using the data collected from 2007 to 2010 by the ANTARES telescope. 
The result of this search shows no excess of neutrino events over the
background for these considered directions.
}


%
\maketitle

\section{Introduction}
Gravitational lensing of electromagnetic radiation from distant astrophysical sources is a
well-known prediction of Einstein's general
relativity~\cite{bibmangano1:Einstein,bibmangano1:Zwicky}, 
now corroborated by numerous 
and sometimes spectacular 
observations.
The gravitational bending of light generated by mass or energy
concentrations along a light path produces magnification, distortion
and multiple images of a background source.    
Since the first detection of multiple images of the gravitationally lensed object,
Q0957+561~\cite{bibmangano1:1979Nature}, more than 100 quasar lenses have been
discovered opening a new branch of astrophysics with a large potential
for future applications.  

Gravitational lensing is also expected to act on cosmic neutrinos, and a
variety of possible configurations for neutrino lensing by massive astrophysical objects 
has been theoretically studied in the literature~\cite{bibmangano1:Gerver,bibmangano1:Elewyck2,bibmangano1:Mena,bibmangano1:Eiroa,bibmangano1:Romero}.
The possibility of neutrino gravitational lensing also offers great interest for 
large-scale Cherenkov detectors currently looking for high-energy
($\sim$ TeV and beyond) cosmic neutrinos. 
Such neutrinos are expected to be emitted along with gamma-rays by
astrophysical sources in processes 
involving the interaction of accelerated hadrons and the subsequent
production and decay of pions and kaons. 
Both gamma-rays and neutrinos will be similarly deflected by any
sufficiently massive object interposed 
between the source and the observer.

While the limited angular resolution of neutrino telescopes prevents
the observation 
of multiple-image patterns,
these instruments could be sensitive 
to an overall enhancement of the neutrino flux from a lensed
astrophysical source. 
The magnification factor so achieved could in fact allow the detection of sources
 whose intrinsic brightness is just below the sensitivity of the
 telescope. 
Known gravitational lenses with large magnification capabilities are
 interesting 
for such telescopes as they would enhance the flux from any neutrino 
source located behind them. Moreover, due to their weak interactions, 
neutrinos suffer much less than photons from absorption while crossing
 the lens. 
Even a source which is a moderate gamma-ray emitter could therefore
 become 
a promising target for neutrino telescopes provided it lies behind a powerful gravitational lens.


The ANTARES neutrino telescope \cite{bibmangano1:Ant1,bibmangano1:Antoine} is located on the
bottom of the Mediterranean Sea at a depth of 2475~m. The main
objective of the experiment is the observation of neutrinos of cosmic
origin in the Southern hemisphere sky.  Sea water is used as the
detection medium of the Cherenkov light induced by relativistic
charged particles resulting from the interaction of neutrinos.
The knowledge of the timing and amplitude of the light pulses recorded by 
the photomultiplier tubes allows to reconstruct the trajectory of the muon and to infer the arrival 
direction of the incident neutrino. The design of ANTARES is optimized for the detection 
of up-going muons produced by neutrinos which have traversed 
the Earth, in order to limit the background from down-going atmospheric muons. Its instantaneous 
field of view is $\sim\, 2 \pi\, \mathrm{sr}$ for neutrino energies 
between about 10~GeV and 100~TeV. 

ANTARES has looked for neutrinos from the direction of 51 candidate
sources~\cite{bibmangano1:AdrianMartinez:2012rp}.
No statistically significant excess has been found. 
In this letter we present a complementary search for point-like sources of cosmic neutrinos in the direction of known gravitational lenses.  As stated above, the assumption of proportionality between neutrino and gamma-ray fluxes can be relaxed in this case. The selection will therefore focus on gravitational lensing systems with a high-amplification lens and a plausible candidate neutrino source.



\section{Gravitational Lensing}

Massive neutrinos do not move along the
same null geodesic as massless photons. 
The non-zero mass brings a time delay and small differences of the angular
deflection of neutrino 
in comparison to photons. However these differences are tiny as shown below.

Massive neutrinos do not travel at the speed of light, 
hence a time delay~\cite{bibmangano1:Beacom:1999bn} with respect to photons $T=0.5s \Big(\frac{D}{10~kpc}\Big) \Big(\frac{m_{\nu}^2}{100~eV^2}\Big) \Big(\frac{100~MeV^2}{E_{\nu}^2}\Big)$, 
where $E_{\nu}$ represents the neutrino energy, $D$ is the source distance to Earth and $m_{\nu}$ is the neutrino mass. For the furthest quasar selected in Table \ref{tablemangano1:Finalunblinedlist}, the neutrino time 
delay in comparison to photons is less than a second for a neutrino mass  $m_{\nu} \sim 1$ eV  and energy  $E_{\nu} \sim 1$ TeV \cite{bibmangano1:Arnett}. Moreover the angular deflection  for a 
relativistic neutrino with mass $m_{\nu}$ that 
passes by a compact lens of mass 
$M$ with an impact parameter $b$ is \cite{bibmangano1:Crocker:2003cw}
\begin{equation}
\alpha(b)=\frac{4GM}{c^2 b}\Big(1+\frac{m_{\nu}^2}{2E_{\nu}^2}\Big),
\label{eqmangano1:lenswithmass}
\end{equation}

where $E_{\nu}$ is the energy of the neutrino, $G$ is 
the gravitational constant and $c$ is the speed of light in vacuum.

For astrophysical neutrinos with $m_{\nu} \sim 1$ eV and 
\mbox{$E_{\nu} \sim 1$ TeV}, 
the value $m_{\nu}^2/2E_{\nu}^2 \ll 1$ and the deflection angle difference 
between photon and neutrino is negligible. 
Therefore Equation \ref{eqmangano1:lenswithmass} simplifies to the deflection angle known from photons 
and the neutrino is treated as a massless particle. 
%
 
Figure \ref{figmangano1:lensing} shows a general lensing configuration~\cite{bibmangano1:Gravlens}, where
a compact lens of mass $M$ lies close to the line of sight of a source at a distance
$D_{OL}$ from the observer O. The angle $\beta$ describes the position of the source S
with respect to the optical axis which in our case is the lens direction. $D_{OS}$ is the distance between
the observer and the source and $D_{LS}$ the distance between the lens and the source. 
The angle $\theta$ describes the apparent position of the source image I with respect to the lens direction. 
Due to the gravitational field of the lens the trajectory of the neutrino is bent by the angle
$\hat{\alpha}$.

\begin{figure}
 \setlength{\unitlength}{1cm}
 \centering
 \begin{picture}(8.0,2.5)
   \put(0.0,0.0){\includegraphics[width=8.0cm]{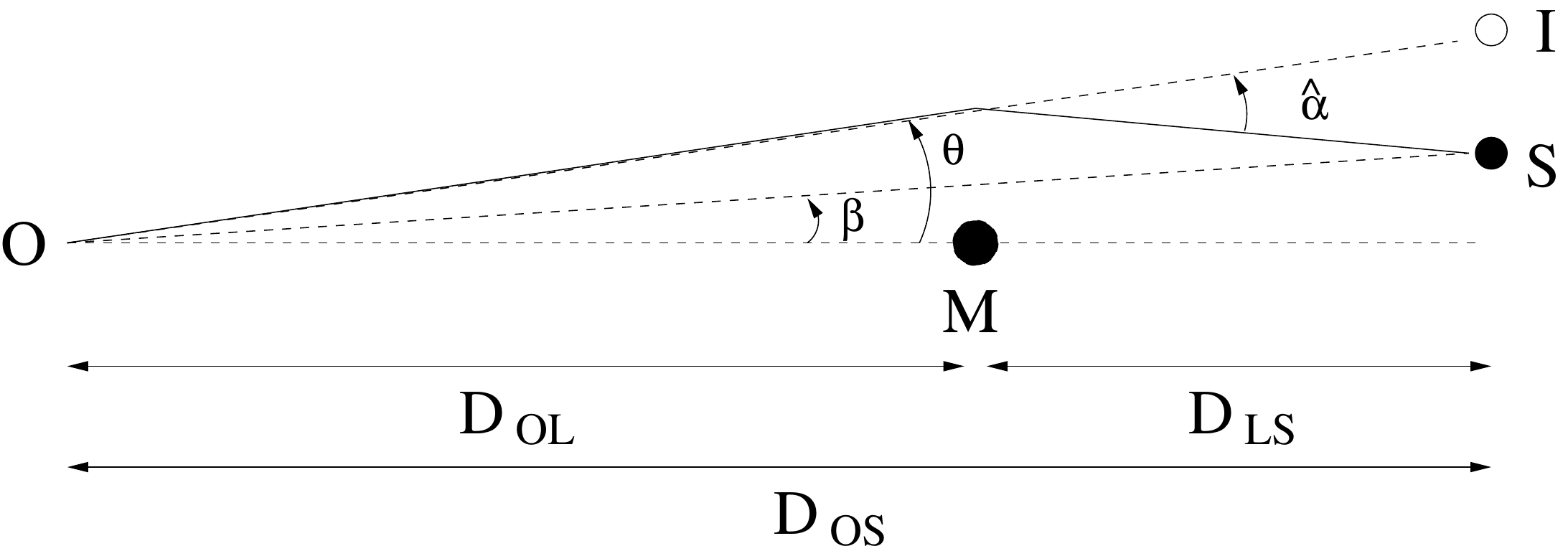}}
 \end{picture}
\vspace{-0.4cm}
\caption[Sc]{\it{The gravitational lens geometry for a compact lens M. O is the location of the observer. S is the position of the source and I is the apparent position of the source.}}
\label{figmangano1:lensing}
\end{figure}

If all the angles are small,
the lens equation~\cite{bibmangano1:Gravlens,bibmangano1:Gravlens2,bibmangano1:Gravlens3} can be written in the following way: 
\begin{equation} 
\beta = \theta - \frac{\theta_E^2} {\theta},
\label{eqmangano1:lensfinal}
\end{equation}
where the Einstein angle $\theta_E$ is 
\begin{equation} 
 \theta_E = \sqrt{\frac{4G}{c^2}} \sqrt{\frac{D_{LS}}{D_{OS}}}\sqrt{\frac{M}{D_{OL}}}.
\label{eqmangano1:einsteinangle}
\end{equation}
Equation \ref{eqmangano1:einsteinangle} can be split in 
three components: a constant term, a term related to the source and a term related to the lens.
In general space-time curvature allows more than one macroscopic path between 
source and telescope and the
source object has multiple images. 
The assumption of the lens to be point-like holds at least roughly if the Einstein angle of the lensing system 
is larger than the physical scale of the lens~\cite{bibmangano1:Crocker:2003cw}.
In particular, if the background neutrino source, the massive lensing object, and the observer lie on a straight 
line ($\beta=0$), 
the original neutrino source will appear as a ring with an angular radius $\theta_E$ around the massive lensing object. 
This kind of images are called Einstein rings.
Moreover, only sources within the Einstein angle are strongly magnified. If the source is  not exactly aligned with the 
lens we obtain from Equation \ref{eqmangano1:lensfinal} 
two solutions. 
The two images lie at opposite positions of the source with an angular separation of 
$\Delta \theta =  \theta_{+}- \theta_{-}$. 
These separations are at most of the order of
a few tens of arcseconds and cannot be resolved by the ANTARES detector. 
However the focusing effect of the gravitational lens means that the intensity of the image is magnified.
In general the magnification is given by the ratio of the solid angle $d\Omega_{\pm}$ in presence of the lens and 
the solid angle  $d\Omega_0$ in absence of the lens. 
The amplification of each image is $ \mu_{\pm} = \left| \frac{d\Omega_{\pm}}{d\Omega_0 } \right|  = \left|  \frac{\theta_{\pm} d\theta_{\pm}}{\beta d\beta} \right|$.
In the case of a double image from a point source the total 
magnification is given by
\begin{equation} 
\mu = \mu_{+} + \mu_{-} = \frac {1+\frac{1}{2} \left (\frac{\beta}{\theta_E}\right)^2} {\frac{\beta}{\theta_E} \sqrt{1+\frac{1}{4}\left(\frac{\beta}{\theta_E}\right)^2}}.
\label{eqmangano1:magnification}
\end{equation}
If $ \beta/\theta_E = 1$ the magnification is $\mu \approx 1.3$, and $\mu \approx 10.0$ if $\beta/\theta_E = 0.1$. 
The magnification is maximized in the limit of perfect alignment ($\beta/\theta_E \rightarrow$ 0).
The observed magnification $\mu_s$ of an extended sources~\cite{bibmangano1:Gravlens3} with surface 
brightness profile $I(\beta)$ is given as 
\begin{equation} 
\mu_s = \frac{1} {\int I(\beta) d^2\beta} \int I(\beta) \mu(\beta) d^2\beta,
\label{eqmangano1:extendedmagnification}
\end{equation}
where $\mu(\beta)$ is the magnification of a point source at position $\beta$. 
For an extended source of radius $R_s$, 
the angular size is $\beta_s=R_s/D_{OS}$.  
The observed magnification $\mu_s$ of a finite size source for small misalignment angle $\beta_s$ using Equation~\ref{eqmangano1:extendedmagnification} is 

\begin{equation} 
\mu_s \approx \frac {\theta_E}{\beta_s} \simeq  \sqrt{\frac{4G}{c^2}}\frac{1}{R_s}\sqrt {D_{LS} D_{OS}} \sqrt{\frac {M} {D_{OL} }}.
\label{eqmangano1:magnificationmax}
\end{equation}




Both the magnification  
(Equation \ref{eqmangano1:magnificationmax}) and the Einstein angle (Equation \ref{eqmangano1:einsteinangle}) are proportional 
to $\sqrt{\frac{M}{D_{OL}}}$.


It should be mentioned that the number of neutrinos reaching the observer from distant sources scales as $\mu/D_{OS}^2 $.
The magnification gives us the possibility to observe sources at larger distances than for unlensed sources.
If $\Phi_{min}$ is the minimal neutrino flux detectable by the ANTARES detector and $\Phi$ is the flux at Earth then we need at least 
$\Phi > \Phi_{min}$ to detect a signal. Considering $\Phi =\Phi_0 (R_0/R)^2$ for a arbitrary source with a flux $\Phi_0$ at a distance $R_0$ then it follows that  
$R < R_0 \sqrt{\Phi_0/\Phi_{min}}$. With the magnification $\mu(\beta)$
the flux is given by $\mu(\beta) \Phi_0$ and thus the condition for a source to be detected is $R < R_0 \sqrt{\Phi_0/\Phi_{min}} \sqrt{\mu(\beta)}$ or equivalently the emission threshold for detection is lowered.

\section{Neutrino Selection}
In the selection of the cosmic neutrino candidates, two kinds 
of background events are considered. First, the muons produced in
the atmosphere by cosmic rays. In order to reject them, only upgoing
events are selected, since muons cannot traverse the Earth. Still, a
small fraction of downgoing atmospheric muons (but large in number,
given the fluxes involved) produces events which are misreconstructed
as upgoing, so additional cuts in the reconstruction quality of the
events are necessary to reject these events. The second kind of
background are the atmospheric neutrinos, which are produced also in
interactions of cosmic rays in the atmosphere. This is a somehow
irreducible background, but which is expected to be diffuse. 

The selection criteria used in this paper were optimized to
search for $E^{-2}$ neutrino fluxes in the data sample taken from
29-1-2007 to 14-11-2010 extending over a livetime of 813 days. The
sample contains around 3000 neutrino candidates with a predicted
atmospheric muon neutrino purity of around 85\%. The estimated angular
resolution is 0.5$\pm$0.1 degrees. In~\cite{bibmangano1:AdrianMartinez:2012rp},
the results of a search for point sources using this sample and
selection criteria were presented, following two different strategies:
an all-sky scan and a search in the direction of 51 particularly
interesting neutrino candidate sources. No statistically significant
excess was found. The selection of those 51 sources was mainly based
on the gamma emission as observed by Fermi and HESS, since it is
expected that neutrino fluxes are proportional to gamma fluxes.  

In this paper, we study eleven additional sources where 
lensing in photons has been observed. These new sources are
located behind gravitational lenses. 
Neutrino fluxes arriving at Earth from
these sources would be enhanced. Gamma fluxes are also enhanced, but
part of them would be absorbed by the lens itself. Therefore, even if
the observed gamma fluxes observed from these sources is moderate
compared with other sources in the list used
in~\cite{bibmangano1:AdrianMartinez:2012rp}, the neutrino fluxes could be
observable, since the proportionality mentioned above would be higher,
due to photon absorption in the lens. 

\section{Potential Neutrino Lensing Systems}
\label{sec:sources}
A list of most promising neutrino lensing systems has been established on the basis of known  systems where lensing has been observed in photons, and where the lensed object is a potential neutrino source. Our sample is taken from the 
Fermi Survey~\cite{bibmangano1:Abdo:2009wu},
Chandra Survey~\cite{bibmangano1:Brandt:2001ur}, 
SDSS Quasar Lens Search (SQLS)~\cite{bibmangano1:Oguri:2006aw}, 
Cosmic Lens All-Sky Survey (CLASS) radio surveys~\cite{bibmangano1:Browne:2002yb} 
and optical CASTLES survey~\cite{bibmangano1:Kochanek}.  

The most promising neutrino lensing systems have been selected from this sample on the basis of the following criteria:
first, the system is in the field of view of ANTARES,
second, the lensed object is a known AGN,
third, the system was detected in X-ray and/or gamma-ray observations and 
fourth, if the object has not been identified as a gamma-ray emitter, a magnification factor (as taken from~\cite{bibmangano1:Mediavilla,bibmangano1:Oguri:2010rh,bibmangano1:Dai:2009bp,bibmangano1:Raychaudhury:2003cf,bibmangano1:Oguri:2012vg}) larger than 20 for at least one of the images is required. 

The criteria above allow to identify nine gravitational lenses, including 
four blazars and five strongly amplified AGNs, where in general
quadruples have the highest magnifications.  
Two galaxy clusters with particularly large Einstein angles and estimated masses 
were also added. 
The sample covers a
source redshift range, $0.6 < z_s < 3.6$, a lens redshift range, $0.2< z_l < 0.9$, and Einstein angle 
from 1 arcsecond to 53 arcseconds. 
These eleven structures are listed in Table \ref{tablemangano1:Finalunblinedlist}. 

\begin{table}[tb]
\small
\begin{center}
\begin{tabular}{|p{2.3cm}|p{0.8cm}|p{0.8cm}|p{0.4cm}|p{0.4cm}|p{0.4cm}|p{0.6cm}|}
\hline
Lens name & $\alpha~[^{\circ}]$     & $\delta~[^{\circ}]$  & Vis. & $z_s$ & $z_l$ & $\theta_E[^{\prime\prime}]$ \\
\hline
PKS 1830-211    &278.42 &-21.06 & 0.61 & 2.51& 0.89  & 1.0  \\
JVAS B0218+357  & 35.27 &+35.94 & 0.26 & 0.96& 0.68  & 0.3   \\
\hline
B1422+231   &216.16&+22.93& 0.38 & 3.62 &0.34 &  1.7 \\
B1030+074 & 158.39& +07.19 &  0.46 &1.54&0.60&   1.7 \\
\hline
RX J1131-1231    &172.97 & -12.53 & 0.56  & 0.66 & 0.30 & 3.8  \\
SDSS 1004+4112      &151.15 & +41.21 & 0.20  &1.73  & 0.68 & 16.0 \\
SDSS 0924+0219      &141.23 & +02.32  & 0.50  & 1.52 & 0.39 & 1.8  \\
RX J0911+0551    &137.86 & +05.85 & 0.46  &2.80  & 0.77 &  2.5  \\
\hline
SDSS 1029+2623      &157.31 & +26.39 & 0.35  &2.20  & 0.55 & 22.5  \\
\hline
A1689           &197.89 & -01.37 & 0.51& - &0.18  & 53 \\
A370            & 39.96 & -01.59 & 0.51& - &0.38  & 37 \\

\hline
\end{tabular}
\vspace{-0.0cm}
\caption{\it{List of selected gravitational lensing systems. We list
    the equatorial coordinates, the ANTARES visibility, the source
    redshift (if applicable), the lens redshift and the Einstein
    angle.  The first nine objects are gravitational lensed
    quasars. PKS 1830 and B0218 are gravitationally lensed quasars
    with bright flat radio spectrum and compact jets and have gamma
    emission.  B1422 and B1030 are BL Lac objects. J1131, J1004, J0924
    and J0911 have quadruple images with large magnifications and
    with X-ray emission.  J1029 is the largest separation image quasar with
    X-ray emission. The galaxy clusters A1689 and A370 are well known
    gravitational lenses with large Einstein angles and large
    estimated masses.}}
\label{tablemangano1:Finalunblinedlist}
\end{center}
\end{table}

According to the unified model of AGNs, blazars are sources with one
of the jets pointing towards the Earth. The blazar emission is predominantly
of non-thermal origin, showing a typical two-hump spectrum at
radio and gamma-ray energies.  The relativistic jets from AGNs can produce
strong shocks which convert a fraction of the kinetic energy into
energy of relativistic particles. The decay products of such particles
can lead to neutrino emission~\cite{bibmangano1:Becker}.  

The blazars PKS 1830-211 and JVAS B0218+357 are gravitationally lensed,
with double images. Both blazars are Fermi detections with pretty
bright flat spectrum. A large absorption in the corresponding lenses
has been measured in both cases~\cite{bibmangano1:Falco:1999jc}.  PKS 1830-211 has
variable gamma ray fluxes and shows substantial
variability only above 10-100 MeV~\cite{bibmangano1:Donna}. The radio-loud blazar
is at a redshift of $z_s=2.51$ and is gravitationally lensed by a spiral
galaxy at $z_l=0.89$.  The JVAS B0218+357 consists of two images of a
background blazar and an extended jet at a redshift of $z_s=0.96$,
while the intervening and lensing spiral galaxy has redshift
$z_l=0.68$. Fermi found recently that JVAS B0218+357 had flaring states. 

The objects B1422+231 and B1030+074 are both in the Multifrequency
Catalog of Blazars~\cite{bibmangano1:Massaro:2010si} (MGL2009) and seem to be BL
Lac objects. BL Lac is a blazar subtype with rapid and large amplitude
flux variability with jet structure. In MGL2009 the B1422+231 blazar
is categorized as a flat-spectrum radio quasars, with an optical
spectrum showing broad emission lines and dominant blazar
characteristics, while B1030+074 is categorized as a blazar of
uncertain type.  B1422+231 is a four-image quasar at redshift $z_s=3.62$
and lensed by a galaxy at redshift $z_l=0.34$, while B1030+074 is viewed
as two spot-like feature with the source at redshift of $z_s=1.54$ and
lens at redshift of $z_l=0.60$.

We select the quadruply lensed quasars RX J1131-1231, SDSS 1004+4112,
SDSS 0924+0219 and RX J0911+0551 and we include SDSS 1029+2623 which is
the quasar with the largest known image separation.  The two large separation quasars
SDSS 1004+4112 and SDSS 1029+2623 are lensed by a galaxy cluster,
while the others quasars are lensed by a galaxy.  The cluster lens
SDSS 1029+2623 at $z_l=0.58$ is the largest known image separation
quasar. It consists of three images at $z_s=2.20$ with a maximum image
separation of about $22.5^{\prime\prime}$.  The other large image separation quasar
SDSS 1004+4112 with four images of a quasar at $z_s=1.74$ is lensed by a
foreground galaxy cluster at $z_l=0.68$.  The maximum separation between
the quasar images is approximately $16^{\prime\prime}$.  RXJ1131-1231 is one of
the nearest gravitationally lensed AGN. 

Finally two well known galaxy clusters are included without any
known neutrino candidate source behind but particularly powerful as
gravitational lenses: A1689 and A370. 
Both are relative close and characterized by large Einstein angles and estimated masses
($M>10^{15}M_{\odot}$). As seen in Equation \ref{eqmangano1:magnificationmax} the magnification is proportional to $\sqrt{M/D_{OL}}$ and these two galaxy clusters 
have large $\sqrt{M/D_{OL}}$ values from the known very massive galaxy
clusters.  A1689 is the largest known gravitational lens with 
some 100 images of 30 multiply lensed background galaxies. The lensed
galaxies have a wide redshift range from 1 to 5.5. This galaxy cluster
has the largest Einstein angle of all known massive
lensing clusters with an angle of 53 arcseconds~\cite{bibmangano1:Broadhurst:2004}.

\section{Search for Neutrino Emission from Gravitational Lensed Sources}
\label{sec:addobjects}

The analysis has been performed using an unbinned maximum likelihood
method as described in~\cite{bibmangano1:AdrianMartinez:2012rp}.  This method uses
the information of the event direction, the number of hits produced
by the track and the angular error estimate. 
For each source, the position of the cluster is fixed as
the direction of the source and the likelihood function is maximized
with respect to the number of signal events ($n_s$).  After the
likelihood maximization, a likelihood ratio is used as a test
statistic.  The test statistic ($\lambda$) is the ratio between the
value of the likelihood given by the maximization and the likelihood
computed for the only-background case. Before unblinding, many
pseudo-experiments are generated both for the case of only background
and for the case of background with some signal added. This allows us
to build the corresponding test statistic distributions. After
unblinding, the observed value of $\lambda_{\rm obs}$ is obtained and
compared with the $\lambda$ distribution for the only-background case,
which provides the $p$-value, i.e. the probability that the background
has produced a value of $\lambda$ as large or larger than
$\lambda_{\rm obs}$. In the absence of a significant excess of
neutrinos above the expected background, the upper limit on the
neutrino flux is obtained, based on the $\lambda$ distributions
generated with different signal levels. These flux upper limits are
calculated at 90\% confidence level and using the approach from
Feldman \& Cousins.
 
The results from the search for neutrino emission from the direction
of the eleven gravitationally lensed directions are presented in
Table~\ref{tablemangano1:Listresults}. No significant excess has been found
around the direction of the gravitational lenses. 
The SDSS 1004+4112
source with equatorial coordinates $\alpha=151.15^{\circ},
\delta=+41.21^{\circ}$ has the highest excess,with a post-trial $p$-value of 49\%. 

\begin{table}[tb]
\small
\begin{center}
\begin{tabular}{|p{2.3cm}|p{0.8cm}|p{0.8cm}|p{0.4cm}|p{0.4cm}|p{0.65cm}|p{0.5cm}|}
\hline
Lens name &  $\alpha~[^{\circ}]$     & $\delta~[^{\circ}]$  &  $\lambda_{\rm obs}$ & $n_s$ & {\tiny $p$-value} &  $\phi_{\nu}^{90\%}$  \\
\hline
SDSS 1004+4112  &151.15 &+41.21 & 0.35  &0.63  & 0.49   & 15.7  \\
RX J0911+0551   &137.86 &+05.85 & 0.09  &0.41  & 0.63   &  8.6  \\
A370            & 39.96 &-01.59 & 0.02  &0.18  & 0.73   &  7.6 \\
A1689           &197.89 &-01.37 & 0.00  &0.00  & 1.00            &  4.4 \\
B1422+231       &216.16 &+22.93 & 0.00  &0.00  & 1.00            &  7.0 \\
SDSS 1029+2623  &157.31 &+26.39 & 0.00  &0.00  & 1.00            &  6.6 \\
B1030+074       &158.39 &+07.19 & 0.00  &0.00  & 1.00            &  5.6 \\
SDSS 0924+0219  &141.23 &+02.32 & 0.00  &0.00  & 1.00            &  4.6 \\
PKS 1830-211    &278.42 &-21.06 & 0.00  &0.00  & 1.00            &  3.8 \\
JVAS B0218+357  & 35.27 &+35.94 & 0.00  &0.00  & 1.00            &  8.2 \\
RX J1131-1231   &172.97 &-12.53 & 0.00  &0.00  & 1.00            &  4.6 \\
\hline

\end{tabular}
\vspace{-0.25cm}
\caption{\it{Results from the search for high-energy neutrinos from
    sources behind gravitational lenses. The equatorial coordinates,
    the observed test statistic value $\lambda_{obs}$, the fitted
    number of signal events $n_s$, the p-value and the 90\% C.L. upper
    limit on the $E^{-2}$ flux in units of $10^{-8} GeV^{-1} cm^{-2} s^{-1}$ are
    given for the eleven selected directions.}}
\label{tablemangano1:Listresults}
\end{center}
\vspace{-0.75cm}
\end{table}

\section {Summary}
The search for point sources with ANTARES data corresponding to the
2007-2010 period has been extended including sources which are
magnified by gravitational lenses. These lenses enhance both the
photon and neutrino fluxes from background sources. Neutrinos,
contrary to photons, are not absorbed in the lens, so the
proportionality factor often assumed between neutrino and photon
fluxes is larger for these cases. In this paper the eleven most
promising gravitational lensed neutrino sources have been selected.
Nine objects are gravitational lensed quasars which are potential
neutrino sources according to theory. In addition we also looked in
the directions of two nearby and massive galaxy clusters with no
known neutrino candidate source behind but which offer particularly
magnification factors.

A likelihood ratio method has been used to search for clusters of
events correlated with the position of these eleven lenses.  No excess of
neutrino events is found at any of the selected locations and upper
limits on the neutrino flux intensity have been given.

{\footnotesize{\bf %
\noindent
Acknowledgment: }I would like to thank J. A. Mu\~noz and E. E. Falco for discussions of experimental aspects of gravitational lensing, and O. Mena and G. E. Romero for discussions of theoretical aspects. I gratefully acknowledge the support of the JAE-Doc postdoctoral programme of CSIC. This work has also been supported by the following Spanish projects: FPA2009-13983-C02-01, MultiDark Consolider CSD2009-00064, ACI2009-1020 of MICINN and Prometeo/2009/026 of Generalitat Valenciana.}

\setcounter{figure}{0}
\setcounter{table}{0}
\setcounter{footnote}{0}
\setcounter{section}{0}
\setcounter{equation}{0}

\newpage
\id{id_schussler2}




\title{\arabic{IdContrib} - 2pt correlation analysis of ANTARES data}
\addcontentsline{toc}{part}{\arabic{IdContrib} - {\sl Fabian Sch\"ussler} : 2pt correlation analysis of {\sc Antares} data
}

\shorttitle{\arabic{IdContrib} - {\sc Antares} 2pt correlation}

\authors{
Fabian Sch\"ussler$^{1}$ for the {\sc Antares} Collaboration.
}

\afiliations{
$^1$ Commissariat \`a l'\'energie atomique et aux \'energies alternatives / Institut de recherche sur les lois fondamentales de l'Univers \\
}

\email{fabian.schussler@cea.fr}

\abstract{Clustering of high energy neutrino arrival directions or correlations with known astrophysical objects would provide hints for their astrophysical origin. The two-point autocorrelation method is sensitive to a large variety of cluster morphologies and, due to its independence from Monte Carlo simulations, provides complementary information to searches for the astrophysical sources of high energy muon neutrinos. Using 4 years of data from the ANTARES neutrino telescope, we present an analysis of the autocorrelation of neutrino candidate events and cross-correlations with catalogues compiled using data of different messengers and wavelengths like high energy gamma rays.}

\maketitle


\section{Introduction}
The key question to resolve the long standing mystery of the origin of cosmic rays is to locate the sources and study the acceleration mechanisms able to produce fundamental particles with energies orders of magnitude above man-made accelerators. Over the last years it has become more and more obvious that multiple messengers will be needed to achieve this task. Fundamental particle physics processes like the production and subsequent decay of pions in interactions of high energy particles predict that the acceleration sites of high energy cosmic rays are also sources of high energy gamma rays and neutrinos. The detection of astrophysical neutrinos and the identification of their sources is one of the main aims of large neutrino telescopes operated at the South Pole (IceCube), in Lake Baikal and in the Mediterranean Sea (ANTARES). 
 
\subsection{The ANTARES neutrino telescope}
\begin{figure}[!h]
  \centering
  \includegraphics[width=0.45\textwidth]{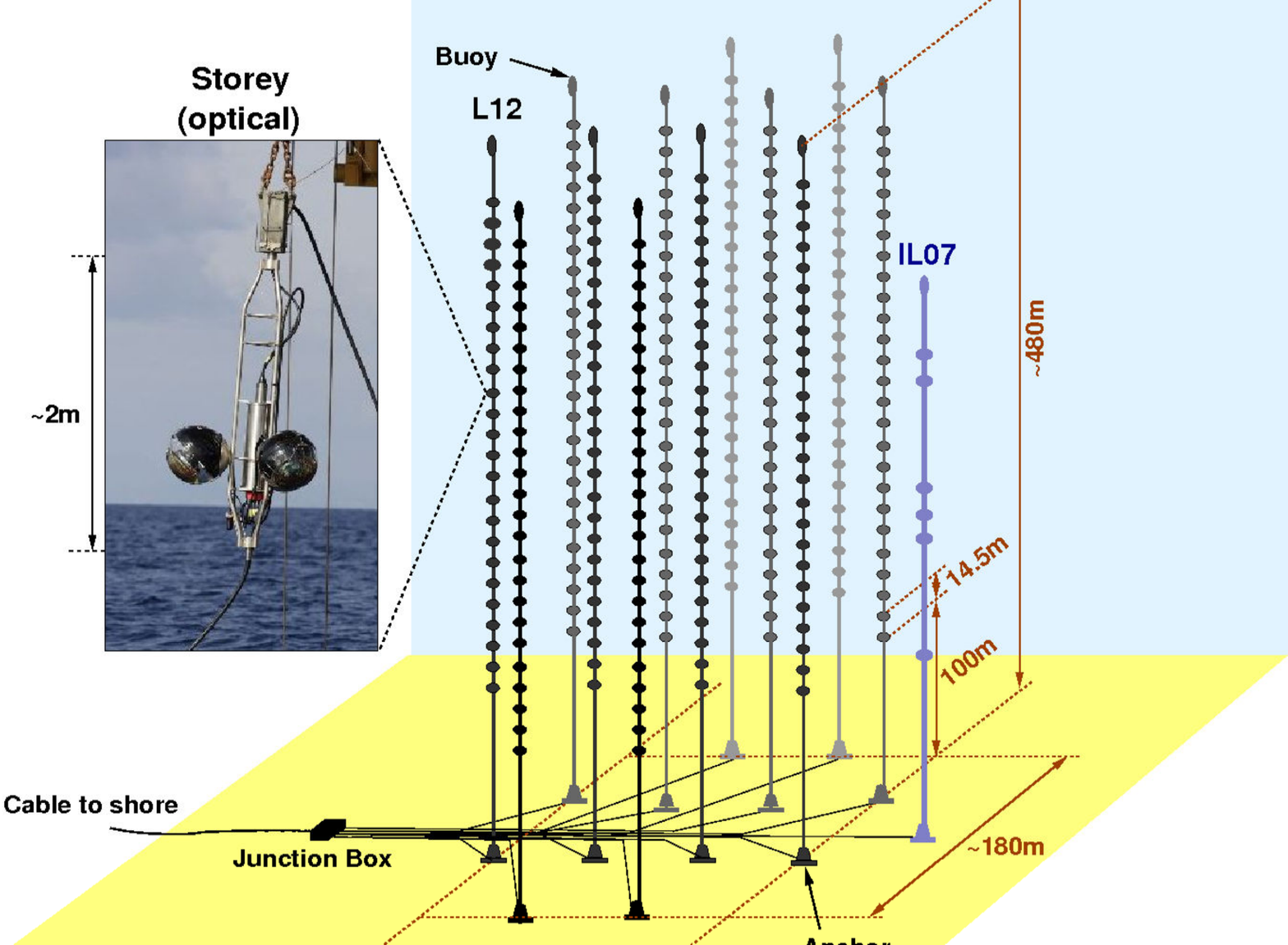}
  \caption{Schematic view of the ANTARES detector.}
\label{figschussler2:layout}	
 \end{figure}
 \noindent
Whereas physics data taking started already during the deployment phase, the ANTARES detector~\cite{bibschussler2:Antares_DetectorPaper} became fully equipped and operational in 2008. The detector is composed of 12 detection lines placed at a depth of 2475m off the French coast near Toulon. The detector lines are about 450m long and hold a total of 885 optical modules (OMs), 17'' glass spheres housing each a 10'' photomultiplier tube. The OMs look downward at $45^\circ$ in order to optimize the detection of upgoing, i.e. neutrino induced, tracks. The geometry and size of the detector makes it sensitive to neutrinos in the TeV-PeV energy range. A schematic layout is shown in Fig.~\ref{figschussler2:layout}.\\
\noindent
The neutrino detection relies on the emission of Cherenkov light by high energy muons originating from charged current neutrino interactions inside or near the instrumented volume. All detected signals are transmitted via an optical cable to a shore station, where a farm of CPUs filters the data for coincident signals or {\it hits} in several adjacent OMs. The muon direction is then determined by maximising a likelihood which compares the times of the hits with the expectation from the Cherenkov signal of a muon track.
\begin{figure*}[!t]
\centerline {
   \includegraphics[width=0.48\textwidth]{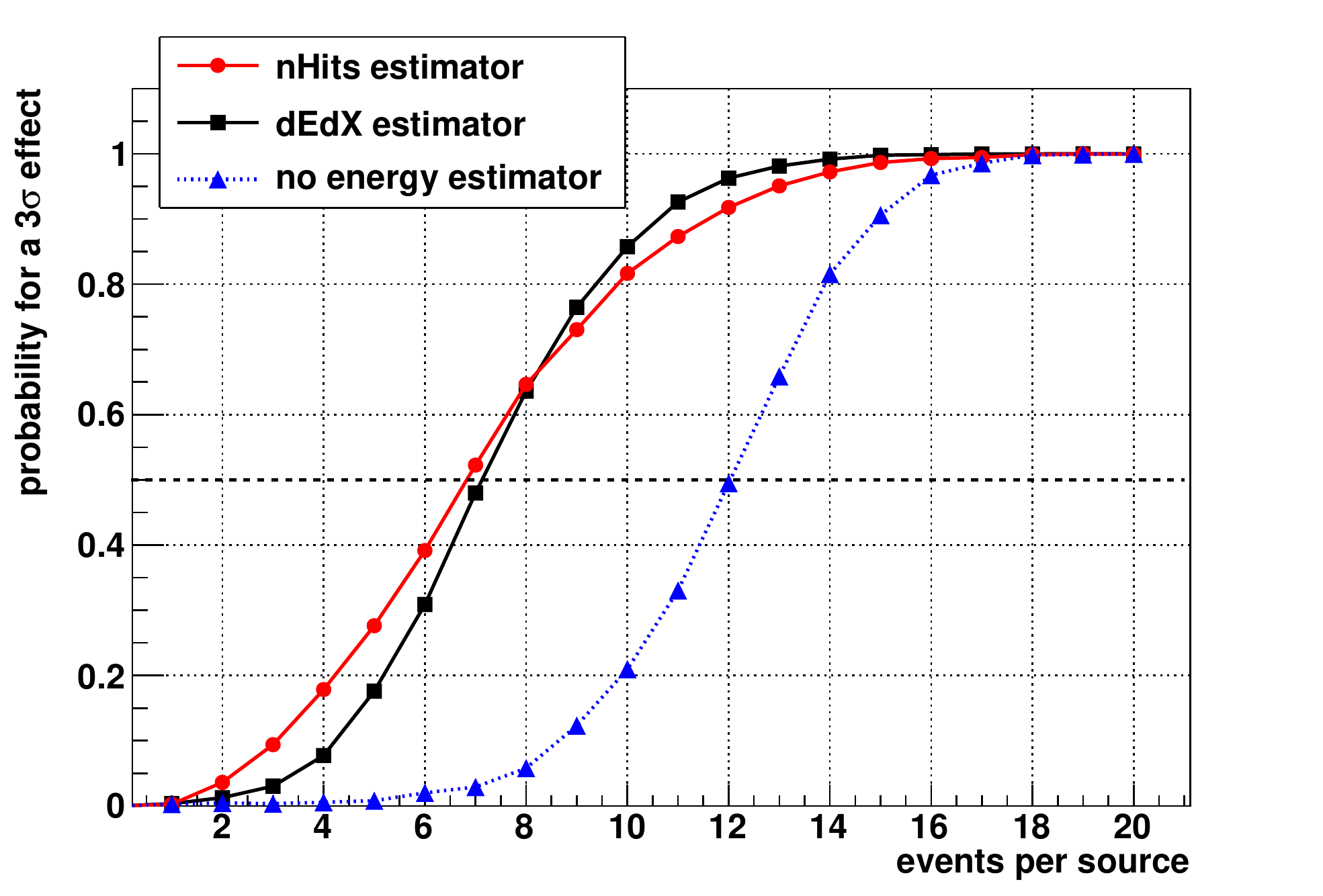}
   \hfill
     \includegraphics[width=0.48\textwidth]{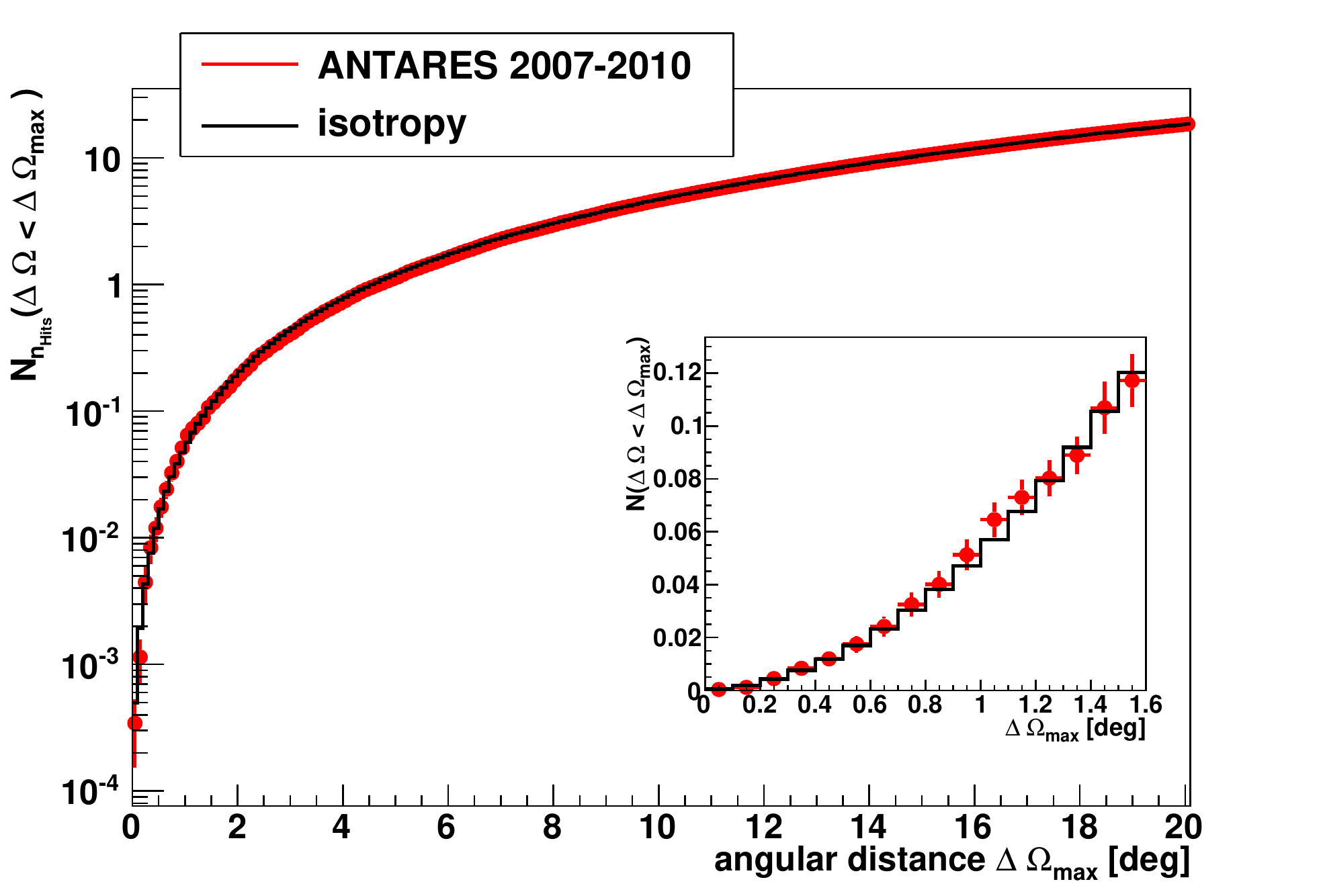}
}
\caption{Left plot: Probability to detect a single point-like source as a function of its luminosity. The blue triangles denote the standard autocorrelation method without an energy estimator. The black squares show the performance including the $\mathrm{d}E/\mathrm{d}X$ estimator~\cite{bibschussler2:ICRC2013_dEdX} and the red circles denote the final method using the $n_\mathrm{Hits}$ estimator. Right plot: Cumulative autocorrelation function of data taken with the ANTARES neutrino telescope in 2007-2010. The red markers denote the ANTARES data and the black histogram represents the reference distribution expected for an isotropic dataset. The inset shows an enlarged view for small angular distances.}
   \label{figschussler2:Performance_AutoCorrelation}
 \end{figure*}
\noindent%

\subsection{Astrophysical neutrinos}
Two main backgrounds for the search for astrophysical neutrinos can be identified: downgoing atmospheric muons which have been mis-reconstructed as upgoing and atmospheric neutrinos originating in cosmic ray induced air showers at the opposite side of the Earth. Depending on the requirements of the analysis both backgrounds can at least partially be discriminated using various parameters like the quality of the event reconstruction or an estimator for the deposited energy~\cite{bibschussler2:ICRC2013_dEdX}.

In addition, analysing the reconstructed arrival directions of the events allows to search for an excess over the uniform atmospheric backgrounds. Despite important efforts, no clear signature for point-like sources of astrophysical neutrinos has been found so far~\cite{bibschussler2:Amanda_PointSources2009, bibschussler2:IceCube_PointSources2011, bibschussler2:ANTARES_PointSources2010, bibschussler2:ICRC2013_ANTARESExcess, bibschussler2:ICRC2013_PointSources}. Both the distribution and morphologies of sources potentially emitting neutrinos in the TeV energy range are yet unknown but are possibly very inhomogeneous with most of them being located in the Galactic disk and spatially extended (e.g. shell type supernova remnants). It seems therefore interesting to study the intrinsic clustering of the arrival directions of neutrino candidates without trying to localize the underlying sources. In this analysis biases are naturally avoided as no prior information about the potential sources is required. Covering a large angular range, i.e. neutrino emission regions of very different sizes, this study complements the searches for point like sources and, if successful, would provide hints for underlying, yet unresolved, source morphologies and source distributions. Here we present an improved autocorrelation method for this task.

 
\section{Autocorrelation analysis}
\subsection{Method}
The most commonly used method to detect intrinsic clusters within a set of $N$ events is the standard 2-point autocorrelation distribution. It is defined as the differential distribution of the number of observed event pairs $N_\mathrm{p}$ in the dataset as a function of their mutual angular distance $\Delta \Omega$. This technique has been applied to ANTARES data in~\cite{bibschussler2:ICRC2011_AutoCorrelation}. Here we extend and improve this method by using an estimator of the event energy. To suppress statistical fluctuations that would reduce the sensitivity of the method, we analyse the cumulative autocorrelation distribution defined as \begin{equation}
\mathcal{N}_\mathrm{p} (\Delta \Omega) = \sum\limits_{i=1}^{N} \sum\limits_{j=i+1}^{N} w_{ij} \times H(\Delta \Omega_{ij} - \Delta \Omega), \label{eqschussler2:autocor}
\end{equation}
where $H$ is the Heaviside step function. The weights $w_{ij}=w_i \times w_j$ are calculated using the individual event weights $w_i= 1 - \int_{\hat{E}_i}^{\inf} f(\hat{E})$, where $f(\hat{E})$ is the normalized distribution of the energy estimator $\hat{E}$ and has been derived from Monte Carlo simulations. Higher values of the estimator $\hat{E_i}$ mean that the event is more likely to originate from an astrophysical neutrino flux in contrast to the atmospheric flux that dominates at lower energies. This is represented by a higher event weight $w_i$. Modifying the standard autocorrelation by these weights leads to a significant increase of sensitivity to detect clustering of events of astrophysical origin. This improvement is illustrated in the left plot of Fig.~\ref{figschussler2:Performance_AutoCorrelation}.

\begin{figure*}[!th]
  \centerline{
  \includegraphics[width=0.48\textwidth]{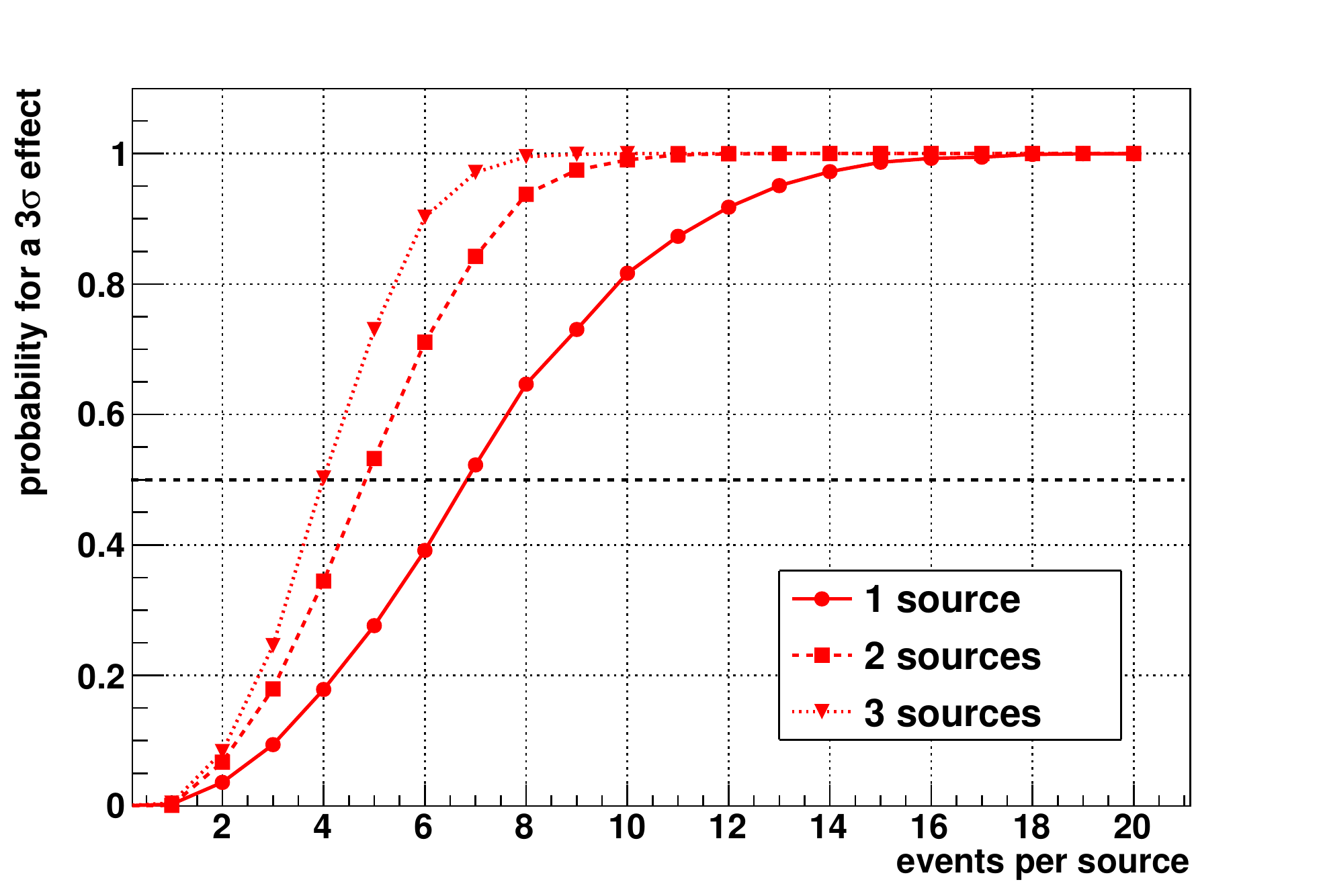}
  \hfill
  \includegraphics[width=0.48\textwidth]{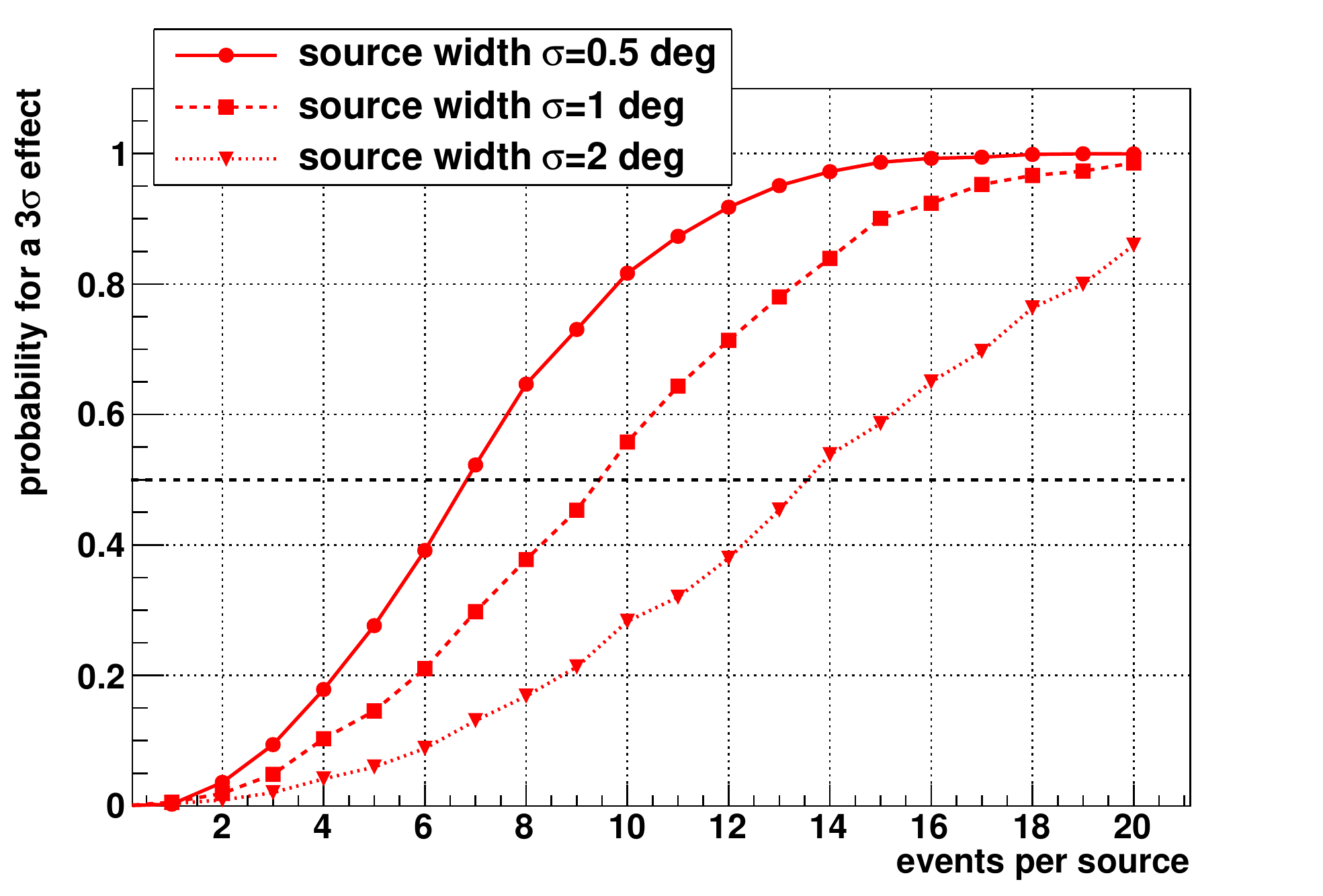}
          }	
  \caption{Probability for a $3~\sigma$ effect using the improved autocorrelation method exploiting the $n_\mathrm{Hit}$ energy estimator. Left plot: Dependence on the number of sources in the visible sky. Right plot: Dependence on the extension of a single source modelled by 2-dimensional Gaussians.}
  \label{figschussler2:nHitsPerformance}
 \end{figure*}

\subsection{Data set}
The analysed data set has been recorded by the ANTARES neutrino telescope between 2007 and 2010. The event selection criteria have been optimized by means of Monte Carlo simulations to yield the best average upper limit on the neutrino flux in the search for point like sources~\cite{bibschussler2:ANTARES_PointSources2010}. They contain a cut on the reconstructed zenith angle $\theta > 90^\circ$, a requirement on the reconstruction quality parameter $\Lambda>-5.2$ as well as a cut on the estimated angular uncertainty of the track reconstruction $\beta < 1^\circ$. A total of 3058 neutrino candidates are found in 813 days of effective lifetime. Following Eq.~\ref{eqschussler2:autocor}, the cumulative autocorrelation distribution of the selected events has been determined. It is shown as the red markers in the right plot of Fig.~\ref{figschussler2:Performance_AutoCorrelation}.

\subsection{Reference autocorrelation distribution}\label{secschussler2:reference}
To detect structures in the sky distribution of the selected events, we need a reference autocorrelation distribution to compare with. This reference is determined by scrambling the data themselves, a method which allows to reduce uncertainties introduced by the use of Monte Carlo simulations. While keeping the pairs of local coordinates zenith/azimuth in order to avoid losing information about possible correlations between them, the detection time is drawn randomly from another event within the same detector configuration to keep track of the changing asymmetry of the detector due to its construction and maintenance. Using all selected events, a randomized sky map with the same coverage as the ANTARES data is constructed. 

This randomized sky is then analysed in exactly the same way as the data to derive the autocorrelation function. The weights are drawn randomly from distributions derived from large statistic Monte Carlo simulations reproducing the actual data taking conditions including for example the time dependent  background fluctuations induced by bioluminescence. This use of simulations is necessary in order to avoid biases induced by the limited statistics of the data sample. The randomization process is performed about $10^6$ times and the derived autocorrelation distributions are averaged in order to suppress statistical fluctuations. 

\subsection{Comparison between data and reference}
Structures in the sky distribution of our data will show up as differences between the autocorrelation distribution of the data and the reference distribution. The comparison between them is performed by using the formalism introduced by Li\&Ma~\cite{bibschussler2:LiMa}. This formalism results in raw significances as a function of the cumulative angular scale. As the comparison is performed bin-by-bin and as we scan over different angular scales, this result has to be corrected for the corresponding trial factor. To limit the number of trial we scan only up to $25~\mathrm{deg}$, a scale which includes most known extended sources and emission regions. In addition we apply the method proposed by Finley and Westerhoff~\cite{bibschussler2:FinleyWesterhoff2004} and perform about $10^5$ pseudo experiments in which the autocorrelation distributions of randomized sky maps are compared with the reference distribution. The final p-value is calculated as the probability to obtain the same or higher raw significance as the one found in the data.\\
 
\subsection{Performance and sensitivity}
The performance of the algorithm has been determined using mock data sets for which we scrambled the selected data events as described above. While keeping the total number of events constant we added predefined source structures with various sizes and source luminosities taking into account the angular resolution of the detector of about $0.5~\deg$. These mock data sets where then analysed in exactly the same way as described above. Results for a single point-like source are shown in the left plot of Fig.~\ref{figschussler2:Performance_AutoCorrelation}. This analysis has been performed using two different energy estimators: the number of hits used during the final step of the event reconstruction as in the search for point like sources~\cite{bibschussler2:ANTARES_PointSources2010}, as well as a recently developed estimator exploiting the correlation between the energy deposit $\mathrm{d}E/\mathrm{d}X$ and the primary energy~\cite{bibschussler2:ICRC2013_dEdX}. Both provide very similar results, but, as shown in the left plot of Fig.~\ref{figschussler2:Performance_AutoCorrelation}, the $n_\mathrm{Hit}$ shows a slightly better performance for weak sources and has therefore been retained for the final analysis.

Compared to a dedicated, likelihood based search for a point like excess in the same dataset~\cite{bibschussler2:ANTARES_PointSources2010} the sensitivity of the autocorrelation analysis is slightly worse for a single source in the sky visible by ANTARES. The 2pt correlation method requires one signal event in addition to obtain a $3~\sigma$ detection with a $50~\%$ probability. On the other hand it outperforms the algorithm optimized for the localization of point-like sources as soon as several, weak sources are available, a fact which further underlines the complementarity of the two methods. Another complementarity is its ability to detect extended emission regions. Both cases are shown in Fig.~\ref{figschussler2:nHitsPerformance}.\\[0.4ex]

\subsection{Results and discussion}
The improved autocorrelation analysis using the $n_\mathrm{Hit}$ energy estimator has been applied to the 3058 selected neutrino candidate events recorded by the ANTARES neutrino telescope between 2007 and 2010. The comparison of the cumulative autocorrelation distribution between data and the expectation from an isotropic distribution is shown in the right plot of Fig.~\ref{figschussler2:Performance_AutoCorrelation}. The maximum deviation between the data and the reference distribution is found for an angular scale $<1.1^\circ$. Correcting for the scanning trial factor this corresponds to a p-value of $9.6~\%$ and is therefore not significant. We conclude that the analysed dataset does not contain significant clusters in addition to the $2.2~\sigma$ point-like excess that has been detected in the dedicated search~\cite{bibschussler2:ANTARES_PointSources2010, bibschussler2:ICRC2013_ANTARESExcess}.

\section{2pt correlation with external catalogues}
\subsection{Introduction}
No astrophysical neutrino source has been identified by current (and past) neutrino telescopes. One way to improve the sensitivity for these detectors is to rely on the connection with other messengers or source scenarios. Here we present a first search for a global correlation between neutrinos and high energy gamma rays and with the matter distribution in the local universe represented by the distribution of galaxies. We extend the autocorrelation function described in Eq.~\ref{eqschussler2:autocor}, to become a 2pt correlation between the $N$ neutrino candidates and an external data set of $n$ objects or sources: 
\begin{equation}
\mathcal{N}_\mathrm{p} (\Delta \Omega) = \sum\limits_{i=1}^{N} \sum\limits_{j=1}^{n} w_{i} \times \hat{w}_{j} \times H(\Delta \Omega_{ij} - \Delta \Omega), \label{eqschussler2:2ptcorr}
\end{equation}
The methods for the calculation of the reference distribution for an isotropic neutrino dataset, the comparison with the data and the correction for trial factors with pseudo experiments is performed in the same way as described above.

\begin{figure*}[!t]
  \centerline{
  \includegraphics[width=0.46\textwidth]{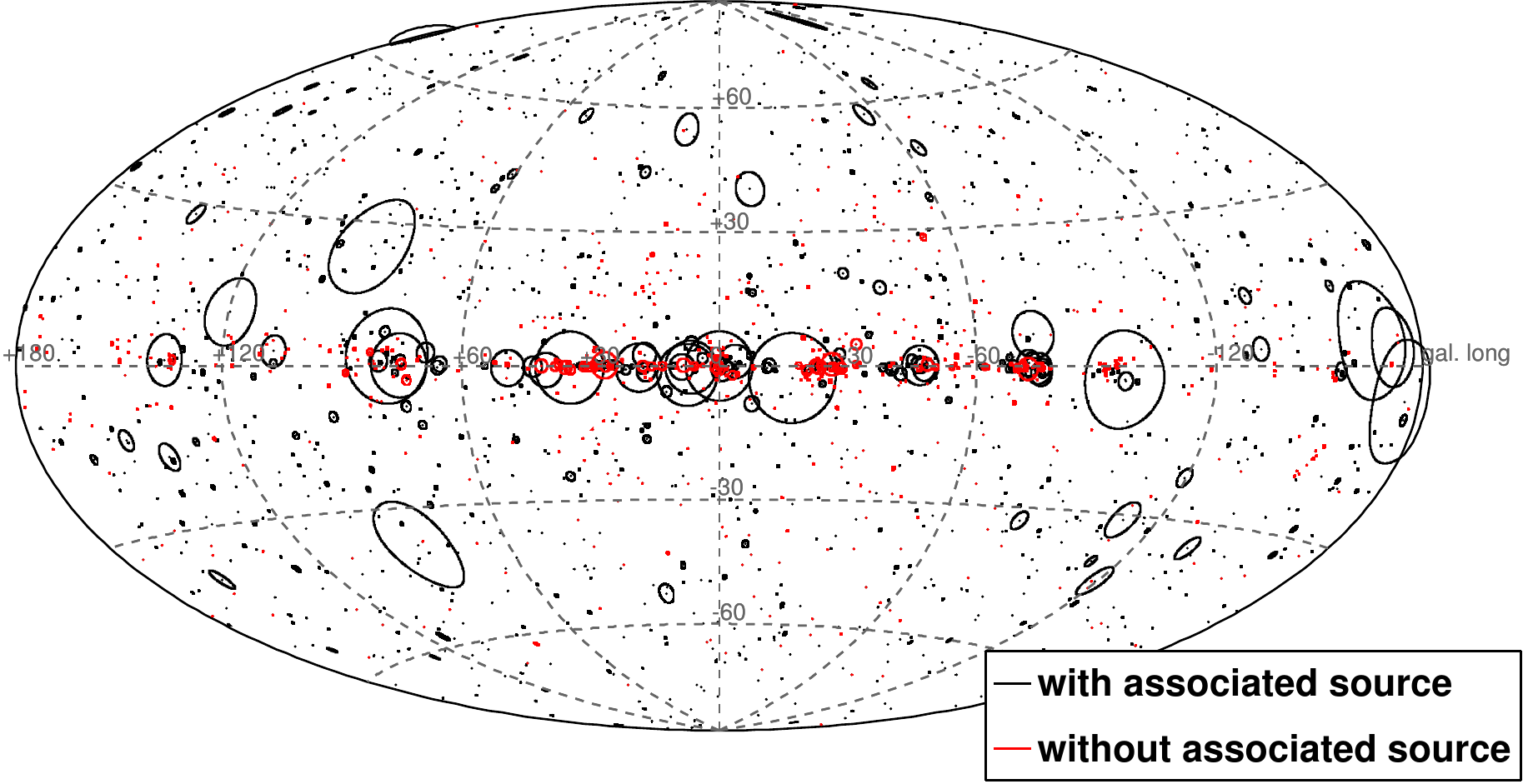}
  \hfill
  \includegraphics[width=0.46\textwidth]{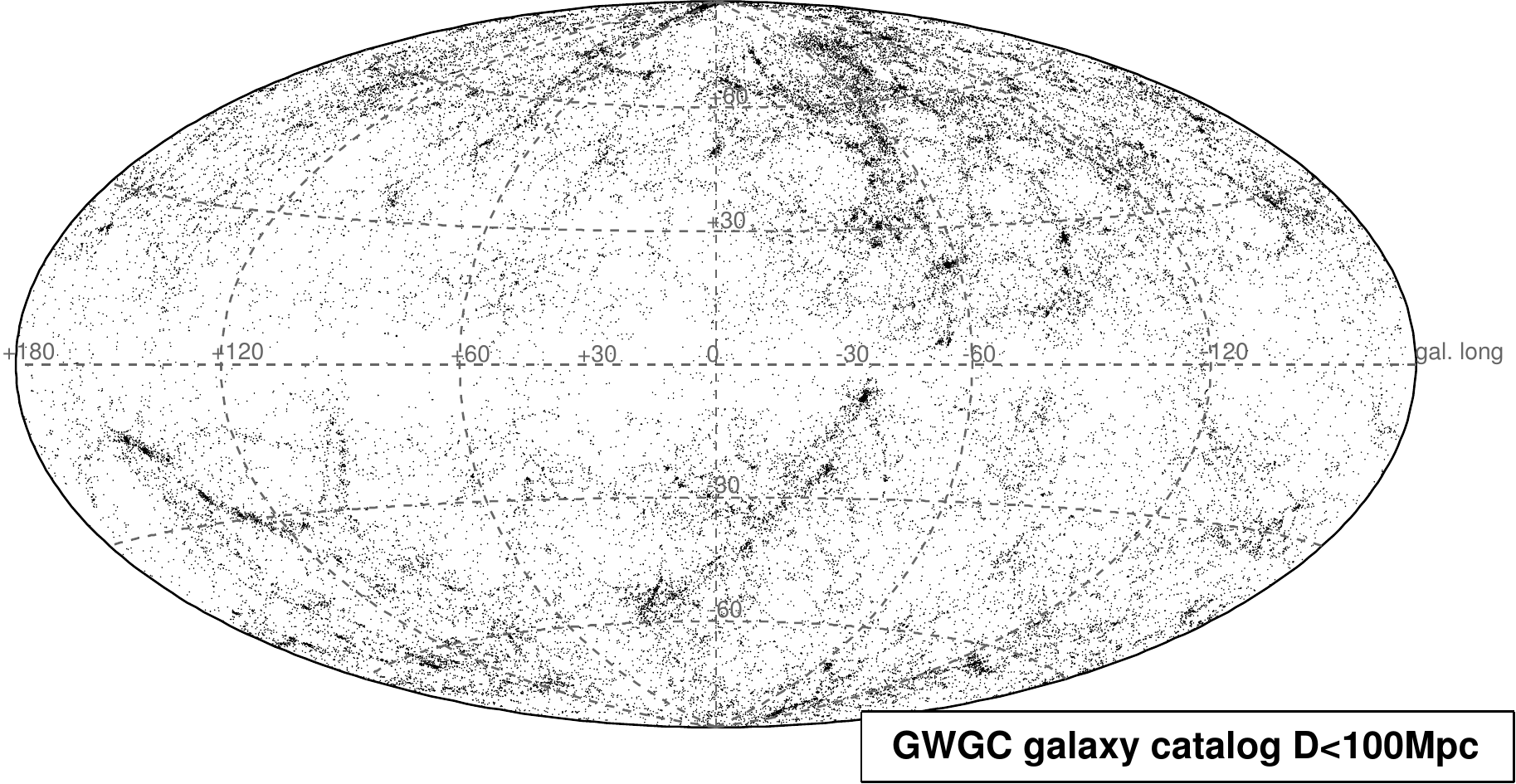}
          }	
  \caption{Left plot: High energy gamma ray sources given in the 2FGL catalogue from Fermi-LAT~\cite{bibschussler2:2FGL}. Sources with (without) an associated counterpart are shown in black (red). The size of the circles indicate the gamma-ray flux in the $[1-100~\mathrm{GeV}]$ energy range. Right plot: Galaxies within $100~\mathrm{Mpc}$ as given in the GWGC catalogue~\cite{bibschussler2:GWGC}.}
  \label{figschussler2:2ptCorrelation}
 \end{figure*}

\subsection{High energy gamma rays}
Data from 2 years of observation with the Fermi-LAT high energy gamma ray satellite have been used to compile the 2FGL point source catalogue~\cite{bibschussler2:2FGL}. It is shown in Fig.~\ref{figschussler2:2ptCorrelation}, where the radii of the circles represent the gamma ray flux in the $1-100~\mathrm{GeV}$ range and associated (non-associated) sources are shown in black (red). As most of the missing associations are due to source confusion within or near the Galactic plane, we use the full catalogue containing 1873 gamma ray sources to correlate their positions with the 3058 selected neutrino candidates. Each 2FGL source is weighted with its gamma ray flux $[1-100~\mathrm{GeV}]$ as given in the Fermi catalogue and the ANTARES events are weighted with the $n_\mathrm{Hit}$ energy estimator. It should be noted that this approach does not take into account changes in the energy spectrum that have been observed for many of the $\mathrm{GeV}$ gamma ray sources. The 2pt correlation analysis did not find any significant correlation. The minimum post-trial p-value of $68~\%$ has been found for angular scales smaller than $0.6^\circ$.\\

\subsection{The local universe}
The cosmic ray accelerators are likely correlated with the matter distribution in the local universe. To exploit this connection, the GWGC catalogue of 53295 galaxies within a distance of $D<100~\mathrm{Mpc}$ has been used as description of the local matter distribution~\cite{bibschussler2:GWGC}. Assuming the simplest case of equal neutrino luminosity from all galaxies, a $D^{-2}$ weighting for the galaxies and the $n_\mathrm{Hit}$ neutrino weights for the neutrino candidates have been used. Limiting the maximum analyzed angular scale to $5~\deg$ due to the large number of sources, the 2pt correlation analysis found the most significant clustering at scales smaller than $0.3~\deg$ with a post-trial p-value of $96~\%$.\\

\section{Summary}
In the search for the sources of high energy cosmic rays, the detection of astrophysical neutrinos sources may play a crucial role. Various experiments are currently taking data or are in a preparatory phase to achieve this goal and the recorded data are scrutinized in numerous ways in order to extract a maximum of information. We presented here a search for intrinsic clustering of data recorded with the ANTARES neutrino telescope using an improved 2pt correlation technique exploiting the energy of the neutrino candidates. The data do neither show evidence for deviations from an isotropic arrival direction distribution expected for the background of atmospheric neutrinos, nor does it correlate with the Fermi 2FGL gamma ray or the GWGC galaxy catalogue.

\clearpage

\setcounter{figure}{0}
\setcounter{table}{0}
\setcounter{footnote}{0}
\setcounter{section}{0}
\setcounter{equation}{0}

\addcontentsline{toc}{part}{{\sc Multi-Messenger searches}%
\vspace{-0.5cm}
}

\newpage
\id{id_james}



\title{\arabic{IdContrib} - A search for neutrinos from long-duration GRBs with the {\sc Antares} underwater neutrino telescope}
\addcontentsline{toc}{part}{\arabic{IdContrib} - {\sl Clancy W. James} : A search for neutrinos from long-duration GRBs%
\vspace{-0.5cm}
}

\shorttitle{\arabic{IdContrib} - {\sc Antares} GRB neutrino search}

\authors{
C.W.James$^{1}$ on behalf of the {\sc Antares} Collaboration.
}

\afiliations{
$^1$ ECAP, University of Erlangen-Nuremberg, Erwin-Rommel-Str 1, 91058 Erlangen, Germany
}

\email{clancy.james@physik.uni-erlangen.de}

\abstract{ANTARES is an underwater neutrino telescope located at a depth of 2475~m off the coast of Toulon, France. In this contribution, a search for neutrino events observed by ANTARES in coincidence with gamma-ray bursts (GRBs) is described. The observed properties of 296 long-duration GRBs visible to ANTARES from Dec.\ 2007 to Dec.\ 2011 are used to construct neutrino-selection criteria which are optimised to discover the expected prompt neutrino emission. In particular, the numerical `Neutrinos from Cosmic Accelerators' (NeuCosmA) method is used for the calculation of the expected neutrino fluxes. Using these predictions, a search is performed using data from the Fermi and Swift satellites to select neutrino candidate events coincident in time and direction with the GRBs. The result of this search is presented, and used to place limits on the predictions of neutrino production from various GRB models.
}



\maketitle

\section{Introduction}

Gamma-ray bursts (GRBs) are intense flashes of gamma rays which are usually divided into two classes: short (duration $\lesssim 2$~s) and long (duration $\gtrsim 2$~s) bursts \cite{bibjames:Kouveliotou93a}. The latter class has been associated with Type 1b/c supernova events \cite{bibjames:Galama1998}, and their emission is commonly treated in terms of the "fireball" model \cite{bibjames:MeszarosRees1993}. In this model, the gamma rays result from a highly relativistic jet formed during the collapse of the massive star. Shocks associated with the jet --- internal shocks, and any resulting from the jet interacting with material previously ejected by the star --- will accelerate electrons, thus producing the observed gamma-ray emission via the synchrotron self-Compton process \cite{bibjames:Meszaros2006}. Any co-acceleration of protons in these shocks would lead to interactions with the local photon field, producing an accompanying neutrino flux via e.g.\ the decay of charged pions \cite{bibjames:Waxman1995a} produced in these interactions. The detection of a neutrino flux associated with GRBs would thus be a clear signature of hadronic acceleration in these sources.

In these proceedings, a search for a neutrino flux in coincidence with GRBs is presented using data from the ANTARES neutrino telescope taken from late 2007 to 2011.
A numerical treatment (`NeuCosmA') of the neutrino flux from GRBs is used in Sec.\ \ref{secjames:model} to model the predicted flux from a sample of 296 candidate GRBs (described in Sec.\ \ref{secjames:data}), which indicates that the expected flux of neutrinos using the fireball model of GRBs is not currently limited by observations. In Sec.\ \ref{secjames:method}, the NeuCosmA predictions are used to optimise the sensitivity of a neutrino-search for each GRB individually. The results of this analysis after unblinding are given in Sec.\ \ref{secjames:results}.

\section{Data selection}
\label{secjames:data}

The ANTARES underwater neutrino telescope is located at a depth of $2475$~m off the coast of Toulon, France \cite{bibjames:Ageron2011,bibjames:AntoineICRC}. It primarily detects charged-current (anti) muon-neutrino interactions by observing the passage of the subsequent relativistic muons through the seawater near the detector. When the resulting Cherenkov light is detected by ANTARES' array of photo-multiplier tubes, the initial direction of the primary neutrino can be estimated. Due to the number of down-going muons coming from cosmic-ray interactions with the atmosphere above the detector, only up-going muons (those coming from below the local horizon) are selected for analysis, and stringent quality-cuts are used to reject mis-reconstructed down-going events (see e.g.\ Ref.\ \cite{bibjames:Ageron2011} for a description of this methodology). The instantaneous sky coverage of ANTARES is therefore $2 \pi$~sr; its latitude of $43^{\circ}$ thus makes it more likely to observe GRBs occurring in the Southern Hemisphere.

For the results presented here, data primarily from {\it Fermi} and {\it Swift} are used to select long-duration GRBs which were visible to ANTARES from Dec.\ 2007 to Dec.\ 2011. A complete description of the data-selection process, and the final table of all relevant GRB parameters used to model these events as described in Sec.\ \ref{secjames:model}, is given in Ref.\ \cite{bibjames:JuliaGRB}. Requiring that the detector was operating under normal conditions for the entire burst duration leads to 296 events located below the ANTARES horizon being chosen for this analysis. From the observed beginning and end times of the bursts, this amounts to a total coincidence window of 6.55 hr. The coordinates of these bursts are shown in Fig.\ \ref{fig_data}.

\begin{figure}[t]
\centering
\includegraphics[width=0.45\textwidth]{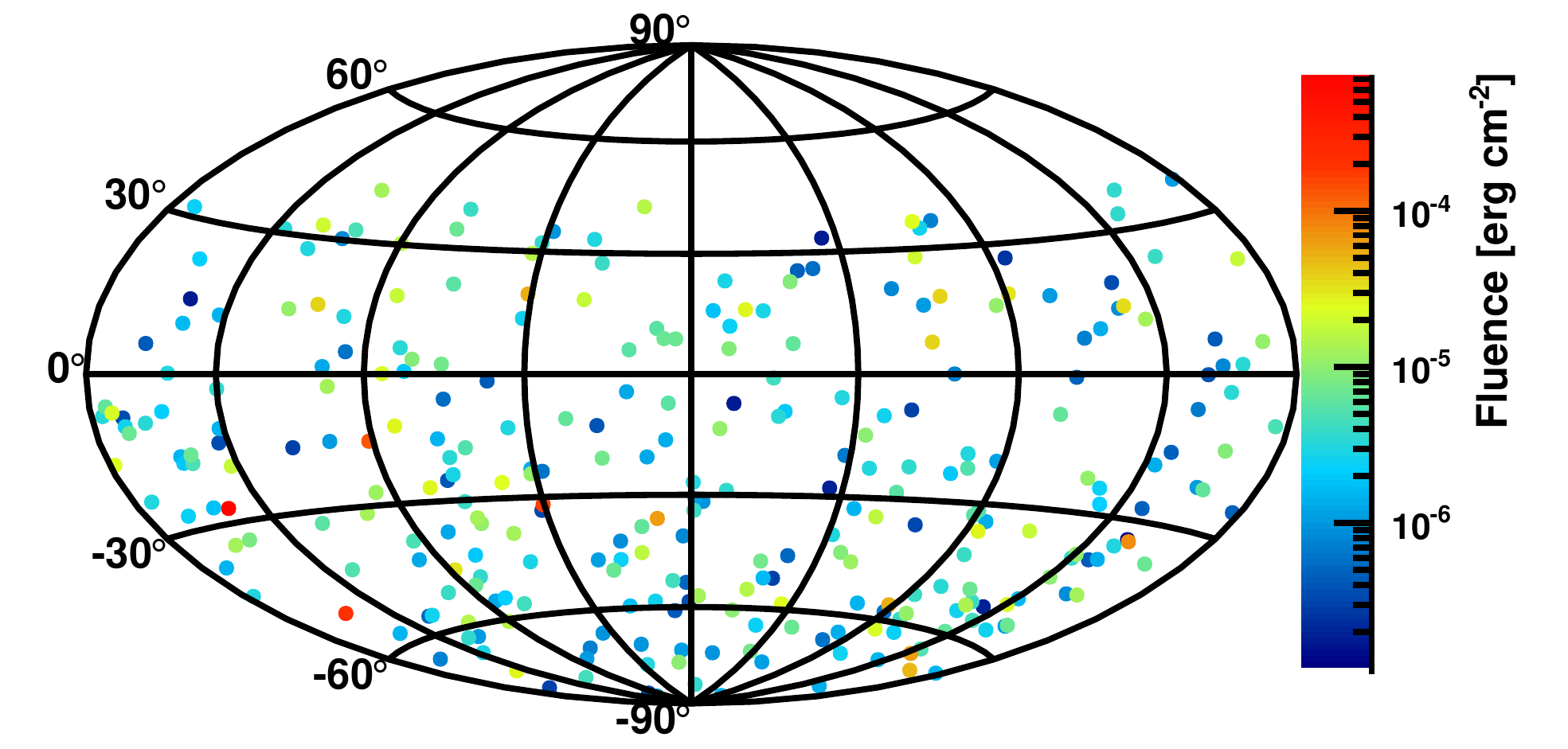}
\caption{Sky distribution of the selected 296 gamma-ray bursts in equatorial coordinates (colour indicates measured gamma-ray fluence).}\label{fig_data}
    \end{figure}

\section{The NeuCosmA model}
\label{secjames:model}

The most commonly used method to estimate the neutrino-emission from long-duration GRBs in the context of the fireball model is that of Waxmann \& Bahcall \cite{bibjames:WaxmannBahcall}, adapted as per Guetta {\it et al.} \cite{bibjames:Guetta2004} to account for individual GRB properties. This calculation models $\Delta$-resonance interactions of Fermi-accelerated protons with the GRB photon field --- neutrinos result from the subsequent decay of charged pions and muons. This method predicts a doubly broken power-law spectrum for the $\nu_{\mu}$ flux. The recently-published limit by the IceCube collaboration on the $\nu_{\mu}+\bar{\nu}_{\mu}$ flux from GRBs lies significantly below the prediction from this two-break model \cite{bibjames:Abbasi2012}.
Accounting for the different maximum break energies for the $\nu_{\mu}$ and $\bar{\nu}_{\mu}$ fluxes within the model of Guetta {\it et al.} produces a combined spectrum that has three break energies. This was the method used for the previously published limits from ANTARES on the neutrino flux from GRBs using 2007 data \cite{bibjames:Adrianmartinez2013}. For comparison, the predictions for the combined flux of $\nu_{\mu}$ and $\bar{\nu}_{\mu}$ from one GRB (GRB110918) from both the `two-break' and `three-break' treatments of the Guetta model are shown in Fig.\ \ref{fig_compare} as the solid and dotted blue curves respectively. It is likely that applying the three-break treatment to the IceCube predictions would still result in the limit lying below the expected flux.

\begin{figure}[t]
\centering
\includegraphics[width=0.45\textwidth]{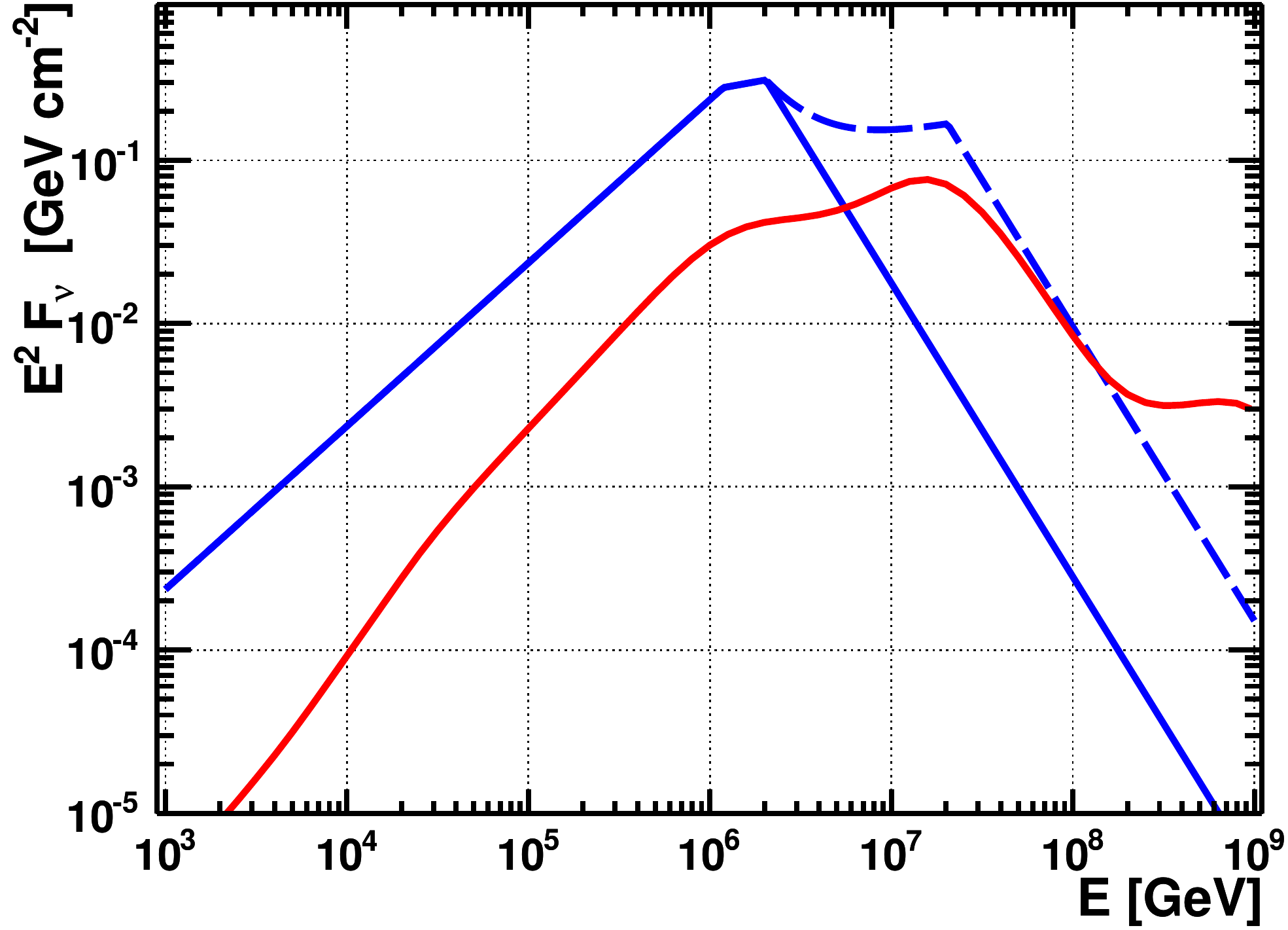}		
\caption{
      Expected  $\nu_{\mu} + \bar{\nu}_{\mu}$ spectra from GRB110918, using both the two-break (solid blue) and three-break (dotted blue) treatments of the model of Guetta {\it et al.}, and the NeuCosmA model (red).} \label{fig_compare}
    \end{figure}

A full account of all the physical processes relevant for neutrino production within the fireball model requires a detailed numerical calculation. The NeuCosmA (Neutrinos from Cosmic Accelerators) code \cite{bibjames:Hummer2010,bibjames:Hummer2012} was developed to include the full photo-hadronic cross-sections as given by the Monte-Carlo code SOPHIA \cite{bibjames:Muecke1998}, the energy-losses of secondary particles produced in such interactions, and the effects of neutrino-mixing in the final calculation of the expected neutrino spectrum from GRBs (among other details). As discussed in H\"{u}mmer {\it et al.} \cite{bibjames:Hummer2010}, the resulting predictions for the GRB neutrino flux lie an order of magnitude below those of the simplified Guetta model, with the addition of a high-energy component due to $K^+$ decays --- see e.g.\ Fig.\ \ref{fig_compare} (red curve) for a comparison in the case of GRB110918. It should be emphasised that predictions from the NeuCosmA model arise simply through a more thorough application of known physics to the same GRB fireball model as used by previous predictions, and thus suffer from no more --- and no less --- error due e.g.\ to uncertainty in GRB parameters such as the jet gamma-factor. Therefore, we used the NeuCosmA model to estimate the $\nu_{\mu}$ and $\bar{\nu}_{\mu}$ flux from the 296 GRBs in the chosen sample for the purposes of optimising data-selection, as described in the next section. The predicted spectrum from each event, and the combined total, is shown in Fig.\ \ref{fig_total}. Observe that GRB110918 contributes approximately half the total predicted flux in the energy-range above $1$~TeV, where ANTARES has the greatest sensitivity.

\begin{figure}[t]
\centering
\includegraphics[width=0.45\textwidth]{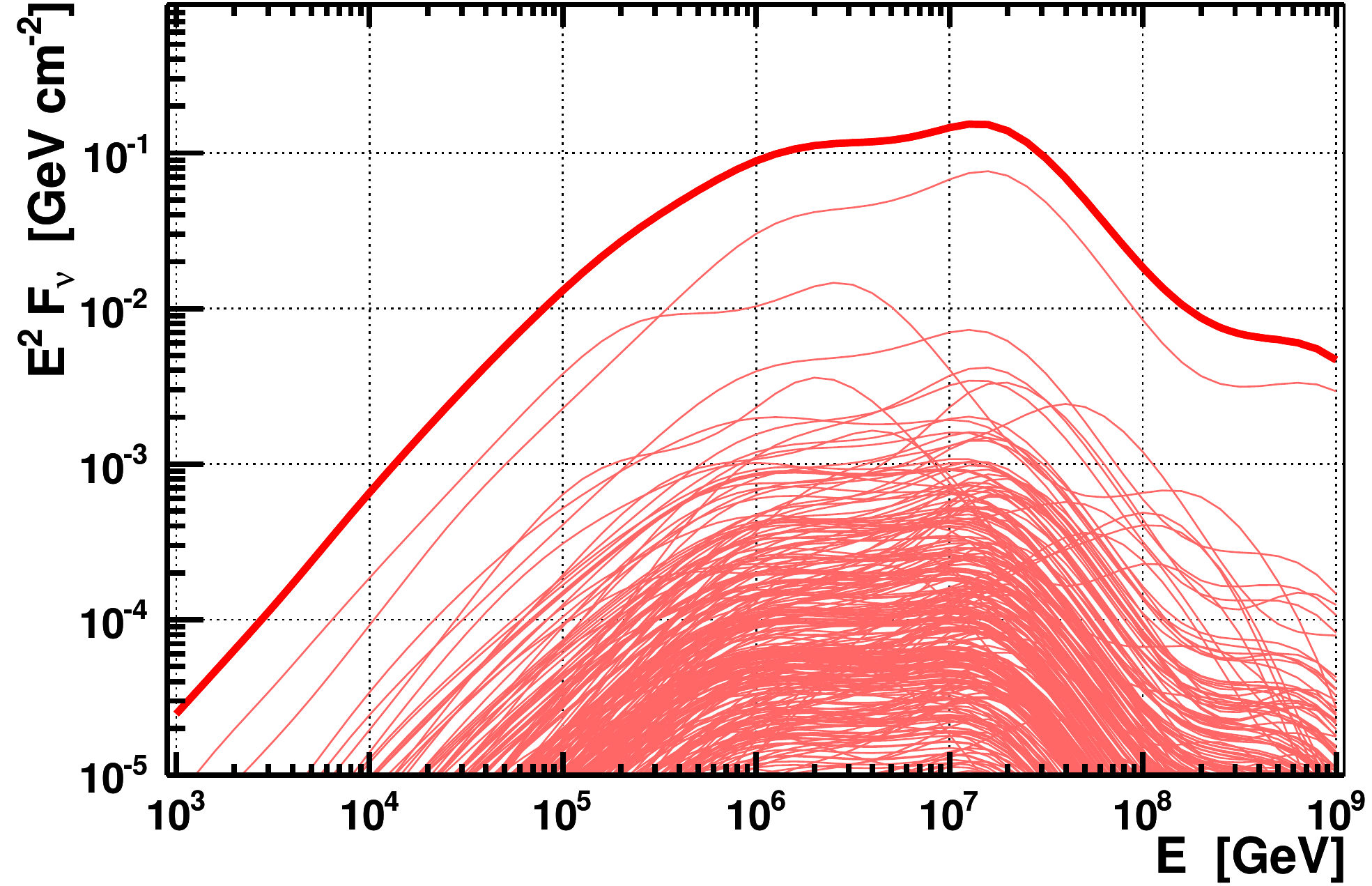}		
\caption{ $\nu_\mu + \bar{\nu}_\mu$ NeuCosmA spectra of the 296 GRBs (thin lines) and their sum (thick line).} \label{fig_total}
\end{figure}

\section{Search methodology}
\label{secjames:method}

The primary criteria used to search for a neutrino flux coming from the sample of GRBs are timing and spatial coincidence of observed up-going muons with the gamma-ray signal. For each GRB, only up-going muons arriving within the observed GRB duration and with reconstructed origin less than $10^{\circ}$ from the GRB direction are considered.
For this purpose, the algorithm used in this analysis to reconstruct the direction of observed muon tracks returns two parameters: the $\Lambda$ parameter, reflecting the quality of the track fit (high values are better), and $\beta$, giving the estimated uncertainty in the reconstructed direction (low values are better). See Ref.\ \cite{bibjames:Adrianmartinez2012} for a more detailed description of the event-reconstruction procedure. A cut of $\beta<1^{\circ}$ was used as per Ref.\ \cite{bibjames:Adrianmartinez2012}, while the values of $\Lambda_{\rm cut}$ used to select events with $\Lambda > \Lambda_{\rm cut}$ vary from source to source and are optimised as follows.

To determine if an observation of neutrinos in coincidence with a given GRB is significant, the `extended maximum-likelihood ratio' \cite{bibjames:Barlow1990} is used to define a test-statistic $Q$. This is the log-likelihood ratio of observing a given spatial-distribution of events with an estimated number of source events $n_s^{\rm est}$ and expected background contribution $\mu_b$ as a function of the observed angular offsets $\delta$ between the events and the source:
\begin{eqnarray}
      Q(n_s^{\rm est}) & = & \max_{\hat{n}_s \in [0,n_\mathrm{tot}]} Q(\hat{n}_s) \label{eqjames:ex_max_likelihood} \\
   Q(\hat{n}_s) & = & \sum\limits_{i=1}^{n_{\rm tot}} \log \frac{\hat{n}_s \cdot S(\delta_i) + \mu_\mathrm{b} \cdot B(\delta_i)}{\mu_\mathrm{b} \cdot B(\delta_i)}  - (\hat{n}_s +\mu_\mathrm{b}) \nonumber
\end{eqnarray}
where $n_s^{\rm est}$ is chosen as the value of $\hat{n}_s$ which maximises $Q$. Here, $S(\delta_i)$ and $B(\delta_i)$ represent the likelihoods of event $i$ (with reconstructed direction $\delta_i$ degrees from the GRB direction) being of signal (GRB) and background (atmospheric cosmic ray) origin respectively. $S$ is thus the ANTARES point-spread function (PSF) in the direction of the GRB, and $S(\delta_i)$ takes a separate value for each of the $n_{\rm tot}$ events passing the selection criteria. From hereon, `$Q$' will be used as shorthand for `$Q(n_s^{\rm est})$'. The distributions $S$ and $B$ are themselves functions of cuts on the $\Lambda$ parameter used to discriminate signal from background; their determination is described below.

\subsection{Signal simulation and background estimation}

\begin{figure}[t]
\includegraphics[width=0.45 \textwidth]{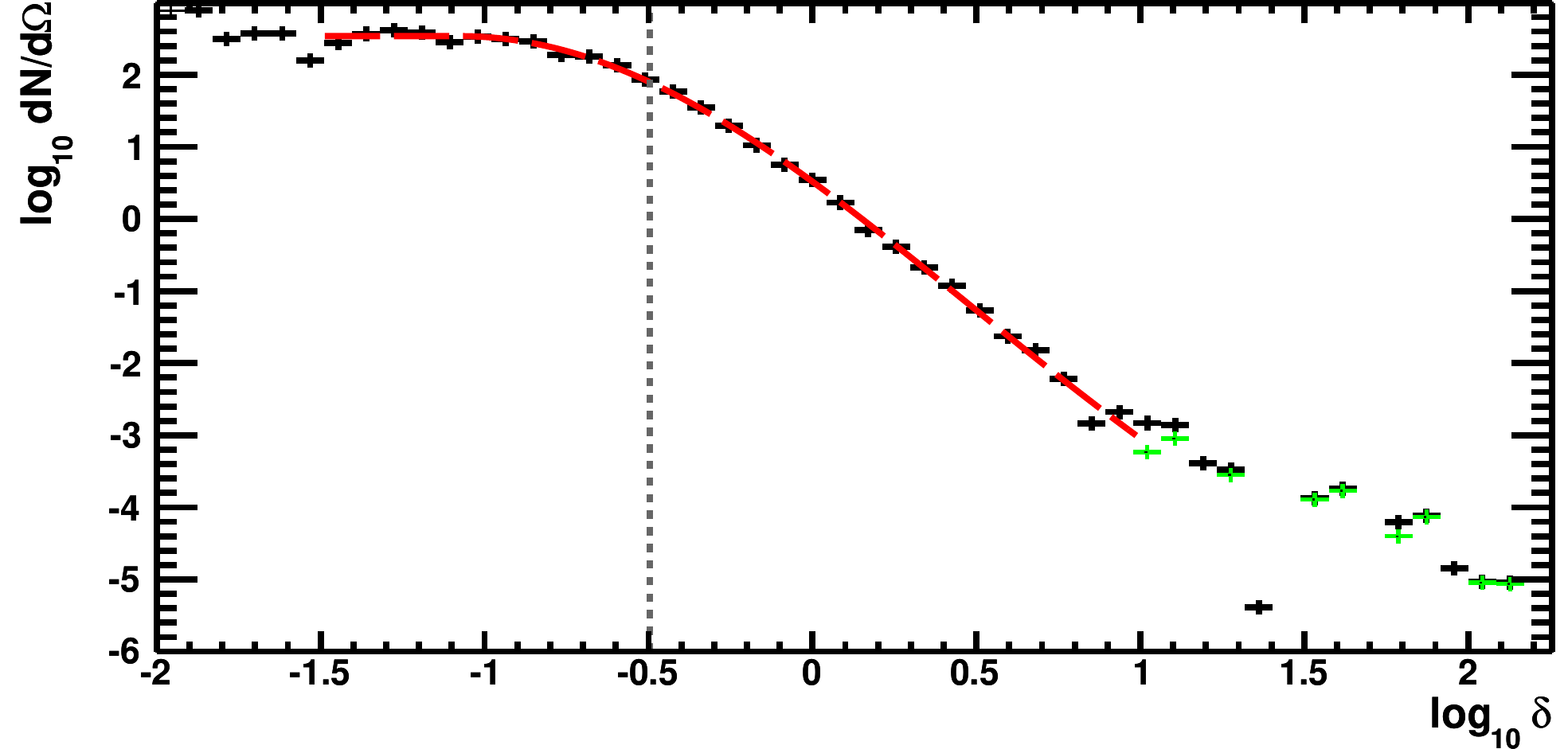}
\caption{Point-spread function $S(\delta)$ (red), $\nu_{\mu}$ and $\bar{\nu}_{\mu}$ events (black), and shower events (green) for the GRB 11091889 with $\Lambda_{\rm cut} = -5.5$.}\label{PSF_fig}
\end{figure}

In order to determine $S$, each GRB is simulated using $4 \cdot 10^{9}$ incident $\nu_{\mu}$ and $\bar{\nu}_{\mu}$ sampled from the spectrum from the NeuCosmA model using the standard ANTARES Monte-Carlo simulation chain \cite{bibjames:Adrianmartinez2012}. A preliminary simulation of $150$ neutral-current `shower' events from neutrinos of all flavours (Fig.\ \ref{PSF_fig}, green points) demonstrates that, using methods currently available, their reconstructed directions correlate poorly with their arrival directions, and thus these events are not used in the fit to $S$. For each simulated event, $\delta$ is the angular offset between the reconstructed direction and the source direction. The distribution $S(\delta)$ (the point-spread function) is modelled for each possible cut value $\Lambda_{\rm cut}$. The fitted function used for $S(\delta)$ is given in Eq.\ \ref{PSF_eqn}:
\begin{eqnarray}
S(\delta) & \equiv & \frac{d N(\delta)}{d \Omega} \label{PSF_eqn} \\
\log_{10} S & = & \left\{ \begin{array}{ c l l}
 C_1 & \delta \le \delta_0 \\
 C_1-C_2 \left( 1 - {\rm e}^{ -\frac{(\log \delta-\log \delta_0)^2}{2 \sigma^2}} \right) & \delta > \delta_0
\end{array} \right. \nonumber
\end{eqnarray}
where $C_1$, $C_2$, $\delta_0$ and $\sigma$ are fitted parameters. An example of this fit, which is performed for each combination of GRB and $\Lambda_{\rm cut}$ value, is given in Fig.\ \ref{PSF_fig}. The grey vertical line gives the median value $\bar{\delta}$.

The distribution of background events $B$ is taken to be constant over the narrow time-windows and small regions of sky about each GRB. The expected number of background events $\mu_b$ is taken from data, and is calculated as per Eq.\ \ref{eqjames:mub}:
\begin{eqnarray}
\mu_b & = & 1.5 \, T_{s} r_{\bar{t}}(\Omega) \, \left(1+\frac{r^{90}_{\bar{\Omega}}(t)}{\bar{r}}\right) \label{eqjames:mub}
\end{eqnarray}
Here, $T_s$ is the duration of the search-time window, $\bar{r}$ is the long-term all-sky mean background rate, $r^{90}_{\bar{\Omega}}(t)$ is the $90$\% upper limit on the observed all-sky background rate around the time $t$ of the GRB, and $r_{\bar{t}}(\Omega)$ is the long-term background rate observed in the direction of the GRB in detector-coordinates $\Omega$. The extra factor of $1.5$ is included to ensure that all observed rates are below the estimates, i.e.\ under the conservative assumption that all fluctuations are real rather than statistical \cite{bibjames:JuliaGRB}.

\subsection{Threshold calculation and MDP optimisation}

The test-statistic $Q$ (Eq.\ \ref{eqjames:ex_max_likelihood}) is defined such that high values are evidence against the background-only (null) hypothesis --- i.e.\ evidence for a neutrino signal from GRBs. The significance of an observation $Q^{\rm obs}$ is characterised by the $p$-value, defined as the probability of observing a $Q$-value $Q>Q^{\rm obs}$ in the case of background only. In general, for any true number of signal events $n_s$ (as opposed to the estimate $n_s^{\rm est}$) and expected number of background events $\mu_b$, $Q$ has the distribution $h_{n_s}(Q)$. The distributions $h_{n_s}(Q)$ are calculated using `pseudo-experiments'. Firstly, $n_b$ background events are Poisson-sampled from the mean $\mu_b$. Then, for these and the $n_s$ signal events, random values of $\delta$ according to the distributions $S$ and $B$ are generated. Since an accurate determination of $h_{0}(Q)$ is critical for determining the discovery threshold $Q^{\rm thresh}$, $10^{10}$ pseudo-experiments are used to estimate each $h_0(Q)$, $10^5$ are used for $h_{n_s > 0}(Q)$, yielding enough statistics for the optimisation procedure.

    \begin{figure}[t]
    \begin{center}
      \includegraphics[width=0.45\textwidth]{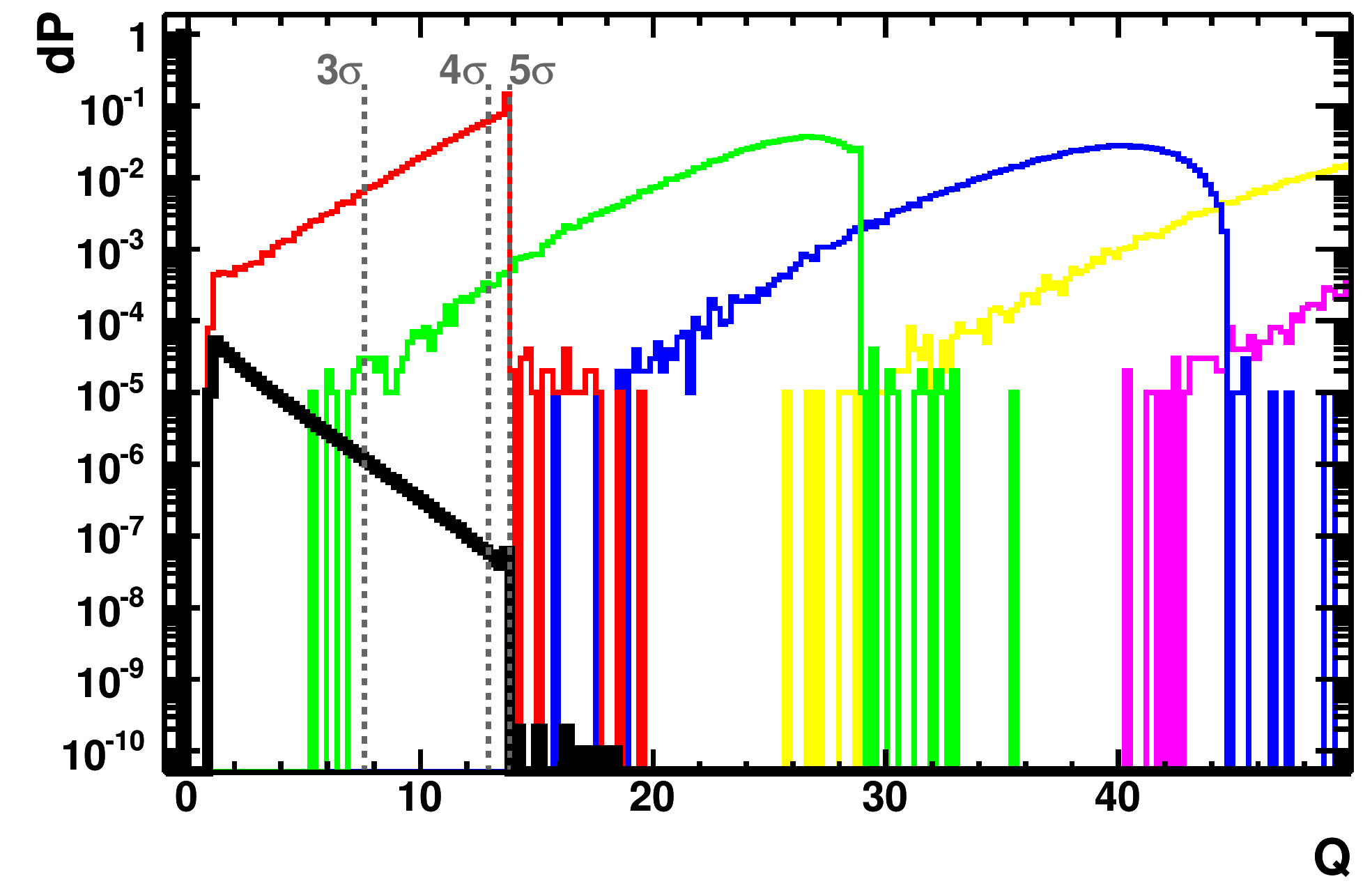}
      \end{center}
      \caption{
      Probability distributions of $Q$-values $ h_{n_s}(Q)$ for (left to right) $n_\mathrm{s}=0, 1,2,3 \ldots$ for GRB110918 and $\Lambda_{\rm cut}=-5.5$; $\mu_b=3.7\cdot 10^{-4}$. Grey vertical lines indicate the threshold values $Q_{p}^{\rm thresh}$ for different levels ($3 \sigma$, $4 \sigma$, $5 \sigma$) of significance after accounting for a trial factor of 296 as calculated from $ h_{0}(Q)$. The three distinct components $h_{0}(Q)$ (black) are due to the random number of background events $n_b$ (Poisson mean $\mu_s$) taking values $0$, $1$, and $2$.}
      \label{figjames:pe_q_hist}
    \end{figure}  

Given $h_0(Q)$, the critical threshold value of $Q^{\rm thresh}$ required for a discovery can then be calculated from a required value of $p$ via:
\begin{equation}
\int_{Q_{p}^{\rm thresh}}^{\infty}  h_{0}(Q) \, {\rm d} Q = p. \label{eqjames:pvalue}
\end{equation}
A global probability of $2.7 \cdot 10^{-3}$ ($3 \sigma$) is chosen for the optimisation --- this corresponds to $p=2.7 \cdot 10^{-3}/296 = 9.1 \cdot 10^{-6}$ per source.
After defining $Q^{\rm thresh}$, the probability to discover a model predicting an expected source flux of $\mu_s$ events (from which $n_s$ are actually observed, according to Poisson statistics) is given by the `model discovery potential' $MDP$:
\begin{eqnarray}
MDP & = & \sum_{i=0}^{\infty} P(Q>Q^{\rm thresh} | n_s) P(n_s | \mu_s) \nonumber \\
& = & \sum_{n_s=0}^{\infty} \int_{Q_{p}^{\rm thresh}}^{\infty} h_{n_s}(Q) {\rm d}Q  \cdot \frac{\mu_s^{n_s} e^{-\mu_s}}{n_s!}. \label{eqjames:mdp}
\end{eqnarray}
The value of $\Lambda_{\rm cut}$ chosen for each source is thus that which maximises the $MDP$. Applying this procedure, a set of $\Lambda_{\rm cut}$, $Q^{\rm thresh}$, $S$, $B$, and $\mu_b$ are determined for all 296 sources. Table \ref{table} gives the expected source and background counts, and the cut parameters, for the ten most-promising GRBs. In total, $0.06$ signal events are expected from the NeuCosmA model, and $0.5$ from the Guetta model, against a background of $0.05$ events.

\begin{table}[htb]
\centering
\begin{small}
\begin{tabular}{r| r | c c | r | r } \hline \hline
GRB$^{\dag}$ & $\Lambda_{\rm cut}$ & $\mu_{\rm b}$ & $\mu_s^{\rm NCA}$ & $\bar{\delta} (^{\circ})$ & $T_{\rm s}$(s)\\
 \hline
110918  & -5.5 & 3.7$\cdot 10^{-4}$ & 3.5$\cdot 10^{-2}$  & 0.32 & 73.4  \\ 
080607  & -5.4 & 5.5$\cdot 10^{-4}$ & 6.5$\cdot 10^{-3}$ & 0.33 & 164.3  \\ 
111008  & -5.5 & 3.6$\cdot 10^{-4}$ & 2.2$\cdot 10^{-3}$ & 0.35 & 75.4  \\ 
101014  & -5.1 & 4.1$\cdot 10^{-4}$ & 1.2$\cdot 10^{-3}$  & 0.89 & 723.1  \\ 
100728  & -5.6 & 2.0$\cdot 10^{-4}$ & 9.6$\cdot 10^{-4}$  & 0.49 & 268.6 \\ 
090201  & -5.4 & 5.4$\cdot 10^{-4}$ & 7.0$\cdot 10^{-4}$  & 0.39 & 126.6  \\ 
111220  & -5.2 & 1.4$\cdot 10^{-4}$ & 6.2$\cdot 10^{-4}$  & 1.13 & 66.5  \\ 
090829  & -5.4 & 1.7$\cdot 10^{-4}$ & 3.9$\cdot 10^{-4}$ & 1.02 & 112.1  \\ 
110622  & -5.4 & 1.7$\cdot 10^{-4}$ & 4.3$\cdot 10^{-4}$ & 1.42 & 116.6  \\ 
081009  & -5.5 & 1.3$\cdot 10^{-4}$ & 3.5$\cdot 10^{-4}$  & 0.94 & 70.2  \\ \hline
All GRBs & & 5.1$\cdot 10^{-2}$ & 6.1$\cdot 10^{-2}$ & &   23500
\end{tabular}
\end{small}
\caption{Sample parameters (see text) of the ten GRBs with the highest discovery probabilities as estimated from the NeuCosmA (`NCA') model, and the total (where relevant) over all 296 GRBs.
} \label{table}
\end{table} 

\section{Results}
\label{secjames:results}

With search parameters defined as in Sec.\ \ref{secjames:method}, the data from Dec.\ 2007 to Dec.\ 2011 were unblinded and analysed for evidence of a neutrino signal correlating with the 296 GRBs in the data-sample. No events passing quality cuts are found within the specified time-intervals and $10^{\circ}$ search windows. Hence, all $Q$-values are zero, and we find no evidence for a neutrino signal from any of the analysed GRBs. A $90$\% confidence limit on the total flux of neutrinos from all GRBs predicted by each of the NeuCosmA and Guetta models is then placed by scaling the flux predictions to obtain $\mu_s=2.3$ (since, from Poisson statistics, $P(n_s>0|\mu_s=2.3)=0.9$). The resulting limits on both the NeuCosmA and Guetta (two-break) fluxes are shown in Fig.\ \ref{fig_limit} (red and blue dotted lines). Neither prediction is constrained by the observations presented here, with the predictions for the NeuCosmA flux lying a factor of $38$ below the limit. While this analysis improves the previous limit set by ANTARES \cite{bibjames:Adrianmartinez2013}, which used a smaller sample of $40$ GRBs and the three-break treatment of the Guetta model, the IceCube limit \cite{bibjames:Abbasi2012} (set using the two-break Guetta model) is not improved-upon due to the larger effective volume of the IceCube detector. Note that both the relative sensitivities to particular GRBs, and the GRB samples used, in these previous searches differ significantly from those presented here, which is particularly relevant given the importance of individual bright events such as GRB110918. We have also shown that the more detailed calculation of the expected neutrino flux from long-duration GRBs using NeuCosmA indicates that current limits on neutrino emission from GRBs are completely consistent with expectations.

 \begin{figure}[t]
 \centering
\includegraphics[width=0.45 \textwidth]{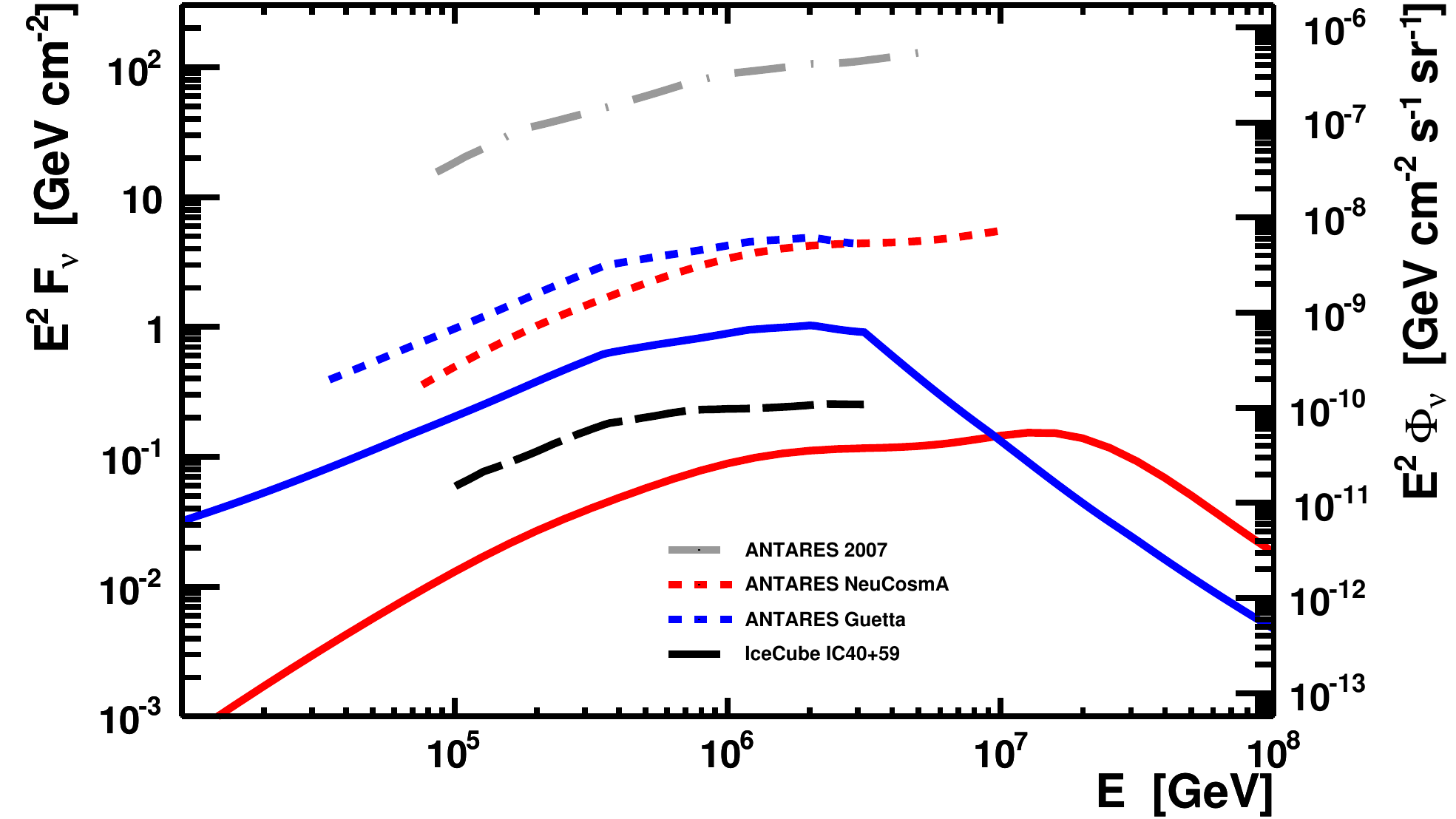}
\caption{ Comparison of $90$\% confidence limits (dashed) with expected fluxes (solid) for the summed $\nu_{\mu}+\bar{\nu}_{\mu}$ spectra of the 296 GRBs used in this analysis, using the NeuCosmA (red) and `two-break' Guetta (blue) models. Limits from a previous IceCube (black dashed) and ANTARES (grey dash-dotted) are also shown for comparison. \label{fig_limit}}
\end{figure}

\vspace*{0.5cm}
{\footnotesize{\bf %
\noindent
Acknowledgment: }We would like to thank Philipp Baerwald for producing the NeuCosmA neutrino-flux predictions, and Walter Winter for helpful discussions and making the NeuCosmA model available. This work is supported by the German government (BMBF) with grant 05A11WEA.}
\setcounter{figure}{0}
\setcounter{table}{0}
\setcounter{footnote}{0}
\setcounter{section}{0}
\setcounter{equation}{0}

\newpage
\id{id_dornic1}
\title{\arabic{IdContrib} - Search for neutrinos from transient sources with the {\sc Antares} telescope and optical follow-up observations}
\addcontentsline{toc}{part}{\arabic{IdContrib} - {\sl Damien Dornic} : Search for neutrinos from transient sources %
and optical follow-up observations%
\vspace{-0.5cm}
}
\shorttitle{\arabic{IdContrib} - The TAToO project}

\authors{Ageron M.$^{1}$, Al Samarai I.$^{1}$, Basa S.$^{2}$, Bertin V.$^{1}$, Brunner J.$^{1}$, Busto J.$^{1}$, Coyle P.$^{1}$, Dornic
D.$^{1}$, Klotz A.$^{3}$, Sch\"ussler F.$^{4}$, Mathieu A.$^{1}$,  Vallage B.$^{4}$ for the {\sc Antares}, TAROT and ROTSE Collaborations }
\afiliations{$^1$ CPPM, CNRS/IN2P3 - Universit\'e de M\'editerran\'ee, 163 avenue de Luminy, 13288 Marseille Cedex 09, France \\
             $^2$ LAM, BP8, Traverse du siphon, 13376 Marseille Cedex 12, France\\
	     $^3$ Universit\'e de Toulouse; UPS-OMP; IRAP; Toulouse, France. CNRS; IRAP; 14, avenue Edouard Belin, F-31400 Toulouse, 
	     France CNRS; Observatoire de Haute-Provence, 04870 Saint Michel l'Observatoire, France\\
             $^4$ CEA-IRFU, centre de Saclay, 91191 Gif-sur-Yvette, France	    	     
	     }
\email{dornic@cppm.in2p3.fr}

\abstract{
The ANTARES telescope is well suited to detect neutrinos produced in astrophysical transient sources as it can observe 
a full hemisphere of the sky at all the times with a duty cycle close to unity. Potential sources include gamma-ray bursts 
(GRBs), core collapse supernovae (SNe), and flaring active galactic nuclei (AGNs). To enhance the sensitivity of ANTARES 
to such sources, a new detection method based on coincident observations of neutrinos and optical signals has been developed. 
A fast online muon track reconstruction is used to trigger a network of small automatic optical telescopes. Such alerts are 
generated twices per month for special events such as two or more neutrinos, coincident in time and direction, or single neutrinos 
of very high energy or in the specific directions of local galaxies. Alert triggers are followed by the TAROT, ROTSE and ZADKO optical 
telescopes and by the SWIFT/XRT telescopes. Results on the optical images analysis to search for GRB in the prompt images and core 
collapse SNe will be presented. 
}

\maketitle

\section{Introduction}

Detection of high-energy neutrinos from an astronomical source would be a direct evidence of the presence of hadronic acceleration 
and provide important information on the origin of the high-energy cosmic rays. Transient astronomical phenomena, such as Gamma-Ray 
Bursts (GRB)~\cite{bibdornic1:GRB} and core collapse supernovae (ccSNe)~\cite{bibdornic1:CCSN}, offer very promising perspectives for the detection of cosmic neutrinos as, due 
to their short duration, the background from atmospheric neutrinos and muons is strongly reduced. 
As neutrino telescopes observe a full hemisphere 
of the sky (even the whole sky if down-going events are considered) at all times, they are particularly well 
suited for the detection of transient phenomena. 

To improve the detection sensibility to transient sources, a multi-wavelength follow-up program, TAToO, operates within the ANTARES 
Collaboration since 2009~\cite{bibdornic1:TAToO}. This method, earlier proposed in~\cite{bibdornic1:Marek}, is based on the optical follow-up of selected neutrino events
very shortly after their detection by the ANTARES neutrino telescope. ANTARES is able to send alerts within one minute after the neutrino detection and with 
a precision of the reconstructed direction better than 0.5 degres at high-energy ($E>1~1TeV$). The optical follow-up is performed by the TAROT~\cite{bibdornic1:Tarot} and 
ROTSE-III~\cite{bibdornic1:Rotse} telescopes. Since February 2009 to December 2012, 83 alerts have been sent, all of them triggered by the two single neutrino 
selection criteria. After a commissioning phase in 2009, more than $80~\%$ of the alerts had an optical follow-up. The main 
advantage of this program is that no hypothesis is required on the nature of the source, only that it produces neutrino and photons. 

In this paper, the first results on the analysis of the early follow-up images associated to eight TAToO alerts are presented. A  
brief description of the ANTARES experiment and the alert system are summarized in section 2 and 3 respectively. The observation strategy and the 
optical data analysis are described in section 4 and 5 respectively. Finally, section 6 outlines the results of this search.

\section{The ANTARES experiment}

The ANTARES experiment \cite{bibdornic1:Antares} aims at searching for neutrinos of astrophysical origin by detecting high-energy muons 
($\geq$100~GeV) induced by their neutrino charged current interaction in the vicinity of the detector. Due to the very large 
background from down-going cosmic ray induced muons, the detector is optimized for the detection of up-going neutrino induced
 muon tracks. 

The ANTARES detector is located in the Mediterranean Sea, 40~km from the coast of Toulon, France, at a depth of 2475~m. It is a 
tridimensional array of photomultiplier tubes (PMTs) arranged on 12 slender detection lines, anchored to the sea bed and kept taught 
by a buoy at the top. Each line comprises up to 25 storeys of triplets of optical modules (OMs), each housing a single 10" PMT. Since 
lines are subject to the sea current and can change shape and orientation, a positioning system comprising hydrophones and compass-tiltmeters 
is used to monitor the detector geometry. PMT signals are digitized offshore and time-stamped using an external GPS signal giving the absolute 
timing at the location of the detector, allowing an absolute time accuracy better than 1~$\mu$s. The onshore data acquisition system collects 
the data from all the individual PMTs of the detector and passes them to the filtering algorithms based on local ``clusters'' which search for a 
collection of signals compatible with a muon track crossing the detector. All hits within a few microseconds around these clusters define 
an ``event'' and are kept for further online and offline reconstructions. Events are typically available less than one minute after the crossing 
of the detector by a high-energy muon. The ANTARES neutrino telescope is fully operational since May 2008\cite{bibdornic1:Antares2}.

\section{ANTARES neutrino alerts}

The criteria for the TAToO trigger are based on the features of the
neutrino signal produced by the expected sources.
Several models predict the production of high energy neutrinos greater than
1 TeV from GRBs~\cite{bibdornic1:GRB} and from Core Collapse
Supernovae~\cite{bibdornic1:CCSN}. Under certain conditions, multiplet of
neutrinos can be expected~\cite{bibdornic1:CCSN1}. A basic requirement for the coincident observation of a neutrino and an optical
counterpart is that the pointing accuracy of the neutrino telescope should be
at least comparable to the field of view of the TAROT and ROTSE 
telescopes ($\approx 2^\circ \times 2^\circ$). 

Atmospheric muons, whose abundance at the ANTARES detector~\cite{bibdornic1:atmu}
is roughly six orders of magnitude larger than the one
of muons induced by atmospheric
neutrinos, are the main background for the alerts and have to be
efficiently suppressed. Among the surviving events, neutrino candidates
with an increased probability to be of cosmic origin are
selected~\cite{bibdornic1:difflux}. 

To select the events which might trigger an alert, a fast and robust algorithm
is used to reconstruct the calibrated data. This algorithm uses an idealized detector
geometry and is independent of the dynamical positioning calibration. 
A detailed description of this algorithm and its performances
can be found in~\cite{bibdornic1:BBfit}. This reconstruction allows to reduce the rate of events from few Hz down to few mHz. This 
algorithm is then coupled to a more precise reconstruction tool~\cite{bibdornic1:AAfit} which allows to confirm the neutrino nature of the event and to 
improve the angular resolution. With this system, ANTARES is able to send alerts in few seconds ($\approx 3-5~s$) after the detection of the neutrinos.

Three online neutrino trigger criteria are currently implemented in the TAToO
alert system~\cite{bibdornic1:vlvnt09}:
\begin{itemize} 
\item the detection of at least two neutrino-induced muons coming from
similar directions ($<3^{o}$) within a predefined time window ($<15 min$);
\item the detection of a single high-energy neutrino induced muon.
\item the detection of a single neutrino induced muon for which the direction points toward a local galaxy.
\end{itemize}

The main performance of these three triggers are described in table~\ref{tabledornic1:perf}. For the highest energy events, the angular 
resolution gets down to less than $0.3^{o}$ (median value). Figure~\ref{figdornic1:ZenithAzimuth} shows the estimate of the point 
spread function for a typical high energy neutrino alert.

\begin{table*}
\caption{Performances of the three alert criteria. The third column corresponds to the fraction of events inside a $2^\circ \times 2^\circ$ field of view assuming a
flux of GRB~\cite{bibdornic1:GRB} and ccSNe~\cite{bibdornic1:CCSN}.}           
\label{tabledornic1:perf}    
\centering
\begin{tabular}{c c c c c}
\hline\hline
Trigger 	 & Angular Resoltion (median)   & Fraction of events in fov     & Muon contamination 	& Mean energy  	 \\        	
\hline
Doublet          &           $\leq~0.7$$^{o}$           &                               &    0~\%               & \~100 GeV  \\
\hline
single HE        &  0.25-0.3$^{o}$              &     96~\% (GRB)   68~\% (SN)  &  $<0.1~\% $           &  \~7 TeV  \\
\hline
single directional   &  0.3-0.4$^{o}$           &     90~\% (GRB)   50~\% (SN)  &  \~2~\%                &  \~1 TeV  \\
\hline
\end{tabular}

\end{table*}

\begin{figure}[ht!]
\centering
\includegraphics[width=0.45\textwidth]{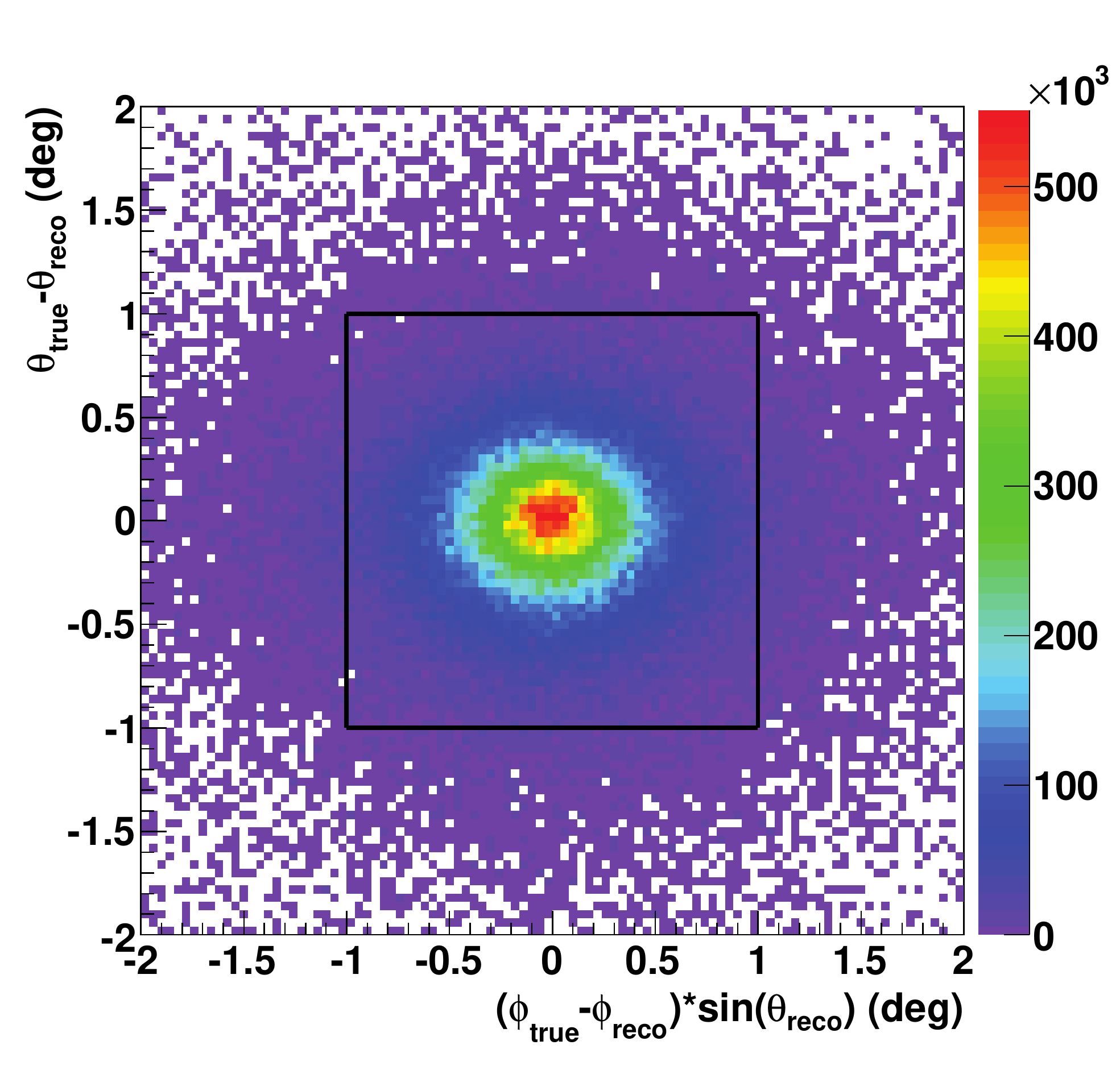}
\caption{Bi-dimensional angular resolution. The black square
corresponds to the TAROT telescope field of view ($\approx 2^\circ \times 2^\circ$).}
\label{figdornic1:ZenithAzimuth}
\end{figure}

\section{Observation strategy of the robotical telescopes}

ANTARES is organizing a follow-up program in collaboration with the TAROT and ROTSE telescopes. The 
TAROT~\cite{bibdornic1:Tarot} network is composed of two 25~cm optical robotic telescopes located at Calern 
(France) and La Silla (Chile). The ROTSE~\cite{bibdornic1:Rotse} network is composed of four 45~cm optical 
robotic telescopes located at Coonabarabran (Australia), Fort Davis (USA), Windhoek (Namibia) and Antalya 
(Turkey). The main advantages of these instruments are the large field of view of about 2 x 2 square degrees 
and their very fast positioning time (less than 10s). These telescopes are perfectly tailored for such a program. 
Thanks to the location of the ANTARES telescope in the Northern hemisphere (42.79 degrees latitude), all the six 
telescopes are used for the optical follow-up program. Depending on the neutrino trigger settings, the alert are sent 
at a rate of about twice per month. With the current settings, the connected telescopes can start taking 
images with a latency of the order of $\approx 20~s$ with respect to the neutrino event (T0) including the telescope slewing.

As already mentionned, this method is sensitive to all transient sources producing high energy 
neutrinos. For example, a GRB afterglow requires a very fast observation strategy in contrary to a core collapse 
supernovae for which the optical signal will appear several days after the neutrino signal. To be sensitive 
to all these astrophysical sources, the observational strategy is composed of a real time observation 
followed by few observations during the following month. For the prompt observation, 6 images 
with an exposure of 3 minutes and 30 images with an exposure of 1~min are taken respectively by the first 
available TAROT and ROTSE telescopes. The integrated time has been defined in order to reach an average magnitude of about 19.
For each delayed observation, six images are taken at T0+1,+2, +3, +4, +5, +6, +7, +9, +15, +27, +45, +60~days after the 
trigger for TAROT (8 images for ROTSE the same days plus T0+16 and T0+28~days). Since one year, the follow-up has been extended 
to the 1~m ZADKO telescope~\cite{bibdornic1:Zadko} located in Australia with the same observational strategy as the one used for TAROT.

\section{Optical image analysis}

Once the images are taken, they are automatically dark subtracted and flat-fielded at the telescope site. 
Once the data is copied from the telescopes, an offline analysis is performed combining the images
from all sites. This off-line program is composed of three main steps: astrometric and photometric 
calibration, subtraction of each image and a reference one and light curve determination for each variable 
candidates. 

Curently, two offline analysis pipelines are used: the ROTSE automated pipeline~\cite{bibdornic1:RotsePipeline} and one specially adapted to the TAROT and ROTSE image quality
based on the Poloka~\cite{bibdornic1:Poloka} program originally developped 
for the supernovae search in the SuperNova Legacy Survey (SNLS) project. Cases like variable PSF due to the atmospheric conditions or the lower quality images on the CCD 
edges have to be optimized in order not to loose any optical information. The choice of the reference is based on 
quality criteria such as the limiting magnitude and the seeing (i.e. mean size of the stars in the image). For the GRB search, the reference is picked among 
the follow-up observations (few days after the alert) where no GRB signal is expected anymore. For SNe 
search, the reference is either the first observation night or an additional image taken few months later to have a better quality in absence of a SN signal. 
It is also planned that the image analysis step will be included at the end of the automatic detection chain.The photometry is done using~\cite{bibdornic1:LePhare}.

\section{Results}

This paper presents the analysis of the data taking from 01/01/2010 to 31/12/2012. ANTARES was running in a reduced configuration (10-11 lines) up to November 2010 
and in the full configuration (12 lines) from this date. During this period, the efficiency of the ANTARES data taking was around 80~\% which includes periods 
of maintenance and calibration of the detector.

During this period a total of 83 alerts were successfully sent to the telescopes, 65 of which benefit from at least 3 observations. For the remaining 18 
alerts were not followed because of the telescope maintenance or due to the Sun position too close to the alert direction. 
Only 12 alerts had a good quality follow-up within less than one day (i.e. prompt follow-up).   
The lack of prompt observations is due to the telescopes observing efficiency upon the reception of the 
alert. 

The two optical image analysis pipeline have been applied to these twelve alerts from which optical images have been recorded during the first 24 hours after 
the neutrino alert sending. The minimum delay between the neutrino detection and the first image is around 20~s. No object has been found for which 
the light curve is compatible with a fast time decreasing signal. 
  
The limitation is most probably due to the sensitivity of the telescope 
and atmospheric conditions during the observation, which reduce the limiting magnitude. We define the limiting magnitude as the mean 
value of sources extracted at a Signal-to-Noise Ratio (SNR) of 5. Table \ref{tabledornic1:limit} shows the obtained limits along with the 
delay of the first image acquisition with respect to the neutrino detection. The limiting magnitude is further reduced when the 
direction is close to the galactic plane, which causes magnitudes to be dimmer and redder than they are (i.e., galactic extinction).

\begin{table*}
\caption{Magnitude limits in R-band ($5\sigma$ threshold) for the neutrino alerts. The second column indicates the time delay between the first image and the
neutrino detection. An estimate of the Galactic extinction is indicated.}           
\label{tabledornic1:limit}    
\centering
\begin{tabular}{c c c c c c}
\hline\hline
Alert  &  Time delay &  Limiting magnitude &  Galactic extinction\\   
       &   (days)   	     &         	 (R mag)   &    (R mag)\\   
\hline
 ANT100123        &  0.64 &         12.0 & 0.2 \\
\hline
ANT100302         &  1.01 &         15.7 &  0.2 \\
\hline
ANT100725         &  8.7e-4 &         14.5 & 0.3 \\
\hline         
ANT100922         &  4.7e-2 &         14.0 & 0.5 \\
\hline
ANT101211         &  0.50 &          15.1 & 0.1 \\
\hline
ANT110409         &  3.0e-3 &         18.1 & 6.7 \\
\hline
ANT110529         &  5.2e-3 &          15.6 & 1.2 \\
\hline
ANT110613        &  7.8e-4 &         17.0 & 2.3 \\
\hline
ANT120730        & 2.4e-4 &          17.6 & 0.4 \\
\hline
ANT120907        & 2.9e-4 &          16.9 & 0.2 \\
\hline
ANT121010        & 2.8e-4 &         18.6 & 0.1 \\
\hline
ANT121206        & 3.1 e-4 &         16.9 & 1.3 \\

\hline
\end{tabular}
\end{table*}

\section{Conclusion}

The method used by the ANTARES collaboration
to implement the search for coincidence between high energy neutrinos and
transient sources followed by small robotic telescopes has been presented. 
Of particular importance for this alert system are the ability
to reconstruct online the neutrino direction and to efficiently reject the
background. With the described ANTARES alert sending capability,
the connected optical telescopes can start taking images with a latency of the order of $\approx 20~s$. 
The precision of the direction of the alert is better than 0.5 degres.

The alert system is operational since February 2009, and as of December 2012,
83 alerts have been sent, all of them triggered by the two single neutrino 
criteria. No doublet trigger has been recorded yet. After a commissioning phase in 2009, almost all alerts had an optical
follow-up since 2010, and the live time of the system over this year
is strictly equal to the one of the ANTARES telescope, 70-80\%.
These numbers are consistent with the expected trigger rate, after accounting for the duty cycle of the neutrino telescope.
The image analysis of the 'prompt' images has not permitted yet to discover any transient sources associated to the selected high energy neutrinos, 
in particular no GRB afterglow. 
The analysis of the rest of the images to look for the light curve of a core collapse SNe is still on-going.

The optical follow-up of neutrino events significantly improves the perspective
for the detection of transient sources. A confirmation by an optical telescope
of a neutrino alert will not only provide information on the nature of
the source but also improve the precision of the source direction determination in order to
trigger other observatories (for example very large telescopes for
redshift measurement). The program for the follow-up of ANTARES neutrino
events is already operational with the TAROT, ROTSE and ZADKO telescopes. 
This technique has been recently extended to the follow-up in X-ray with the Swift/XRT
telescope to further improve the sensitivity to fast transient sources, like GRBs.

\vskip 0.5cm
{\footnotesize{\bf %
\noindent
Acknowledgment: }
This work has been financially supported by the GdR PCHE in France. 
}

\setcounter{figure}{0}
\setcounter{table}{0}
\setcounter{footnote}{0}
\setcounter{section}{0}
\setcounter{equation}{0}

\newpage
\id{id_dornic2}

\title{\arabic{IdContrib} - Search for neutrino emission of gamma-ray flaring blazars with the {\sc Antares} telescope}
\addcontentsline{toc}{part}{\arabic{IdContrib} - {\sl Damien Dornic} : Search for neutrino emission of gamma-ray flaring blazars%
\vspace{-0.5cm}
}
\shorttitle{\arabic{IdContrib} - Time-dependent neutrino point source search}
\authors{
Dornic D.$^{1}$,
S\'{a}nchez-Losa A. $^{2}$, 
Coyle P.$^{1}$ 
for the {\sc Antares} Collaboration
}
\afiliations{
$^1$ CPPM, Aix-Marseille Universit\'{e}, CNRS/IN2P3, Marseille, France \\
$^2$ IFIC - Instituto de F\'{i}sica Corpuscular, Edificios Investigaci\`{o}n de Paterna, CSIC - Universitat de Valencia, Apdo. de Correos 22085, 46071 Valencia, Spain	
}
	     
\email{dornic@cppm.in2p3.fr}

\abstract{
The ANTARES telescope is well suited to detect neutrinos produced in astrophysical transient 
sources as it can observe a full hemisphere of the sky at all the times with a duty cycle 
close to unity. The background and point-source sensitivity can be drastically reduced by 
selecting a narrow time window around the assumed neutrino production period. Radio-loud 
active galactic nuclei with their jets pointing almost directly towards the observer, the 
so-called blazars, are particularly attractive potential neutrino point sources, since they 
are among the most likely sources of the observed ultra high energy cosmic rays and therefore, 
neutrinos and gamma-rays may be produced in hadronic interactions with the surrounding medium. 
The gamma-ray light curves of blazars measured by the LAT instrument on-board the Fermi 
satellite reveal important time variability information. A strong correlation between the 
gamma-ray and the neutrino fluxes is expected in this scenario.

An unbinned method based on the minimization of a likelihood ratio was applied to a subsample 
data collected between 2008 and 2011. By looking for neutrinos detected in the high state period 
of the AGN light curve, the sensitivity to these sources has been improved by a factor 2-3 with 
respect to a standard time-integrated point source search. The assumed neutrino time distribution 
is directly extracted from the gamma-ray light curve. The typical width for a flare ranges from 
1 to 100 days depending on the source. The results of this analysis are presented.
}

\maketitle

\section{Introduction}
The production of high-energy neutrinos has been proposed for several kinds
of astrophysical sources, such as active galactic nuclei, gamma-ray bursters, supernova
remnants and microquasars, in which the acceleration of hadrons may occur. 
Neutrinos are unique messengers to study the high-energy universe as there are neutral and 
stable, interact weakly and travel directly from their point of creation in the source without absorption. 
Neutrinos could play an important role in understanding the mechanisms of cosmic ray acceleration and their 
detection from a source would be a direct evidence of the presence of hadronic acceleration in that source.

Radio-loud active galactic nuclei with their jets pointing almost directly towards
the observer, the so-called blazars, are particularly attractive potential neutrino point
sources, since they are among the most likely sources of the observed ultra high energy
cosmic rays and therefore, neutrinos and gamma-rays may be produced in hadronic
interactions with the surrounding medium~\cite{bibdornic2:AGNhadronic}. The gamma-ray light curves of blazars
measured by the LAT instrument on-board the Fermi satellite reveal important time
variability information on timescale of hours to several weeks, with intensities
always several times larger than the typical flux of the source in its quiescent state~\cite{bibdornic2:FermiLATAGNvariability}. 
A strong correlation between the gamma-ray and the neutrino fluxes is expected in this scenario. 

In this paper, the results of the time-dependent search for cosmic neutrino sources in
the sky visible to the ANTARES telescope using data taken from 2008 to 2011 are presented. 
This analysis is the extension of a previous ANTARES analysis~\cite{bibdornic2:flare} using only the last four months of 2008. 
The data sample used in
this analysis is described in Section 2,
together with a discussion on the systematic uncertainties. The point source search
algorithm used in this time-dependent analysis is explained in Section 3. The results are presented in
Section 4 for a search on a list of ten selected candidate sources.

\section{ANTARES}

The ANTARES collaboration has completed the construction of a neutrino
telescope in the Mediterranean Sea with the connection of its twelfth detector line
in May of 2008~\cite{bibdornic2:Antares}. The telescope is located 40 km on the southern coast of France
(42$^{o}$48'N, 6$^{o}$10'E) at a depth of 2475 m. It comprises a three-dimensional array of
photomultipliers housed in glass spheres (optical modules), distributed along twelve
slender lines anchored at the sea bottom and kept taut by a buoy at the top. 
Each line comprises up to 25 storeys of triplets of optical modules (OMs), each housing a single 10" PMT. 
Since lines are subject to the sea current and can change shape and orientation, a positioning system 
comprising hydrophones and compass-tiltmeters is used to monitor the detector geometry. The main goal of the experiment 
is to search for neutrinos of astrophysical origin 
by detecting high energy muons ($>$100~GeV) induced by their neutrino charged current interaction in the vicinity 
of the detector~\cite{bibdornic2:Antares2}. 

The arrival time and intensity of the Cherenkov light on the OMs are digitized into hits and transmitted to shore, 
where events containing muons are separated from the optical backgrounds due to natural radioactive decays and bioluminescence, 
and stored on disk. A detailed description of the detector and the data acquisition is given in~\cite{bibdornic2:Antares}~\cite{bibdornic2:antaresdaq}. The arrival times 
of the hits are calibrated as described in~\cite{bibdornic2:TimeCalib}. The online event 
selection identifies triplets of OMs that detect multiple photons. At least 5 of these are required throughout the detector, 
with the relative photon arrival times being compatible with the light coming from a relativistic particle. Independently, 
events were also selected which exhibit multiple photons on two sets of adjacent, or next to adjacent floors. 

The data used in this analysis corresponds to the period from September 6th, 2008 up to 
December 31st, 2011 (54720-55926 modified Julian day), taken with the full detector. Some filtering has been applied 
in order to exclude periods in which the bioluminescence-induced optical background was high. The resulting effective lifetime is 750 days.
Atmospheric neutrinos are the main source of background in the search for astrophysical neutrinos. These neutrinos 
are produced from the interaction of cosmic rays in the Earth's atmosphere. Only charged 
current interactions of neutrinos and antineutrinos were considered. An additional source of background 
is due to the mis-reconstructed atmospheric muons. 
The track reconstruction algorithm derives the muon track parameters that maximize a likelihood function built from the 
difference between the expected and the measured arrival time of the hits from the Cherenkov photons emitted along the muon 
track. This maximization takes into account the Cherenkov photons that scatter in the water and the additional photons 
that are generated by secondary particles (e.g. electromagnetic showers created along the muon trajectory). The algorithm used is outlined in~\cite{bibdornic2:AAfit}.
The value of the log-likelihood per degree of freedom ($\Lambda$) from the track reconstruction fit is a measure of the track fit 
quality and is used to reject badly reconstructed events, such as atmospheric muons that are mis-reconstructed as upgoing tracks. 
Neutrino events are selected by requiring that tracks are reconstructed as upgoing and have a good reconstruction quality. In addition, the error 
estimate on the reconstructed muon track direction obtained from the fit is required to be less than 1$^{o}$. This additional cut allows to further reduce the bad-reconstructed muon
contamination.

The angular resolution can not be determined directly in the data and has to be estimated from simulation. However, the comparison of 
the data and MonteCarlo simulations from which the time accuracy of the hits has been degraded has yielded to a constrain on the 
uncertainty of the angular resolution of the order of 0.1$^{o}$~\cite{bibdornic2:AAfitps}. Figure~\ref{figdornic2:Angres} shows the cumulative distribution of the angular difference between 
the reconstructed muon direction and the neutrino direction with an assumed spectrum proportional to $E_{\nu}^{-2}$, where $E_{\nu}$ is 
the neutrino energy. For this period, the median resolution is estimated to be 0.4 +/- 0.1 degree. 

\begin{figure}[ht!]
\centering
\includegraphics[width=0.4\textwidth]{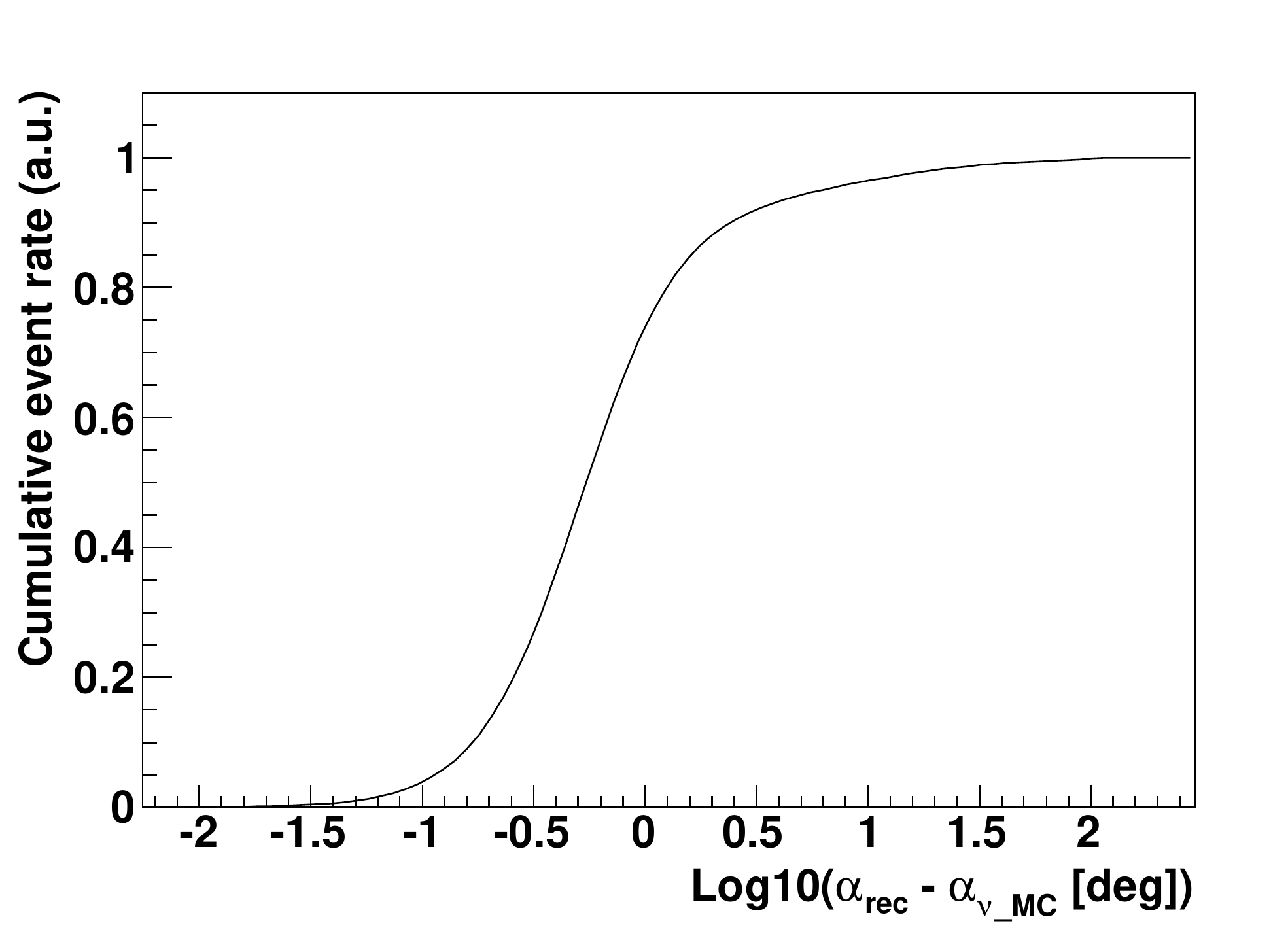}
\caption{Cummulative distribution of the angle between the true Monte Carlo neutrino direction and the reconstructed
muon direction for E$^{-2}$ upgoing neutrino events selected for this analysis.}
\label{figdornic2:Angres}
\end{figure}

\section{Time-dependent search algorithm}
This time dependent point source analysis is performed using an unbinned method based on a
likelihood ratio maximization. The data is parameterized as a two components
mixture of signal and background. The signal and the background are described by two probability distribution function (PDF), $S_{i}$ and $B_{i}$ respectively. These PDF 
are described by the product of three components: one for the direction, one for the energy and one for the timing. The signal PDF is:

\begin{equation}
S_{i} = S_{i}^{space}(\Psi_{i}(\alpha_{s},\delta_{s})).S_{i}^{energy}(Nhit_{i})S_{i}^{time}
\label{eqdornic2:EQ_likelihood}
\end{equation}
where $S_{i}^{space}$ is a parametrization of the point spread function, i.e., the probability density function of reconstruction an event i at an angular distance $\Psi_{i}$ from 
the true source location
($\alpha_{s}$,$\delta_{s}$).The energy PDF is pararametrized based on the distribution of the number of hits, $Nhit_{i}$, of an event according to an energy spectrum of $E^{-2}$. The shape of the 
time PDF, $S_{i}^{time}$, for the signal event is extracted directly from the gamma-ray light curve assuming 
 the proportionality between the gamma-ray and the neutrino fluxes. A possible lag up to +/- 5 days has been introduced in the Likelihood to test this proportionality.

The background PDF is:

\begin{equation}
B_{i} = B_{i}^{space}(\alpha_{i},\delta_{i}).B_{i}^{energy}(Nhit_{i})B_{i}^{time}
\label{eqdornic2:EQ_likelihood2}
\end{equation}

The directional PDF $B_{i}^{space}$, the energy PDF $B_{i}^{energy}$ and the time PDF $B_{i}^{time}$ for the background are derived from the data using respectively the observed 
declination distribution of selected events 
in the sample, the distribution of the number of hits and the observed time distribution of all the reconstructed muons. 
Figure~\ref{figdornic2:TimeDistri} shows the time distribution of all the reconstructed events. Once normalized, 
the distribution is used directly as the time PDF for the background.
Null values indicate the absence of data taken during these periods (ie detector in maintenance) or data with a very poor quality (high bioluminescence or bad calibration).

\begin{figure}[ht!]
\centering
\includegraphics[width=0.4\textwidth]{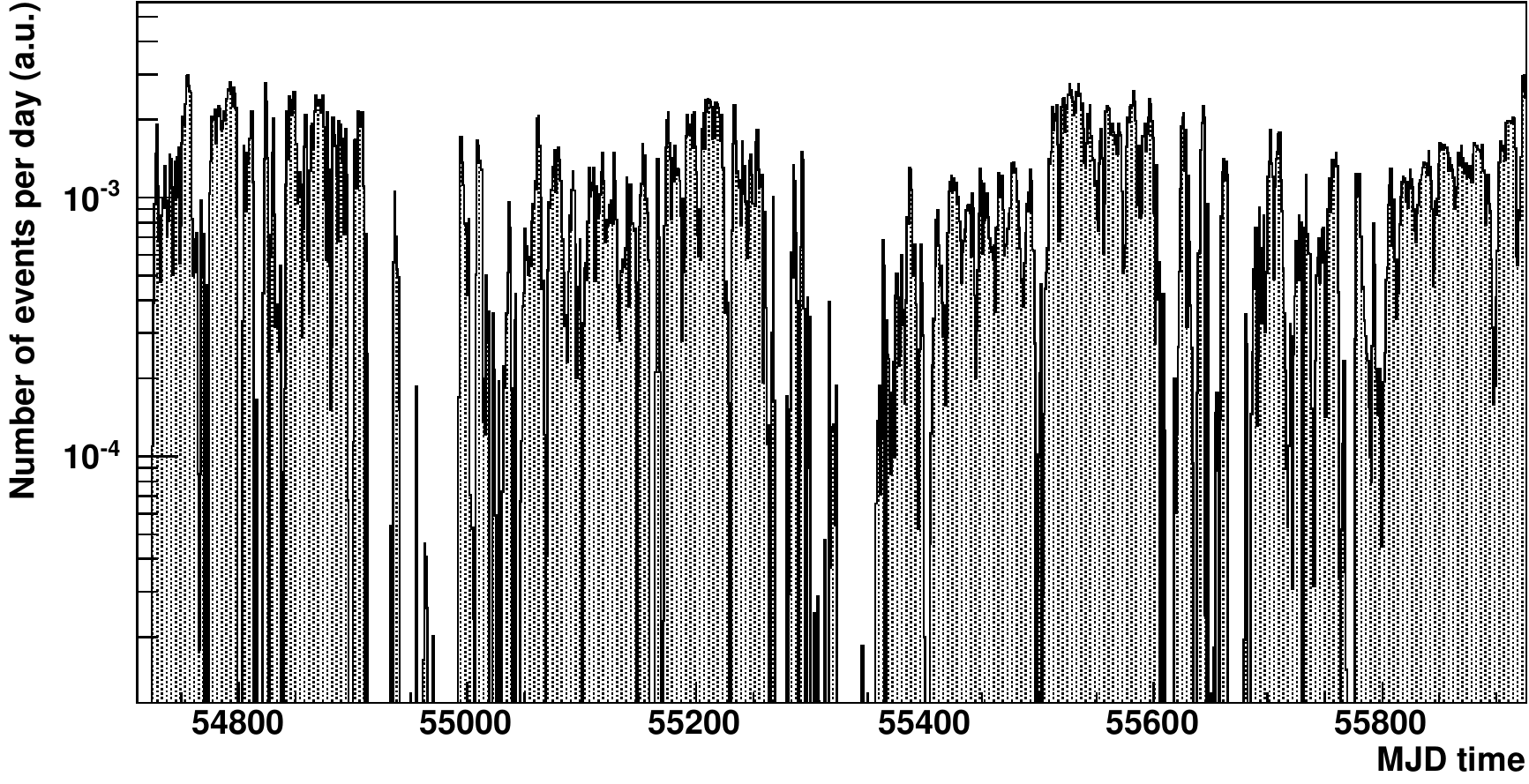}
\caption{Time distribution of the well reconstructed events. 
}
\label{figdornic2:TimeDistri}
\end{figure}

The goal is to determine, at a given point in the sky and at a given
time, the relative contribution of each component and to calculate the probability to have a signal
above a given background model. This is done via the test statistic, $\lambda$, which is the ratio of the probability density for the hypothesis 
of background and signal ($H_{sig+bkg}$) over the probability density of only background ($H_{bkg}$):

\begin{equation}
\lambda=\sum_{i=1}^{N} log\frac{P(x_{i}|H_{sig+bkg}(n_{sig}))}{P(x_{i}|H_{bkg})} 
\label{eqdornic2:TS}
\end{equation}
where $n_{sig}$ and N are respectively the unknown number of signal events and the total number of events in the considered data sample. 

The null hypothesis is given by $n_{sig}=0$. The obtained value of $\lambda_{data}$ on the data is then 
compared to the distribution of $\lambda$ given the null hypothesis. Large values of $\lambda_{data}$ compared to the distribution of $\lambda$ for the 
background only reject the null hypothesis with a confident level equal to the fraction of the scrambled trials above $\lambda_{data}$. This fraction of 
trials above $\lambda_{data}$ is referred to as the p-value. The discovery potential is then defined as the average number of signal events required to 
achieve a p-value lower than 5$\sigma$ in 50~$\%$ of trials. Figure~\ref{figdornic2:Nev5sigma} shows the average number of events required for a 5$\sigma$ discovery (50~\% C.L.) 
for a single source located at a declination of -40$^{o}$ as 
a function of the total width of the flare periods. These numbers are compared to the one obtained without using the timing information. Using the timing information yields 
an improvement of the discovery potential of a factor 2-3 with respect to a standard time-integrated point source search~\cite{bibdornic2:AAfitps}. Moreover, 
adding the energy information yields an additional improvement of about 25~$\%$.

\begin{figure}[ht!]
\centering
\includegraphics[width=0.4\textwidth]{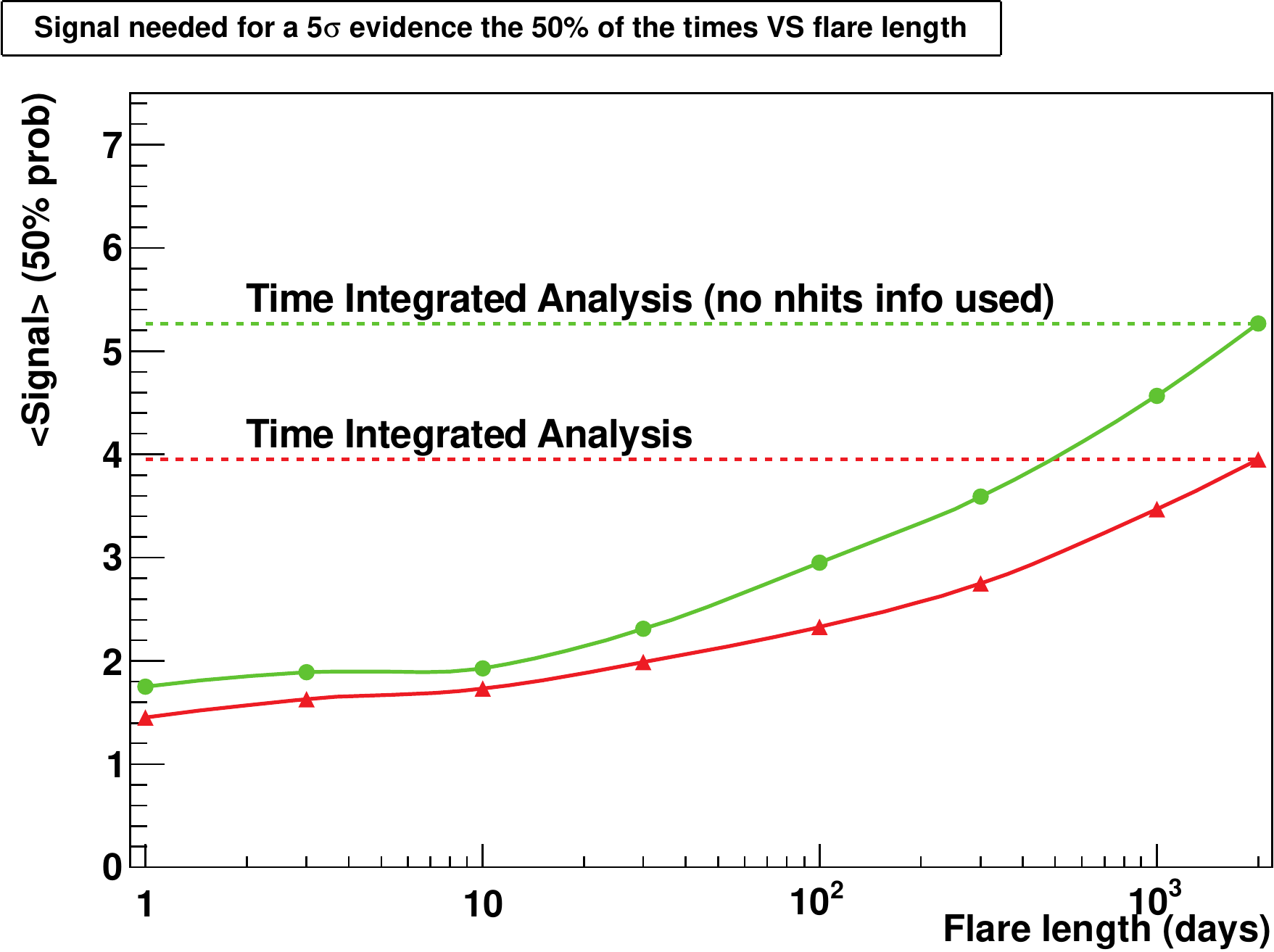}
\caption{Average number of events required for a 5$\sigma$ discovery (50~$\%$ C.L.) for a source located at a
declination of -40$^{o}$ as a function of the width of the flare period. These numbers are compared to the one obtained without
using the timing information (flat dashed lines).These distributions are computed with (red) and without (green) using the energy information in the likelihood.}
\label{figdornic2:Nev5sigma}
\end{figure}

\section{Search for neutrino emission from gamma-ray flare}

This time-dependent analysis has been applied to bright and variable Fermi blazar sources reported in the second year Fermi LAT catalogue~\cite{bibdornic2:Fermicatalogue} and in the LBAS catalogue 
(LAT Bright AGN sample~\cite{bibdornic2:FermicatalogueAGN}). The sources located in the part of the sky visible by ANTARES ($\delta$$<$35$^{o}$) from which the flux is greater than 10$^{-9}$
photons.cm$^{-2}$.s$^{-1}$ above 1~GeV, a detection significance $TS>25$ in the studied time period and with a significant time variability are selected. This list is completed by adding
sources reported as flaring in the Fermi Flare advocates in 2011~\cite{bibdornic2:FermiAdvocates}). This list includes a total of 136 sources. 

The light curves are produced using the Fermi Public Release Pass 7 data using the diffuse class event selection and the Fermi Science Tools v9r23p1 package~\cite{bibdornic2:FermiData}. Light curves are computed from the 
photon counting in a two degrees cone around the studied source direction corrected by the total exposure. This light curve corresponds to 
the one-day binned time evolution of the average gamma-ray flux above a threshold of 100~MeV from August 2008 to December 2011. Finally, a maximum likelihood block (MLB) algorithm is used to denoise 
the lightcurve by iterating over the data points to select periods from which data are consistent with a constant flux taking into account statistical errors. 

The high state periods are defined using a simple and robust method. After the determination of the value of the steady state, a fixed threshold on the gamma-ray flux is applied to select the high
state period (5 sigma above the steady state). Finally, an additional delay of 0.5 day is added before and after the flare in order to take into account that the precise time of the flare is not 
known (1-day binned LC). With this definition, a flare has a width of at least two days. Figure~\ref{figdornic2:4C2135} shows the time distribution of the Fermi LAT
gamma-ray light curve of 4C+21.35 for the studied and the determined high state periods (blue histogram). With the hypothesis that the neutrino emission follows the gamma-ray emission, the signal time PDF is 
simply the normalized denoised light curve. The final list includes 40 bright and variable Fermi blazars, 32 Flat Spectrum Radio Quasars, 6 BL-Lacs and 2 unknowned identifications. The list is 
composed by PKS1510-089, 3C279, PKS1502+106, PKS2326-502, 3C273, AO0235+164, PKS0426-380, 4C+28.07, PKS0454-234, PKS1329-049, PKS0537-441, 4C+14.23, PMNJ 0531-4827,
PKS0402-362, PKS1124-186, Ton~599, PKS2142-75, PKS0208-512, PKS0235-618, PKS1830-211, PKS2023-07, PKSB1424-418, PMNJ2345-1555, OJ~287, PKS0440-00, PKS0250-225,
B22308+34, B21520+31, PKS1730-13, PKS0301-243, PKS0405-385, CTA~102, OG~050, PMNJ2331-2148, TXS0506+056, PKS2320-035, PKS0521-36 and PKS2227-08. The selected 
periods are mostly dominated by three main sources: 3C354.3, 4C +21.35 and PKS1510-089. 

\begin{figure}[ht!]
\centering
\includegraphics[width=0.4\textwidth]{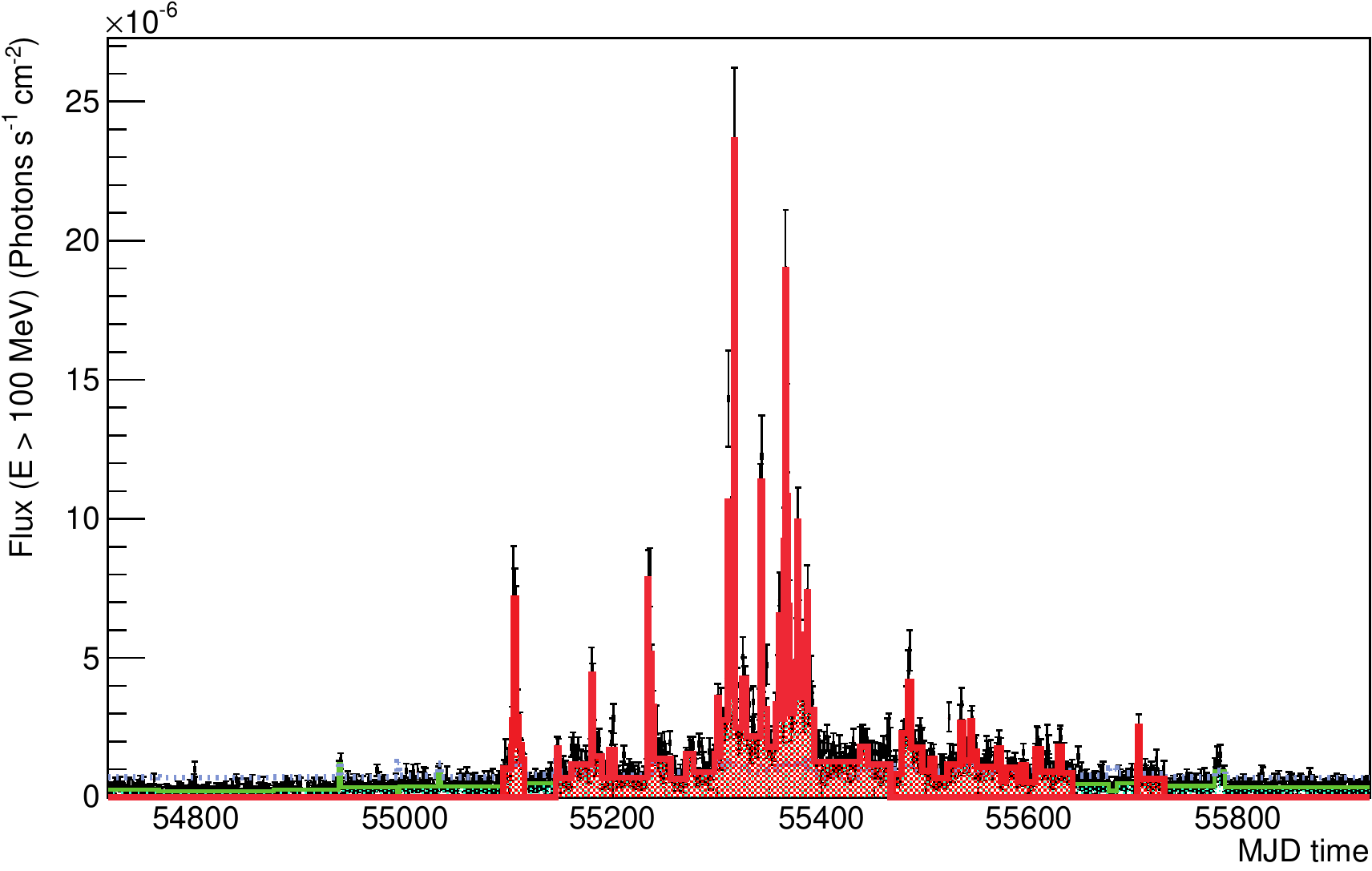}
\caption{Gamma-ray light curve (black dots) of the blazar 4C+21.35 measured by the LAT instrument onboard the Fermi satellite above 100~MeV for more than 3.5 years 
of data. The green line shows the denoise light curve while the red histogram displays the selected high state periods. 
}
\label{figdornic2:4C2135}
\end{figure}

\begin{figure}[ht!]
\centering
\includegraphics[width=0.4\textwidth]{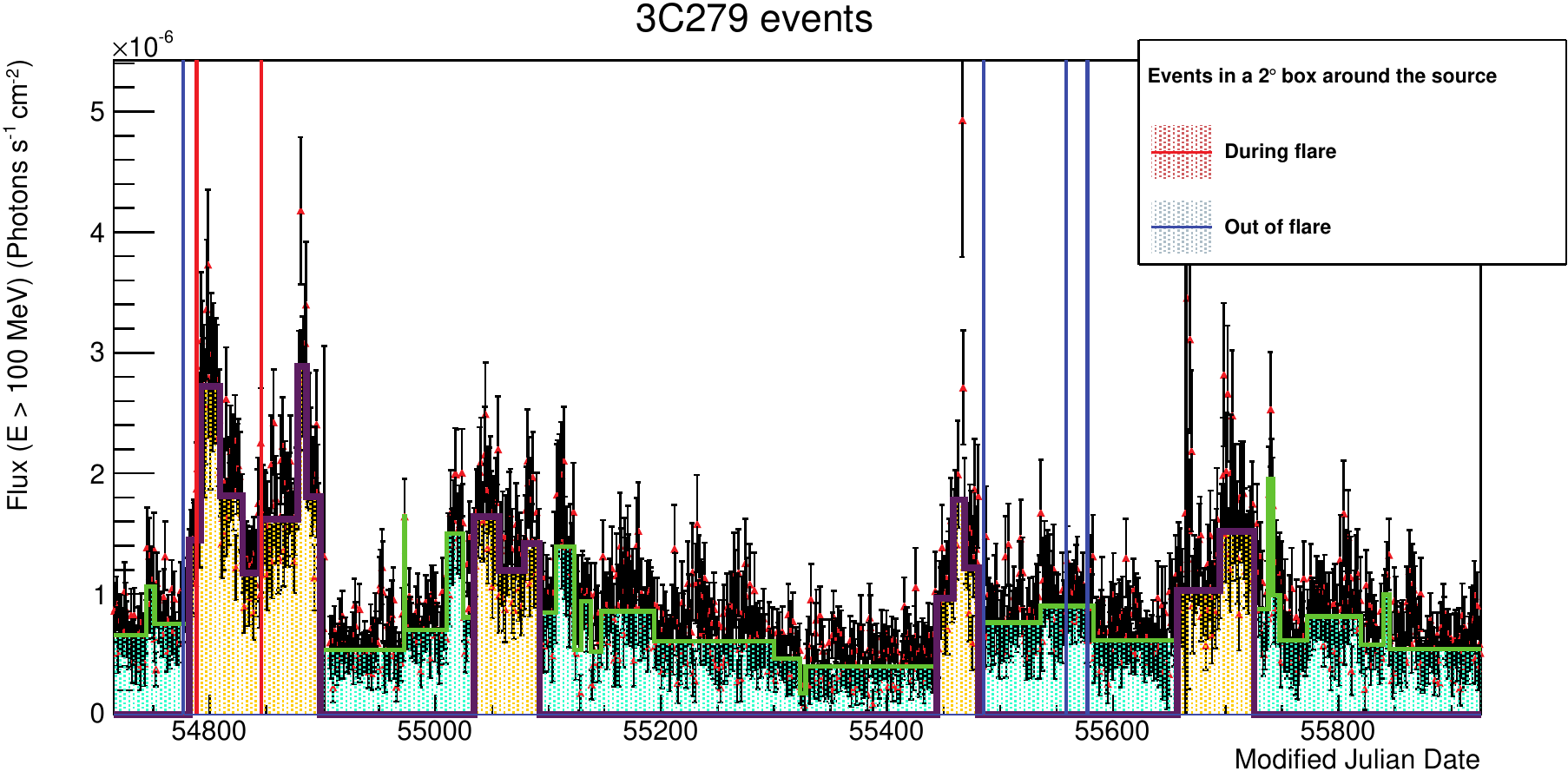}
\caption{Gamma-ray light curve (black dots) of the blazar 3C279 measured by the LAT instrument above 100~MeV. The green and yellow histogram show the denoise light curve and the selected flare periods
respectively. 
The red and blue lines display the time of the two ANTARES events associated with the source and all the events in a two degres box around the source position.
}
\label{figdornic2:3C279results1}
\end{figure}

\begin{figure}[ht!]
\centering
\includegraphics[width=0.4\textwidth]{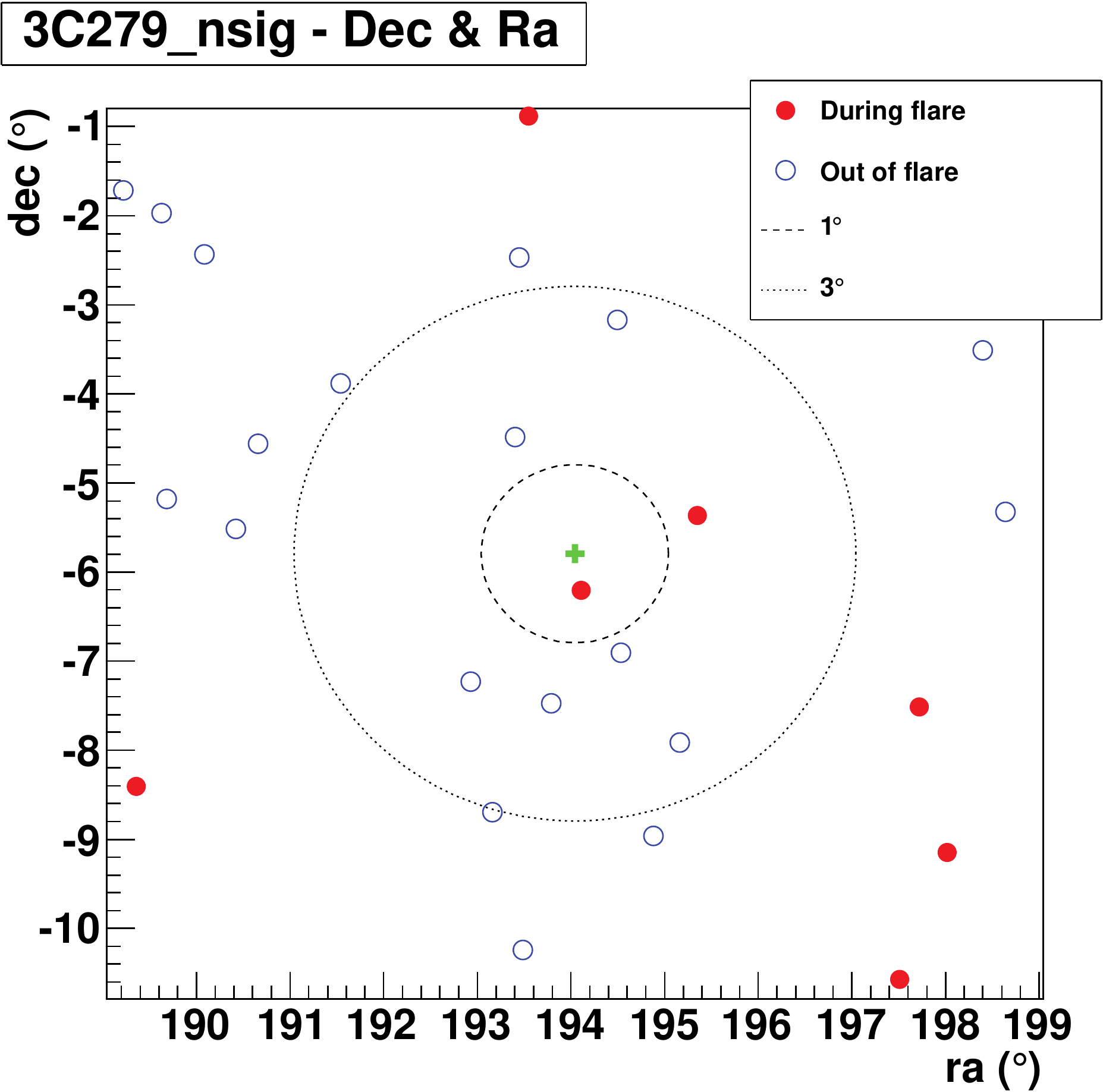}
\caption{Event map around the direction of 3C279 indicated by the green cross with two events compatible in time and direction (red dots). The blue empty dots
represent the direction of all the selected events outside the flaring periods.
}
\label{figdornic2:3C279results2}
\end{figure}

The results of these searches are summarized in the table~\ref{tabledornic2:results}. Only for six sources (3C279, 3C454.3, 4C+21.35, PKS1830-211, PKS1124-186 and CTA~102), 
a neutrino signal has been found in coincidence. The lower
p-value is obtained for the source 3C279 where two events are coincident with a gamma-ray flare detected by the Fermi/LAT. The pre-trial p-value is 0.17~$\%$.
Figure~\ref{figdornic2:3C279results1} and ~\ref{figdornic2:3C279results2} show respectively the time distribution of the Fermi gamma-ray light curve of 3C279 with the time of
the coincident neutrino events and the angular distribution of the events around the position of this source. The post-trial probability computed taking into
account the forty searches is 9.9~$\%$ and thus is compatible with background fluctuations.

\begin{table}[ht!]
\caption{Results of the search of neutrino coincidences with Fermi blazars. The second column indicates the total selected flare duration for each source. The third and
fourth column show the number of fitted signal events with the likelihood and the corresponding pre-trial probability respectively.}           
\label{tabledornic2:results}    
\centering
\begin{tabular}{c c c c}
\hline\hline
Source 	 & Flares (days)  & Fitted signal     & p-value  	 \\        	
\hline
3C279   &  279                    & 1.08      & 0.2~$\%$   \\
\hline
PKS1124-186   &  73                    & 1.04      & 1.1~$\%$   \\
\hline
PKS1830-211   &  63                    & 0.93      & 1.4~$\%$   \\
\hline
3C454.3   &  830                    & 0.77      & 3.5~$\%$   \\
\hline
4C21.35   &  515                    & 0.73      & 3.7~$\%$   \\
\hline
CTA 102   &  16                    & 0.62      & 4.6~$\%$   \\
\hline
\end{tabular}

\end{table}

\section{Summary}
This paper discusses the extended time-dependent search for cosmic neutrinos using the data taken with the full 12 lines ANTARES detector between 2008 and 2011. 
Time-dependent searches are significantly more sensitive than standard point-source searches to variable sources thanks to the large reduction of the background of 
atmospheric muons and neutrinos over short time scales. This search has been applied to 40 very bright and variable Fermi LAT blazars. The most significant
corelation was found with a flare of the blazar 3C279 for which two neutrino events were detected in time/direction coincidence with the famma-ray emmission. The
post-trial probability is about 10~$\%$. Upper-limits were then obtained on the neutrino fluence for the selected sources.




\setcounter{figure}{0}
\setcounter{table}{0}
\setcounter{footnote}{0}
\setcounter{section}{0}
\setcounter{equation}{0}

\newpage
\id{id_thierry}

%
\def\ant{{\sc Antares}}
\def\vo{{\sc Virgo}}
\def\lo{{\sc LIGO}}
\title{\arabic{IdContrib} - Searches for coincident High Energy Neutrinos and Gravitational Wave Bursts
using the {\sc Antares} and {\sc VIRGO/LIGO} detectors}
\addcontentsline{toc}{part}{\arabic{IdContrib} - {\sl Thierry PRADIER} : 
Coincident High Energy Neutrinos and Gravitational Wave Bursts
}

\shorttitle{\arabic{IdContrib} - GWHEN searches using \ant~and \lo/\vo~data in 2007 and 2009-2010}

\authors{
T. PRADIER$^{1}$
for the {\sc ANTARES Collaboration}, the {\sc LIGO Scientific Collaboration} and the {\sc Virgo Collaboration}.
}

\afiliations{
$^1$ IPHC - University of Strasbourg and CNRS/IN2P3, 23 rue du Loess, BP 28, 67037
Strasbourg Cedex 2, France - pradier@in2p3.fr
}


\abstract{
Cataclysmic cosmic events can be plausible sources of both gravitational waves (GW) and high energy neutrinos
(HEN). Both GW and HEN are alternative cosmic messengers that may traverse very dense media and travel 
unaffected over cosmological distances, carrying information from the innermost regions of the astrophysical 
engines. Such messengers could also reveal new, hidden sources that have not been observed by conventional 
photon astronomy.
The \ant~Neutrino Telescope can determine accurately the time and direction of HEN events, 
and the \vo/\lo~network of GW interferometers can  provide timing/directional information 
for GW bursts. Combining these informations obtained from totally independent detectors 
provide a novel way of constraining the processes at play in the sources, and also help confirming the 
astrophysical origin of a HEN/GW signal in case of concomitant observation. 

This contribution describes the first joint GW+HEN search performed using concomitant data taken with the 
\ant, \vo~and \lo~detectors in 2007, during the \vo~VSR1 and \lo~S5 science runs, while \ant~was operating 
in a 5-line configuration, approximately half of its final size. No coincident GW/HEN event 
was observed, which allowed for the first time to place upper limits on the density of joint GW+HEN emitters, 
which can be compared to the densities of mergers and core-collapse events in the local universe. 
More stringent limits will be soon available by performing a new and optimized search, described in this 
contribution, using the data of the full \ant~telescope in 2009-2010, concomittant with the S6 \lo~science 
run and VSR2-3 \vo~science runs, where all the involved interferometers took data with improved sensitivities.
}


\maketitle

\section{Introduction}

A new generation of detectors offer unprecedented opportunities to observe 
the universe through all kind of cosmic radiations. In particular, both high-energy (TeV) neutrinos (HEN) and gravitational waves (GW), 
which have not yet been directly observed from astrophysical sources, are considered as promising tools for the development of a multi-messenger 
astronomy (see e.g.~\cite{bibthierry:pradiermm} for a recent review of HEN-related searches). Both HEN and GW can escape 
from the core of the sources and travel over large distances through magnetic fields and matter without being altered. 
They are therefore expected to provide important information about the processes taking place in the core of the production sites and they 
could even reveal the existence of sources opaque to hadrons and photons, that would have remained undetected so far.
The detection of coincident signals in both these channels would then be a landmark event and sign the first observational evidence that GW 
and HEN originate from a common source. The most plausible astrophysical emitters of GW+HEN are presented in Section~\ref{secthierry:sources}.

The concomitant operation of GW and HEN detectors is summarized in the 
time chart of Fig. \ref{figthierry:calendar}.  Section~\ref{secthierry:det} briefly describes the detection principles and the performances achieved by 
the \ant~neutrino telescope \cite{bibthierry:antares} as well as by the GW inter\-fe\-ro\-meters \vo~\cite{bibthierry:virgo} and \lo~\cite{bibthierry:ligo}, 
that are currently part of this joint search program. 
As both types of detectors have completely independant sources of backgrounds,  the correlation between HEN and GW significances can 
also be exploited to enhance the sensitivity of the joint channel, even in the absence of detection. The combined false alarm rate is 
indeed severely reduced by the requirement of space-time consistency between both channels.  In Sections~\ref{secthierry:ana} and \ref{secthierry:ana2}, the strategies 
being developed for joint GW+HEN searches between \ant~and the network of GW interferometers using the currently 
available datasets are presented. The results of the first GWHEN search have recently been published \cite{bibthierry:gwhen2007}. 

\begin{figure}[ht!]
 \centering
 \includegraphics[width=0.5\textwidth,clip]{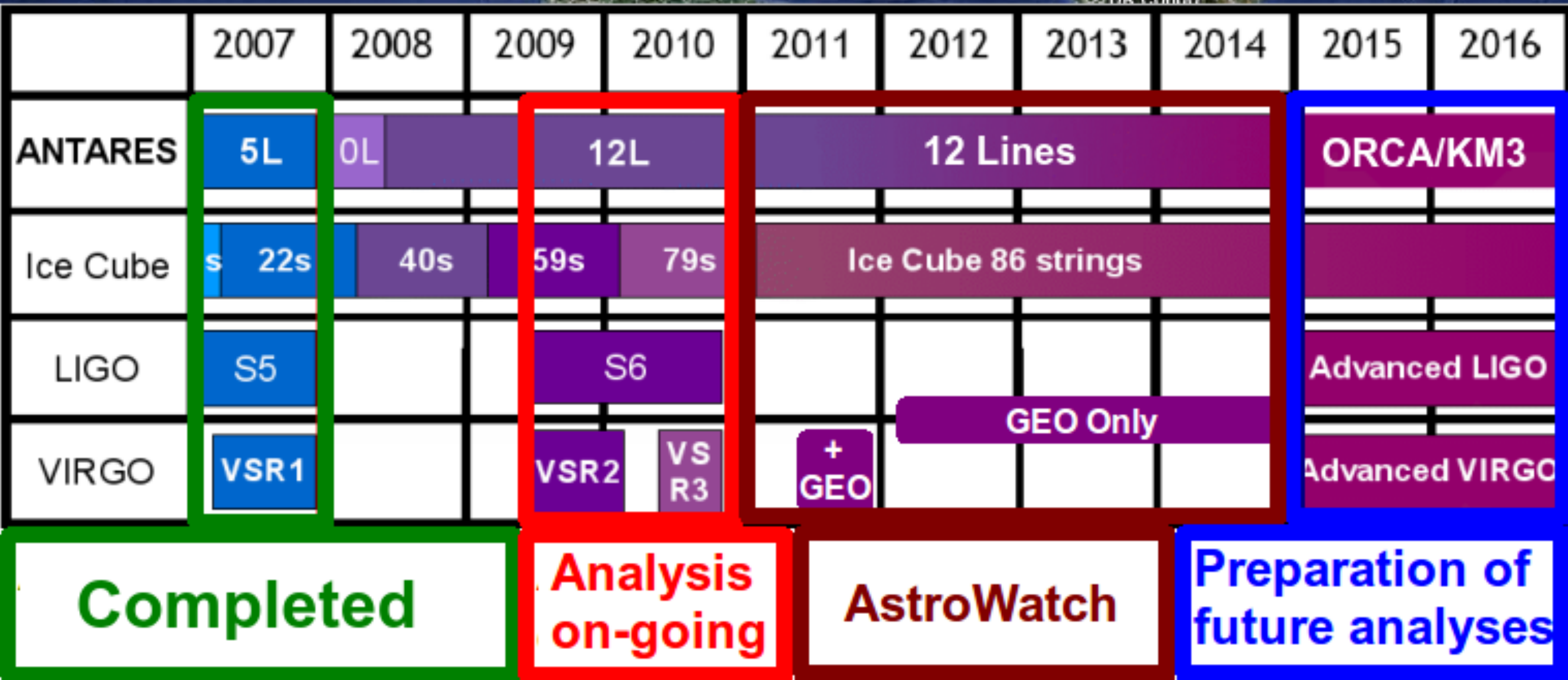}      
  \caption{Time chart of the data-taking periods for the ANTARES, VIRGO and LIGO experiments, indicating the respective upgrades of the detectors (as described in the text). 
  The deployment of the KM3NeT neutrino telescope is expected to last three to four years, 
  during which the detector will be taking data with an increasing number of PMTs before reaching its final configuration \cite{bibthierry:km3}.}
  \label{figthierry:calendar}
\end{figure}


\section{Joint GW-HEN astrophysical emitters}
\label{secthierry:sources}

Potential sources of GWs and HENs are likely to be very energetic and to exhibit bursting activity. 
The most promising class of known extragalactic bursting sources are surely Gamma-Ray Bursts (GRBs), most frequent and better modelled - for 
details on these sources and other possible emitters, refer to \cite{bibthierry:gwhen2007}.
In the prompt and afterglow phases, 
HEN ($10^5 - 10^{10}$ GeV) are expected to be produced by accelerated protons in relativistic shocks and several models predict detectable 
fluxes in km$^3$-scale detectors~\cite{bibthierry:GRBnu1,bibthierry:GRBnu2,bibthierry:GRBnu3}, although no evidence for GRB neutrinos has been observed yet by IceCube 40~\cite{bibthierry:ice3nu}.  
While gamma-ray and HEN emission from GRBs are related to the mechanisms driving the relativistic outflow, GW emissions are closely connected 
to the central engine and hence to the GRB progenitor. 
{\bf Short-hard GRBs} are thought to originate from coalescing binaries involving  black holes and/or neutron stars; such mergers could emit 
 GW detectable from relatively large distances, with significant associated HEN fluxes \cite{bibthierry:GRBshort1,bibthierry:GRBshort2}. 
 {\bf Long-soft GRBs} are most probably  induced by  "collapsars",  i.e. collapses of massive stars into black holes, with the formation of 
 an accretion disk and a jet that emerges from the stellar envelope~\cite{bibthierry:GRBlong}.
{\bf Low-luminosity GRBs}, with $\gamma$-ray luminosities a few orders of magnitude smaller, are believed to originate from 
  particularly energetic Type Ib/c core-collapse supernovae. They could produce stronger GW signals together with significant high- and low-energy neutrino emission; moreover they are more frequent 
  than typical long GRBs and often discovered at shorter distances \cite{bibthierry:razzaque}.
  Finally, {\bf choked GRBs} are thought to be associated with supernovae driven by mildly relativistic, baryon-rich and optically thick jets, 
  so that no $\gamma$-rays escape~\cite{bibthierry:choked}. Such ``hidden sources'' could be among the most promising emitters of GW and HEN, as current 
  estimates predict a relatively high occurrence rate in the volume probed by current GW and HEN detectors \cite{bibthierry:ando}.

\section{Detectors and concomittant data taking}
 \label{secthierry:det}

  The \ant~detector \cite{bibthierry:antares} is the first undersea neutrino telescope;  its deployment at a depth of 2475m in the Mediterranean 
  Sea near Toulon was completed in May 2008 \cite{bibthierry:antps2012}. It consists in a three-dimensional array of 885 photomultiplier tubes (PMTs) distributed on 12 lines 
  anchored to the sea bed and connected to the shore through an electro-optical cable. Before reaching this final (12L) setup, \ant~has been 
  operating in various configurations with increasing number of lines, from one to five (5L) and ten (10L) lines.
  
   \ant~detects the Cherenkov radiation emitted by charged leptons (mainly muons, but also electrons and taus) induced by cosmic neutrino 
   interactions with matter inside or near the instrumented volume. The knowledge of the timing and amplitude of the light pulses recorded by 
   the PMTs allows to reconstruct the trajectory of the muon and to infer the arrival direction of the incident neutrino. The current reconstruction 
   algorithms achieve an angular resolution (defined as the median angle between the neutrino and the reconstructed muon) of about $0.4^\circ$ 
   for neutrinos above 10 TeV~\cite{bibthierry:aart}. The design of \ant~is optimized for the detection of up-going muons produced by neutrinos which have 
   traversed the Earth and interacted near the detector; its field of view is $\sim\, 2 \pi\, \mathrm{sr}$ for neutrino energies 
   between 100 GeV and 100 TeV.
   Above this energy, the sky coverage is reduced because of neutrino absorption in the Earth; but it can be partially 
recovered by looking for horizontal and downward-going neutrinos, which can be more easily identified at these high energies where 
the background of atmospheric muons and neutrinos is fainter. \ant, especially suited for the search 
of astrophysical point sources, and transients in particular \cite{bibthierry:tatoo}, is intended as the first step towards a km$^3$-sized neutrino telescope 
in the Mediterranean Sea \cite{bibthierry:km3}. 

The GW detectors \vo~\cite{bibthierry:virgo}, with one site in Italy, and \lo~(see e.g. \cite{bibthierry:ligo}), with two sites in the United States, 
are Michelson-type laser interferometers. They consist of two light storage arms enclosed in vacuum tubes oriented at $90^\circ$ from each other, able to detect 
the differential strain in space produced by the GW. 
Suspended, highly reflective mirrors play the role of test masses. Current detectors are sensitive to relative displacements of the order of $10^{-20}$ to $10^{-22}$ Hz$^{-1/2}$. Their detection horizon is about 15 Mpc for standard binary sources.


The first concomitant data-taking phase with the whole \vo/\lo~network VSR1/S5 was carried out in 2007, while \ant~was operating in 
5L configuration (see  Fig.~\ref{figthierry:calendar}). A second data-taking phase was conducted between mid-2009 and end 2010 with upgraded detectors, 
S6/VSR2 and VSR3, 
in coincidence with the operation of \ant~12L (see section \ref{secthierry:ana2}). Another major upgrade for both classes of detectors is scheduled for the upcoming decade: 
the Advanced \vo/Advanced \lo~and {\sc KM3NeT} projects should gain a factor of 10 in sensitivities with respect to the presently operating instruments.  
The \vo/\lo~network monitors a good fraction of the sky in common with \ant: the instantaneous overlap of visibility maps is about 4~sr, or $\sim 30\%$ of the sky \cite{bibthierry:pradier}.

\section{First joint GW+HEN search}
\label{secthierry:ana}

GW interferometers and HEN telescopes share the challenge to look for faint and rare signals on top of abundant noise or background events. 
Preliminary studies on the feasibility of such searches \cite{bibthierry:pradier,bibthierry:aso} indicated that, even if the constituent 
observatories provide several triggers a day, the false alarm rate for the combined detector network can be kept at a very low level (e.g. $\sim$1/600 yr$^{-1}$ in \cite{bibthierry:aso}).

\subsection{Coincidence Time Window}
\label{secthierry:tw}

An important ingredient of these searches is the definition of an appropriate coincidence time window between HEN and GW signals hypothetically 
arriving from the same astrophysical source. A case study that considered the duration of different emission processes in long GRBs, based on BATSE, 
Swift and Fermi observations, allowed to derive a conservative upper bound  $t_{GW} - t_{HEN} \in [-500s, +500s]$ on this time window~\cite{bibthierry:baret}. 
For short GRBs, this time-delay could be as small as a few seconds. For other sources, this delay is poorly constrained.

\subsection{Analysis Strategy}

The strategy chosen for the 2007 search consists in an event-per-event search for a GW signal  
correlating in space and time with a given HEN event considered as an external trigger. 
Such a search is rather straightforward to implement 
as it allows to make use of existing analysis pipelines developed e.g. for GRB searches. 
It has been applied to the concomitant set of data 
taken between January 27 and September 30, 2007 with \ant~5L-VSR1/S5. 
Such a triggered GW search has been proven to be more efficient than a
classical all-sky analysis, because of the knowledge of the direction and time of arrival of the signal.
More details on this analysis can be found in \cite{bibthierry:gwhen2007}.

The \ant~5L data were filtered according to quality 
requirements similar to those selecting the well-reconstructed events that are used for the standalone searches 
for HEN point sources. The list of 
candidate HEN includes their arrival time, direction on the sky, and an event-by-event estimation of the angular accuracy,  
which serves to define 
the angular search window for the GW search. 
For the purpose of this joint search, the angular accuracy is defined as the 90\% quantile (and not the median) 
of the distribution of the error on the reconstructed neutrino direction, obtained from Monte Carlo studies. 
The on-source time window defined in Section \ref{secthierry:tw}.

The list of HEN triggers is then transmitted to the X-pipeline, an algorithm which performs coherent searches 
for unmodelled bursts of 
GWs on the combined stream of data coming from all active interferometers (ITFs). 
The background estimation and the optimization of the selection 
strategy are performed using time-shifted data from the off-source region in order to avoid contamination by a potential GW signal. 
Once the search 
parameters are tuned, the analysis is applied to the on-source data set. 
If a coincident event is found, its significance is
obtained by comparing to the distribution of accidental events obtained with Monte-Carlo simulations using time-shifted data streams from the 
off-source region ; this is particularly efficient to look for strong signals but one can also look for an accumulation of weakest signals, by
performing a dedicated statistical test, as will be shown later.
 
\subsection{HEN events and error box for GW search}

The HEN candidates have been selected using the BBFit reconstruction \cite{bibthierry:gwhen2007}. A total of 414 events,
among which 198 reconstructed with 2 lines, with 2 azimuthal possible solutions, and 18 more energetic events
reconstructed with more than 2 lines (with a unique solution), were selected. Finally, when taking into account the fact that
2 or more ITFs are needed in order to reconstruct a possible GW arrival direction on the sky, 144 2-line events
and 14 3-line events were analyzed for a possible GW counterpart.

The angular accuracy with which the HEN arrival direction is reconstruted depends on the energy of the 
event and its direction. The space-angle error distribution between the true neutrino direction and the reconstructed
muon direction has been parametrized using a log-normal law in intervals of declination and energy.
The parameters of the function has been used in the GW analysis to estimate the consistency of a
reconstructed signal with the HEN arrival direction. This is the 90\% quantile of this distribution 
which is used as a angular window for the GW search, seen in Figure~\ref{figthierry:space}.
%
\begin{figure}[ht!]
 \centering
\includegraphics[width=0.5\textwidth]{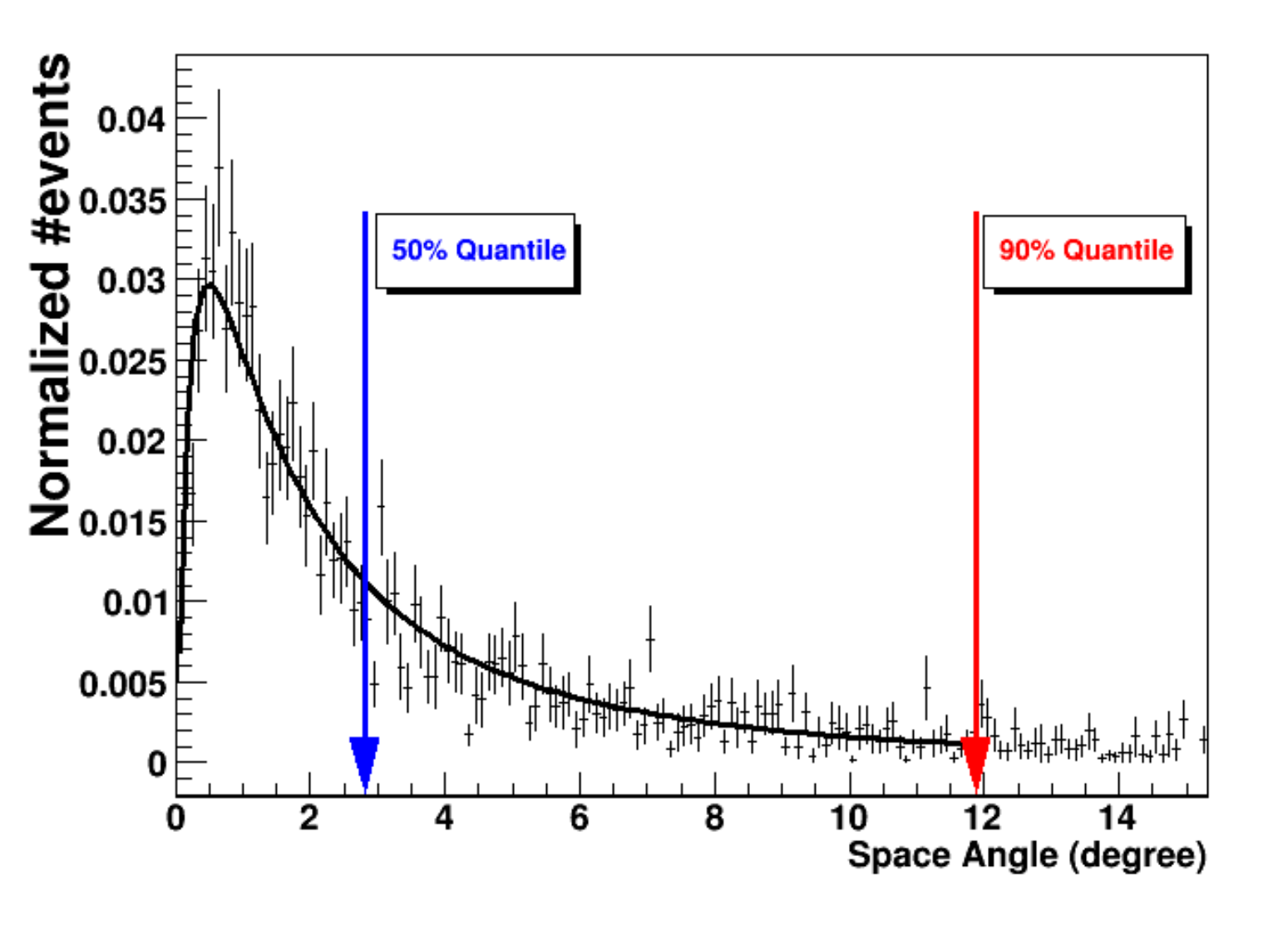}          
  \caption{Space angle error between the true neutrino direction and the reconstructed
muon direction, together with the log-normal parametrization.}
  \label{figthierry:space}
\end{figure}

\subsection{Results of the GW search}

A low-frequency search, with a cut-off frequency at 500 Hz, was performed
for all the HEN events. An additional high-frequency search up to 2kHz, more time-consuming,
was performed for the 3 line events, more energetic and more likely to be of astrophysical origin.

No GW candidate was observed. This allowed to extract GW exclusion distances for typical signal scenarios.
For binary merger signals, expected in the case of short GRBs, the null observation means that no merger
of this type has occured within $\sim 10$ Mpc. The exclusion distances obtained are similar for
collapse-like signals, which are to be expected in the case of long GRBs for instance.

A binomial test has been performed to look for an accumulation of weak GW signals. 
Its results are negative for both the low and high frequency searches - the post-trial
significance of the largest deviation from the null hypothesis is 66\%.

\subsection{Astrophysical interpretation of the search}

The non-observation of a GW+HEN coincidence during the $\sim 100$ days of concomittant data taking
allows to set that the actual number of coincidences verified $N_{\mathrm{GWHEN}} = \rho_{\mathrm{GWHEN}} V_{\mathrm{GWHEN}} T_{\mathrm{obs}} \leq 2.3$ 
at the $90\%$ confidence level. Here $\rho_{\mathrm{GWHEN}}$ is the density of objects aimed at with 
the present analysis, typically the collapse or coalescence of compacts stars, GW emitters, followed by a jet, 
in which HEN are produced, in the local universe. This is a novel way to test the non-constrained
gravitational origin of astrophysical jets formation. 

$V_{\mathrm{GWHEN}}$ is the effective volume
of universe probed by the search, which depends on the horizon of the involved experiments
for typical signals. The GW horizon has been estimated to be $\sim 10$ Mpc for mergers, and $\sim$ 20
Mpc for collapses. The HEN horizons are weaker for the \ant~5 line detector, of the order of 5 Mpc
for mergers (computed using typical short GRB models), and 10 Mpc for long GRBs. The variation of 
the detection efficiencies of both experiments with distance have to be taken into account to have a realistic
estimate of the effective volume.

Converting the null observation into a density yields a limit ranging from $10^{-2}~
\mathrm{Mpc}^{-3}.\mathrm{yr}^{-1}$ for short GRB-like signals down to $10^{-3}~
\mathrm{Mpc}^{-3}.\mathrm{yr}^{-1}$ for long GRB-like emissions. The comparisons with existing estimates
of occurence rates for short/long GRBs or other objects of interest is made in Figure \ref{figthierry:astro}.

\begin{figure}[ht!]
 \centering
\includegraphics[width=0.5\textwidth]{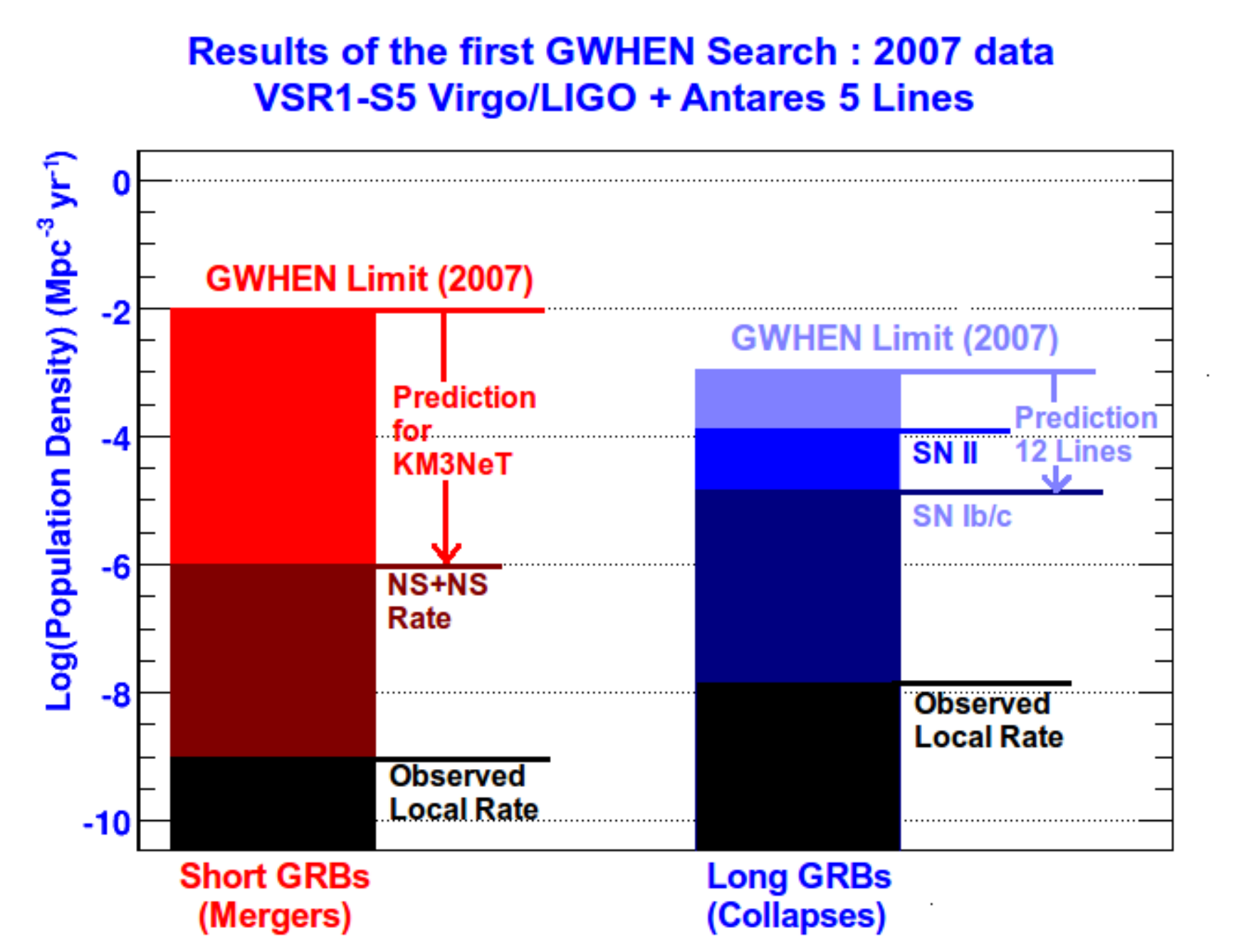}          
  \caption{GWHEN 2007 astrophysical limits are compared with local short/long GRB rates, merger
  rates, and SN II and SN Ib/c rates. 
  Also shown are the potential reach of ongoing or future analyses.}
  \label{figthierry:astro}
\end{figure}


\section{Joint Analysis using 2009-2010 data}
\label{secthierry:ana2}

Data taken with the 9-12 lines \ant~detector in 2009-2010, concomitant with the \vo~VSR2 and
\lo~S6 joint runs, with GW upgraded detectors, yield 129 days of joint operations to be analyzed. 
A new HEN reconstruction algorithm has been used to reduce the HEN angular error~\cite{bibthierry:aart}. 
A different GW software \cite{bibthierry:cwb} has also been developed specifically for this search to perform joint simulations 
- a necessary step to optmize the joint analysis.
The need for such an optimization can be understood with the following arguments, already discussed 
in \cite{bibthierry:pradier} : 
the false-coincidence rate of the GW+HEN search indeed depends on the individual false-alarm rates 
$f_{\mathrm{HEN}}$ and $f_{\mathrm{GW}}$. For instance, if $f_{\mathrm{HEN}}$ is high, 
because of loose selection cuts to obtain a high HEN detection efficiency,
$f_{\mathrm{GW}}$ has to be reduced to conserve the same significance in case of a detection, reducing
the joint GWHEN detection figure of merit defined as : 

\begin{equation}
 \eta_{\textrm{GWHEN}} = \frac{\epsilon_{\textrm{HEN}}}{\rho_{\textrm{GW}}}
\end{equation}

Here $\epsilon_{\textrm{HEN}}$ represents the HEN efficiency,
which depends on the selection cuts, and $\rho_{\textrm{GW}}$ stands for the threshold signal-to-noise ratio
in the GW detection process. Reciprocally, if the HEN selection criteria are too tight, 
resulting in a low number of selected HEN events, the HEN detection horizon is dramatically reduced, resulting in a 
smaller $\eta_{\textrm{GWHEN}}$, even if the GW search is more powerful because of the lower allowed thresholds. 
Of course, this optimization depends on the, e.g., HEN spectrum index, and the GW assumed signals. This optimization
has been performed to find the optimal HEN selection cuts. Figure~\ref{figthierry:optim} shows a summary of the differences
between this new \ant-\vo/\lo~search with respect to the one presented in Section \ref{secthierry:ana}.

\begin{figure}[t!]
 \centering
\includegraphics[width=0.5\textwidth]{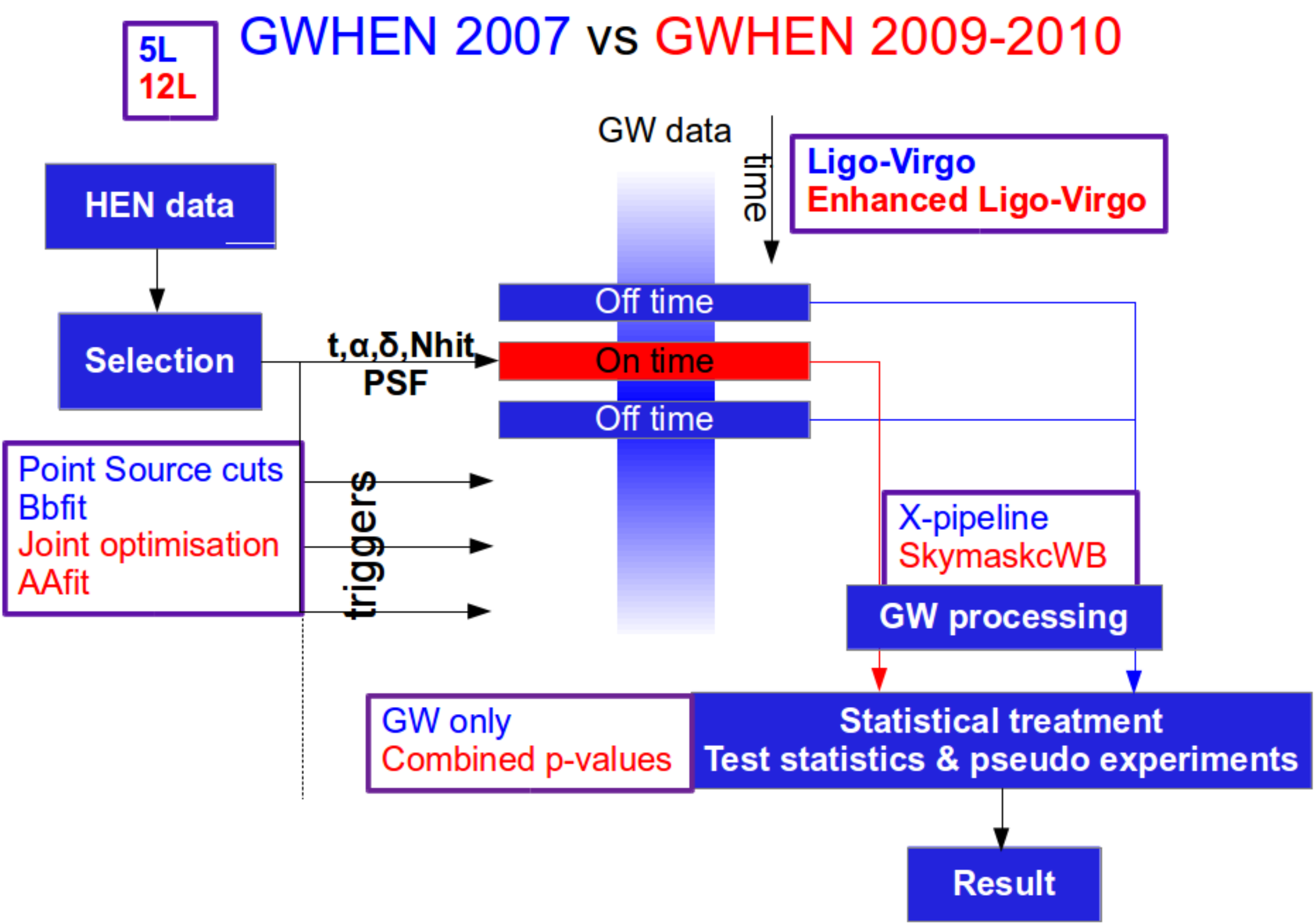}  
  \caption{Strategy of the 2009-2010 \ant-\vo/\lo~joint search compared to the 2007 analysis.
  A new HEN reconstruction method has been used, with improved angular resolution, and a new GW software, suited for joint simulations,
  has been developed for the search.
  }
  \label{figthierry:optim}
\end{figure}

According to preliminary estimates, the enhanced sensitivity of the GW interferometers, combined
with the increased joint live time and the improvement of a factor 3 in the \ant~effective area above 100 TeV
should result in an increase by a factor 5 of the volume of universe probed by the joint search : this ongoing search could be able to constrain for the
first time the fraction of star collapses followed by the ejection of a hadronic jet.

\section{Conclusions}

These pioneering GW+HEN searches, developed in \cite{bibthierry:gwhen2007} and \cite{bibthierry:gwhen2012}, opens the way towards a new multi-messenger astronomy.
Beyond the benefit of a potential high-confidence discovery, future analyses, particularly the one 
involving a km$^3$ HEN telescope \cite{bibthierry:km3} and advanced interferometers \cite{bibthierry:advitf}, could be able
to constrain the density of joint sources down to astrophysically-meaningful levels - hence contrain
for the first time the fraction of binary mergers followed by the emission of a relativistic jet.


\setcounter{figure}{0}
\setcounter{table}{0}
\setcounter{footnote}{0}
\setcounter{section}{0}
\setcounter{equation}{0}

\addcontentsline{toc}{part}{{\sc Exotic Physics, Environmental parameters and new detection techniques}%
\vspace{-0.5cm}
}

\newpage
\id{id_lambard}



\title{\arabic{IdContrib} - Indirect search of dark matter with the {\sc Antares} neutrino telescope}
\addcontentsline{toc}{part}{\arabic{IdContrib} - {\sl Guillaume Lambard} : Indirect search of dark matter%
\vspace{-0.5cm}
}

\shorttitle{\arabic{IdContrib} - Dark Matter search with {\sc Antares}}

\authors{
Juan Jos\'e Hern\'andez--Rey$^{1}$, 
Guillaume Lambard$^{1}$ 
for the {\sc Antares}  Collaboration.}

\afiliations{ 
$^1$ IFIC - Instituto de F\'{\i}sica Corpuscular (CSIC- Universitat de
Val\`encia), E-46980 Paterna, Spain}

\email{Juan.J.Hernandez@ific.uv.es}

\abstract{The results of a search for high-energy neutrinos coming
  from the direction of the Sun using the data recorded by the ANTARES
  neutrino telescope during 2007 and 2008 are presented.  The number
  of neutrinos observed is found to be compatible with background
  expectations and upper limits for the spin-dependent and
  spin-independent WIMP-proton cross-sections are derived and compared
  to predictions of the CMSSM.  These limits are comparable to those
  obtained by other neutrino telescopes and are more stringent than
  those obtained by direct search experiments for the spin-dependent
  WIMP-proton cross-section assuming the self-annihilation proceeds
  through hard channels, i.e. via {\rm $W^{+}W^{-}$} and
  {$\tau^{+}\tau^{-}$}.}


\maketitle

\section{Introduction}

The existence of dark matter, i.e. non-baryonic matter that does not
interact electromagnetically, has been educed from a variety of
gravitational effects. The rotation curves of galaxies and galaxy
clusters and the observations of weak lensing point to the existence
of dark matter. It is also required to understand the features of the
cosmic microwave background and the large scale structure formation
within our present accepted models of the evolution of the
Universe~\cite{biblambard:bertone}.

One hypothesis is that dark matter is formed by weakly interacting
massive particles (WIMPs), candidates for which are copiously provided
by different extensions of the Standard Model, in particular those based on
Supersymmetry (SUSY). Within SUSY models the lightest neutralino is one of 
the favoured candidates for dark matter WIMP. 

WIMPs can be captured in massive astrophysical objects, like the Sun,
where they can self-annihilate giving rise to high-energy
neutrinos. Neutrino telescopes are well suited to look for such a
signal coming from sources like the Sun or the centre of the
Galaxy. This type of indirect search complements direct searches and
has a variety of advantages compared to others. For the case of the
Sun, for instance, such a WIMPigenic neutrino flux depends only mildly
on the WIMP velocity distribution. On the other hand, it has a strong
dependence on the WIMP-proton cross-section, in particular the part
that involves spin, making it very sensitive to this important
magnitude. Moreover, unlike other indirect searches (e.g. using
gamma-rays or positrons) a signal would be very clean, since it is
very difficult to imagine astrophysical processes that would produce
high-energy neutrinos coming from the Sun and those which are possible
can only yield negligible fluxes.

\section{ANTARES}

ANTARES is an undersea neutrino telescope located in the Mediterranean
Sea, $40$ km offshore from Toulon (France)~\cite{biblambard:antares}.  The
telescope consists of $12$ detection lines anchored to the seabed at
2475 m depth. Each line has 25 storeys. A standard storey includes
three optical modules (OMs)~\cite{biblambard:OM} each housing a 10-inch
photomultiplier~\cite{biblambard:PMT} and a local control module that contains
the electronics~\cite{biblambard:frontend, biblambard:DAQ}. The OMs are directed
45$^{\circ}$ downwards so as to optimise their acceptance to upgoing
light and to avoid the effect of sedimentation and
biofouling~\cite{biblambard:biofouling}. The length of a line is 450 m and the
horizontal distance between neighbouring lines is 60-75 m. In one of
the lines, the upper storeys put up a test system for acoustic
neutrino detection~\cite{biblambard:amadeus}.  Other acoustic devices are
installed in an additional line that contains instrumentation aimed at
measuring environmental parameters~\cite{biblambard:instrumentation}. The
location of the active components of the lines is known better than
10~cm by a combination of tiltmeters and compasses in each storey and
a series of acoustic transceivers in certain storeys along the line
and surrounding the telescope~\cite{biblambard:alignment}. A common time
reference is maintained in the full detector by means of a 25 MHz
clock signal broadcast from shore. The time offsets of the individual
optical modules are determined in dedicated calibration facilities
onshore and regularly monitored in situ, and corrected if needed, by
means of optical beacons distributed throughout the lines and which
emit short light pulses through the water~\cite{biblambard:OBs}. This allows to
reach a sub-nanosecond accuracy on the relative timing~\cite{biblambard:timing}.

The first 5 lines of the detector were operational in 2007 and the full 
detector with 12 lines was completed in May 2008.

\section{Event selection}

The search presented here uses the data taken during 2007 and 2008 by
the ANTARES neutrino telescope.  Good detector and environmental
conditions, in particular a low level of bioluminescent light, are
required for a run to be kept for further analysis. For that period, a
total of $2693$ runs are found to have the appropiate conditions. This
corresponds to a livetime of 134.6, 38.0, 39.0 and 83.0 days for the
periods in which the dectector consisted of $5$, $9$, $10$ and $12$
lines, respectively.

 \begin{figure}[tb]
  \centering
  \includegraphics[width=0.55\textwidth]{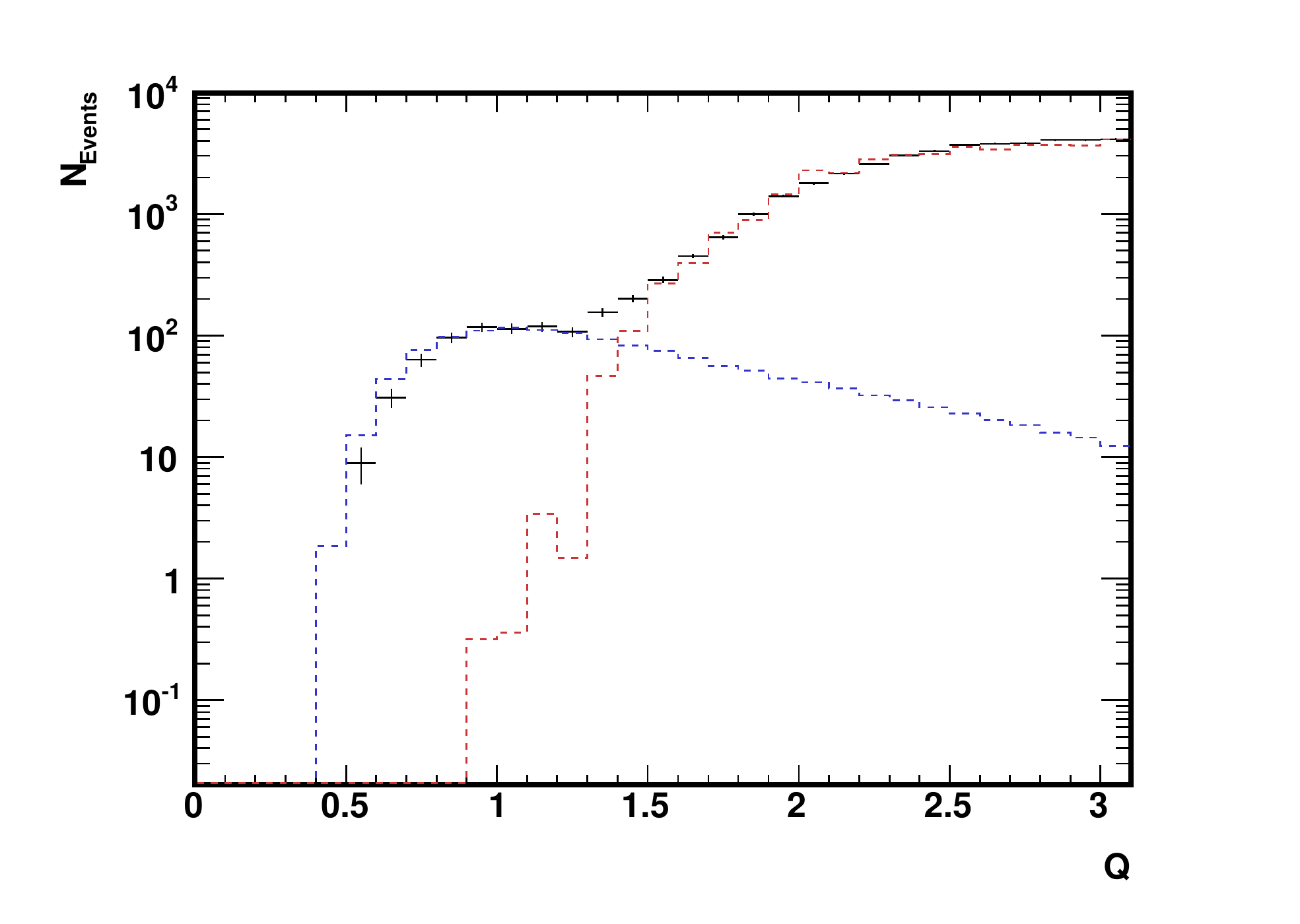}
  \caption{Number of events as a function of the quality parameter of
    the track fit, $\rm Q$. The blue and red dashed lines correspond
    to the atmospheric neutrino and muon events, respectively,
    according to simulation, and the black crosses are the 2007-2008
    data.}
  \label{figlambard:quality}
 \end{figure}

The muon tracks are reconstructed from the position and time of the
hits of the Cherenkov photons in the OMs.  The reconstruction
algorithm is based on the minimisation of a $\rm \chi^{2}$-like
quality parameter, $\rm Q$, which uses the difference between the
expected and measured times of the detected photons plus a correction
term that takes into account the effect of light
absorption~\cite{biblambard:bbfit}.  The distribution of the number of events as
a function of this quality parameter, $\rm Q$, shows a good agreement
between data and simulated events, as can be seen in
figure~\ref{figlambard:quality}. The angular resolution provided by this
algorithm for upgoing neutrinos depends on the configuration of the
detector and of the neutrino energy. In figure~\ref{figlambard:angres}, the
angular resolution for upgoing neutrinos as a function of the neutrino
energy is shown for the different detector configurations.

 \begin{figure}[tb]
 \begin{center}
  \includegraphics[width=0.55\textwidth]{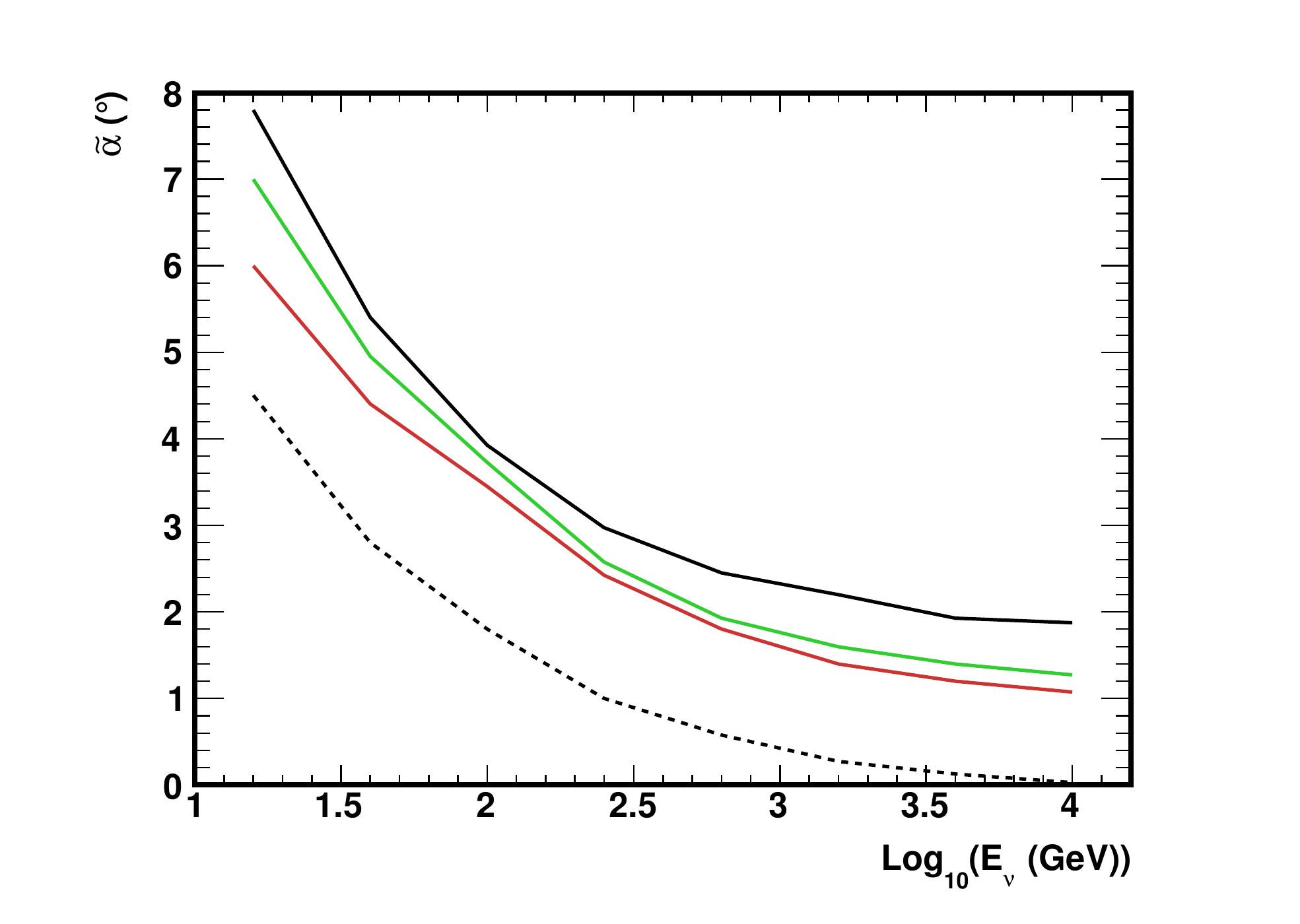}
  \caption{Angular resolution for upgoing neutrinos for the detector
    configuration with 5 (black), 9 (green) and 10 and 12 (red) lines. The
    black dashed line is the resolution due solely to the angle between the
    muon and the neutrino at the interaction vertex.}
  \label{figlambard:angres}
 \end{center}
 \end{figure}

The fit of the muon track is required to use a number of hits greater
than five in at least two lines. This yields a non-degenerate
5-parameter fit that provides a good angular resolution. Only events
that go upwards, according to the fitted angle, are kept in order to
reject most of the atmospheric muon background.  The quality parameter
of the track, $\rm Q$, and its angular distance to the Sun, $\Psi$,
are then used to further reduce the background. They are chosen so as
to optimise the model rejection factor~\cite{biblambard:mrf}.  For each WIMP mass
and each annihilation channel, the selected values, $\rm Q_{cut}$ and
$\rm \Psi_{cut}$, are specifically those that minimise the average
90\% confidence level (CL) upper limit on the $\rm
\nu_{\mu}+\bar{\nu}_{\mu}$ flux, $\rm
\overline{\Phi}_{\nu_{\mu}+\bar{\nu}_\mu}$, defined as:

\begin{equation}
\rm{\overline{\Phi}_{\nu_{\mu}+\bar{\nu}_\mu} = \frac{\bar{\mu}^{90\%}}{\sum\limits_{i} \bar{A}_{eff}^{i}(M_{WIMP}) \times T_{eff}^{i}}} \, ,
\label{eqlambard:mrfeq}
\end{equation}

\noindent where the index $\rm i$ denotes the periods with different
detector configurations (5, 9, 10 and 12 detection lines), $\rm
\bar{\mu}^{90\%}$ is the average upper limit of the background at
$90$\% CL computed using a Poisson distribution with the
Feldman-Cousins prescription~\cite{biblambard:feldmancousins}, $\rm {T_{eff}^{i}}$
is the livetime for each detector configuration and
$\rm{\bar{A}_{eff}^{i}(M_{\rm WIMP})}$ is the effective area averaged
over the neutrino energy.

To perform this optimisation, simulated events both for the signal and
the background were generated. Downgoing atmospheric
muons were simulated using Corsika~\cite{biblambard:corsika} and upgoing
atmospheric neutrinos using the Bartol flux~\cite{biblambard:bartol}. A possible
background coming from the interaction of cosmic rays with the Sun's
corona was neglected, since it was found to be less than 0.4\% of the
total atmospheric background in the Sun's direction.

The signal was simulated assuming that the high-energy neutrinos from
the Sun were produced by the annihilation of neutralinos. The flux of
neutrinos as a function of their energy arriving at the Earth's
surface from the Sun's core was computed using the software package
WimpSim~\cite{biblambard:wimpsim} without assuming any concrete model. The
neutrinos resulting from the neutralino self-annihilation channels to
$q\bar{q}$, $l\bar{l}$, $W^{+}W^{-}$, $ZZ$, Higgs doublets
$\phi\phi^{*}$ and $\nu\bar{\nu}$ were simulated for $17$ different
WIMP masses ranging from 10~GeV to 10~TeV.  Three main
self-annihilation channels were chosen as benchmarks for the lightest
neutralino, $\rm \tilde{\chi}_{1}^{0}$, namely: a soft neutrino
channel, $\rm \tilde{\chi}_{1}^{0}\tilde{\chi}_{1}^{0}\rightarrow
b\bar{b}$, and two hard neutrino channels, $\rm
\tilde{\chi}_{1}^{0}\tilde{\chi}_{1}^{0} \rightarrow W^{+}W^{-}$ and
$\rm \tilde{\chi}_{1}^{0}\tilde{\chi}_{1}^{0} \rightarrow
\tau^{+}\tau^{-}$. A 100\% branching ratio was assumed for all of them
in order to explore the widest theoretical parameter space.
Oscillations among the three neutrino flavours (both in the Sun and
during their flight to Earth) were taken into account, as well as
$\nu$ absorption and $\tau$ lepton regeneration in the Sun's medium.

The optimisation procedure gave a stable value of 1.4 for $\rm
Q_{cut}$, independently of the cut on the angle to the Sun's
direction. The optimised values for the latter, $\rm \Psi_{cut}$, were
typically between 3.5$^{\circ}$ and 5.5$^{\circ}$.  The extreme upper
and lower values for this cut, namely 8.4$^{\circ}$ and 3.2$^{\circ}$,
occurred, respectively, for low mass in the soft-channel (M$_{WIMP}$=
50~GeV and $b \bar{b}$) and for high masses in the hard channels
(M$_{WIMP} >$ 1~TeV and $W^+ W^-$, $\tau^+ \tau^-$), as expected.

\section{Results}

After unblinding, a total of 27 events are found within a 20$^\circ$
angular separation from the Sun's direction.  In figure~\ref{figlambard:angular}
the distribution of the angular distance for the selected event
tracks is shown. The triangles are the data with the one sigma
statistical uncertainty and the straight line is the expected
background obtained scrambling the data in right ascension. No
significant excess is found.

 \begin{figure}[tb]
  \centering
  \includegraphics[width=0.55\textwidth]{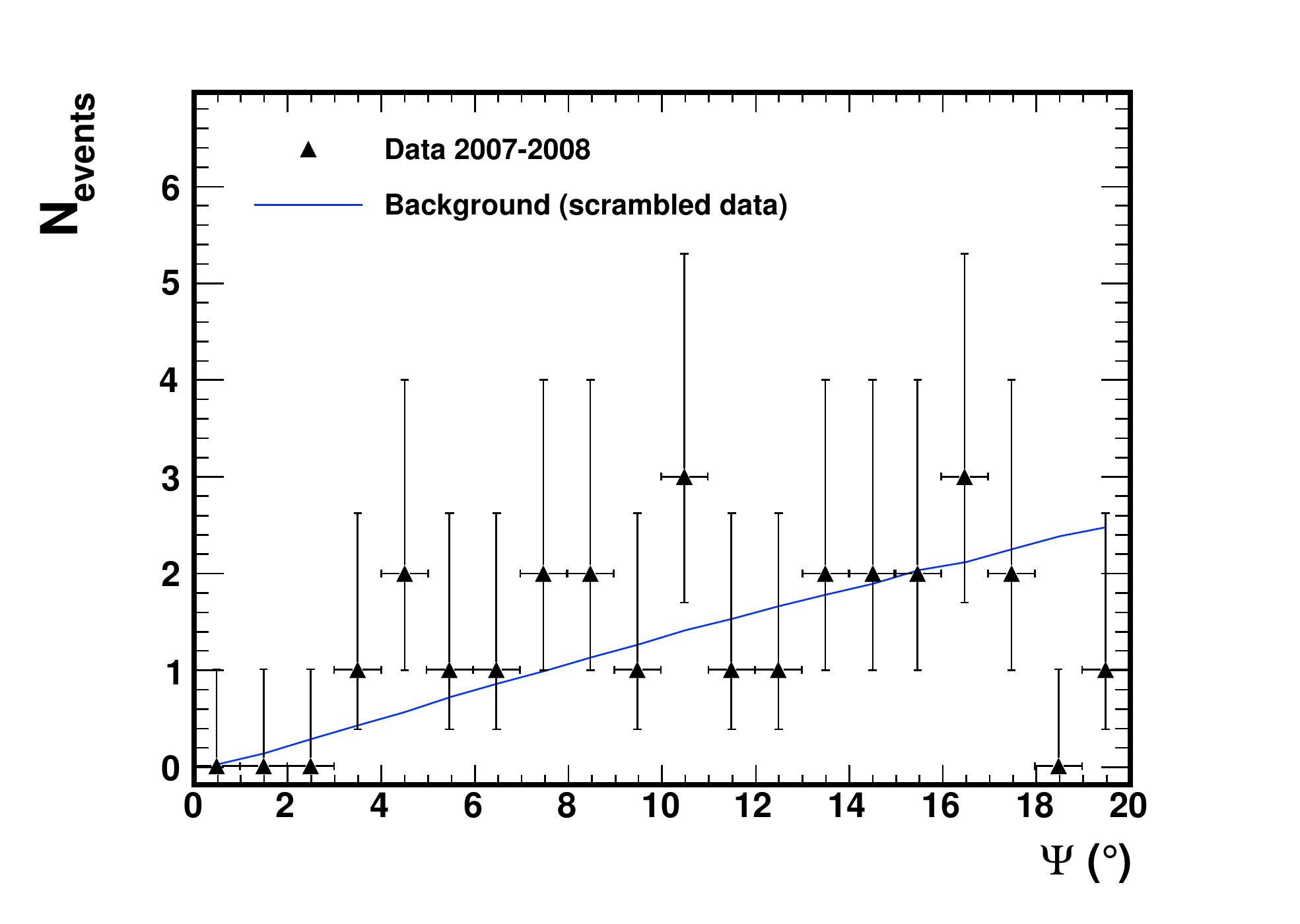}
  \caption{Number of events as a function of the angle between the
    track and the Sun's direction. The straight line is the expected
    background as obtained scrambling real data. The triangles are the
    data. }
  \label{figlambard:angular}
 \end{figure}

 Using the values for the cuts obtained in the optimisation procedure,
 $90$\% CL limits on the $\rm \nu_{\mu}+\bar{\nu}_{\mu}$ flux, $\rm
 \Phi_{\nu_{\mu}+\bar{\nu}_\mu}$, can be extracted from the data
 according to Equation~\ref{eqlambard:mrfeq}, where now $\rm \mu^{90\%}$ is not
 the average but the actual $90$\% CL limit on the number of observed
 events. The limits on the muon flux are calculated using a conversion
 factor between the neutrino and the muon fluxes computed using the
 package DarkSUSY~\cite{biblambard:gondolo}. Figure~\ref{figlambard:phimulimit} shows the
 90\% CL muon flux limits, $\rm \Phi_{\mu}$, for the channels
 $b\bar{b}$, $W^{+}W^{-}$ and $\tau^{+}\tau^{-}$. The latest results
 from Baksan~\cite{biblambard:baksan}, Super-Kamiokande~\cite{biblambard:superk} and
 IceCube-$79$~\cite{biblambard:icecube} are also shown for comparison.

\begin{figure}[tb]
  \centering
  \includegraphics[width=0.5\textwidth]{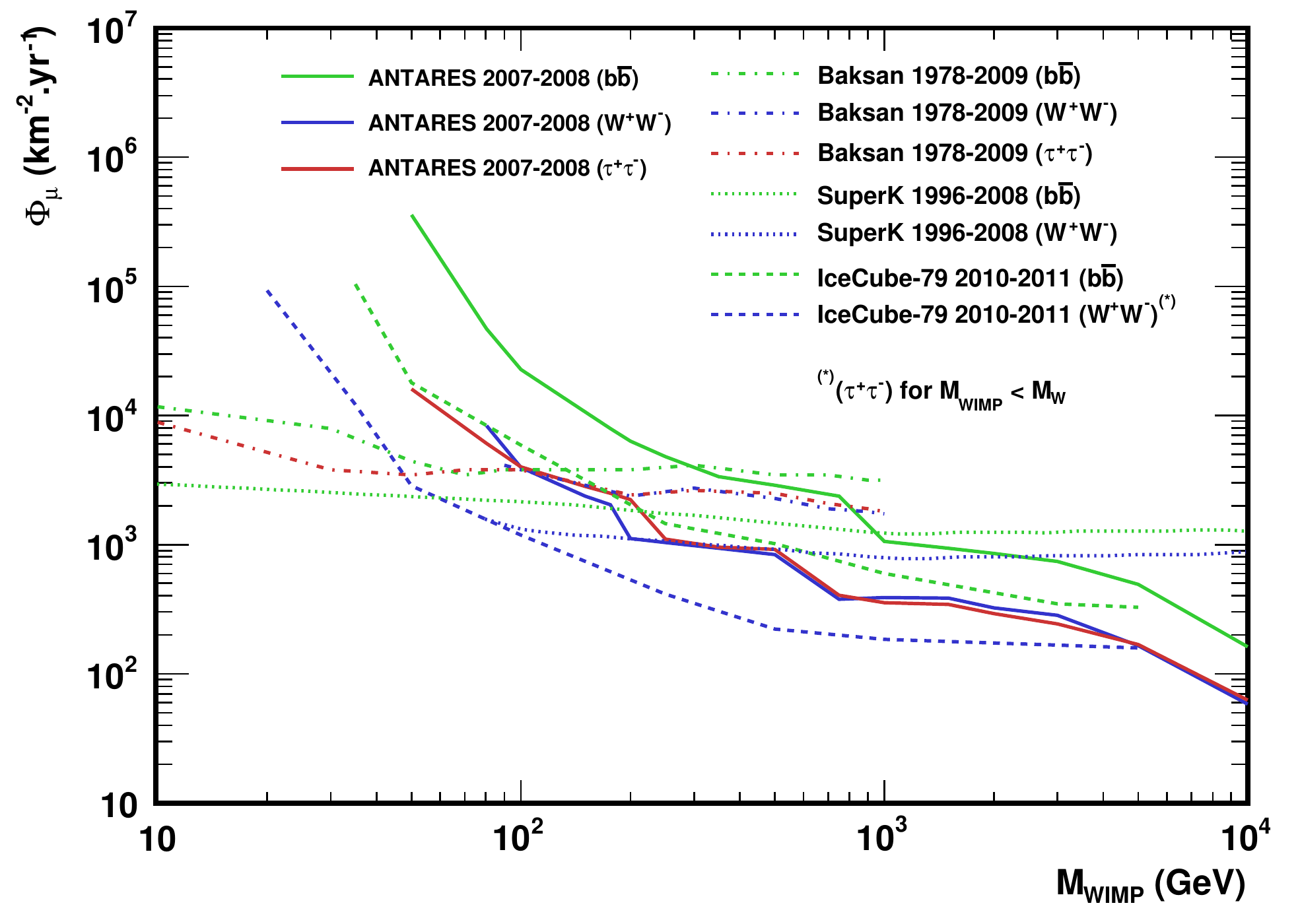}

  \caption{$90$\% CL upper limit on the muon flux as a function of the
    WIMP mass for the neutralino self-annihilation channels to
    $b\bar{b}$ (green), $W^{+}W^{-}$ (blue) and $\tau^{+}\tau^{-}$
    (red). The results from Baksan $1978-2009$~\cite{biblambard:baksan}
    (dash-dotted lines), Super-Kamiokande $1996-2008$~\cite{biblambard:superk}
    (dotted lines) and IceCube-$79$ $2010-2011$~\cite{biblambard:icecube} (dashed
    lines) are also shown.}
\label{figlambard:phimulimit}
\end{figure}

\begin{figure}[tb]
\begin{center}
\begin{minipage}[c]{.8\linewidth}
\includegraphics[width=\linewidth]{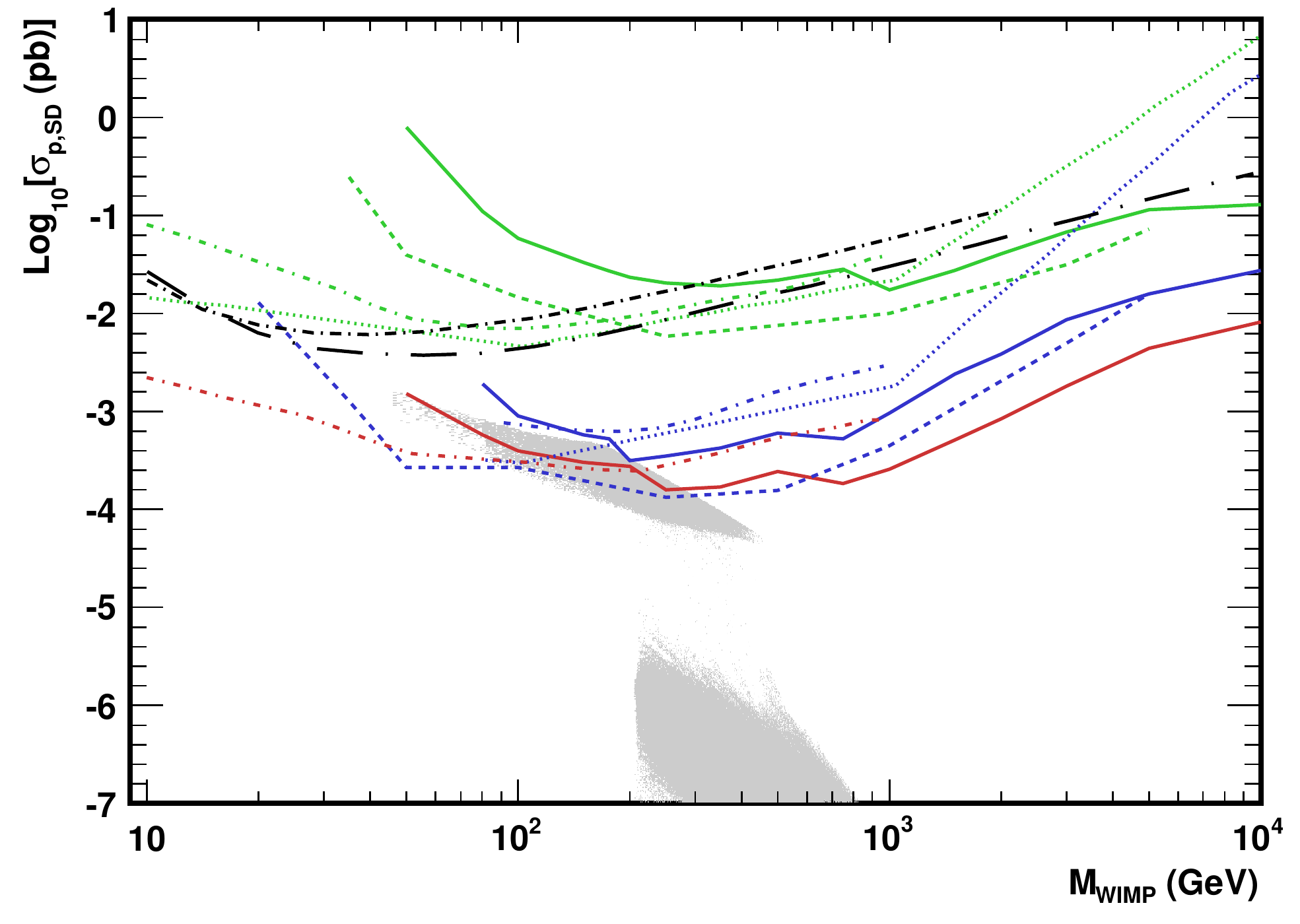}
\\ \includegraphics[width=\linewidth]{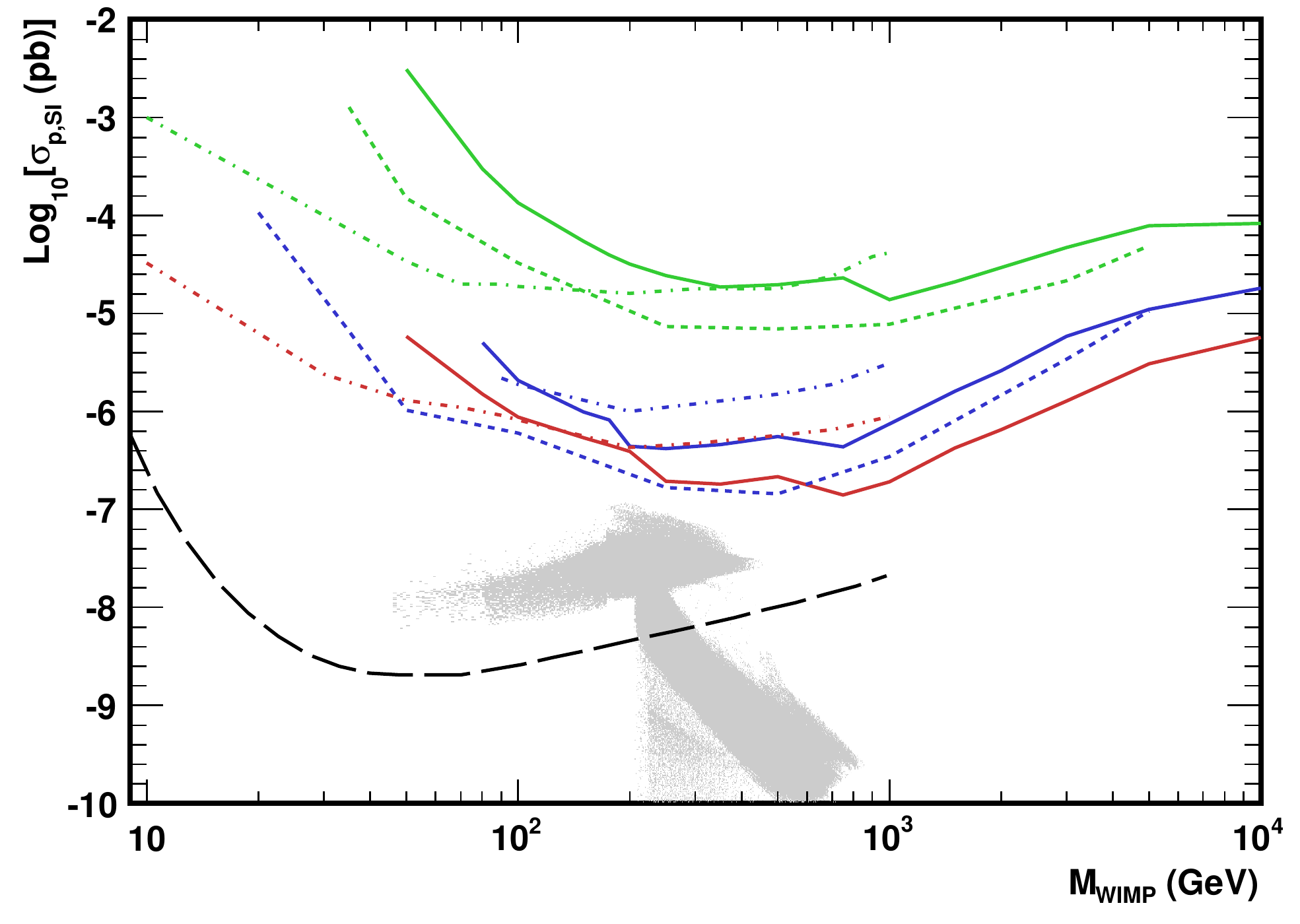}
\end{minipage}
 \caption{$90$\% CL upper limits on the SD and SI WIMP-proton
   cross-sections (upper and lower plots, respectively) as a function
   of the WIMP mass, for the three self-annihilation channels:
   $b\bar{b}$ (green), $W^{+}W^{-}$ (blue) and $\tau^{+}\tau^{-}$
   (red), for ANTARES 2007-2008 (solid lines) compared to the results
   of other indirect search experiments: Baksan
   $1978-2009$~\cite{biblambard:baksan} (dash-dotted lines), Super-Kamiokande
   $1996-2008$~\cite{biblambard:superk} (dotted lines) and IceCube-$79$
   $2010-2011$~\cite{biblambard:icecube} (dashed lines) and the result of the
   most stringent direct search experiments (black): SIMPLE
   $2004-2011$~\cite{biblambard:simple} (short dot-dashed line in upper plot),
   COUPP $2010-2011$~\cite{biblambard:coupp} (long dot-dashed line in upper plot)
   and XENON100 $2011-2012$~\cite{biblambard:xenon} (dashed line in lower
   plot). The results of a grid scan of the CMSSM are included (grey
   shaded areas) for the sake of comparison.}
\label{figlambard:sdsilimit}
\end{center}
\end{figure}

\begin{table}[h]
\begin{center}
\begin{tabular}{c}
\hline
\\[-0.1in]
CMSSM Parameter Range \\
\\[-0.1in]
\hline
\\[-0.1in]
 $50$ GeV$\rm <m_{0}<4$ TeV\\
 $500$ GeV$\rm <m_{1/2}<2.5$ TeV\\
 $\rm 5< tan \, \beta <62$\\
 $-5$ TeV$\rm <A_{0}<5$ TeV\\
 $\rm sgn(\mu) > 0$\\
\hline
\end{tabular}
 \caption{Range of parameters scanned for the CMSSM model. $\rm m_{0}$
   and $\rm m_{1/2}$ are the common scalar and gaugino masses,
   respectively; $\rm tan \, \beta$ is the ratio of vevs of the Higgs
   field; $\rm A_{0}$ is the common trilinear coupling and $\rm
   sgn(\mu)$ the sign of the Higgs mixing.}
\label{tablelambard:cmssmlimits}
\end{center}
\end{table}

Limits on the spin-dependent (SD) and spin-independent (SI)
WIMP-proton scattering cross-sections --when one or the other is
dominant-- can also be obtained. To this end, the following
assumptions are made. First, equilibrium between the WIMP capture and
self-annihilation rates in the Sun is assumed. Secondly, the Sun is
considered to be free in the galactic halo. Thirdly, a local dark
matter density of $0.3$ GeV cm$^{-3}$ and a Maxwellian velocity
distribution of the WIMP with a dispersion of $270$ km s$^{-1}$ are
used.  Finally, it is assumed that no additional dark matter disk that
could enhance the local dark matter density exists.

The $90$\% CL limits for the SD, $\rm \sigma_{p,SD}$, and SI, $\rm
\sigma_{p,SI}$, WIMP-proton cross-sections extracted from the
self-annihilation channels to $b\bar{b}$, $W^{+}W^{-}$ and
$\tau^{+}\tau^{-}$ are presented in Figure~\ref{figlambard:sdsilimit}.
Systematic uncertainties were included in the evaluation of the limits
following the approach described in reference~\cite{biblambard:conrad}. The total
systematic uncertainty on the detector efficiency is around 20\% and
comes essentially from the uncertainties on the average quantum
efficiency and the angular acceptance of the PMTs and the sea
water absorption length (a $10$\% uncertainty for each one). This
systematic uncertainty worsens the upper limit between 3\% and 6\%,
depending on the WIMP mass.  

Also shown in Figure~\ref{figlambard:sdsilimit} are the latest results from
Baksan~\cite{biblambard:baksan}, Super-Kamiokande~\cite{biblambard:superk} and
IceCube-$79$~\cite{biblambard:icecube} as well as the limits from the direct
search experiments SIMPLE~\cite{biblambard:simple}, COUPP~\cite{biblambard:coupp} and
XENON100~\cite{biblambard:xenon}.

The allowed values from the CMSSM model according to the results from
an adaptative grid scan performed with DarkSUSY are also shown.  The
parameters of this model are moved within the ranges indicated in
Table~\ref{tablelambard:cmssmlimits}.  All the limits are computed with a muon
energy threshold at $\rm E_\mu = 1$ GeV. The shaded regions show a
grid scan of the model parameter space taking into account the latest
constraints for various observables from accelerator-based experiments
and in particular the results on the Higgs boson mass from ATLAS and
CMS, $\rm M_{h} = 125 \pm 2$ GeV~\cite{biblambard:buchmeller}.  A relatively
loose constraint on the neutralino relic density $\rm 0 <
\Omega_{CDM}h^{2} < 0.1232$ is used to take into account the possible
existence of other types of dark matter particles.

Since the capture rate of WIMPs in the core of the Sun depends very
much on the SD WIMP-proton cross-section, the neutrino flux coming
from WIMP annihilation strongly depends on it. This makes this type of
indirect search very competitive for this magnitude.  This is not the
case for the SI WIMP-proton cross-section, where the limits coming
from direct search experiments like XENON100 are better thanks to the
composition of their target materials.

\section{Conclusions}

An indirect search for dark matter towards the Sun has been
carried out  using the first two years of data recorded by the ANTARES neutrino
telescope. The number of neutrino events coming from the Sun's
direction is compatible with the expectation from the atmospheric
backgrounds. The derived limits are comparable to those obtained by
other neutrino observatories and are more stringent than those
obtained by direct search experiments for the spin-dependent
WIMP-proton cross-section. The present ANTARES limits start to
constraint the parameter space of the CMSSM model.

\vspace*{0.5cm}
{\footnotesize{\bf %
\noindent
Acknowledgment: }
This work was supported by the Spanish Ministry of Science and
  Innovation and the Generalitat Valenciana under grants:  
FPA2009-13983-C02-01, FPA2012-37528-C02-01, PROMETEO/2009/026 and the
Consolider MultiDark, CSD 2009-00064}

\setcounter{figure}{0}
\setcounter{table}{0}
\setcounter{footnote}{0}
\setcounter{section}{0}
\setcounter{equation}{0}

\newpage
\id{id_mangano2}



\title{\arabic{IdContrib} - Measurement of Velocity of Light in Deep Sea Water at the Site of {\sc Antares}}
\addcontentsline{toc}{part}{\arabic{IdContrib} - {\sl Salvatore Mangano} : Measurement of Velocity of Light in Deep Sea Water at the Site of {\sc Antares}%
\vspace{-0.5cm}
}

\shorttitle{\arabic{IdContrib} - Velocity of light in water}

\authors{Salvatore Mangano$^{1}$ and Juan Jos\'e Hern\'andez-Rey$^{1}$ for the {\sc Antares} Collaboration.
}

\afiliations{
$^1$ IFIC - Instituto de F\'{\i}sica Corpuscular (CSIC- Universitat de
Val\`encia), E-46980 Paterna, Spain}

\email{manganos@ific.uv.es}

\abstract{
The ANTARES experiment is currently the largest underwater neutrino
telescope and is taking high quality date since 2007. Sea water is used 
as the detection medium of the Cherenkov light emitted by relativistic
charged particles resulting from the interaction of neutrinos.  The particle
trajectory is reconstructed from the measured arrival times of the detected 
photons. The propagation of Cherenkov light depends on the optical 
properties of the sea water and their understanding is crucial to reach the optimal performance of the detector.  The group velocity depends on the wavelength 
of the photons and this is usually referred to as chromatic dispersion. This group velocity of light has been measured at eight different wavelengths between 385 nm and 532 nm in the Mediterranean Sea at a depth of about 2 km with the ANTARES optical beacon systems. A parametrization of the dependence of the velocity of light on wavelength based on the salinity, pressure and temperature of the sea water at the ANTARES site is in good agreement with these measurements.
}


\maketitle

\section{Introduction}
\label{secmangano2:introduction}
The ANTARES neutrino telescope is located at the bottom of the
Mediterranean Sea at a depth of 2475~m, roughly \mbox{40~km} offshore
from Toulon in France. Sea water is used as the detection medium of
the Cherenkov light emitted by relativistic charged particles resulting
from the interaction of neutrinos around the detector. The particle
direction is reconstructed from the measured arrival times of the
detected photons through a three dimensional array of 885
photomultiplier tubes (PMTs) arranged in twelve vertical lines. Along
each line with a vertical separation of 14.5~m, three PMTs are
oriented with their axis pointing downward at an angle of $45^o$ with
respect to the vertical line direction. The horizontal separation
between lines is about 70~m. Further details can be found in~\cite{bibmangano2:Ant1}.

Charged particles crossing sea water induce the emission of Cherenkov
light whenever the condition $\beta > 1/n_p$ is fulfilled. 
$\beta$ is the speed of the particle in sea water with respect to the
speed of light in vacuum and $n_p$ is the phase refractive index,
i.e. the ratio between the phase speed of light in the water and that
in vacuum. The Cherenkov photons are emitted at an angle with respect
to the particle track given by $cos \, \theta_c = \frac{1}{\beta n_p}$, 
and travel in the medium at the group velocity, therefore the relevant
quantity for light transmission is the group refractive index,
$n_g$. Both the phase and group refractive indices depend on the
wavelength of the photons, an effect usually referred to as chromatic 
dispersion and has the effect of making the emission
angle and the speed of light wavelength dependent.  The group
refractive index is related to its phase refractive index through
\begin{equation}
n_g=\frac{n_p}{1+\frac{\lambda}{n_p}\frac{d n_p}{d \lambda}},
\label{eqmangano2:n_g}
\end{equation}
\noindent where $\lambda$ is the wavelength of light.


The present measurement of the refractive index has been made using
the Optical Beacon system~\cite{bibmangano2:Timecalibration} of ANTARES. This system consists 
of a set of pulsed light sources
(LEDs and lasers) which are 
distributed throughout the detector and illuminate the PMTs with 
short duration flashes of light. The refractive index is
deduced from the recorded time of flight distributions of photons at
different distances from the source and has been measured for eight
different wavelengths between 385~nm and 532~nm. 

As the PMTs are color blind, the variation of the photon emission
angle and of the group velocity due to chromatic dispersion cannot be
accounted for on a photon by photon basis. Nevertheless, a good
knowledge of this wavelength dependence enables to make the optimum
average corrections and to estimate the uncertainty introduced~\cite{bibmangano2:Coll2010}.
 
Several measurements similar to the one presented in this
paper have been performed in the past~\cite{bibmangano2:Agui,bibmangano2:Lubs,bibmangano2:opt}.

\section{Experimental Setup}
\label{secmangano2:expsetup}
The PMTs of ANTARES are sensitive to photons in the wavelength range
between 300~nm and 600~nm. They have a peak quantum efficiency of
about 22\% between 350~nm and 450~nm. The PMT measures the arrival
time and charge amplitude of the detected photons.  The single
photoelectron transit time spread of the PMT is around 3.5~ns 
full width at half maximum (FWHM)~\cite{bibmangano2:Timecalibration}.

The refractive index has been measured using the ANTARES Optical
Beacon system. This system of short pulse light sources was primarily
designed to perform time calibration in situ~\cite{bibmangano2:Timecalibration,bibmangano2:Ageron}, 
but is also being used to measure the optical
properties of water. There are two types of Optical Beacons (OBs), the
LED Beacons (LOBs) and the Laser Beacons (LBs). There are four LOBs
per line and two LBs at the bottom of the two central lines.

An LOB contains 36 individual LEDs distributed over six vertical faces
shaping a hexagonal cylinder (Figure~\ref{figmangano2:opticalb}, left).  On
each face, five LEDs point radially outwards and one upwards. All the
LEDs emit light at an average wavelength of 469~nm, except two LEDs
located on the lowest LOB of line 12 which emit light at an average
wavelength of 400~nm. In order to extend the measurement of the
optical properties, a modified LOB was deployed in
2010 taking advantage of the redeployment of Line 6. This modified
LOB has three LEDs per face instead of six, all of them pointing
upwards.  The three LEDs of each
face emit at the same wavelength.  The
average wavelength of the light from the six faces of the modified LOB are
385, 400, 447, 458, 494 and
518~nm. The LEDs emit 
light with a maximum intensity of $\sim$160~pJ and a
pulse width of \mbox{$\sim$4 ns} (FWHM). The intensity of the emitted light
can be varied changing the voltage feeding the LEDs.

\begin{figure}
 \setlength{\unitlength}{1cm}
 \centering
 \begin{picture}(8.0,3.5)
   \put(0.0, 0.0){\includegraphics[width=3.8cm]{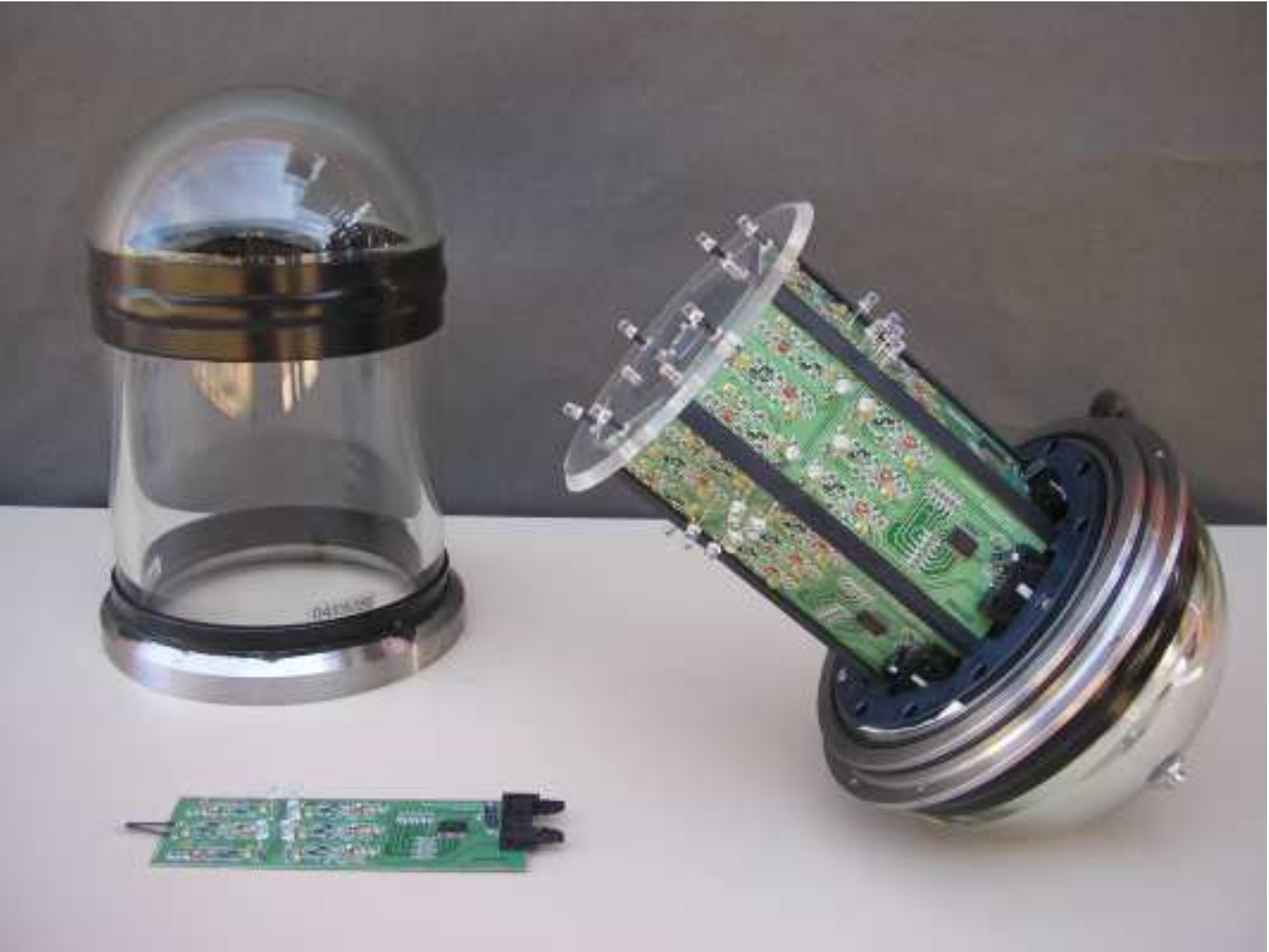}}
   \put(4.0,-0.0){\includegraphics[width=4.0cm]{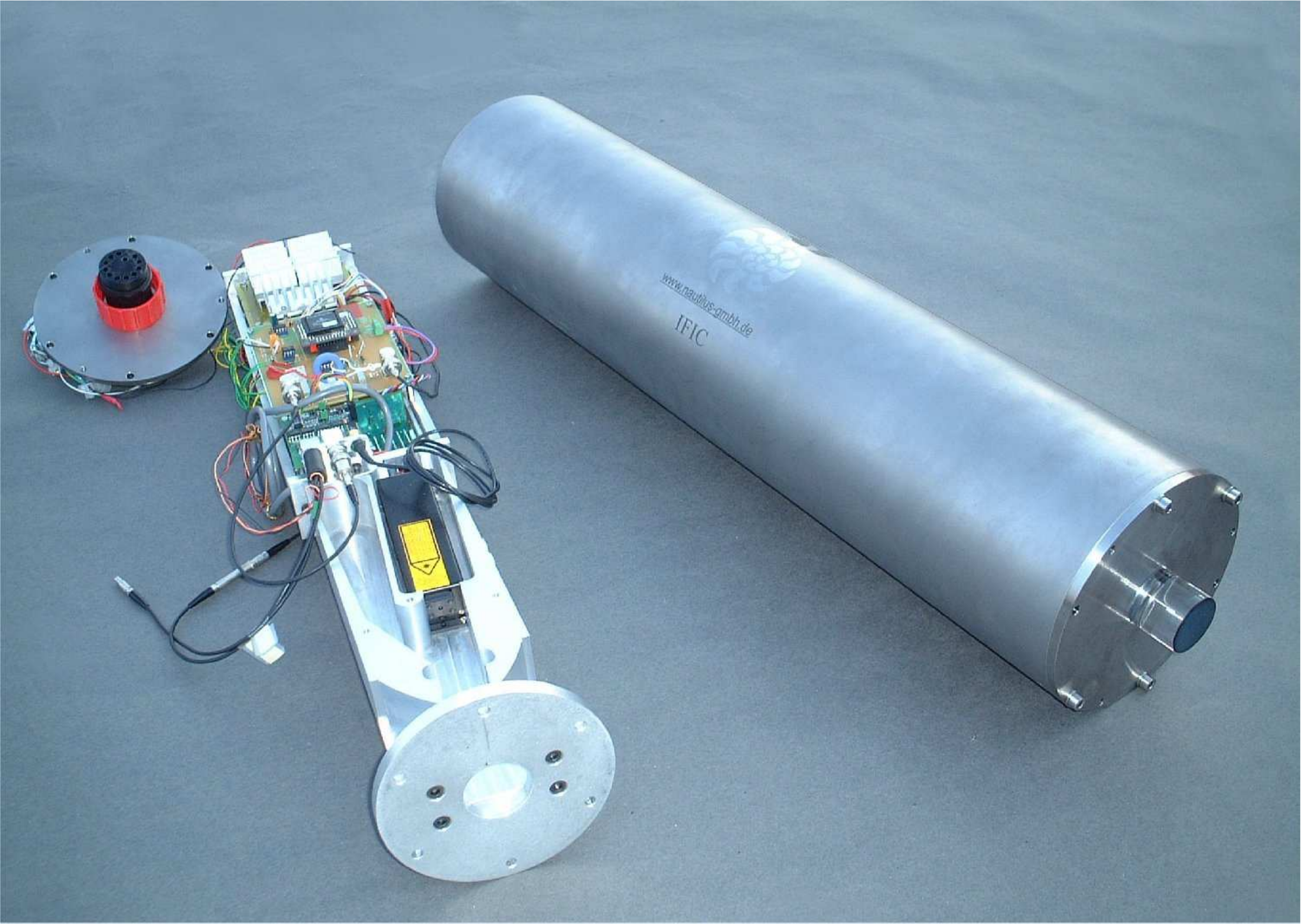}}
 \end{picture}
\caption[Sc]{\textit{Left: Picture of a LED Optical Beacon. Right: Picture of a Laser Beacon.}}
\label{figmangano2:opticalb}
\end{figure}

The laser, a Nd-YAG solid state laser, is a more powerful device and
emits pulses of light with an intensity of $\sim$1~$\mu$J and pulse
width of $\sim$0.8~ns (FWHM) at an average wavelength of 532~nm
(Figure~\ref{figmangano2:opticalb}, right).  The LEDs and lasers flash at a
frequency of 330~Hz provided by the DAQ system~\cite{bibmangano2:DAQ}. Further
details about the OBs system can be found in~\cite{bibmangano2:Timecalibration,bibmangano2:Ageron}.

The light spectrum of the eight sources used for this \mbox{analysis} (seven
LEDs with wavelengths between 385~nm and 518~nm and one Laser with
wavelength of \mbox{532~nm}) were measured using a calibrated
spectrometer.  The spectrometer was
first crosschecked with the green Nd-YAG Laser (532~nm) and then used
to measure the spectral emission of the LEDs in pulsed mode operation
(see Figure~\ref{figmangano2:specmeas}, upper). As can be seen the width of the wavelength spectra
is around 10~nm for all the light sources except for the LED
with average value of 518~nm, which is larger, and the laser, which is
much smaller ($\sim$2~nm). 

Due to the wavelength dependence of the absorption of light in water,
the spectra change as a function of the distance traveled by the light.
The expected wavelength distributions as a
function of distance have been estimated by Monte Carlo simulation
using the dependence of absorption length on wavelength given by~\cite{bibmangano2:abs}. In the lower panel of Figure~\ref{figmangano2:specmeas}
the estimated
wavelength distributions at 120~m are shown for the different LEDs.
Most of the wavelength distributions show small changes, except the
light sources with wavelengths above 500~nm where the
absorption length spectrum is steeper and the absorption has a larger effect.
Notice that the distributions have been normalized to unity in each peak
and therefore the relative effect of absorption between sources is not 
observed. This renormalization is performed in order to 
show the change in the shapes of the distributions, which 
influences the velocity measurement. The evolution of 
these wavelength distributions
along the light path is taken into account in the final 
results (Section \ref{secmangano2:data}). In particular the 
uncertainty assigned to the wavelengths has been 
taken to be the root mean square of the wavelength 
distribution given by the simulation.

\begin{figure}
 \setlength{\unitlength}{1cm}
 \centering
 \begin{picture}(8.0,5.)
   \put(0.0, 2.5){\includegraphics[width=9.3cm]{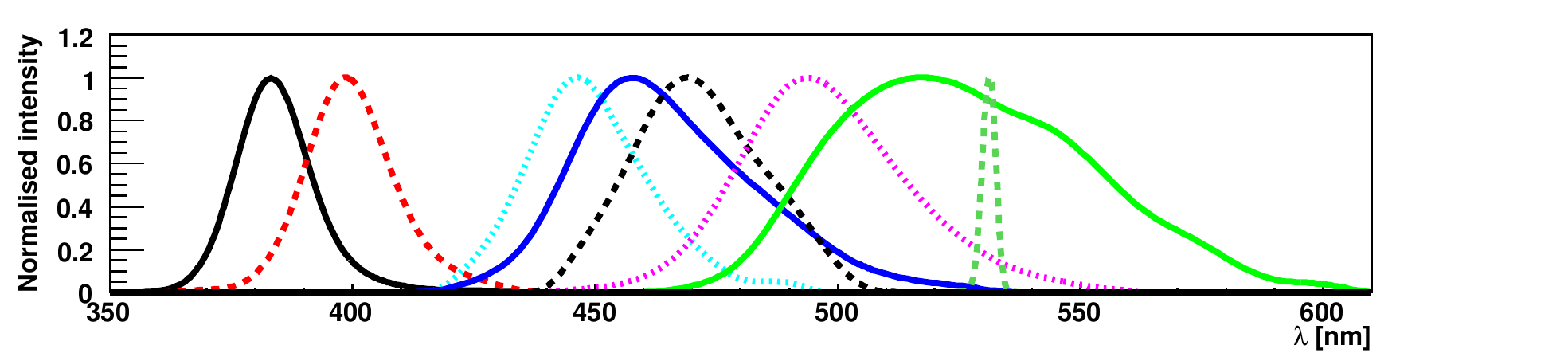}}
   \put(0.0,-0.0){\includegraphics[width=9.3cm]{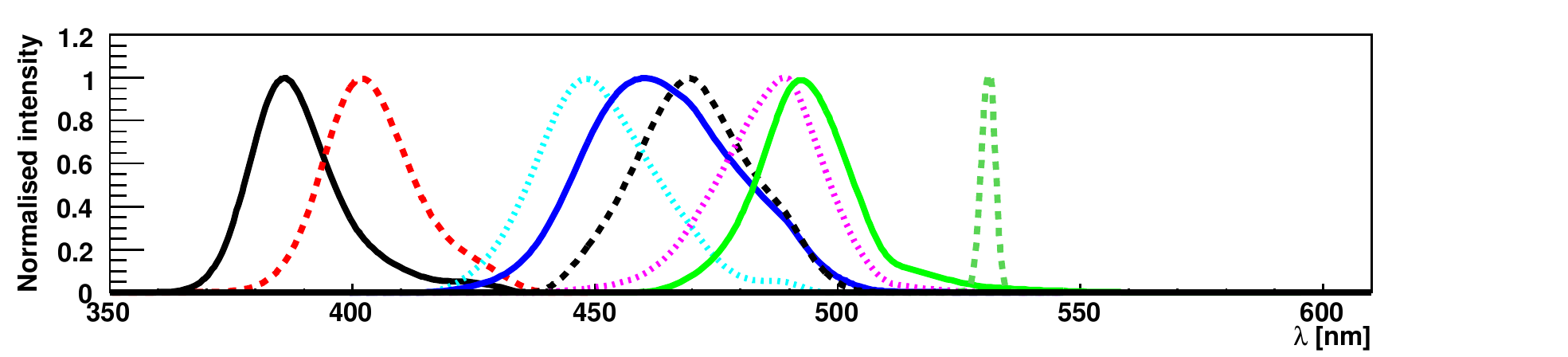}}
 \end{picture}
\caption[Sc]{\textit{Upper: Light source distributions measured with the spectrometer (normalized to unit intensity). The spectrometer measurement are made in air. The data points have been smoothed  and the highest value of each wavelength distribution is set equal to one. Lower: Simulated spectral distribution at a distance of 120 m in sea water with the highest value normalized to one. The difference between the upper and lower spectral distributions are due to the variation of the absorption length as a function of the wavelength.}}
\label{figmangano2:specmeas}
\end{figure}

\section{Data Acquisition and Analysis}
\label{secmangano2:dataacqu}
In order to measure the optical properties of the deep sea water, such
as the refractive index and the absorption and scattering lengths,
special dedicated runs were performed using the OB system. During these 
runs, one single upward looking LED of
the lowest OB in a line was flashed (see Figure \ref{figmangano2:linearfit}, left). 
Only the signals recorded by the
PMTs along the same line where the flashing OB was located are used in
the analysis, so that the influence due to the line movements are
negligible and no corrections to the nominal PMT positions have to be
applied.

The runs used in this analysis were taken from May 2008 to April 2011.
Each run contained typically more than 10$^5$ light flashes. For each
flash, its time of emission as given by the small PMT inside the OB
was recorded. The arrival time of the photons to the PMTs further
above the line were recorded within a time window spanning from 1500~ns
before the flash to 1500~ns afterwords. The integrated charge of the
analogue pulses of the PMTs were also recorded.  
Only runs were used with rates of
background light below
100~kHz and stable in time.

Figure \ref{figmangano2:time} shows the distribution of the time of arrival of
photons to a PMT located 100~m above the flashing LED optical 
beacon ($\lambda$ = 469~nm) with a clear peak at
around 470~ns. The zero time is the time of emission of the flash as
recorded by the internal PMT of the LOB. The peak at t=470~ns corresponds 
to the shortest propagation time of light.
The tail of delayed photons
is due to light scattering.  The flat distribution before and after
the peak is the optical background due to $^{40}$K decays and
bioluminescence.

\begin{figure}
 \setlength{\unitlength}{1cm}
 \centering
 \begin{picture}(8.0,5.5)
   \put(-0.5,0.0){\includegraphics[width=9.5cm]{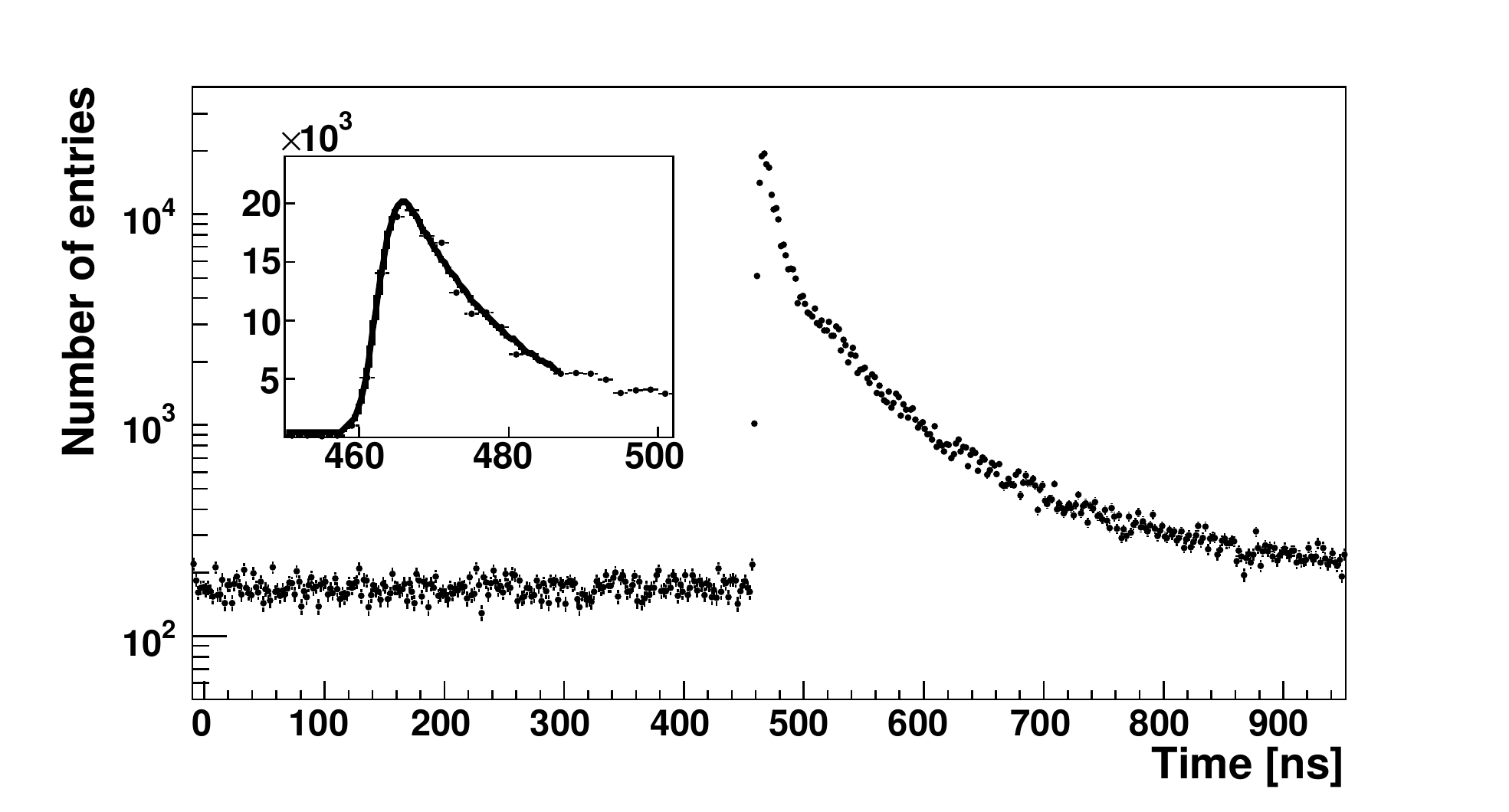}}
 \end{picture}
\caption[Sc]{\textit{Arrival time distribution of LED light detected by a PMT
  at 100 m.  The inset shows a zoom around the signal
  region. Superimposed is the result of the fit to the convolution of
  a Gaussian and an exponential distribution (see text).}}
\label{figmangano2:time}
\end{figure}

Since the distance traveled by the light from the emitting OB to the
different OMs along the line is of the order of hundreds of
meters the effect of scattering has to be taken into account.  The
arrival time distributions are fitted to a convolution of a Gaussian
and an exponential distribution on top of a flat background. The Gaussian distribution models the
transit time spread of the PMTs, the time width of the pulsed emitted
by the LED optical source and the effect of the chromatic dispersion
in water. The exponential distribution takes into account the effect of the scattering of photons in water.  The fit function can be formulated as:
\begin{equation}
  f(t)=b + h \cdot e^{-\frac{t-\mu}{\tau}} \times \textrm{erfc} \Bigg(\frac{1}{\sqrt{2}} \Big(\frac{\sigma}{\tau} - \frac{t-\mu}{\sigma}\Big)\Bigg),
\label{eqmangano2:ciro}
\end{equation}
where $t$ is the arrival time of the photons. The fit parameters
are the optical \mbox{background $b$}, the height of the fit
function $h$, the mean $\mu$ and width $\sigma$ of the Gaussian distribution
and the exponential decay constant
$\tau$. In Equation \ref{eqmangano2:ciro} $\textrm{erfc}(t)$ is the complementary error function
distribution, i.e.  erfc(t)= $\frac {2}{\sqrt{\pi}}\int_{t}^{\infty}
e^{-t^2} dt $.  An example of the fit is shown in the inset of
Figure~\ref{figmangano2:time} where the peak region has been enlarged. The
fit is made in the range from 200~ns before the most populated bin and
20~ns after.  The arrival time at each PMT is given by the fitted mean
value of the Gaussian distribution.  The fit results are stable to
changes in the fit range and histogram binning.

An example of the measured arrival times versus the distances between the
corresponding OB and the PMTs are shown in
Figure~\ref{figmangano2:linearfit} right. Since at each storey there are three PMTs
there can be up to three measurements for each distance.  The nearest
PMTs to the OB are excluded from the fit since they receive a very
high amount of light and have a bias in the time estimate caused by
the early photon effect~\cite{bibmangano2:Timecalibration}.  The minimum distance
for the fit range is chosen in such a way that the average collected
charge amplitude per flash is below 1.5 photoelectrons (p.e.).  The
maximum distance is chosen so that the signal is at least seven
sigmas above the average background. The slope of a linear fit through
the measured points gives the inverse of the measured velocity of
light in water ($v_{measured}$).  The measured refractive index is
defined as
\begin{equation}
n=c/v_{measured}
\label{eqmangano2:refindexmes}
\end{equation}
with c = $2.9979 \times 10^8$~m $\cdot$ s$^{-1}$.  The error of the
refractive index given in Figure \ref{figmangano2:linearfit} right is the error
estimated by the linear fit.

\begin{figure}
 \setlength{\unitlength}{1cm}
 \centering
 \begin{picture}(8.0,5.5)
   \put(0.0,0.5){\includegraphics[width=2.0cm]{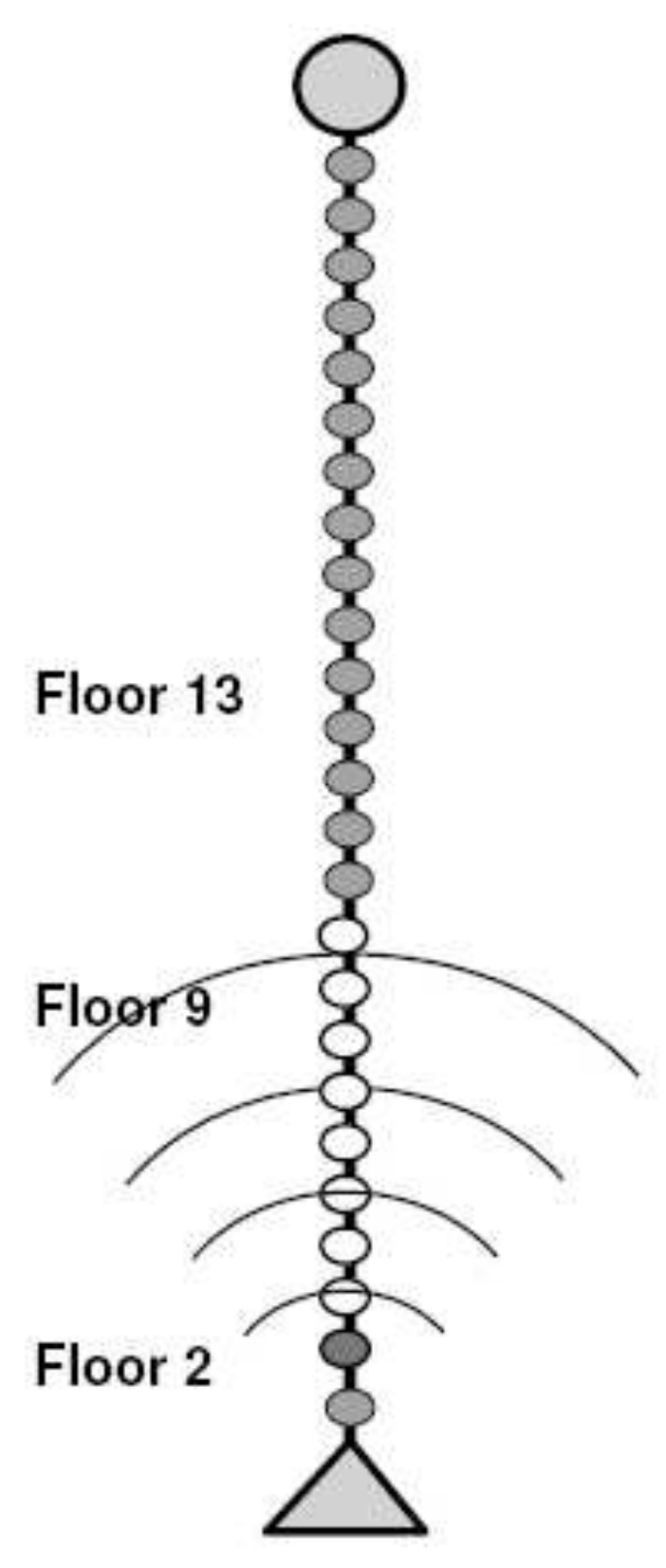}}
   \put(2.0,0.0){\includegraphics[width=6.0cm]{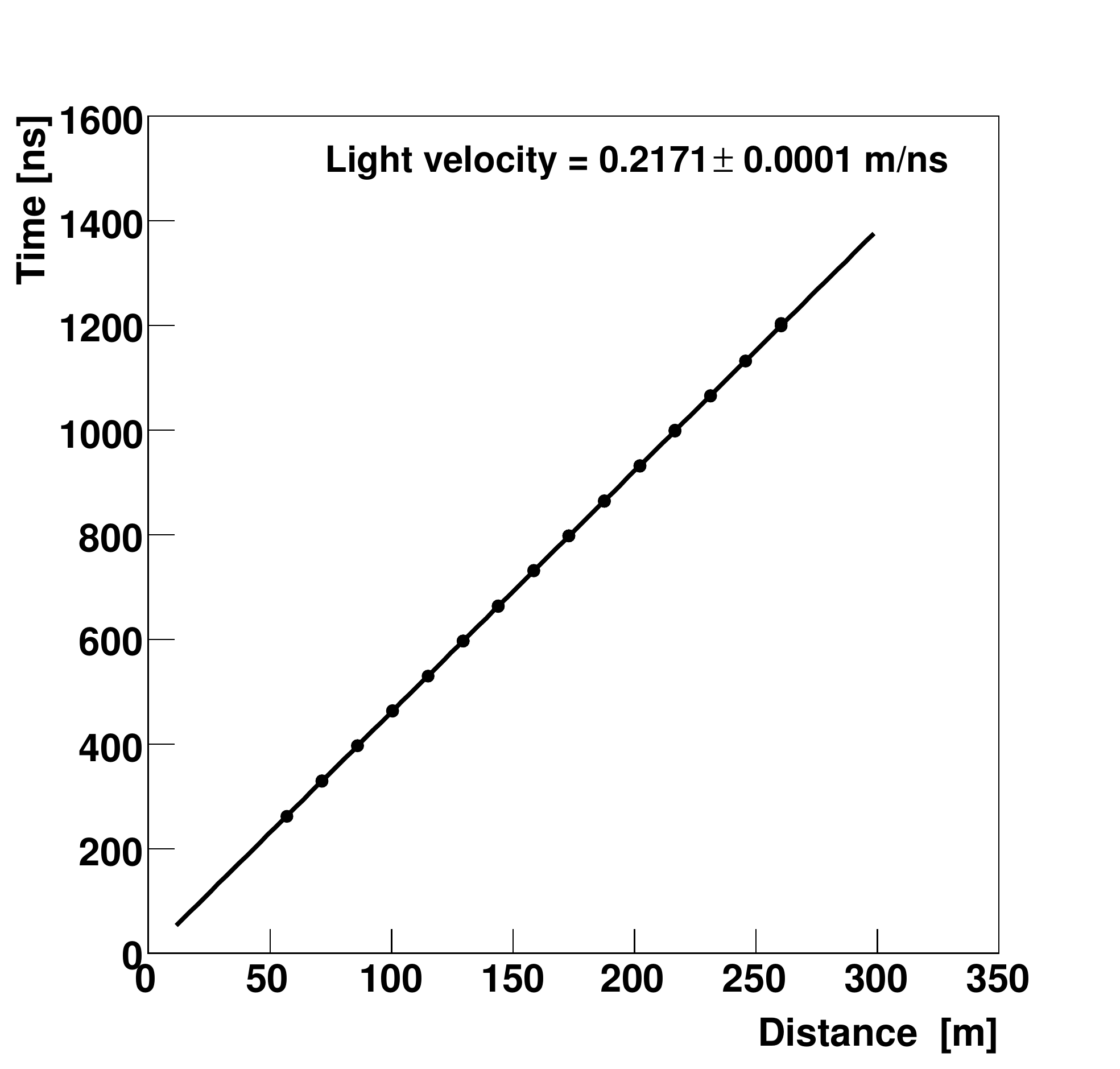}}
 \end{picture}
\caption[Sc]{\textit{Left: Schematic view of an upwards LED flashing light located in the lowest optical beacon. Only signals recorded by the PMTs along the same line are used. Right: The Figure shows the arrival time as a function of the distance between the LED and the different PMTs. The slope of a linear fit to the arrival time versus distance gives the inverse of the light velocity.}}
\label{figmangano2:linearfit}
\end{figure}

Monte Carlo simulations have been used to  check the stability of the analysis
and to study the systematic effects.
Monte Carlo runs taking into account the geometry of the detector and
modeling the light propagation in water have been used to analyze the
time distributions and calculate the refractive index for different
optical parameters.  The
analysis method was first validated with Monte Carlo samples for the wavelength values of 400~nm, 470~nm and 532~nm.  A
variation in the absorption length between 30~m and 120~m produces a
variation in the refractive index lower than 0.1\%.  The main
systematic uncertainty is given by the variation of the scattering
length between 20~m and 70~m.  The measured refractive index increases
up to 0.3\% using the scattering length of 20~m.
Adding in quadrature the individual errors one obtains
a systematic error that varies from 0.3\% to 0.4\% depending on the
wavelength.

\section{Refractive Index and Data Results}
\label{secmangano2:data}
The refractive index of sea water at a given wavelength depends
on the temperature, the salinity of the water and the pressure.  
A parametric formula for the phase
refractive index proposed by~\cite{bibmangano2:Quanfry}, based
on data from~\cite{bibmangano2:Austin}, has been modified with
appropriate pressure corrections~\cite{bibmangano2:Agui}.  The phase refractive
index for sea water as a function of wavelength at temperature $T=12.9 \pm 0.1^{\circ}$ and
salinity $S = 3.848 \pm 0.001\%$ and pressure $p$ is given by
\begin{eqnarray}
n_p (\lambda,p)= 1.32292+(1.32394-1.32292) \times \nonumber  \\
\times \frac{p-200}{240-200}+\frac{16.2561}{\lambda}-\frac{4382}{\lambda^2}+\frac{1.1455\times10^6}{\lambda^3}.
\label{eqmangano2:nphase}
\end{eqnarray}
The parametrization of the refractive indexes $n_g$ 
(using Equation \ref{eqmangano2:n_g} and Equation \ref{eqmangano2:nphase}) is shown
in \mbox{Figure~\ref{figmangano2:refdat}} for the given values of temperature
and salinity, and for a pressure between 200~atm and 240~atm.

The environmental parameters can be considered stable in time so that,
from measurements of temperature, salinity and pressure, the
refractive index for a particular wavelength and at a given depth can
be calculated with a precision better than $4 \times 10^{-5}$.

\begin{figure}
 \setlength{\unitlength}{1cm}
 \centering
 \begin{picture}(8.0,6.)
   \put(1.0,0.0){\includegraphics[width=6.5cm]{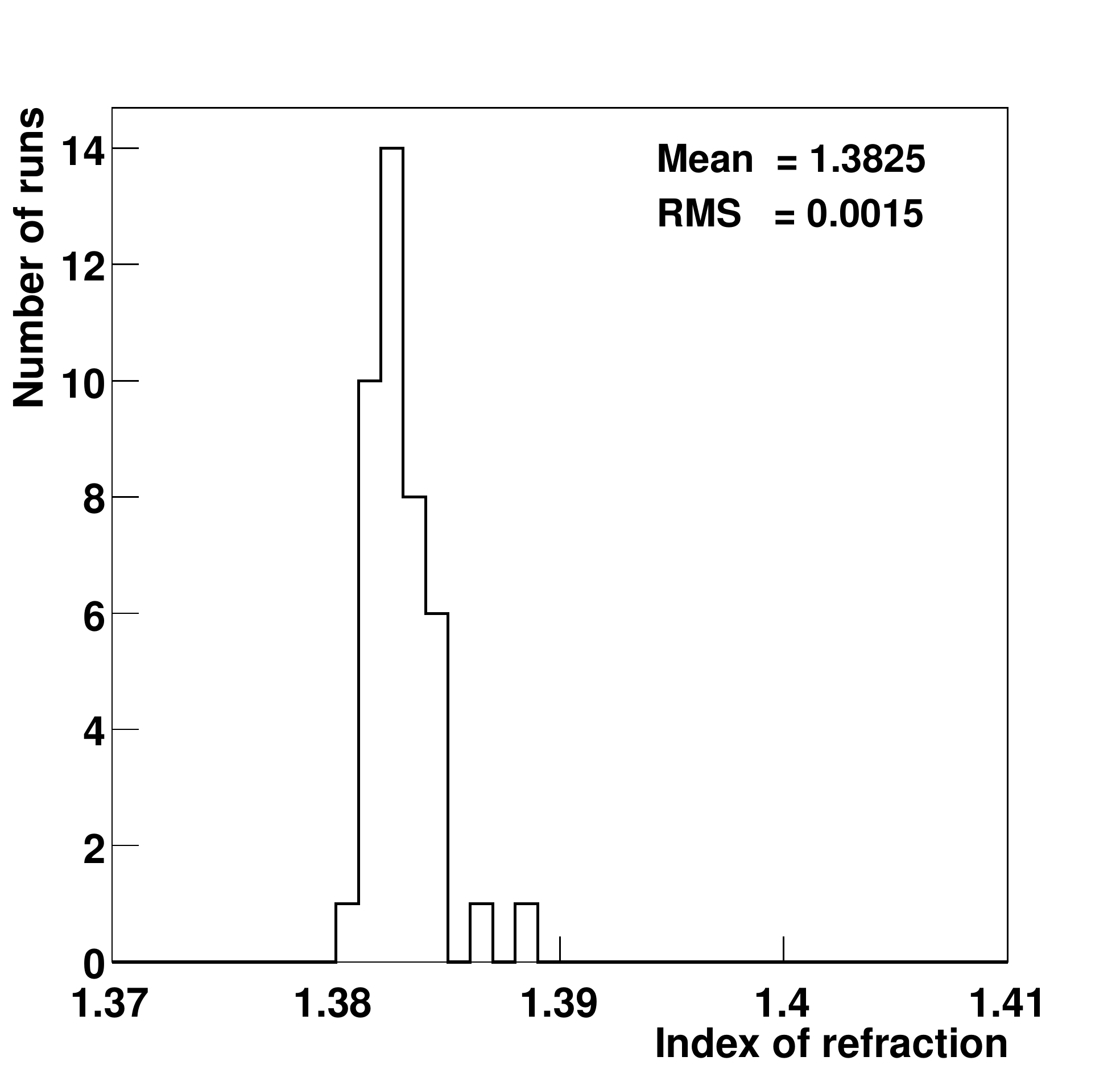}}
 \end{picture}
\caption[Sc]{\textit{Distribution of the measured refractive index for a total of 42 runs for a LED with $\lambda$=469~nm taken between May 2008 and March 2010.}}
\label{figmangano2:fitresults}
\end{figure}

Between May 2008 and March 2010, a total of 42 runs taken with an
average wavelength of 469~nm, 14 runs with an average wavelength of
400 nm and 13 runs with an average wavelength of \mbox{532 nm} were
collected and analyzed according to the methods explained in
Section~\ref{secmangano2:dataacqu}.  The runs were taken with three different
LED intensities. Depending on the LED light intensity, the minimum and
maximum fit range values varies.  For a typical high, middle or low
intensity run at a wavelength of \mbox{469~nm} the fit range is
between 50~m and 250~m, 40~m and 220~m, and 10~m and 130~m
respectively. The measured refractive index values of the 42 runs with a  wavelength of 469~nm are shown in Figure \ref{figmangano2:fitresults}. 
Between November 2010 and April 2011 eight runs with the
modified OB have been collected, extending the measurements with six additional 
 wavelengths.   
The measured  refractive indices with their systematic errors
estimated in Section~\ref{secmangano2:dataacqu} are compared with the Equation~\ref{eqmangano2:nphase} in Figure~\ref{figmangano2:refdat}.  
As mentioned in Section~\ref{secmangano2:expsetup}, 
the uncertainties in the wavelengths
have been taken to be the root mean square of the corresponding distribution at the middle of
 the distance ranges.
As can be seen from Figure~\ref{figmangano2:refdat} the results are compatible with the theoretical prediction.

As mentioned in Section~\ref{secmangano2:introduction}, the PMTs are unable to distinguish
the wavelength
of the incoming photons, so the effect of this chromatic dependence
can only be taken into account on average. 
The spread of the arrival time residuals with
respect to the expected arrival time of a 460~nm photon have been
computed by means of a standalone Monte Carlo simulation using the
phase velocity for the emission angle and the group velocity (as given
by the Equations \ref{eqmangano2:nphase} and Equation \ref{eqmangano2:n_g}) for the arrival time.
This simulation indicates that the spread of the time residual is
0.6~ns at 10~m, 1.6~ns at 40~m, 2.7~ns at 100~m and 3.6~ns at 200~m. 
The
time uncertainty introduced by this spread is unavoidable and is taken
into account in the ANTARES official simulation program~\cite{bibmangano2:KM3,bibmangano2:Bru}.
Even though the exact influence of the medium depends on the particular Cherenkov
photon (wavelength, distance to the hit PMT) and therefore requires a
full simulation, a rough estimate of the average effect can be
obtained assuming that a majority of hits are between 40~m and 100~m
from the track, which gives a value of $\sim$2~ns for the uncertainty
introduced by the transmission of light in sea water, including
chromatic dispersion. This value is to be compared with $\sim$1.3~ns
coming from the PMTs transit time spread and to $\sim$1~ns from time
calibration.

\begin{figure}
 \setlength{\unitlength}{1cm}
 \centering
 \begin{picture}(18.5,6.)
   \put(0.0,0.){\includegraphics[width=9.0cm]{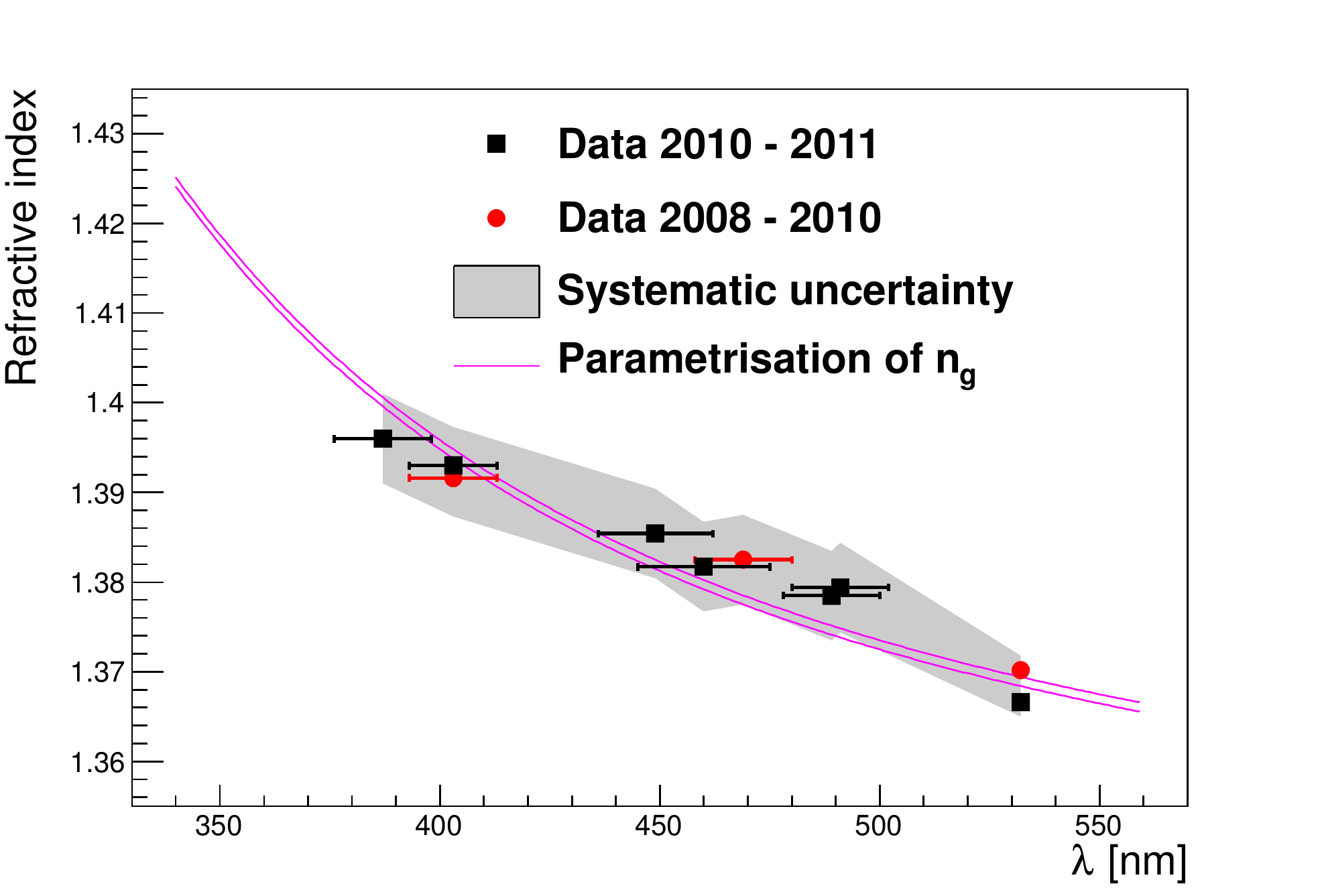}}
 \end{picture}
\caption[Sc]{Group and phase refractive index as a function of the wavelength for a given temperature and salinity with measured data points and its systematic error bars.}
\label{figmangano2:refdat}
\end{figure}

\section {Conclusion}
\label{conclusion}
Pulsed light sources with wavelengths between 385~nm and 532~nm
shining through sea water have been used to measure the refraction
index using the time of light distribution detected by the PMTs at
distances between few tenth and few hundred meters from the
source. Dedicated Monte Carlo simulations have been used to validate the
analysis method and to evaluate the systematic errors.  The data
results obtained for the dependence of the group refractive index on wavelength are compatible with the
parametrization as a function of salinity, pressure
and temperature of sea water at the ANTARES site.

\vspace*{0.5cm}
{\footnotesize{\bf %
\noindent 
Acknowledgment: }I gratefully acknowledge the support of the JAE-Doc postdoctoral program of CSIC. This work has also been supported by the following Spanish projects: FPA2009-13983-C02-01, MultiDark Consolider CSD2009-00064, ACI2009-1020 of MICINN and Prometeo/2009/026 of Generalitat Valenciana.}

\setcounter{figure}{0}
\setcounter{table}{0}
\setcounter{footnote}{0}
\setcounter{section}{0}
\setcounter{equation}{0}

\newpage
\id{id_lahmann}



\newcommand{\dd}{\mathrm{d}}

\title{\arabic{IdContrib} - Recent Results of the Acoustic Neutrino Detection Test System AMADEUS}
\addcontentsline{toc}{part}{\arabic{IdContrib} - {\sl Robert~Lahmann} : Recent Results of the Acoustic Neutrino Detection Test System AMADEUS%
\vspace{-0.5cm}
}

\shorttitle{\arabic{IdContrib} - AMADEUS Results}

\authors{R.~Lahmann for the {\sc Antares} Collaboration}
\afiliations{Friedrich-Alexander-Universit\"{a}t Erlangen-N\"{u}rnberg, Erlangen Centre for Astroparticle Physics, Erwin-Rommel-Str. 1, 91058 Erlangen, Germany}

\email{robert.lahmann@physik.uni-erlangen.de}

\abstract{
The technique of acoustic neutrino detection is a promising approach
for future large-scale ultra-high-energy neutrino detectors in water. 
To investigate this technique in the deep sea, the AMADEUS system has
been integrated into the ANTARES neutrino telescope in the
Mediterranean Sea. Installed at a depth of more than 2000\,m, the 36
acoustic sensors of AMADEUS are based on piezo-ceramics elements for
the continuous broadband recording of signals with frequencies ranging up to 125\,kHz.
In order to assess the background for acoustic neutrino
detection in the deep sea, the characteristics of transient
signals, which can mimic the acoustic signature of a neutrino interaction, and of the ambient noise have been investigated. 
In this context, 
an offline analysis package 
was developed, including signal classification and acoustic source
reconstruction algorithms.
%
In addition, 
a complete simulation chain was developed. 
It comprises the generation
of acoustic pulses produced by neutrino interactions and their
propagation to the sensors within the detector, ambient and transient
noise models for the Mediterranean Sea and the response functions of
the AMADEUS data acquisition hardware. 
In this article, the AMADEUS system will be introduced and the  procedures
for offline event selection and Monte Carlo simulations will be described.
Recent AMADEUS results 
will be discussed and first conclusions concerning the feasibility of 
acoustic detection of ultra-high-energy neutrinos in the Mediterranean 
Sea presented.
}


\maketitle

\section{Introduction}
\label{seclahmann:intro}

Measuring acoustic pressure pulses in huge underwater acoustic arrays
is a promising approach for the detection of 
ultra-high-energy (UHE, $E_{\nu} \gtrsim 10^9$\,GeV) neutrinos.
These
are expected to be produced in interactions of cosmic rays with the cosmic microwave background~\cite{biblahmann:Berezinsky-1969}.
The pressure signals are produced by the
particle showers that evolve when neutrinos interact with nuclei in
water.
The resulting energy deposition within a cylindrical volume of a few
centimetres in radius and several metres in length leads to a local
heating of the medium which is instantaneous with respect to the
hydrodynamic time scales.  This temperature change induces an
expansion or contraction of the medium depending on its volume
expansion coefficient.  According to the thermo-acoustic
model~\cite{biblahmann:Askariyan1979267,biblahmann:PhysRevD.19.3293}, the accelerated expansion of the
heated volume---a micro-explosion---forms a pressure pulse of bipolar
shape which propagates in the surrounding medium.
Coherent superposition of the elementary sound waves, produced over the
volume of the energy deposition, leads to a propagation within a flat
disk-like volume (often referred to as {\em pancake})
in the direction perpendicular to the axis of the particle shower.
After propagating several hundreds of metres in sea water, the pulse
has a characteristic frequency spectrum that is expected to peak
around 10\,kHz~
\cite{biblahmann:S.Bevan:2007wd,biblahmann:Bevan:2009rr,biblahmann:Bertin_Niess}.
As the attenuation length in sea water in the relevant frequency range
is about
 one to two orders of magnitude larger than that for visible light,
a potential acoustic neutrino detector would require
a less dense instrumentation of a given volume
than an optical neutrino telescope.

The AMADEUS project~
\cite{biblahmann:collaboration:2010fj}
was conceived to perform a feasibility study for a
potential future large-scale acoustic neutrino detector. For this purpose, 
a dedicated array of acoustic sensors was integrated into the
ANTARES neutrino telescope~\cite{biblahmann:ANTARES-paper}. 

\section{The ANTARES Detector}
\label{seclahmann:antares_detector}

The ANTARES neutrino telescope was designed to detect neutrinos by
measuring the Cherenkov light emitted along the tracks of relativistic
secondary muons generated in neutrino interactions.
A sketch of the detector, with the AMADEUS modules highlighted, is
shown in Fig.~\ref{fig:ANTARES_schematic_all_storeys}.  The detector
is located in the Mediterranean Sea at a water depth of 2475\,m,
roughly 40\,km south of the town of Toulon at the French coast at the
geographic position of 42$^\circ$48$'$\,N, 6$^\circ$10$'$\,E.  ANTARES was
completed in May 2008 and comprises 12 vertical structures, the {\em
  detection lines}.  Each detection line holds up to 25 {\em storeys}
that are arranged at equal distances of 14.5\,m along the line,
starting at about 100\,m above the sea bed and interlinked by
electro-optical cables.  A standard storey consists of a titanium
support structure, holding three {\em optical modules}
(each one consisting of a photomultiplier tube inside a
water-tight pressure-resistant glass sphere) and one 
cylindrical electronics container.

\begin{figure}[tb]
\centering
\includegraphics[width=0.93\columnwidth]{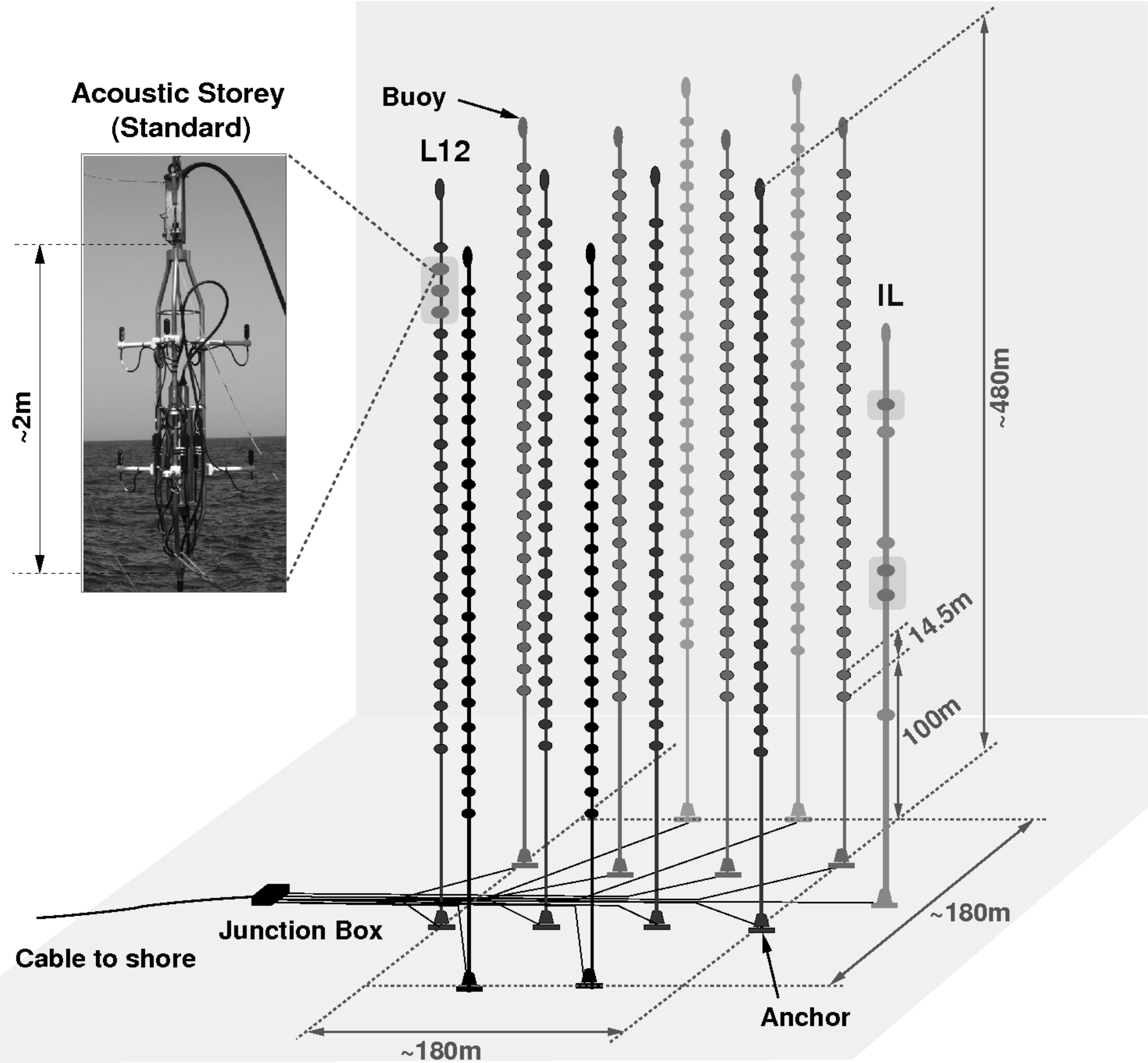}
\caption{A sketch of the ANTARES detector.  The six acoustic storeys
  are highlighted and a photograph of a storey in standard
  configuration is shown.  L12 and IL denote the 12th detection line
  and the Instrumentation Line, respectively.}
\label{fig:ANTARES_schematic_all_storeys}
\end{figure}

A 13th line, called {\em Instrumentation Line (IL)}, is equipped with
instruments for monitoring the environment. It holds six storeys.
For two pairs of consecutive storeys in the IL, the vertical distance
is increased to 80\,m.

Each line is fixed on the sea floor by an anchor equipped with
electronics and held taut by an immersed buoy.  An interlink cable
connects each line to the {\em Junction Box} from where the main
electro-optical cable provides the connection to the shore station.

\section{The AMADEUS System}
\label{seclahmann:amadeus}
\label{seclahmann:data_processing}
Within the AMADEUS system~
\cite{biblahmann:collaboration:2010fj}, acoustic sensing is
integrated in the form of {\em acoustic storeys} that are modified
versions of standard ANTARES storeys, in which the optical modules are
replaced by custom-designed acoustic sensors.  Dedicated electronics
is used for the amplification, digitisation and pre-processing of the
analogue signals.  Figure~\ref{fig:acou_storey_drawing} shows the
design of a standard acoustic storey.  Six acoustic
sensors per storey were implemented, arranged at distances of roughly
1\,m from each other.  

\begin{figure}[ht]
\centering
\hspace*{-10mm}\includegraphics[width=6.0cm]{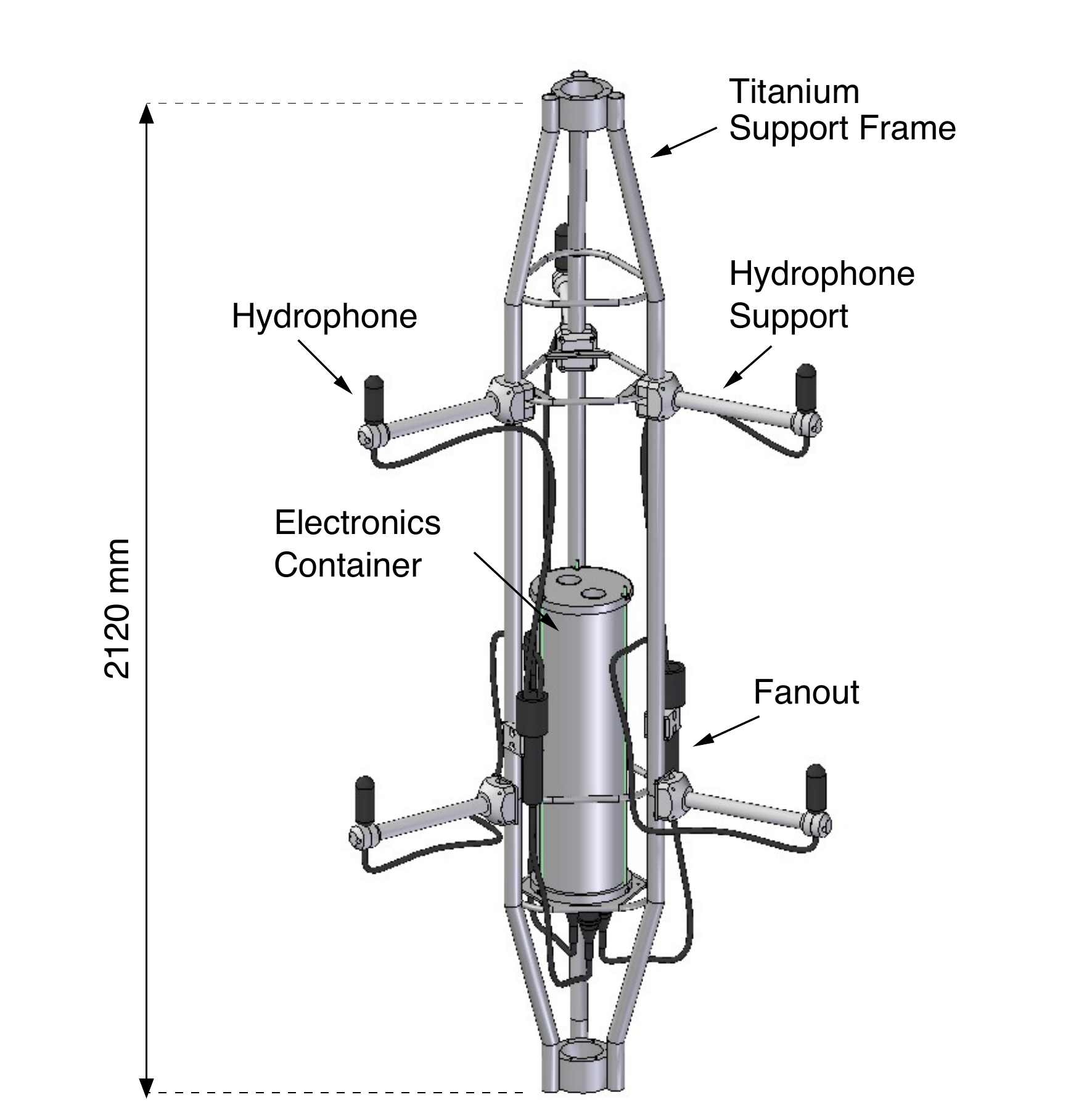}
\caption{
Drawing of a standard acoustic storey. 
\label{fig:acou_storey_drawing}
}
\end{figure}

The AMADEUS system comprises a total of six acoustic storeys: three on
the IL
and three on the
12th detection line (Line 12).
%
The analysis presented in this article was done with the standard acoustic storeys
which are equipped with hydrophones.
These acoustic sensors employ piezo-electric elements, coated in polyurethane, for
the broadband recording of signals with frequencies ranging up to
125\,kHz. 
The lowest acoustic storey on Line 12 is equipped with alternative sensing devices, so-called
{\em acoustic modules} which are described elsewhere~\cite{biblahmann:collaboration:2010fj}.

All data from the 36 acoustic sensors
is transmitted to the shore station.
Here an adjustable software filter selects events from the data stream 
for storage on disk and further offline analysis.
Currently, three filter schemes are in 
operation~
\cite{biblahmann:collaboration:2010fj}:
A minimum bias trigger which records data continuously for about 10\,s
every 60\,min, a threshold trigger which is activated when the signal
exceeds a predefined amplitude, and a pulse shape recognition
trigger. For the latter, a cross-correlation of the signal with a
predefined bipolar signal, as expected for a neutrino-induced shower,
is performed. The trigger condition is met if the output of the
cross-correlation operation exceeds a predefined threshold.  
For the latter two triggers,
the thresholds are automatically 
adjusted to the prevailing
ambient noise and the condition must be met in at least four sensors of a 
storey.

\section{Acoustic Background in the Deep Sea}
To assess the feasibility of acoustic neutrino detection in a natural body of water, 
 transient and ambient noise
at the site of the installation
have to be investigated.
The ambient noise is broadband and is mainly caused by agitation of the sea surface\,\cite{biblahmann:urick1986ambient}, i.e.~by wind, breaking waves, spray, and cavitations. Thus it is correlated to the weather conditions, mainly to the wind speed, see e.g.~\cite{biblahmann:NeffVLVnT11}. 
It is predominantly the ambient background that determines the 
energy threshold for neutrino detection.
Transient noise signals have short duration and an amplitude that exceeds the ambient noise level. These signals can mimic bipolar pulses from neutrino interactions. Sources of transient signals can be 
anthropogenic, such as shipping traffic, or
marine mammals. In particular dolphins emit short signals with a spectrum similar to that of acoustic
emissions from neutrino interactions. 
Given the expected low rate of cosmogenic neutrinos of the order of 1 per year and km$^3$, the 
transient background must be {\em completely} suppressed, which poses a major challenge.

\section{Suppression of Transient Background}
\label{seclahmann:apps}
To suppress signals from transient sources which are not consistent with 
acoustic signals of neutrino interactions, a classification strategy is employed\,\cite{biblahmann:Neff:2011pd}. 
This strategy stems from machine learning algorithms trained and tested with data from a simulation (see Sec.~\ref{seclahmann:sim_chain}). Random Forest and Boosted Trees algorithms have achieved the best results for individual sensors and clusters of sensors. For individual sensors, the classification error is of the order of 10\,\% for a well trained model. The combined results of the individual sensors in an acoustic storey are used as new input for training. This method obtains a classification error below 2\,\%.

Using the six hydrophones of an acoustic storey, direction reconstruction of point sources 
is possible with an accuracy of $1.6^\circ \pm 0.2^\circ$ in azimuth and $0.6^\circ \pm 0.1^\circ$ in zenith~\cite{biblahmann:Neff_phd}. 
If the directions were reconstructed for at least two of the acoustic storeys, the best approximation of the intersection point of the rays starting from the sensor clusters and pointing into the reconstructed direction is searched for. 
In principal, a powerful method to reduce ambient background is to restrict 
neutrino searches to a fiducial volume that excludes the topmost part of the sea above
a depth of about 500\,m. 
This eliminates 
signals from ships and from dolphins, which do not dive below 500\,m. 
However, given the geometry of AMADEUS, the small angular errors of the zenith angle reconstruction
 translate into large uncertainties of the position reconstruction at large distances.
%
Hence the more effective approach of suppressing sources using a clustering algorithm was chosen: since neutrino interactions are rare and
isolated events, a number of events clustering temporally and spatially can be assumed to stem from a single source such as a ship or a sea mammal. 
To avoid the aforementioned problems with the position reconstruction, 
for each event the intersection of the reconstructed direction with the sea surface  was used as input
for the clustering algorithm. 

Figure~\ref{fig:r_z} shows the distribution of events over an integrated measurement time of about 156
days without any offline suppression and after applying the signal classification and clustering
cuts described above. The background from transient signals is reduced from 
$15\times10^3$\,km$^{-3}$\,yr$^{-1}$ to 100\,km$^{-3}$\,yr$^{-1}$. 
Figure~\ref{fig:r_z} shows the limitations of the reconstruction of the $z$-coordinate, which is smeared out and even extends to positive values which would imply a source above the sea surface.

The background rate was reduced significantly by applying the cuts described above, but 
as mentioned in the previous section,
basically a complete reduction of the transient background is required. 
The characteristic propagation of the signal within a flat disk-like shape
(the ``pancake'') is another important feature to classify neutrinos. 
This feature however cannot be 
exploited with AMADEUS due to its small size and two-dimensional arrangement of acoustic storeys. 
With the software framework developed and the data collected with the AMADEUS detector, 
designs of large acoustic neutrino detectors
can be investigated using Monte Carlo techniques in order to optimise the detector layout with respect to the potential for the reconstruction of the 
source position and the three-dimensional shape of the acoustic emission pattern.

\begin{figure}[tb]
\centering
\includegraphics[width=0.99\columnwidth]{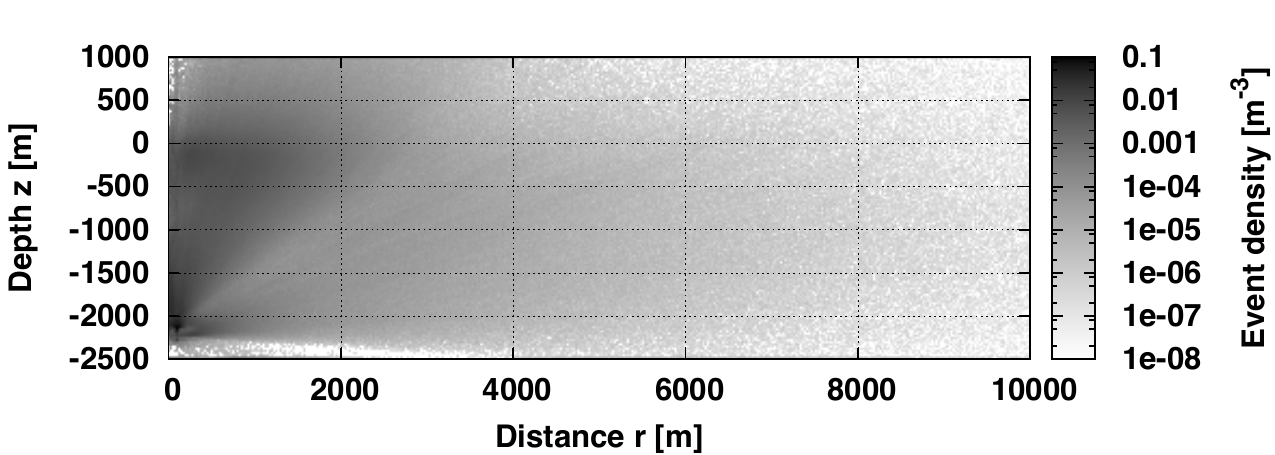}
\includegraphics[width=0.99\columnwidth]{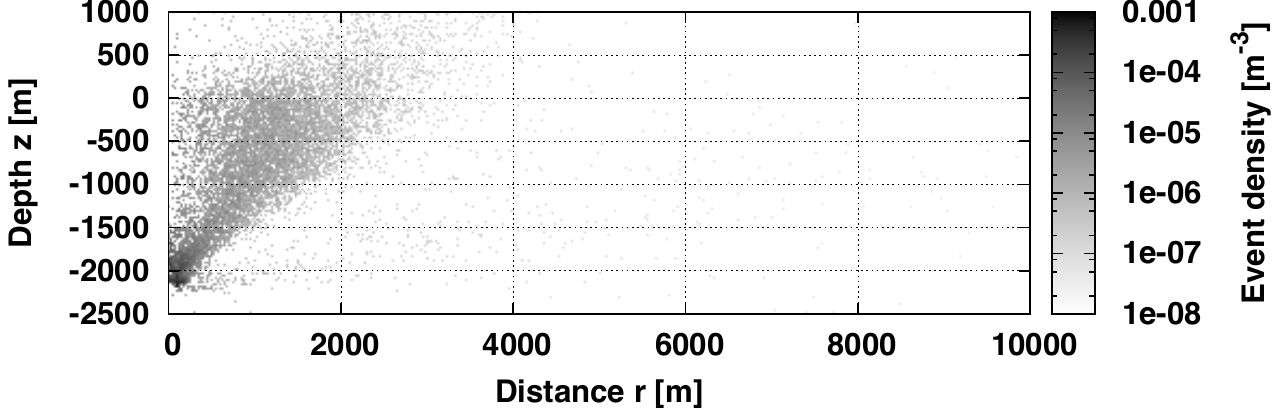}
\caption{
Spatial density of transient signals as a function of the depth $z$ and the distance $r$ from the
centre of the AMADEUS detector. Top:
the density of transient signals as selected by the AMADEUS online filter; bottom:
events remaining after the offline cuts described in the text. 
\label{fig:r_z}}
\end{figure}

\section{Simulation Chain}
\label{sim}
\label{seclahmann:sim_chain}

The simulation chain~\cite{biblahmann:NeffVLVnT11} consists of the following modules, which build upon each other to create a simulated neutrino event: 
\begin{enumerate}
\itemsep 0pt
\item
An interaction vertex is located at a random position in a given volume around the detector and the energy and direction of the incident neutrino are set randomly within predefined ranges.
\item
The Monte Carlo hadronic shower is produced from a parametrisation, 
valid up to a total shower energy of $10^{12}$\,GeV,
which is based on work by the ACoRNE collaboration~\cite{biblahmann:S.Bevan:2007wd,biblahmann:Bevan:2009rr}. 
The fraction of energy of the neutrino energy that is deposited in the shower
is calculated according to parametrisations from~\cite{biblahmann:Connolly-etal-inelast}.

After the cascade has been simulated, the acoustic pulse and its propagation to the sensors within the detector are calculated following~\cite{biblahmann:S.Bevan:2007wd,biblahmann:Bevan:2009rr}. 
This comprises the disk-like propagation pattern 
and the bipolar pulse recorded by the sensor.
\item
A realistic model of both the ambient and transient background, reproducing the characteristics measured with AMADEUS 
have been implemented. 
Signal and 
noise are then superimposed.
\item
The inherent noise of the sensor is added and 
the resulting waveform is convoluted with the system transfer function of the sensor.
Subsequently, the same steps of adding noise and applying the system transfer function
are applied for the read-out electronics. 
\item
The output is directed to the simulation of the online filter system
described in Sec.~\ref{seclahmann:amadeus}.
\end{enumerate}

The complete chain allows for the investigation of the neutrino detection efficiency
of AMADEUS or any other existing or potential acoustic neutrino detection device.

\section{Effective Volume}
A measure of the sensitivity of the AMADEUS device to neutrino interactions is its
effective volume. While the AMADEUS layout will not allow for an
effective volume sufficient to set a competitive limit on the flux
of ultra-high-energy neutrinos, an estimate of the energy detection threshold can be derived
and the effects of various cuts and environmental conditions can be quantified.
The simulation and analysis chains described above were used to simulate the data required for this study. An effective volume $V_{\mathrm{eff}}$ for the {AMADEUS} detector can be defined as:
\begin{equation}
V_{\mathrm{eff}}(E_{\nu}) = \frac{\sum_{N_{\mathrm{gen}}} \delta_{\mathrm{sel}} p(E_{\nu},\mathbf{r},\mathbf{e_{p}})}{N_{\mathrm{gen}}}V_{\mathrm{gen}}\text{,}
\label{chap4:seclahmann:eq:veff}
\end{equation}
where $N_\mathrm{gen}$ is the number of generated neutrino interactions in a volume $V_\mathrm{gen}$ and $p(E_{\nu},\mathbf{r},\mathbf{e_{p}})$ is the probability that the neutrino reaches the interaction vertex set in the simulation. $\delta_{\mathrm{sel}} \in \{0,1\}$ accounts for the fact that the probability only contributes to the effective volume 
 if the pressure pulse corresponding to the neutrino interaction was selected by the online filter within a time window of 128\,${\mathrm{\mu}}$s around the expected arrival time. The probability that the neutrino reaches the interaction vertex is given by:  
\begin{equation}
p(E_{\nu},\mathbf{r},\mathbf{e_{p}}) = e^{-d_{\mathrm{WE}}(\mathbf{r},\mathbf{e_{p}})/\lambda_{\mathrm{water}}(E_{\nu})}\,\text{,}
\label{chap4:seclahmann:eq:prob}
\end{equation}
where $\mathbf{r}$ is the position of the interaction vertex, $\mathbf{e_{p}}$ is the unit vector of the direction of the flight trajectory. The mean free path $\lambda_{\mathrm{water}}(E_{\nu})$ of the neutrino in water is anti-proportional to the total cross section of the neutrino. The total cross section as a function of the neutrino energy $E_{\nu}$ was parametrised using values from\,\cite{biblahmann:Cooper-Sarkar:2011fk}. The distance $d_{\mathrm{WE}}$ is the water equivalent of the distance travelled through matter of varying density encountered by the neutrino along its flight path. For the determination of the density distribution along the flight path, the {PREM}\footnote {Preliminary Earth Reference Model}\,\cite{biblahmann:Dziewonski1981297} was used to model the density profile of the earth. In addition, it is assumed that the earth is covered by water of 2.5\,km depth and the detector is placed on the sea floor.
\begin{figure}[tb]
\centering
\includegraphics[width=0.99\columnwidth]{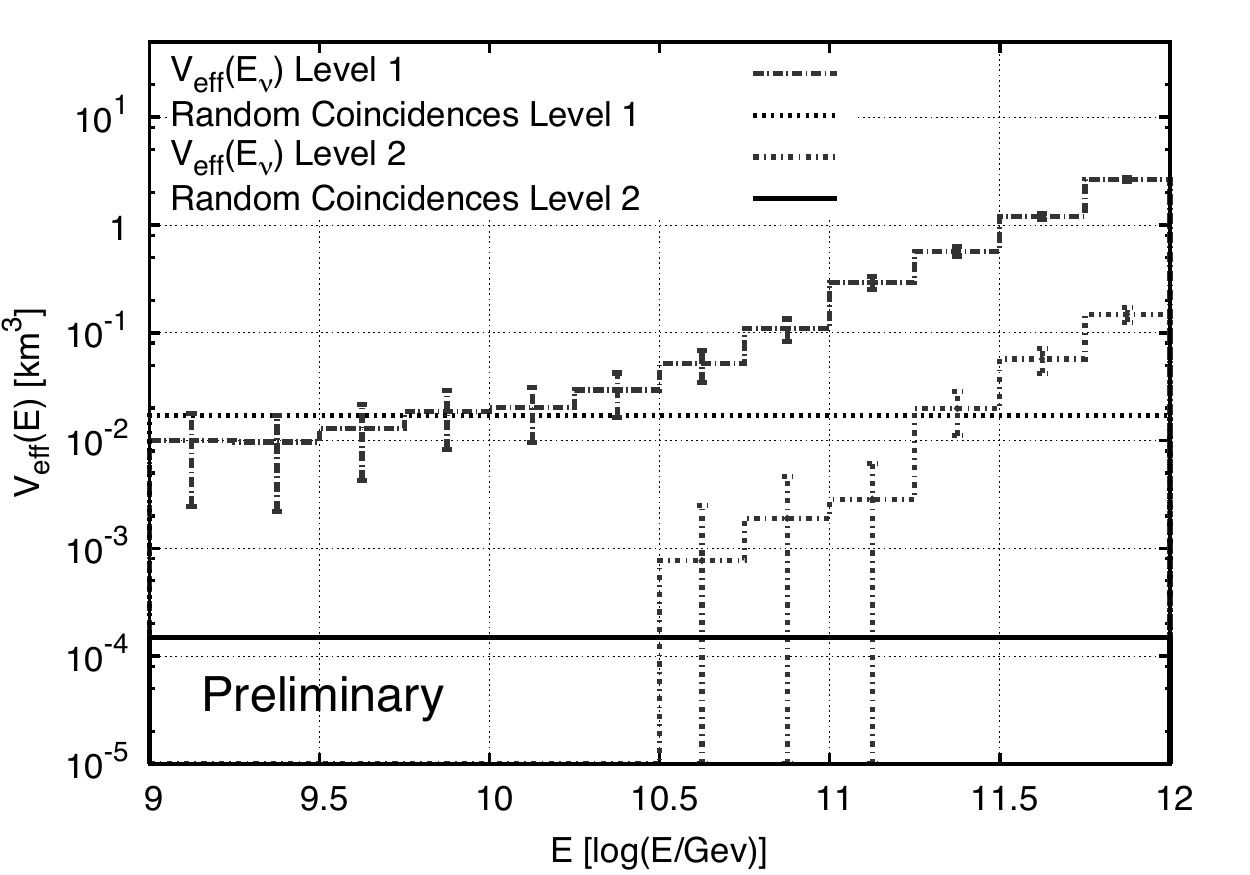}
\caption{
The effective volume of the AMADEUS detector as a function of the logarithmic neutrino energy for the two different levels, as described in the text. Also shown are the random coincidence rates for the levels~1 and 2.
\label{fig:veff}}
\end{figure}
For the calculation of the effective volume, $10^7$ neutrinos with energies uniformly distributed between $10^{9}$\,GeV and $10^{12}$\,GeV
were simulated. With the uniform energy distribution, a sufficient number of events over the entire energy range is available. The interaction vertexes of these neutrinos were chosen in a cylindrical volume of 1200\,km$^3$ around the {AMADEUS} detector. The heading of the flight path was ranging from $0\--360\,^{\circ}$ in azimuth and from $0\--100\,^{\circ}$ in zenith\footnote{A zenith angle of $0\,^{\circ}$ corresponds to a $\nu$ coming from above.}. Neutrinos entering the volume $V_\text{gen}$ from below the horizon will traverse an increasing amount of matter. For a zenith angle greater than $100\,^{\circ}$, the probability of a neutrino in the energy range under consideration to reach the interaction vertex is practically zero. To determine random coincidences formed by the ambient noise, a separate set of simulated data was created not containing any signals. 
The effective volume has been calculated for two different ``levels'' describing increasingly realistic conditions: 
\begin{description}
\itemsep 0pt
\item[Level~1:]{Ambient noise is assumed to be minimal, always corresponding to a perfectly calm sea, 
and the coincidence requirement for the filter simulation is that at least two sensors on one storey need to respond.}
\item[Level~2:]The complete ambient noise model and the standard online filter of {AMADEUS} are used, requiring a signal in at least four sensors on two storeys each.
\end{description}
The results of this study are shown in Fig.\,\ref{fig:veff}. 
For levels~1 and 2, the effective volume is 
at least one sigma above the ``fake flux'' from
random coincidences for energies exceeding $1.8\times 10^{10}$\,GeV and  $1.8\times 10^{11}$\,GeV, 
respectively.
The requirements of level~1 are minimal, so this can be seen as an idealised detection threshold of the {AMADEUS} detector. The effective volume for level~1 exceeds 2\,km$^3$ at $10^{12}$\,GeV, while for level~2, it is about 0.1\,km$^3$ 
at this energy. 

Note that for a potential future large-scale acoustic neutrino detector with a three-dimensional arrangement of acoustic sensors, the 
effective volume cannot be easily derived from the results for AMADEUS by simple scaling.
For neutrinos with energies near the detection threshold, the effective volume will be mainly determined by the instrumented volume 
 (assuming sufficiently dense instrumentation).
For rising energies, 
the water volume monitored around the detector
will become increasingly important.
The energy threshold as estimated in this study presumably can be decreased
if a larger number of sensors is used to trigger events, as the effect of random coincidences will be reduced. Applying pattern recognition methods on the triggered signals might further reduce
the threshold.


\section{Conclusion and Outlook}
\label{seclahmann:c&o}
Recent results from the acoustic neutrino detection test system AMADEUS, an integral part of the ANTARES detector in the Mediterranean Sea, have been presented. 
The suppression of the transient background was discussed and a full simulation chain, starting from
the interaction of the neutrino in water and ending with the electric signal registered in the 
acoustic sensors, was applied to calculate the effective volume of AMADEUS.
For realistic conditions, the effective volume is around 0.1\,km$^3$ at $10^{12}$\,GeV 
and the detection threshold was  estimated as about $1.8 \times 10^{11}\,$GeV. 
The latter is expected to decrease for a larger device.
The 
data recorded with AMADEUS 
can be used for Monte Carlo simulations to optimise the design of a potential future large-scale acoustic neutrino detector.

{\footnotesize{\bf %
\noindent
Acknowledgment: }
This work is supported by the German government (BMBF) with grants 05A08WE1 and 05A11WE1.
}


\newpage
{\small
\begin{thebibliography}{}

\bibitem{bibak:wb} E. Waxman \& J. Bahcall, \Journal{\PRD}{64}{023002-1}{2001}
\bibitem{bibak:hooper} D. Hooper \& J.Silk, \Journal{New Journal of Physics}{6}{23} {2004}
\bibitem{bibak:detector} J.A. Aguilar et al., NIM A 656 (2011) 11-38
\bibitem{bibak:amadeus} J.A. Aguilar et al., NIM A 626-627 (2011) 128-143
\bibitem{bibak:calibration} J.A. Aguilar et al., Astropart. Phys 34 (2011)
\bibitem{bibak:osc} S.~Adri\'an-Mart\'inez et al., Physics Letters B 714 (2012) 224-230
\bibitem{bibak:minos} P. Adamson et al., Phys. Rev. Lett. 101, 131802 (2008)
\bibitem{bibak:ICosc} Ch. Wiebusch for IceCube, 33rd ICRC, id 848
\bibitem{bibak:ARS} E. Akhmedov et al., JHEP 02 (2013) 082
\bibitem{bibak:MH} D. Franco et al., JHEP 04 (2013)
\bibitem{bibak:ORCA} P. Kooijman for KM3NeT, 33rd ICRC, id 164
\bibitem{bibak:antares_nu_spectrum} L.A Fusco for {\sc antares}, 33rd ICRC, Article~\ref{id_fusco} p.~\pageref{id_fusco}
\bibitem{bibak:bartol1} G.D. Barr et al., Phys. Rev. D 70, 023006 (2004)
\bibitem{bibak:bartol2} G.D. Barr et al., Phys. Rev. D 74, 094009 (2006)
\bibitem{bibak:amandanu} R. Abbasi et al. Astropart. Phys. 34 (2010)
\bibitem{bibak:icecubenu} R. Abbasi et al., Phys. Rev. D 83, 012001 (2011)
\bibitem{bibak:antaresdiffuse} J.A. Aguilar et al., Phys. Lett. B696:16-22, 2011
\bibitem{bibak:IC40} R. Abbasi et al., Physical Review D84 (2011) 082001
\bibitem{bibak:IC59} G.W. Sullivan, proc. of Neutrino 2012, to appear in Nucl. Phys.B, Proc. Supp.
\bibitem{bibak:mrf} G. C. Hill \& K. Rawlins, Astropart. Phys. 19 (2003) 393-402
\bibitem{bibak:FBflux} M. Su, T. Slatyer, D. Finkbeiner, ApJ 724 (2010) 1044
\bibitem{bibak:fb} R. Crocker \& F. Aharonian, Phys.Rev.Lett.106:101102, 2011
\bibitem{bibak:fbkm3net} S. Adri\'an-Mart\'inez et al., Astropart. Phys. 42, 7 (2013)
\bibitem{bibak:antaresps} S. Adri\'an-Mart\'inez et al., ApJ, 760, 53, 2012
\bibitem{bibak:psicrc} S. Shulte for {\sc antares}, 33rd ICRC, Article~\ref{id_schulte} p.~\pageref{id_schulte}
\bibitem{bibak:hotspot} F. Sch\"ussler et al., 33rd ICRC, id 547
\bibitem{bibak:2pt_ps} F. Sch\"ussler for {\sc antares}, 33rd ICRC, Article~\ref{id_schussler2} p.~\pageref{id_schussler2}
\bibitem{agn_jet1} F.~Halzen \& E.~Zas, \Journal{Astrophys.~J.}{488}{669}{1997}
\bibitem{agn_jet2} K.~Mannheim, \Journal{Astropart.~Phys.} {3}{295}{1995}
\bibitem{galactic_sources} W.~Bednarek et al.,\Journal{New~Astron.~Rev.} {49}{1}{2005}
\bibitem{bibak:antareslensing} J.J. Hernandez-Rey for {\sc antares}, 33rd ICRC, Article~\ref{id_mangano1} p.~\pageref{id_mangano1}
\bibitem{nu_grb} E.~Waxman \& J.~Bahcall, \Journal{\PRL}{78}{2292}{1997}
\bibitem{antares_grb} S. Adri\'an-Mart\'inez et al., JCAP03 (2013) 006
\bibitem{bibak:cosmA} H\"{u}mmer et al.,  \Journal{\PRL}{108}{231101}{2012}
\bibitem{bibak:GRB130427A} J. Schmid for {\sc antares}, TeVPA 2013 proceedings, and Article~\ref{id_james} p.~\pageref{id_james}
\bibitem{bibak:alerts} M. Ageron et al., Astropart. Phys. 35 (2012) 530-536
\bibitem{bibak:icrtatoo} D. Dornic for {\sc antares}, 33rd ICRC, Article~\ref{id_dornic1} p.~\pageref{id_dornic1}
\bibitem{bibak:blazars} S. Adri\'an-Mart\'inez et al., Astropart. Phys. 36 (2012) 204-210
\bibitem{bibak:icrc_blazars} D. Dornic for {\sc antares}, 33rd ICRC, Article~\ref{id_dornic2} p.~\pageref{id_dornic2}
\bibitem{bibak:microquasars} C. Distefano et al., Astrophys.J. 575 (2002) 378-383
\bibitem{bibak:GWHEN} The {\sc antares}, VIRGO \& LSC collaborations,  JCAP06(2013)008
\bibitem{bibak:icrcdm} J.J Hernandez-Rey for {\sc antares}, 33rd ICRC, Article~\ref{id_lambard} p.~\pageref{id_lambard}
\bibitem{bibak:mssm-7} L. Bergstr\"om \& P. Gondolo, Astropart. Phys. 5 (1996) 263-278.
\bibitem{bibak:antaresmonopoles} S. Adri\'an-Mart\'inez et al., Astropart. Phys. 35 (2012) 634ﾐ640
\bibitem{bibak:parker} E.N. Parker, Astrophys. J. 160 (1970) 383
\bibitem{bibak:nuclearites} A. De Rujula \& S.L. Glashow, Nature 312 (1984) 734
\bibitem{bibak:km3net} P. Coyle for KM3NeT, 33rd ICRC, id 891


\end{thebibliography}
}


\newpage
{\small
\begin{thebibliography}{}

\bibitem{bibschulte:becker} J. K. Becker, Physics Reports, 2008, \textbf{458} : 173-246.

\bibitem{bibschulte:Antoine} A. Kouchner [ANTARES Collaboration], Recent results from the ANTARES neutrino telescopes, 33rd ICRC, Article~\ref{id_ak} p.~\pageref{id_ak}

\bibitem{bibschulte:Electronics} J. A. Aguilar \emph{et al}, Nucl. Instr. Meth., 2010, {\bf A622}: 59-73.

\bibitem{bibschulte:Aart} A. Heijboer, 2004, PhD Thesis, http://antares.in2p3.fr/Publications.

\bibitem{bibschulte:Bartol} V. Agrawal, T. K. Gaisser, P. Lipari, T. Stanev Phys. Rev. D, 1996, {\bf 53}: 1314-1323.

\bibitem{bibschulte:MUPAGE}  G. Carminati, M. Bazzotti, A. Margiotta and M. Spurio, Comp. Phys. Comm., {\bf 179}: 915.

\bibitem{bibschulte:ICRC2011}
  C.~Bogazzi [ANTARES Collaboration],
  arXiv:1112.0478 [astro-ph.HE].

\bibitem{bibschulte:Icecube} R. Abbasi \emph{et al}, Astrophys. J., 2011, {\bf 732}: 18.


\bibitem{bibschulte:Kappes_2006fg}
  A.~Kappes, J.~Hinton, C.~Stegmann and F.~A.~Aharonian,
  Astrophys.\ J.\  {\bf 656} (2007) 870
   [Erratum-ibid.\  {\bf 661} (2007) 1348]

\bibitem{bibschulte:AdrianMartinez_2012rp}
  S.~Adrian-Martinez {\it et al.}  [ANTARES Collaboration],
  Astrophys.\ J.\  {\bf 760} (2012) 53

\end{thebibliography}
}


{\small
\begin{thebibliography}{}
\bibitem{bibschussler1:Antares_DetectorPaper} M. Ageron et al. (ANTARES Collaboration), NIM A 656, 11-38, 2011%

\bibitem{bibschussler1:Bartol} V. Agrawal, T. K. Gaisser, P. Lipari, T. Stanev, Phys. Rev. D 53 (1996) 1314-1323%

\bibitem{bibschussler1:ICRC2013_Spectrum} L.A. Fusco et al. on behalf of the ANTARES Collaboration, Article~\ref{id_fusco} p.~\pageref{id_fusco}

\bibitem{bibschussler1:Moriond2013_DiffuseFlux} S. Biagi et al. on behalf of the ANTARES Collaboration, Rencontres de Moriond (2013), arXiv:1305.6442

\bibitem{bibschussler1:IceCube_dEdX} R. Abbasi et al. [IceCube Collaboration], NIM A 703, 1 March 2013, 190-198

\bibitem{bibschussler1:ICRC2011_DiffuseFlux} F. Sch\"ussler on behalf of the ANTARES Collaboration, Proceedings $32^{\mathrm{nd}}$ ICRC (2011), arXiv:1112.0478

\bibitem{bibschussler1:Antares_PointSources2010} S. Adri\'an-Martínez et al. [ANTARES Collaboration], APJ 760, (2012), 53

\bibitem{bibschussler1:KM3NeT} KM3NeT Technical Design Report, available on
http://www.km3net.org/TDR/TDRKM3NeT.pdf 

\end{thebibliography}
}


\newpage
{\small
\begin{thebibliography}{}

\bibitem{bibfusco:antares}  M.Ageron et al. (the ANTARES Collaboration), Nucl. Inst. and Meth. in Phys. Res. \textbf{A656}:11 (2011), doi:10.1016/j.nima.2011.06.103

\bibitem{bibfusco:2012PS} S. Adri\'an-Martinez et al. (the ANTARES Collaboration), The Astrophysical Journal \textbf{760}:53 (2012), doi:10.1088/0004-637X/760/1/53

\bibitem{bibfusco:antoine} A. Kouchner (for the ANTARES Collaboration), Article~\ref{id_ak} p.~\pageref{id_ak}

\bibitem{bibfusco:bartol}  G. D. Barr et al., Phys. Rev. \textbf{D70}:023006 (2004) doi:10.1103/PhysRevD.70.023006

\bibitem{bibfusco:honda} M. Honda et al., Phys. Rev. \textbf{D75}:043006 (2007), doi:10.1103/PhysRevD.75.043006

\bibitem{bibfusco:martin} A.D. Martin, M.G. Ryskin \& A.M. Stasto, Acta Phys. Polon. \textbf{B34}:3237 (2003), doi:10.1016/S0370-2693(03)00656-7

\bibitem{bibfusco:enberg} R. Enberg, M.H. Reno \& I. Sarcevic, Phys. Rev. \textbf{D78}:043005 (2008), doi:10.1103/PhysRevD.78.043005

\bibitem{bibfusco:amanda}  R. Abbasi et al. (the IceCube Collaboration), Astropart. Phys. \textbf{34}:48 (2010), doi:10.1016/j.astropartphys.2010.05.001

\bibitem{bibfusco:ic40}  R. Abbasi et al. (the IceCube Collaboration), Phys. Rev. \textbf{D83}:012001 (2011), doi:10.1103/PhysRevD.83.012001

\bibitem{bibfusco:dimitris} D. Palioselitis, PhD thesis, University of Amsterdam (2012).

\bibitem{bibfusco:fabian} F. Sch\"{u}ssler (for the ANTARES Collaboration), Article~\ref{id_schussler1} p.~\pageref{id_schussler1}

\bibitem{bibfusco:regul} A.N. Tikhonov, Sov. Math. \textbf{4}:1035 (1963).

\bibitem{bibfusco:svd} A. Hocker \& V. Kartvelishvili, Nucl. Instr. \& Meth. in Phys. Res. \textbf{A372}:469 (1996), doi:10.1016/0168-9002(95)01478-0

\bibitem{bibfusco:bayes} G. D'Agostini, Nucl. Instr. \& Meth. in Phys. Res. \textbf{A362}:487 (1995), doi:10.1016/0168-9002(95)00274-X

\bibitem{bibfusco:roounfold} T. Adye, Unfolding algorithms and tests using RooUnfold, arXiv:1105.1160v1

\end{thebibliography}
}


{\small
\begin{thebibliography}{00}
\bibitem{bibvladimir:Su}
M. Su, T. Slatyer, D. Finkbeiner, Astrophys. J. 724 (2010) 1044.
\bibitem{bibvladimir:Dobbler}
G. Dobbler, arXiv:1109.4418v2 [astro-ph.GA], 2011.
\bibitem{bibvladimir:ROSAT}
S.L. Snowden et al., ApJ 485 (1997)125.
\bibitem{bibvladimir:radio}
E. Carretti et al., Nature 493(2013) 66.
\bibitem{bibvladimir:CrockerAharonian}
R.Crocker, F.Aharonian,  Phys. Rev. Lett. 106 (2011) 101102.
\bibitem{bibvladimir:Lacki}
B. C. Lacki, arXiv:1304.6137 [astro-ph.HE] 2013.
\bibitem{bibvladimir:Thoudam}
S. Thoudam, arXiv:1304.6972 [astro-ph.HE] 2013.
\bibitem{bibvladimir:VillanteVissani}
F.L.Villante, F.Vissani, Phys.Rev. D 78, 103007 (2008).
\bibitem{bibvladimir:ANTARES}
ANTARES Collaboration, J.A. Aguilar et al., Nucl. Instrum. Meth. A 656 (2011) 11-38 doi:10.1016/j.nima.2011.06.103. See also A. Kouchner, for the ANTARES collaboration, Article~\ref{id_ak} p.~\pageref{id_ak}.
\bibitem{bibvladimir:timing}
ANTARES Collaboration, J.A. Aguilar et al., Astropart. Phys. 34 (2011) 539.
\bibitem{bibvladimir:Positioning}
ANTARES Collaboration, S. Adri\'an-Mart\'inez et al., JINST 7 (2012) T08002.
\bibitem{bibvladimir:pointsearch}
ANTARES Collaboration, S. Adri\'an-Mart\'inez et al., ApJ 760 (2012) 53. 
\bibitem{bibvladimir:bbfit}
ANTARES Collaboration, J.A. Aguilar et al., Astropart. Phys. 34 (2011) 652-662.
\bibitem{bibvladimir:Jutta}
J.Schnabel, Nucl. Instrum. Meth. A, NIMA55282, doi:10.1016/j.nima.2012.12.109.
\bibitem{bibvladimir:Brunner}
Brunner, J. 2003, in VLVnT Workshop (Amsterdam), ANTARES simulation tools, ed. E. de Wolf (Amsterdam:NIKHEF), \url{http://www.vlvnt.nl/proceedings.pdf}.
\bibitem{bibvladimir:Bartol}
Bartol group, V. Argawal et al., Phys. Rev. D, 53 (1996) 1314-1323.
\bibitem{bibvladimir:MRF}
G.C. Hill and K. Rawlins, Astropart. Phys. 19 (2003) 393-402.
\bibitem{bibvladimir:Conrad}
J. Conrad et al., Phys. Rev. D. 67 012002 (2003).
\bibitem{bibvladimir:LiMa}
T. Li, Y. Ma, Astrophys. J. 272 (1983) 317-323.
\bibitem{bibvladimir:km3}
KM3NeT Collaboration, S. Adri\'an-Mart\'inez et al., Astropart. Phys., 42 (2013) 7-14. See also R.Coniglione et~al., for the KM3NeT collaboration, this conference.
\end{thebibliography}
}


\vspace*{-0.5cm}
{\small
\begin{thebibliography}{}

\bibitem{bibmangano1:Einstein} A. Einstein, Science 84 (1936) 506-507.

\bibitem{bibmangano1:Zwicky} F. Zwicky, Astrophysical Journal 86 (1937) 217-246.

\bibitem{bibmangano1:1979Nature} D. Walsh et al, Nature 279 (1979) 381-384.

\bibitem{bibmangano1:Gerver} J.L. Gerver, Phys. Lett. A127 (1988) 301-303.

\bibitem{bibmangano1:Elewyck2}  R. Escribano et al, Phys. Lett. B512 (2001) 8-17.

\bibitem{bibmangano1:Mena}  O. Mena et al, Astropart. Phys. 28 (2007) 248-356. 

\bibitem{bibmangano1:Eiroa}  E.F. Eiroa et al,  Phys. Lett. B663 (2008) 377-381.

\bibitem{bibmangano1:Romero}  G.E. Romero, {\tiny Int. J. Mod. Phys. Proc. Suppl.} 3 (2011) 475-481.

\bibitem{bibmangano1:Ant1} M. Ageron et al,  Nucl. Instrum. Methods A 656 (2011) 11-38.

\bibitem{bibmangano1:Antoine} A. Kouchner et al, Article~\ref{id_ak} p.~\pageref{id_ak}



\bibitem{bibmangano1:AdrianMartinez:2012rp} S. Adri\'an-Martinez et al, Astrophys. J. 760 (2012) 53.

\bibitem{bibmangano1:Beacom:1999bn} J.F. Beacom, hep-ph/9901300 (1999).

\bibitem{bibmangano1:Arnett} W.D. Arnett, Phys. Rev. Lett.  58 (1987) 1906.

\bibitem{bibmangano1:Crocker:2003cw} R.M. Crocker, Phys. Rev.  D69 (2004) 063008.

\bibitem{bibmangano1:Gravlens} S. Mollerach and E. Roulet, Text book (2002).  

\bibitem{bibmangano1:Gravlens2} P. Schneider, J. Ehlers and E.E. Falco, Text book (1992).  

\bibitem{bibmangano1:Gravlens3} P. Schneider \etal, Text book (2006).  

\bibitem{bibmangano1:Abdo:2009wu} A.A. Abdo et al, Astrophys. J. 700 (2009) 597-622.

\bibitem{bibmangano1:Brandt:2001ur}  W.N. Brandt et al, Astron. J. 122 (2001) 2810-2832.

\bibitem{bibmangano1:Oguri:2006aw}  M. Oguri et al, Astron. J. 132 (2006) 999-1013.

\bibitem{bibmangano1:Browne:2002yb} I.W.A. Browne  et al, {\tiny Mon. Not. Roy. Astron. Soc.} 341 (2003) 13.

\bibitem{bibmangano1:Kochanek} C.S. Kochanek, AIP Conf. Proc. 470 (1998) 163-175.

\bibitem{bibmangano1:Mediavilla} E.  Mediavilla et al,Astrophys. J. 706 (2009) 1451-1462.

\bibitem{bibmangano1:Oguri:2010rh} M. Oguri, Publ. Astron. Soc. Jap. 62 (2010) 1017-1024.

\bibitem{bibmangano1:Dai:2009bp} X. Dai et al, Astrophys. J. 709 (2010) 278-285.

\bibitem{bibmangano1:Raychaudhury:2003cf} S. Raychaudhury et al, Astron. J. 126 (2003) 29.

\bibitem{bibmangano1:Oguri:2012vg}  M. Oguri et al, {\tiny Mon. Not. Roy. Astron. Soc.} 429 (2013) 482-493.

\bibitem{bibmangano1:Becker} J. K. Becker, Phys. Rept. 458 (2008) 173-246.

\bibitem{bibmangano1:Falco:1999jc} E.E. Falco et al, Astrophys. J. 523 (1999) 617. 

\bibitem{bibmangano1:Donna} I. Donnarumma et al, Astrophys. J. 736 (2011) L30.

\bibitem{bibmangano1:Massaro:2010si} E. Massaro et al, arXiv:1006.0922 (2010). 

\bibitem{bibmangano1:Broadhurst:2004} T.J. Broadhurst et al, Astrophys. J. 621 (2005) 53-88.


\end{thebibliography}
}


\vspace{-0.3cm}
{\small
\begin{thebibliography}{}

\bibitem{bibschussler2:Antares_DetectorPaper} M. Ageron et al. (ANTARES Collaboration), ANTARES: the first undersea neutrino telescope, NIM A 656, 11-38, 2011\\
A. Kouchner on behalf of ANTARES, Article~\ref{id_ak} p.~\pageref{id_ak}

\bibitem{bibschussler2:ICRC2013_dEdX} F. Sch\"ussler on behalf of the ANTARES Collaboration, Energy reconstruction in neutrino telescopes, Article~\ref{id_schussler1} p.~\pageref{id_schussler1}

\bibitem{bibschussler2:Amanda_PointSources2009} R. Abbasi et al. (AMANDA Collaboration), Search for point sources of high energy neutrinos with final data from AMANDA-II, PRD 79, 062001, 2009%
 
\bibitem{bibschussler2:IceCube_PointSources2011} R. Abbasi et al. (IceCube Collaboration), Time-integrated Searches for Point-like Sources of Neutrinos with the 40-string IceCube Detector, APJ 732, 18-34, 2011%

\bibitem{bibschussler2:ANTARES_PointSources2010} S. Adri\'an-Mart\'inez et al. (ANTARES Collaboration), Search for cosmic neutrino point sources with four years of data from the ANTARES telescope, APJ 760, 53, 2012%

\bibitem{bibschussler2:ICRC2013_ANTARESExcess} F. Sch\"ussler et al. on behalf of the H.E.S.S. Collaboration, Multiwavelength study of the region around the ANTARES neutrino
   excess, these proceedings, ID 547%

\bibitem{bibschussler2:ICRC2013_PointSources} S. Schulte on behalf of the ANTARES Collaboration, Article~\ref{id_schulte} p.~\pageref{id_schulte}


\bibitem{bibschussler2:ICRC2011_AutoCorrelation} F. Sch\"ussler on behalf of the ANTARES Collaboration, Autocorrelation analysis of ANTARES data, Proceedings $32^\mathrm{nd}$ ICRC (2011), arXiv:1112.0478%

\bibitem{bibschussler2:LiMa} T.-P. Li and Y.-Q. Ma, Analysis methods for results in gamma-ray astronomy, APJ 272, 317-324, 1983%

\bibitem{bibschussler2:FinleyWesterhoff2004} C. Finley and S. Westerhoff, On the evidence for clustering in the arrival directions of AGASA's ultrahigh energy cosmic rays, APP 21, 359-367, 2004%

\bibitem{bibschussler2:2FGL} P. L. Nolan et al. (Fermi-LAT Collaboration), Fermi Large Area Telescope Second Source Catalog, APJ Supplement Series 199, 31, 2012%

\bibitem{bibschussler2:GWGC}  White D.J.,  Daw E.J., Dhillon V.S., A List of Galaxies for Gravitational Wave Searches, Class. Quantum Grav. 28, 085016, 2011%


\end{thebibliography}
}


{\small
\begin{thebibliography}{}

\bibitem{bibjames:Kouveliotou93a} C. Kouveliotou {\it et al.}, ApJ Letters 413 (1993) 101--104 doi:10.1086/186969.

\bibitem{bibjames:Galama1998} T.J. Galama {\it et al.}, Nature 395 (1998) 670--672 doi:10.1038/27150.

\bibitem{bibjames:MeszarosRees1993} P. Meszaros and M.J. Rees, ApJ 405 (1993) 278--284 doi:10.1086/172360.

\bibitem{bibjames:Meszaros2006} P. {M{\'e}sz{\'a}ros}, Reports on Progress in Physics 69 (2006) 2259--2321 doi:10.1088/0034-4885/69/8/R01.

\bibitem{bibjames:Waxman1995a} E. Waxman, Phys.Rev.Lett.\ 75 (1995) 386--389 doi:10.1103/PhysRevLett.75.386.

\bibitem{bibjames:Ageron2011} M. Ageron {\it et al.}, NIMPR A 656 (2011) 11--38, doi:10.1016/j.nima.2011.06.103.

\bibitem{bibjames:AntoineICRC} A. Kouchner (The ANTARES Collaboration), Article~\ref{id_ak} p.~\pageref{id_ak}

\bibitem{bibjames:JuliaGRB} S. Adri{\'a}n-Mart{\'{\i}}nez {\it et al.} (in preparation).

\bibitem{bibjames:WaxmannBahcall} E. Waxmann, J. Bahcall, Phys.Rev.Lett 78 (1997) 2292--2295, doi:10.1103/PhysRevLett.78.2292.

\bibitem{bibjames:Guetta2004} D. Guetta, D. Hooper, J. Alvarez-Mu\~niz, F. Halzen, E. Reuveni, Astropart.Phys. 20 (2004) 429--455 doi:10.1016/S0927-6505(03)00211-1.

\bibitem{bibjames:Abbasi2012} R. Abbasi {\it et al.}, Nature 484 (2012) 351--354 doi:10.1038/nature11068.

\bibitem{bibjames:Adrianmartinez2013} S. Adri{\'a}n-Mart{\'{\i}}nez {\it et al.}, JCAP 3 (2013) 6 doi:10.1088/1475-7516/2013/03/006.

\bibitem{bibjames:Hummer2010} S. H{\"u}mmer, M. R{\"u}ger, F. Spanier and W. Winter, ApJ 721 (2010) 630--652 doi:10.1088/0004-637X/721/1/630.

\bibitem{bibjames:Hummer2012} S. H{\"u}mmer, P. Baerwald and W. Winter, PRL 23 (2012) 231101 doi:10.1103/PhysRevLett.108.231101.

\bibitem{bibjames:Muecke1998} A. M{\"u}cke, R.R. Engel, R.J. Protheroe, J.P. Rachen, T. Stanev, Nuc.Phys.B Proc.Supp 80 (1998) C809.

\bibitem{bibjames:Adrianmartinez2012} S. Adri{\'a}n-Mart{\'{\i}}nez {\it et al.}, ApJ 760 (2012) 53 doi:10.1088/0004-637X/760/1/53.

\bibitem{bibjames:Barlow1990} R. Barlow, NIMPR A 297 (1990) 496--506 doi:10.1016/0168-9002(90)91334-8


\end{thebibliography}
}


\newpage
{\small
\begin{thebibliography}{}
\bibitem{bibdornic1:GRB} E.~Waxman, J.~Bahcall, Phys. Rev. Lett. 78 (1997) 2292;
P.~Meszaros, E.~Waxman, Phys. Rev. Lett. {\bf 87} (2001) 171102;
C.~Dermer, A.~Atoyan, Phys. Rev. Lett. {\bf 91} (2003) 071102;
S.~Razzaque, P.~Meszaros, E.~Waxman, Phys. Rev. Lett. {\bf 90} (2003) 241103.
\bibitem{bibdornic1:CCSN} S.~Ando, J.~Beacom,
Phys. Rev. Lett. {\bf 95} (2005) 061103.
\bibitem{bibdornic1:TAToO} Ageron M.  et al., Astropart. Phys. 35 (2012) 530-536
\bibitem{bibdornic1:Marek} M.~Kowalski, A.~Mohr, Astropart. Phys. {\bf 27} (2007) 533.
\bibitem{bibdornic1:Tarot} M.~Boer {\it et al.},
Astron. Astrophys. Suppl. Ser. {\bf 138} (1999) 579.
\bibitem{bibdornic1:Rotse} C.W.~Akerlof {\it et al.},
Public. Astron. Soc. Pac. {\bf 115} (2003) 132.
\bibitem{bibdornic1:Antares} E. Aslanides {\it et al.}, ANTARES Collaboration, 
astro-ph/9907432.\\
M. Ageron  et al., ANTARES Collaboration, astro-ph.IM/1104.1607v1.
\bibitem{bibdornic1:Antares2} Kouchner A., ANTARES Collaboration, Article~\ref{id_ak} p.~\pageref{id_ak}
\bibitem{bibdornic1:CCSN1} S.~Razzaque, P.~Meszaros, E.~Waxman, 
Phys. Rev. Lett. {\bf 94} (2005) 109903
\bibitem{bibdornic1:atmu} J.A.~Aguilar {\it et al.}, ANTARES Collaboration,
Astropart. Phys. {\bf 34} (2010) 179
\bibitem{bibdornic1:difflux} J.A.~Aguilar {\it et al.}, ANTARES Collaboration,
Phys. Lett. B{\bf 696} (2011) 16
\bibitem{bibdornic1:BBfit} J.A.~Aguilar {\it et al.}, ANTARES Collaboration,
Astropart. Phys. {\bf 34} (2011) 652.
\bibitem{bibdornic1:AAfit} S. Adrian-Martinez {\it et al.}, ANTARES Collaboration,
Astrophysics J. 760:53(2012)
\bibitem{bibdornic1:AAFit} A.~Heijboer,
http://antares.in2p3.fr/Publications/thesis/2004/Aart-Heijboer-phd.pdf,
PhD thesis, Universiteit van Amsterdam, Amsterdam, The Netherlands.
\bibitem{bibdornic1:vlvnt09} D.~Dornic {\it et al.},
{Nucl. Instrum. Meth. A} {\bf 626-627} (2011) S183.
S.~Basa {\it et al.},
{Nucl. Instrum. Meth. A} {\bf 602} (2009) 275.
\bibitem{bibdornic1:Zadko} Coward D.M.{\it et al.}, Publications of the Astronomical Society of Australia, 2010, 27, 331, arXiv:1006.3933  
\bibitem{bibdornic1:RotsePipeline} Yuan, F., Akerlof, C. W. 2008, Astrophys. J., 677, 808
\bibitem{bibdornic1:Poloka} http://supernovae.in2p3.fr/poloka/
\bibitem{bibdornic1:LePhare}    Arnouts, S.; Cristiani, S.; Moscardini, L., Matarrese, S., Lucchin, F.  et al., 1999, MNRAS,  310, 540\\
Ilbert, O.; Arnouts, S.; McCracken, H. J.; Bolzonella, M.; Bertin, E et al., 2006, A\&A, 457, 841 
\end{thebibliography}
}


{\small
\begin{thebibliography}{}

\bibitem{bibdornic2:AGNhadronic} Gaisser T.K., Halzen F., Stanev T., Phys. Rep. 258 (1995) 173;
Learned J.G., Mannheim K., Ann. Rev. Nucl. Part. Sci. 50 (2000) 679;
Halzen F., Hooper D., Rep. Prog. Phys. 65 (2002) 1025.

\bibitem{bibdornic2:FermiLATAGNvariability}  Abdo A. A. {\it et al}. 2010, ApJ, 722, 520

\bibitem{bibdornic2:Antares} Ageron M. {\it et al.}, ANTARES Collaboration, NIMA-D-10-00948R2, astro-ph.IM/1104.1607v1.

\bibitem{bibdornic2:Antares2} Kouchner A., ANTARES Collaboration, Article~\ref{id_ak} p.~\pageref{id_ak}

\bibitem{bibdornic2:antaresdaq} Aguilar J.A. {\it et al.}, ANTARES Collaboration,
{Nucl. Instrum. Meth. A} {\bf 570} (2007) 107.

\bibitem{bibdornic2:TimeCalib} Aguilar J.A. {\it et al.}, ANTARES Collaboration,
Astropart. Phys. {\bf 34} (2011) 539.

\bibitem{bibdornic2:flare} S. Adrian-Martinez et al., Astropart. Phys. 36 (2012) 204-210

\bibitem{bibdornic2:AAfit} Heijboer A.,
http://antares.in2p3.fr/Publications/thesis/2004/Aart-Heijboer-phd.pdf,
PhD thesis, Universiteit van Amsterdam, Amsterdam, The Netherlands.

\bibitem{bibdornic2:Neyman} Neyman, J. 1937, Phil. Trans. Royal Soc. London, Series A, 236, 333

\bibitem{bibdornic2:AAfitps} S. Adrian-Martinez et al., The Astrophysical Journal 760:53(2012)

\bibitem{bibdornic2:Fermicatalogue} Abdo, A. A. {\it et al.} 2010, ApJS, 188, 405

\bibitem{bibdornic2:FermicatalogueAGN} Abdo, A. A. {\it et al.} 2010, ApJ, 715, 429

\bibitem{bibdornic2:FermiAdvocates} Ciprini S. et al, 2011 Fermi Symposium proceedings - eConf C110509, arXiv:1111.6803. 
http://fermisky.blogspot.fr/

\bibitem{bibdornic2:FermiData} http://fermi.gsfc.nasa.gov/cgi-bin/ssc/LAT/LATDataQuery.cgi

\end{thebibliography}
}


{\small
\begin{thebibliography}{}
\bibitem{bibthierry:pradiermm} T. Pradier, Class. Quant. Grav. (2010) 27, 194004

\bibitem{bibthierry:antares} A. Kouchner, for \ant, Article~\ref{id_ak} p.~\pageref{id_ak}
\bibitem{bibthierry:virgo} F. Acernese et al., Class. Quant. Grav. (2008) 25, 184001
\bibitem{bibthierry:ligo} D. Sigg et al., Class. Quant. Grav. (2008) 25, 114041
\bibitem{bibthierry:gwhen2007} S. Adri\'an-Martinez et al., JCAP (2013) 06, 008  arxiv:1205.3018
\bibitem{bibthierry:km3} J. J. Hernandez-Rey, J. Phys. Conf. Ser. (2009) 171, 012047
\bibitem{bibthierry:GRBnu1} E. Waxman \& J. N. Bahcall, Phys. Rev. Lett. (1997) 78, 2292
\bibitem{bibthierry:GRBnu2} J. P. Rachen \& P. Meszaros, Phys. Rev. D (1998) 58, 123005
\bibitem{bibthierry:GRBnu3} J. Alvarez-Muniz et al., Phys. Rev. D (2000) 62, 093015
\bibitem{bibthierry:ice3nu} R. Abbasi et al., Phys. Rev. Lett. (2011) 106, 141101

\bibitem{bibthierry:GRBshort1} S. Kochanek \& T. Piran, Astrophys. Journ. (1993) 417, L17
\bibitem{bibthierry:GRBshort2} E. Nakar, Phys. Rept. (2007) 442, 166
\bibitem{bibthierry:GRBlong} S. E. Woosler \& J. S. Bloom, Ann. Rev. Astron. Astroph. (2008) 44, 507
\bibitem{bibthierry:razzaque} S. Razzaque et al., Phys. Rev. Lett. (2004) 93, 181101
\bibitem{bibthierry:choked} P. Meszaros \& E. Waxman, Phys. Rev. Lett. (2001) 87, 17
\bibitem{bibthierry:ando} S. Ando \& J. Beacom, Phys. Rev. Lett. (2005) 95, 061103
\bibitem{bibthierry:antps2012} S. Adri\'an-Martinez et al., Astrophys. Journ. (2012) 760, 53
\bibitem{bibthierry:aart} S. Adri\'an-Martinez et al., Astrophys. Journ. Lett. (2011) 743, 14
\bibitem{bibthierry:tatoo} D. Dornic et al., Article~\ref{id_dornic1} p.~\pageref{id_dornic1}
\bibitem{bibthierry:pradier} T. Pradier, Nucl. Instrum. Meth. A (2009) 602, 268
\bibitem{bibthierry:aso} Y. Aso et al., Class. Quant. Grav. (2008) 25, 114039
\bibitem{bibthierry:baret} B. Baret et al., Astropart. Phys. (2011) 35, 1

\bibitem{bibthierry:cwb} B. Bouhou, PhD Thesis, University Pierre et Marie Curie, Paris (2012)
\bibitem{bibthierry:gwhen2012} B. Baret et al., Phys. Rev. D 85 (2012) 103004
\bibitem{bibthierry:advitf} G. M. Harry et al., Class. Quant. Grav. (2010) 27, 084006





\end{thebibliography}
}


\newpage
{\small
\begin{thebibliography}{}

 \bibitem{biblambard:bertone}G. Bertone, D. Hooper, J. Silk, Phys. Rept., 2005,
   {\bf 405}: pp. 279-390.

 \bibitem{biblambard:antares} M. Ageron et al., ANTARES Collaboration,
   \textit{ANTARES: the first undersea neutrino telescope},
   Nucl. Inst. and Meth. in Phys. Res. A {\bf 656} (2011) 11-38.

\bibitem{biblambard:OM}  P. Amram, ANTARES Collaboration, \textit{The ANTARES optical module}, 
Nucl. Inst. and Meth. in Phys. Res. A {\bf 484} (2002) 369.

 \bibitem{biblambard:PMT} J.A. Aguilar et al., ANTARES Collaboration,
   Nucl. Inst. and Meth. in Phys. Res. A {\bf 555} (2005) 132.

 \bibitem{biblambard:frontend} J.A. Aguilar et al., ANTARES Collaboration,
   Nucl. Inst. and Meth. in Phys. Res. A {\bf 622} (2010) 59.

 \bibitem{biblambard:DAQ} J.A. Aguilar et al., ANTARES Collaboration,
   Nucl. Inst. and Meth. in Phys. Res. A {\bf 570} (2007) 107.

 \bibitem{biblambard:biofouling} P. Amram, ANTARES Collaboration,
   Astropart. Phys. {\bf 19} (2003) 253.

 \bibitem{biblambard:amadeus} J.A. Aguilar et al., ANTARES Collaboration,
   Nucl. Inst. and Meth. in Phys. Res. A {\bf 626-627} (2011) 128.

 \bibitem{biblambard:instrumentation} J.A. Aguilar et al., ANTARES Collaboration,
   Astropart. Phys. {\bf 26} (2006) 314.

 \bibitem{biblambard:alignment} S. Adri\'an-Mart\'{\i}nez et al., ANTARES
   Collaboration, JINST {\bf 7} (2012) T08002.

 \bibitem{biblambard:OBs} M. Ageron et al., ANTARES Collaboration,
   Nucl. Inst. and Meth. in Phys. Res. A {\bf 578} (2007) 498.

 \bibitem{biblambard:timing} J.A. Aguilar et al., ANTARES Collaboration,
  Astropart. Phys. {\bf 34} (2011) 539.

 \bibitem{biblambard:bbfit} J.A. Aguilar et al., Astropart. Phys. {\bf 34} (2011)
   652.

  \bibitem{biblambard:mrf} G.C. Hill, K. Rawlins, Astropart. Phys. {\bf 19} (2003)
    393-402.
 
 \bibitem{biblambard:feldmancousins}G.J. Feldman, R.D. Cousins, Phys. Rev. {\bf D
   57} (1998) 3873-3889.

 \bibitem{biblambard:corsika} D. Heck et al., Report FZKA 6019 (1998),
   Forschungszentrum Karlsruhe; D. Heck and J. Knapp, Report FZKA 6097
   (1998), Forschungszentrum Karlsruhe.

 \bibitem{biblambard:bartol} G. Barr et al., Phys. Rev. D {\bf 39}, 3532 (1989);
   V. Agrawal et al., Phys. Rev. D {\bf 53}, 1314 (1996).

 \bibitem{biblambard:wimpsim} J. Edsj\"{o}, http://www.physto.se/~edsjo/wimpsim/.

 \bibitem{biblambard:gondolo} P. Gondolo et al., J. Cosm. and Astropart. Phys.,
   JCAP07, 008 (2004).

 \bibitem{biblambard:baksan} M.M. Boliev et al., Baksan Collaboration, [astro-ph/1301.1138].

 \bibitem{biblambard:superk} T. Tanaka et al., Super-Kamiokande Collaboration,
   Astrophys. J. {\bf 742}, 78 (2011).

 \bibitem{biblambard:icecube} M.G. Aartsen et al., IceCube
   Collaboration,Phys. Rev. Lett. 110, (2013) 131302.


\bibitem{biblambard:simple} M. Felizardo et al., SIMPLE Collaboration, 
Phys. Rev. Lett. {\bf 108} (2012) 201302. 

\bibitem{biblambard:coupp} E. Behnke et al., COUPP Collaboration,  Phys. Rev. D
  {\bf 86} (2012) 052001. 


\bibitem{biblambard:xenon} E. Aprile et al., XENON Collaboration, [astro-ph/1207.5988].

\bibitem{biblambard:buchmeller} O. Buchmueller et al., [hep-ph/1207.7315].

\bibitem{biblambard:conrad}F. Tegenfeldt, J. Conrad, Nucl. Inst. and Meth. in
  Phys. Res. A {\bf 539} (2005) 407-413; J. Conrad et al.,
  Phys. Rev. D {\bf 67}, 012002 (2003); J. Conrad, [astro-ph/0612082].


\end{thebibliography}
}


{\small
\begin{thebibliography}{}

\bibitem{bibmangano2:Ant1} M. Ageron et al,  Nucl. Instrum. Methods A 656 (2011) 11-38.

\bibitem{bibmangano2:Timecalibration} J.A. Aguilar et al, Astropart. Phys. 34 (2011) 539-549.


\bibitem{bibmangano2:Coll2010} J.A. Aguilar et al, Astropart. Phys. 34 (2010) 179-184.

\bibitem{bibmangano2:Agui} J.A. Aguilar et al, Astropart. Phys. 23 (2005) 131-155. 

\bibitem{bibmangano2:Lubs} B.K. Lubsandorzhiev et al, Nucl. Instrum. Meth. A502 (2003) 168-171.

\bibitem{bibmangano2:opt}  S. Mangano, ICRC, 0100 HE2.3 (2011). 


\bibitem{bibmangano2:Ageron} M. Ageron et al, Nucl. Instrum. Meth. A578 (2007) 498-509.

\bibitem{bibmangano2:DAQ} J.A. Aguilar et al, Nucl. Instrum. Meth. A570 (2007) 107-116.

\bibitem{bibmangano2:abs} R.C. Smith and K.S. Baker, Appl. Opt., 20 (1981) 177-184.

\bibitem{bibmangano2:Quanfry} X. Quan and E. Fry, Appl. Opt., 34:18 (1995)  3477-3480.

\bibitem{bibmangano2:Austin}  R.W. Austin and G. Halikas, Scripps Inst. Oceanogr., SIO Ref. 76-1 (1976).

\bibitem{bibmangano2:KM3} D.J.L. Bailey, PhD thesis, Oxford University (2002).

\bibitem{bibmangano2:Bru} J. Brunner, Proceedings of the VLVnT Workshop, Amsterdam, (2003) 109-113.

\end{thebibliography}
}


{\small

\begin{thebibliography}{10}

\expandafter\ifx\csname url\endcsname\relax
  \def\url#1{\texttt{#1}}\fi
\expandafter\ifx\csname urlprefix\endcsname\relax\def\urlprefix{URL }\fi
\expandafter\ifx\csname href\endcsname\relax
  \def\href#1#2{#2} \def\path#1{#1}\fi
\bibitem{biblahmann:Berezinsky-1969}
{V.S.~Berezinsky and G.T.~Zatsepin}, 
Phys.\ Lett. B\,28 (1969) 423,
\path{doi:10.1016/0370-2693(69)90341-4}.

\bibitem{biblahmann:Askariyan1979267}
G.A.~Askariyan \etal, 
Nucl. Instrum. Meth. A 164 (1979) 267 -- 278.
\newblock \href {http://dx.doi.org/10.1016/0029-554X(79)90244-1}
  {\path{doi:10.1016/0029-554X(79)90244-1}}.

\bibitem{biblahmann:PhysRevD.19.3293}
J.G.~Learned, 
  Phys. Rev. D 19 (1979) 3293--3307.
  {\path{doi:10.1103/PhysRevD.19.3293}}.

\bibitem{biblahmann:S.Bevan:2007wd}
S.~Bevan~\etal, 
  Astropart. Phys. 28 (2007) 366--379,
\path{doi:10.1016/j.astropartphys.2007.08.001}.
	
\bibitem{biblahmann:Bevan:2009rr}
S.~Bevan~\etal, 
  Nucl. Instrum. Meth. A 607 (2009) 398--411,
\path{doi:10.1016/j.nima.2009.05.009}.

\bibitem{biblahmann:Bertin_Niess}
{V.\ Niess and V.\ Bertin}, 
  Astropart.\ Phys. 26 (2006) 243, 
\path{doi:10.1016/j.astropartphys.2006.06.005}.

\bibitem{biblahmann:collaboration:2010fj}
J.A.~Aguilar~\etal [ANTARES Collaboration], 
  Nucl. Instrum. Meth. A 626 (2011) {128--143},
\path{doi:10.1016/j.nima.2010.09.053}.

\bibitem{biblahmann:ANTARES-paper}
M.~Ageron~\etal [ANTARES Collaboration], 
Nucl. Instrum. Meth. A656 (2011) 11--38,
  {\path{doi:10.1016/j.nima.2011.06.103}}.

\bibitem{biblahmann:urick1986ambient}
R.J.~Urick, {\it Ambient noise in the sea}, Peninsula
, 1986.

\bibitem{biblahmann:NeffVLVnT11}
M.~Neff~\etal. 
  in: Proceedings of the VLVnT 2011, in press,
\path{doi:10.1016/j.nima.2012.11.147}.



\bibitem{biblahmann:Neff:2011pd}
M.~Neff \etal 
in: Proceedings of the ARENA 2010, Nucl. Instrum. Meth. A 662, Supplement 1 (2012), S242 -- S245.
\newblock \href {http://dx.doi.org/10.1016/j.nima.2010.11.016}
  {\path{doi:10.1016/j.nima.2010.11.016}}.

\bibitem{biblahmann:Neff_phd}
{M.~Neff}, {\it {Studies on the Selection of Neutrino-like Signals for the
  Acoustic Detection Test Device AMADEUS}}, Ph.D. thesis, Univ.\
  Erlangen-N{\"u}rnberg ({2013}).


\bibitem{biblahmann:Connolly-etal-inelast}
{A.~Connolly, R.~Thorne, and D.~Waters}, 
Phys.\ Rev.\ D 83, (2011) 113009, 
\path{doi:10.1103/PhysRevD.83.113009}.

\bibitem{biblahmann:Cooper-Sarkar:2011fk}
A.~Cooper-Sarkar, P.~Mertsch, and S.~Sarkar, 
J. High Energ. Phys.~8 (2011) 042, 
\path{doi:10.1007/JHEP08(2011)042}.
	
\bibitem{biblahmann:Dziewonski1981297}
A.M.~Dziewonski and D.L.~Anderson, 
Phys. Earth Planet. Inter. 25~(4) (1981) 297 -- 356,
  {\path{doi:10.1016/0031-9201(81)90046-7}}.

\end{thebibliography}
}

\setcounter{figure}{0}
\setcounter{table}{0}
\setcounter{footnote}{0}
\setcounter{section}{0}

\newpage
\shorttitle{}
\begin{onecolumn}
\section*{Acknowledgements}

\large
The authors acknowledge the financial support of the funding agencies:
Centre National de la Recherche Scientifique (CNRS), Commissariat
\`{a} l'\'{e}nergie atomique et aux \'{e}nergies alternatives (CEA), Agence
National de la Recherche (ANR), Commission Europ\'{e}enne (FEDER fund
and Marie Curie Program), R\'{e}gion Alsace (contrat CPER), R\'{e}gion
Provence-Alpes-C\^{o}te d'Azur, D\'{e}parte-ment du Var and Ville de
La Seyne-sur-Mer, France; Bundesministerium f\"{u}r Bildung und
Forschung (BMBF), Germany; Istituto Nazionale di Fisica Nucleare
(INFN), Italy; Stichting voor Fundamenteel Onderzoek der Materie
(FOM), Nederlandse organisatie voor Wetenschappelijk Onderzoek (NWO),
the Netherlands; Council of the President of the Russian Federation
for young scientists and leading scientific schools supporting grants,
Russia; National Authority for Scientific Research (ANCS), Romania;
Ministerio de Ciencia e Innovaci\'{o}n (MICINN), Prometeo of Generalitat
Valenciana (GVA) and MultiDark, Spain. We also acknowledge the
technical support of Ifremer, AIM and Foselev Marine for the sea
operation and the CC-IN2P3 for the computing facilities.

\end{onecolumn}

\end{document}